\RequirePackage{fix-cm}
\documentclass[smallextended]{svjour3}       
\smartqed  

\newif\ifdraft
\drafttrue

\newif\ifconf
\conffalse

\newif\ifrebut
\rebutfalse

\usepackage{graphicx}
\usepackage{natbib}
\usepackage[usenames]{xcolor}
\usepackage[justification=centering]{caption}
\usepackage{amsfonts}
\usepackage{amsmath}
\usepackage{hyperref}
\usepackage{tabularx,booktabs}
\usepackage{multicol}
\usepackage{multirow}
\usepackage{subcaption}
\usepackage{placeins}
\usepackage[title]{appendix}
\usepackage{hyperref}
\usepackage{longtable}
\usepackage[framemethod=tikz]{mdframed}

\newcommand{\customsize}{1}

\newmdenv[innerlinewidth=0.5pt, roundcorner=4pt,linecolor=red,innerleftmargin=6pt,
innerrightmargin=6pt,innertopmargin=6pt,innerbottommargin=6pt]{mybox}

\newcommand{\ans}[1]{{\leavevmode\color{black}{{#1}}}}
\newcommand{\que}[1]{{\leavevmode\color{blue}\textbf{{{#1}}}}}
\newcommand{\multipagetable}[1]{\begin{scriptsize}\begin{longtable}{lp{6.5cm}}#1 \end{longtable}
\end{scriptsize}}
\newcommand{\popdatatable}[1]{\begin{scriptsize}\begin{tabular}{p{2.5cm}p{2.5cm}p{2.5cm}p{2.5cm}}#1\end{tabular}\end{scriptsize}}
\newcommand{\tasktable}[1]{\begin{scriptsize}    \begin{tabular}{p{4.5cm}p{6.5cm}}#1\end{tabular}\end{scriptsize}}
\newcommand{\ack}[1]{\begin{normalsize}#1\end{normalsize}}

\newcommand{\novel}[1]{\ifrebut{\leavevmode\color{blue}{{#1}}}\else{#1}\fi}

\begin{document}

\title{Algorithmic Fairness Datasets: the Story so Far 
}


\author{Alessandro Fabris \and Stefano Messina \and Gianmaria Silvello \and
Gian Antonio Susto}

\authorrunning{Fabris, Messina, Silvello, Susto}


\institute{Dipartimento di Ingegneria dell'Informazione \\
Università di Padova \\
Via Giovanni Gradenigo 6B -- 35131 Padova, Italy \\
\email{fabrisal@dei.unipd.it, silvello@dei.unipd.it, gianantonio.susto@unipd.it}}

\date{Received: date / Accepted: date}

\maketitle

\begin{mybox}
\textbf{Outdated preprint.} Please read and cite the published version  of this manuscript:

\vspace{0.2cm}

\noindent Alessandro Fabris, Stefano Messina, Gianmaria Silvello, and Gian Antonio Susto. Algorithmic fairness datasets: the story so far. Data Mining and Knowledge Discovery. 2022.

\noindent \url{https://doi.org/10.1007/s10618-022-00854-z} 
\end{mybox}

\begin{abstract}
Data-driven algorithms are studied and deployed in diverse domains to support critical decisions, directly impacting people's well-being. As a result, a growing community of researchers has been investigating the equity of existing algorithms and proposing novel ones, advancing the understanding of risks and opportunities of automated decision-making for historically disadvantaged populations. Progress in fair Machine Learning (ML) and equitable algorithm design hinges on data, which can be appropriately used only if adequately documented. Unfortunately, the algorithmic fairness community, as a whole, suffers from a collective data documentation debt caused by a lack of information on specific resources (\emph{opacity}) and scatteredness of available information (\emph{sparsity}). In this work, we target this data documentation debt by surveying over two hundred datasets employed in algorithmic fairness research, and producing standardized and searchable documentation for each of them.
Moreover we rigorously identify the three most popular fairness datasets, namely Adult, COMPAS, and German Credit, for which we compile in-depth documentation.

This unifying documentation effort supports multiple contributions. Firstly, we summarize the merits and limitations of Adult, COMPAS, and German Credit, adding to and unifying recent scholarship, calling into question their suitability as general-purpose fairness benchmarks. Secondly, we document hundreds of available alternatives, annotating their domain and supported fairness tasks, along with additional properties of interest for fairness practitioners and researchers, including their format, cardinality, and the sensitive attributes they encode. We summarize this information, zooming in on the tasks, domains, and roles of these resources.
Finally, we analyze these datasets from the perspective of five important data curation topics: anonymization, consent, inclusivity, labeling of sensitive attributes, and transparency. We discuss different approaches and levels of attention to these topics, making them tangible, and distill them into a set of best practices for the curation of novel resources.

\end{abstract}

\section{Introduction}

Following the widespread study and application of data-driven algorithms in contexts that are central to people's well-being, a large community of researchers has coalesced around the growing field of algorithmic fairness, investigating algorithms through the lens of justice, equity, bias, power, and harms. A line of work gaining traction in the field, intersecting with critical data studies, human-computer interaction, and computer-supported cooperative work, focuses on 
data transparency and standardized documentation processes to describe key characteristics of datasets \citep{gebru2018datasheets,holland2018dataset,bender2018data,geiger2020garbage,jo2020lessons,miceli2021documenting}. Most prominently, \citet{gebru2018datasheets} and \citet{holland2018dataset} proposed two complementary documentation frameworks, called \emph{Datasheets for Datasets} and \emph{Dataset Nutrition Labels}, to improve data curation practices and favour more informed data selection and utilization for dataset users. Overall, this line of work has contributed to an unprecedented attention to dataset documentation in Machine Learning (ML), including a novel track focused on datasets at the Conference on Neural Information Processing Systems (NeurIPS), an initiative to support dataset tracking in repositories for scholarly articles,\footnote{\url{https://medium.com/paperswithcode/datasets-on-arxiv-1a5a8f7bd104}} and dedicated works producing retrospective documentation for existing datasets \citep{bandy2021addressing,garbin2021structured}, auditing their properties \citep{prabhu2020large} and tracing their usage \citep{peng2021mitigating}.

Data documentation is important and caters to different goals. It increases transparency, favouring an improved understanding of the data and resulting models \citep{hutchinson2021towards}, it reduces chances of data misuse \citep{gebru2018datasheets} and supports accountability in dataset and model creation \citep{hutchinson2021towards}, it helps connect the data with its context to guide scientific inquiry \citep{paullada2020data}, and it makes the values influencing dataset curation explicit \citep{scheuerman2021datasets}. Technical debt is a cost incurred in software development when speed of execution is prioritized over quality \citep{hutchinson2021towards}. In recent work, \citet{bender2021:dangers} propose the notion of \emph{documentation debt}, in relation to training sets that are undocumented and too large to document retrospectively, which compounds over time with serious consequences on dataset understanding and use. We extend this definition to the collection of datasets employed in a given field of research. We see two components at work contributing to the documentation debt of a research community. On one hand, \emph{opacity} is the result of poor documentation affecting single datasets, contributing to misunderstandings and misuse of specific resources. On the other hand, when relevant information exists but does not reach interested parties, there is a problem of documentation \emph{sparsity}. One example that is particularly relevant for the algorithmic fairness community is represented by the German Credit dataset \citep{hofmann1994:sg}, a popular resource in this field. Many works of algorithmic fairness, including recent ones, carry out experiments on this dataset using sex as a protected attribute \citep{he2020geometric,yang2020fairness,baharlouei2020renyi,lohaus2020too,martinez2020minimax,wang2021fair,perrone2021fair,sharma2021fairn}, while existing yet overlooked documentation shows that this feature cannot be reliably retrieved \citep{gromping2019:sg}. Moreover, the mere fact that a dataset exists and is relevant to a given task or a given domain may be unknown. The BUPT Faces datasets, for instance, were presented as the second existing resource for face analysis with race annotations \citep{wang2020mitigating}. However several resources were already available at the time, including Labeled Faces in the Wild \citep{han2014age}, UTK Face \citep{zhifei2017age}, Racial Faces in the Wild \citep{wang2019racial}, and Diversity in Faces \citep{merler2019diversity}.\footnote{Hereafter, for brevity, we only report dataset names. The relevant references and additional information can be found in Appendix \ref{sec:databriefs}.}

To tackle the documentation debt of the algorithmic fairness community, we survey the datasets used in over 500 articles on fair ML and equitable algorithmic design, presented at seven major conferences, considering each edition in the period 2014--2021, and more than twenty domain-specific workshops in the same period. We find over 200 datasets employed in studies of algorithmic fairness, for which we produce compact and standardized documentation, called \emph{data briefs}. Data briefs are intended as a lightweight format to document fundamental properties of data artifacts used in algorithmic fairness, including their purpose, their features, with particular attention to sensitive ones, the underlying labeling procedure, and the envisioned ML task, if any.
To favor domain-based and task-based search from dataset users, data briefs also indicate the domain of the processes that produced the data (e.g., radiology) and list the fairness tasks studied on a given dataset (e.g. fair ranking). For this endeavour, we have contacted creators and knowledgeable practitioners identified as primary points of contact for the datasets. We received feedback (incorporated into the final version of the data briefs) 
from 79 curators and practitioners, whose contribution is acknowledged at the end of this article. Moreover, we identify and carefully analyze the three datasets most often utilized in the surveyed articles (Adult, COMPAS, and German Credit), retrospectively producing a datasheet and a nutrition label for each of them. From these documentation efforts, we extract a summary of the merits and limitations of popular algorithmic fairness benchmarks, a categorization of domains and fairness tasks for existing datasets, and a set of best practices for the curation of novel resources.

Overall, we make the following contributions.
\begin{itemize}
    \item \textbf{\novel{Unified} analysis of popular fairness benchmarks}. We produce \novel{\emph{datasheets} and \emph{nutrition labels}} for Adult, COMPAS, and German Credit, from which we extract a summary of their merits and limitations. We \novel{add to and unify recent scholarship on these datasets, calling} into question their suitability as general-purpose fairness benchmarks due to contrived prediction tasks, noisy data, severe coding mistakes, and age.
    \item \textbf{Survey of existing alternatives}. We compile standardized and compact documentation for over two hundred resources used in fair ML research, annotating their domain, the tasks they support, and the roles they play in works of algorithmic fairness. By assembling sparse information on hundreds of datasets into a single document, we aim to support  multiple goals by researchers and practitioners, including domain-oriented and task-oriented search by dataset users. Contextually, we provide a novel categorization of tasks and domains investigated in algorithmic fairness research (summarized in Tables \ref{tab:domainwise} and \ref{tab:taskwise}).
    \item \textbf{Best practices for the curation of novel resources}. We analyze different approaches to anonymization, consent, inclusivity, labeling, and transparency across these datasets. By comparing existing approaches and discussing their advantages, we make the underlying concerns visible and practical, and extract best practices to inform the curation of new datasets and post-hoc remedies to existing ones.
\end{itemize}

\noindent \textbf{Roadmap}. Readers looking for alternative fairness datasets should prioritize Section \ref{sec:alternatives}, Appendix \ref{sec:databriefs}, and take account of the web app under development (see Footnote \ref{fn:web_app}). Overall, this work is organized as follows. Section \ref{sec:related} introduces related works. Section \ref{sec:methodology} presents the methodology and inclusion criteria of this survey. Section \ref{sec:popular} analyzes the perks and limitations of the most popular datasets, namely Adult (\autoref{sec:popular_adult}), COMPAS (\autoref{sec:popular_compas}), and German Credit (\autoref{sec:popular_german}), and provides an overall summary of their merits and limitations as fairness benchmarks (\autoref{sec:popular_summary}). Section \ref{sec:alternatives} discusses alternative fairness resources from the perspective of the underlying domains (\autoref{sec:domain}), the fair ML tasks they support (\autoref{sec:tasks}), and the roles they play (\autoref{sec:roles}). Section \ref{sec:best_practices} presents important topics in data curation, discussing existing approaches and best practices to avoid re-identification (\autoref{sec:reidentification}), elicit informed consent (\autoref{sec:consent}), consider inclusivity (\autoref{sec:inclusivity}), collect sensitive attributes (\autoref{sec:sensitive_attribute}), and document datasets (\autoref{sec:transparency}). Section \ref{sec:relevance} summarizes the broader benefits of our documentation effort and envisioned uses for the research community. Finally, Section \ref{sec:conclusion} contains concluding remarks and recommendations. Interested readers may find the data briefs in Appendix \ref{sec:databriefs}, followed by the detailed documentation produced for Adult (\autoref{sec:adult}), COMPAS (\autoref{sec:compas}), and German Credit (\autoref{sec:german}).

\novel{
\section{Related Work}
\label{sec:related}

\ifconf
\else{
\subsection{Algorithmic fairness surveys}

Multiple surveys about algorithmic fairness have been published in the literature \citep{mehrabi2019survey,caton2020fairness,pessach2020algorithmic}. These works typically focus on describing and classifying important measures of algorithmic fairness and methods to enhance it. Some articles also discuss sources of bias \citep{mehrabi2019survey}, software packages and projects which address fairness in ML \citep{caton2020fairness}, or describe selected sub-fields of algorithmic fairness \citep{pessach2020algorithmic}. Datasets are typically not emphasized in these works, which is also true of domain-specific surveys on algorithmic fairness, focused e.g.\ on ranking  \citep{pitoura2021fairness}, Natural Language Processing (NLP) \citep{sun2019mitigating} and computational medicine \citep{sun2019mitigating}. As an exception, \citet{pessach2020algorithmic} and \citet{zehlike2021fairness} list and briefly describe 12 popular algorithmic fairness datasets, and 19 datasets employed in fair ranking research, respectively.
}\fi 

\subsection{Data studies}
\ifconf
\else{
The work most closely related (and concurrently carried out) to ours is \citet{lequysurvey2022}. The authors perform a detailed analysis of 15 tabular datasets used in works of algorithmic fairness, listing important metadata (e.g. domain, protected attributes, collection period and location), and carrying out an exploratory analysis of the probabilistic relationship between features. Our work complements it by placing more emphasis on (1) a rigorous methodology for the inclusion of resources, (2) a wider selection of (over 200) datasets spanning different data types, including text, image, timeseries, and tabular data, (3) a fine-grained evaluation of domains and tasks associated with each dataset\ifconf{.}\else{, and (4) the analysis and distillation of best practices for data curation.}\fi 
Different goals of the research community, such as selection of appropriate resources for experimentation and data studies, can benefit from the breadth and depth of both works.

Other works analyzing multiple datasets along specific lines have been published in recent years. \citet{crawford2021excavating} focus on resources commonly used as training sets in computer vision, with attention to associated labels and underlying taxonomies. \citet{fabbrizzi2021survey} also consider computer vision datasets, describing types of bias affecting them, along with methods for discovering and measuring bias, while \citet{scheuerman2021datasets} analyze the values encoded in their documentation. 
\citet{koch2021reduced} study the data employed in machine learning research and show a concentration of work on a small number of benchmark datasets curated at few well-resourced institutions. \citet{peng2021mitigating} analyze ethical concerns in three popular face and person recognition datasets, stemming from derivative datasets and models, lack of clarity of licenses, and dataset management practices.  
\citet{geiger2020garbage} evaluate transparency in the documentation of labeling practices employed in over 100 datasets about Twitter. \citet{leonelli2020data} study practices of collection, cleaning, visualization, sharing, and analysis across a variety of research domains. \citet{romei2014multidisciplinary} survey techniques and data for discrimination analysis, focused on measuring, rather than enforcing, equity in human processes.

A different, yet related, family of articles provides deeper analyses of single datasets. \citet{prabhu2020large} focus on Imagenet (ILSVRC 2012) which they analyze along the lines of consent, problematic content, and individual re-identification. \citet{kizhner2020digital} study issues of representation in the Google Arts and Culture project across countries, cities and institutions. Some works provide datasheets for a given resource, such as CheXpert \citep{garbin2021structured} and the BookCorpus \citep{bandy2021addressing}. Among popular fairness datasets, COMPAS has drawn scrutiny from multiple works, analysing its numerical idiosyncrasies \citep{barenstein2019propublica} and sources of bias \citep{bao2021COMPASlicated}. \citet{hardt2021facing} study numerical idiosyncrasies in the Adult dataset, and propose a novel version, for which they provide a datasheet. \citet{gromping2019:sg} discuss issues resulting from coding mistakes in German Credit.

Our work combines the breadth of multi-dataset and the depth of 
single-dataset studies.
On one hand, we survey numerous resources used in works of algorithmic fairness, analyzing them across multiple dimensions. On the other hand, we identify the most popular resources, compiling their \emph{datasheet} and \emph{nutrition label}, and summarize their perks and limitations. Moreover, by making our data briefs available, we hope to contribute a useful tool to the research community, favouring further data studies and analyses, as outlined in Section \ifconf{\ref{sec:conclusion}.}\else{\ref{sec:relevance}.}\fi 
}\fi

\subsection{Documentation frameworks}

Several data documentation frameworks have been proposed in the literature; three popular ones are described below. \emph{Datasheets for Datasets} \citep{gebru2018datasheets} are a general-purpose qualitative framework with over fifty questions covering key aspects of datasets, such as motivation, composition, collection, preprocessing, uses, distribution, and maintenance. Another qualitative framework is represented by \emph{Data statements} \citep{bender2018data}, which is tailored for NLP, requiring domain-specific information on language variety and speaker demographics. \emph{Dataset Nutrition Labels} \citep{holland2018dataset} describe a complementary, quantitative framework, focused on numerical aspects such as the marginal and joint distribution of variables. More broadly, recent initiatives focused on ML and AI documentation strongly emphasize data documentation \citep{arnold2019factsheets,pai2022:about}.

Popular datasets require close scrutiny; for this reason we adopt these frameworks, producing three datasheets and nutrition labels for Adult, German Credit, and COMPAS. This approach, however, does not scale to a wider documentation effort with limited resources. For this reason, we propose and produce \emph{data briefs}, a lightweight documentation format designed for algorithmic fairness datasets. Data briefs, described in Appendix \ref{sec:databriefs}, include fields specific to fair ML, such sensitive attributes and tasks for which the dataset has been used in the algorithmic fairness literature.
}

\section{Methodology}
\label{sec:methodology}
In this work, we consider (1) every article published in the proceedings of domain-specific conferences such as the ACM Conference on Fairness, Accountability, and Transparency (FAccT), and the AAAI/ACM Conference on Artificial Intelligence, Ethics and Society (AIES); (2) every article published in proceedings of well-known machine learning and data mining conferences, including the IEEE/CVF Conference on Computer Vision and Pattern Recognition (CVPR), the Conference on Neural Information Processing Systems (NeurIPS), the International Conference on Machine Learning
(ICML), the International Conference on Learning Representations (ICLR), the ACM SIGKDD International Conference on Knowledge Discovery and Data Mining (KDD); (3) every article available from Past Network Events and Older Workshops and Events of the FAccT network.\footnote{\url{https://facctconference.org/network/}} We consider the period from 2014, the year of the first workshop on Fairness, Accountability, and Transparency
in Machine Learning, \novel{to June 2021, thus including works presented at FAccT, ICLR, AIES, and CVPR in 2021.}\footnote{We are working on an update covering more recent work, including articles presented at the ACM conference on Equity and Access in Algorithms, Mechanisms, and Optimization.}

To target works of algorithmic fairness, we select a subsample of these articles whose titles contain either of the following strings, where the star symbol represents the wildcard character: \texttt{*fair*} (targeting e.g. fairness, unfair), \texttt{*bias*} (biased, debiasing), \texttt{discriminat*} (discrimination, discriminatory), \texttt{*equal*} (equality, unequal), \texttt{*equit*} (equity, equitable), disparate (disparate impact), \texttt{*parit*} (parity, disparities). These selection criteria are centered around equity-based notions of fairness, typically operationalized by measuring disparity in some algorithmic property across individuals or groups of individuals. Through manual inspection by two authors, we discard articles where these keywords are used with a different meaning. \novel{Discarded works, for instance, include articles on handling pose distribution bias \citep{zhao2021camera}, compensating selection bias to improve accuracy without attention to sensitive attributes \citep{kato2018learning}, enhancing desirable discriminating properties of models \citep{chen2018virtual}, or generally focused on model performance \citep{li2018learning,zhong2019unequaltraining}. This leaves us with 558 articles.}

From the articles that pass this initial screening, we select datasets treated as important data artifacts, either being used to train/test an algorithm or undergoing a data audit, i.e., an in-depth analysis of different properties. We produce a data brief for these datasets by (1) reading the information provided in the surveyed articles, (2) consulting the provided references, and (3) reviewing scholarly articles or official websites found by querying popular search engines with the dataset name.  \ifconf\else{ 
We discard the following: 
\begin{itemize}
    \item Word Embeddings (WEs). We only consider the corpora they are trained on, provided WEs are trained as part of a given work and not taken off the shelf;
    \item toy datasets, i.e., simulations with no connection to real-world processes, unless they are used in more than one article, which we take as a sign of importance in the field;
    \item auxiliary resources that are only used as a minor source of ancillary information, such as the percentage of US residents in each state;
    \item datasets for which the available information is insufficient. This happens very seldom when points (1), (2), and (3) outlined above result in little to no information about the curators, purposes, features, and format of a dataset. For popular datasets, this is never the case.
\end{itemize}
For each of the 226 datasets satisfying the above criteria, we produce a data brief, available in Appendix \ref{sec:databriefs} with a description of the underlying coding procedure. }\fi From this effort, we rigorously identify the three most popular resources, whose perks and limitations are summarized in the next section.

\section{Most Popular Datasets}
\label{sec:popular}


\begin{figure*}
  \includegraphics[width=\customsize\textwidth]{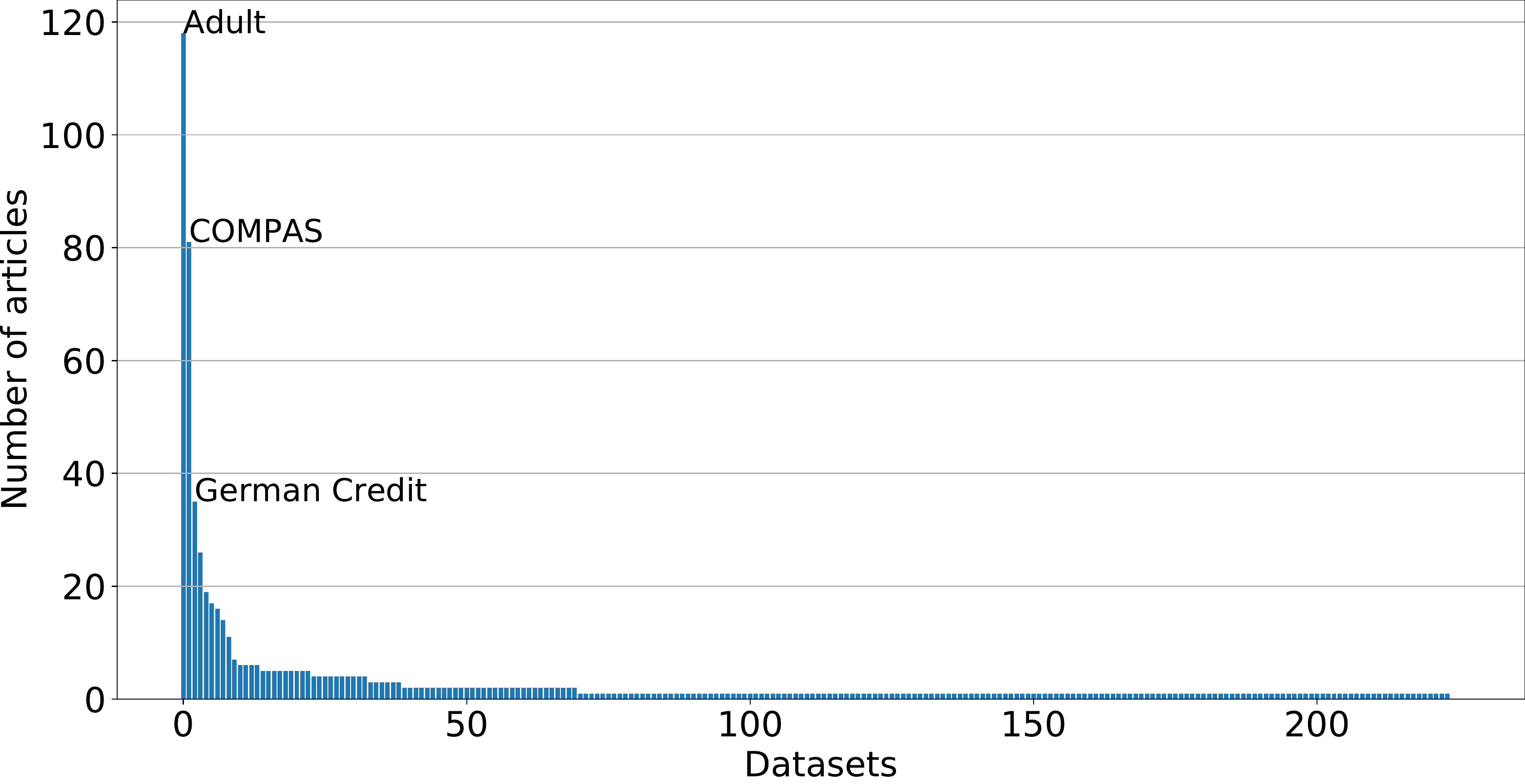}
\caption{Utilization of datasets in fairness research follows a \novel{long tail} distribution.}
\label{fig:bar}
\end{figure*}

Figure \ref{fig:bar} depicts the number of articles using each dataset, showing that dataset utilization in surveyed scholarly works follows a \novel{long tail} distribution, reflecting findings of data use in computer vision \citep{koch2021reduced}. Over 100 datasets are only used once, also because some of these resources are not publicly available. Complementing this long tail is a short head of nine resources used in ten or more articles. These datasets are Adult (118 usages), COMPAS (81), German Credit (35), Communities and Crime (26), Bank Marketing (19), Law School (17), CelebA (16), MovieLens (14), and Credit Card Default (11). The tenth most used resource is the toy dataset from \citet{zafar2017fairness}, used in 7 articles. In this section, we summarize positive and negative aspects of the three most popular datasets, namely Adult, COMPAS, and German Credit, informed by extensive documentation in Appendices \ref{sec:adult}, \ref{sec:compas}, and \ref{sec:german}.

\subsection{Adult}
\label{sec:popular_adult}

The Adult dataset was created as a resource to benchmark the performance of machine learning algorithms on socially relevant data. \ifconf\else{Each instance is a person who responded to the March 1994 US Current Population Survey, represented along demographic and socio-economic dimensions, with features describing their profession, education, age, sex, race, personal, and financial condition. The dataset was extracted from the census database, preprocessed, and donated to UCI Machine Learning Repository in 1996 by Ronny Kohavi and Barry Becker. A binary variable encoding whether respondents' income is above \$50,000 was chosen as the target of the prediction task associated with this resource.

Adult inherits some positive sides from the best practices employed by the US Census Bureau. Although later filtered somewhat arbitrarily, the original sample was designed to be representative of the US population. Trained and compensated interviewers collected the data. Attributes in the dataset are self-reported and provided by consensual respondents. Finally, the original data from the US Census Bureau is well documented, and its variables can be mapped to Adult by consulting the original documentation \citep{usdeptcomm1995current}, except for a variable denominated \texttt{fnlwgt}, whose precise meaning is unclear.

}\fi A negative aspect of this dataset is the contrived prediction task associated with it. Income prediction from socio-economic factors is a task whose social utility appears rather limited. Even discounting this aspect, the arbitrary \$50,000 threshold for the binary prediction task is high, and model properties such as accuracy and fairness are very sensitive to it \citep{hardt2021facing}. \novel{Furthermore, there are several sources of noise affecting the data. Roughly 7\% of the data points have missing values, plausibly due to issues with data recording and coding, or respondents’ inability to recall information.   Moreover, the tendency in household surveys for respondents to under-report their income is a common concern of the Census Bureau \citep{moore2000income}. Another source of noise is top-coding of the variable ``capital-gain'' (saturation to \$99,999) to avoid the re-identification of certain individuals \citep{usdeptcomm1995current}. Finally, the dataset is rather old; sensitive attribute ``race'' contains the outdated ``Asian Pacific Islander'' class. It is worth noting that a set of similar resources was recently made available, allowing more current socio-economic studies of the US population \citep{hardt2021facing}.}

\begin{table*}[h]
  \caption{\novel{Limitations of popular algorithmic fairness datasets.}}
  \label{tab:popular}
  \popdatatable{
    \toprule
    & \multicolumn{1}{c}{Adult} & \multicolumn{1}{c}{COMPAS} & \multicolumn{1}{c}{German Credit} \\
    \midrule
    Age& Old (1994)& Recent (2013--2016)& Very old (1973--1975)\\
    Prediction task& Contrived (income $>50$K\$) & Realistic (recidivism)& Realistic (creditworthiness)\\
    Sensitive attributes&Outdated racial categories&Outdated racial categories&Sex cannot be retrieved\\
    Sources of noise& Top-coding; tendency to under-report income & Data leakage; label bias; clerical errors & Incorrect code table \\
    Sample representativeness & US working population & Convenience sample (Broward County) & Artificial sample (credit granted, negative class oversampled)\\
    Preprocessing needed&Handling missing values (7\%)&Handling missing values (80\%); removing redundant features; ground truth on detainment& None\\
    Additional concerns & Accuracy and fairness are sensitive to arbitrary $50$K\$ threshold & Potential for misguided discussion on criminal justice & Interpretability and exploratory analyses are invalid \\ 
  \bottomrule
  }
\end{table*}

\subsection{COMPAS}
\label{sec:popular_compas}

This dataset was created for an external audit of racial biases in the Correctional Offender Management Profiling for Alternative Sanctions (COMPAS) risk assessment tool developed by Northpointe (now Equivant), which estimates the likelihood of a defendant becoming a recidivist. \ifconf\else{ Instances represent defendants scored by COMPAS in Broward County, Florida, between 2013--2014, reporting their demographics, criminal record, custody and COMPAS scores. Defendants' public criminal records were obtained from the Broward County Clerk’s Office website matching them based on date of birth, first and last names. The dataset was augmented with jail records and COMPAS scores provided by the Broward County Sheriff’s Office. Finally, public incarceration records were downloaded from the Florida Department of Corrections website. Instances are associated with two target variables (is\_recid and is\_violent\_recid), indicating whether defendants were booked in jail for a criminal offense (potentially violent) that occurred after their COMPAS screening but within two years.

}\fi On the upside, this dataset is recent and captures some relevant aspects of the COMPAS risk assessment tool and the criminal justice system in Broward County. On the downside, it was compiled from disparate sources, hence clerical errors and mismatches are present \citep{larson2016how}. Moreover, in its official release \citep{propublica2016compas}, the COMPAS dataset features redundant variables and data leakage due to spuriously time-dependent recidivism rates \citep{barenstein2019propublica}. For these reasons, researchers must perform further preprocessing in addition to the standard one by ProPublica. More subjective choices are required of researchers interested in counterfactual evaluation of risk-assessment tools, due to the absence of a clear indication of whether defendants were detained or released pre-trial \citep{mishler2021fairness}. The lack of a standard preprocessing protocol beyond the one by ProPublica \citep{propublica2016compas}, which is insufficient to handle these factors, may cause issues of reproducibility and difficulty in comparing methods. Moreover, according to Northpointe's response to the ProPublica's study, several risk factors considered by the COMPAS algorithm are absent from the dataset \citep{dieterich2016compas}. \novel{As an additional concern, race categories lack Native Hawaiian or Other Pacific Islander, while Hispanic is redefined as race instead of ethnicity \citep{bao2021COMPASlicated}.}
Finally, defendants' personal information (e.g. race and criminal history) is available in conjunction with obvious identifiers,
making re-identification of defendants trivial. 

\novel{
\ifconf\else{ 
COMPAS also represents a case of a broad phenomenon which can be termed \emph{data bias}. With terminology from \citet{friedler2021impossibility}, when it comes to datasets encoding complex human phenomena, there is often a disconnect between the \emph{construct space} (what we aim to measure) and the \emph{observed space} (what we end up observing). This may be especially problematic if the difference between construct and observation is uneven across individuals or groups. COMPAS, for example, is a dataset about criminal offense. Offense is central to the prediction target $Y$, aimed at encoding recidivism, and to the available covariates $X$, summarizing criminal history. However, the COMPAS dataset (observed space) is an imperfect proxy for the criminal patterns it should summarize (construct space). The prediction labels $Y$ actually encode re-arrest, instead of re-offense \citep{larson2016how}, and are thus clearly influenced by spatially differentiated policing practices \citep{Fogliato:2021mb}. This is also true of criminal history encoded in COMPAS covariates, again mediated by arrest and policing practices which may be racially biased \citep{bao2021COMPASlicated,mayson2018bias}. As a result, the true fairness of an algorithm, just like its accuracy, may differ significantly from what is reported on biased data. For example, algorithms that achieve equality of true positive rates across sensitive groups on COMPAS are deemed fair under the \emph{equal opportunity} measure \citep{hardt2016equality}. However, if both the training set on which this objective is enforced and the test set on which it is measured are affected by race-dependent noise described above, those algorithms are only ``fair'' in an abstract observed space, but not in the real construct space we ultimately care about \citep{friedler2021impossibility}.
}\fi
}

Overall, these considerations paint a mixed picture for a dataset of high social relevance that was extremely useful to catalyze attention on algorithmic fairness issues, displaying at the same time several limitations in terms of its continued use as a flexible benchmark for fairness studies of all sorts. In this regard, \citet{bao2021COMPASlicated} suggest avoiding the use of COMPAS  to demonstrate novel approaches in algorithmic fairness, as considering the data without proper context may lead to misleading conclusions, which could misguidedly enter the broader debate on criminal justice and risk assessment.

\subsection{German Credit}
\label{sec:popular_german}

The German Credit dataset was created to study the problem of computer-assisted credit decisions at a regional Bank in southern Germany. Instances represent loan applicants from 1973 to 1975, who were deemed creditworthy and were granted a loan, bringing about a natural selection bias. \novel{Within this sample, bad credits are oversampled to favour a balance in target classes \citep{gromping2019:sg}.} The data summarizes applicants' financial situation, credit history, and personal situation, including housing and number of liable people. A binary variable encoding whether each loan recipient punctually payed every installment is the target of a classification task. Among the covariates, marital status and sex are jointly encoded in a single variable. Many documentation mistakes are present in the UCI entry associated with this resource \citep{hofmann1994:sg}. 
A revised version with correct variable encodings, called South German Credit, was donated to \citet{gromping2019:sg2} with an accompanying report \citep{gromping2019:sg}.

The greatest upside of this dataset is the fact that it captures a real-world application of credit scoring at a bank. On the downside, the data is half a century old, significantly limiting the societally useful insights that can be gleaned from it. Most importantly, the popular release of this dataset \citep{hofmann1994:sg} comes with highly inaccurate documentation which contains wrong variable codings. For example, the variable reporting whether loan recipients are foreign workers has its coding reversed, so that, apparently, fewer than 5\% of the loan recipients in the dataset would be German. Luckily, this error has no impact on numerical results obtained from this dataset, as it is irrelevant at the level of abstraction afforded by raw features, with the exception of potentially counterintuitive explanations in works of interpretability and exploratory analysis \citep{lequysurvey2022}. This coding error, along with others discussed in \citet{gromping2019:sg} was corrected in a novel release of the dataset \citep{gromping2019:sg2}. Unfortunately and most importantly for the fair ML community, retrieving the sex of loan applicants is simply not possible, unlike the original documentation suggested. This is due to the fact that one value of this feature was used to indicate both women who are divorced, separated, or married, and men who are single, while the original documentation reported each feature value to correspond to same-sex applicants (either male-only or female-only). This particular coding error ended up having a non-negligible impact on the fair ML community, where many works studying group fairness extract sex from the joint variable and use it as a sensitive attribute, even years after the redacted documentation was published \citep{wang2021fair,lequysurvey2022}. These coding mistakes are part of a documentation debt whose influence continues to affect the algorithmic fairness community. 

\subsection{Summary}
\label{sec:popular_summary}

\ifconf\else{
Adult, COMPAS, and German Credit are the most used datasets in the surveyed algorithmic fairness literature, despite the limitations summarized in Table  \ref{tab:popular}. Their status as de facto fairness benchmarks is probably due to their use in seminal works \citep{pedreshi2008discriminationaware,calders2009building} and influential articles \citep{angwin2016machine} on algorithmic fairness. Once this fame was created, researchers had clear incentives to study novel problems and approaches on these datasets, which, as a result, have become even more established benchmarks in the algorithmic fairness literature \citep{bao2021COMPASlicated}. }\fi On close scrutiny, the fundamental merit of these datasets lies in originating from human processes, encoding protected attributes, and having different base rates for the target variable across sensitive groups. Their use in recent works on algorithmic fairness can be interpreted as a signal that the authors have basic awareness of default data practices in the field and that the data was not made up to fit the algorithm. Overarching claims of significance in real-world scenarios stemming from experiments on these datasets should be met with skepticism. Experiments that claim extracting a sex variable from the German Credit dataset should be considered noisy at best. As for alternatives, \citet{bao2021COMPASlicated} suggest employing well-designed simulations.
A complementary avenue is to seek different datasets that are relevant for the problem at hand. We hope that the two hundred data briefs accompanying this work will prove useful in this regard, favouring both domain-oriented and task-oriented searches, according to the classification discussed in the next section.

\section{Existing Alternatives}
\label{sec:alternatives}
In this section, we discuss existing fairness resources from different perspectives. In section \ref{sec:domain} we describe the different domains spanned by fairness datasets. In section \ref{sec:tasks} we provide a categorization of fairness tasks supported by the same resources. \ifconf\else{ In section \ref{sec:roles} we discuss the different roles played by these datasets in fairness research, such as supporting training and benchmarking.}\fi

\subsection{Domain}
\label{sec:domain}

\begin{figure*}[tb]
  \includegraphics[width=\customsize\textwidth]{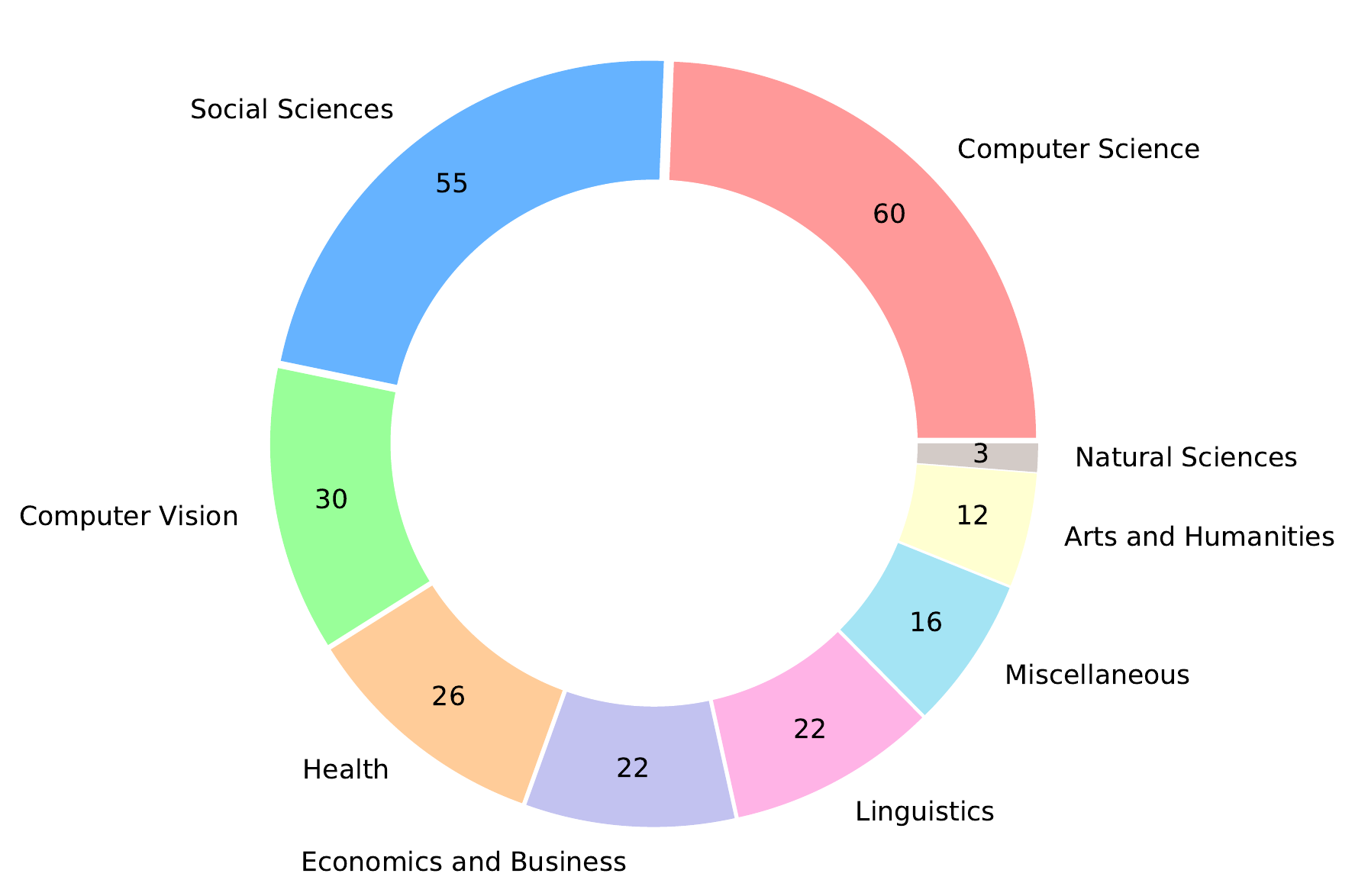}
\caption{Datasets employed in fairness research span diverse domains. See Table \ref{tab:domainwise} for a detailed breakdown.}
\label{fig:pie}
\end{figure*}

\ifconf\else{ 
Algorithmic fairness concerns arise in any domain where Automated Decision Making (ADM) systems may influence human well-being. Unsurprisingly, the datasets in our survey reflect a variety of areas where ADM systems are studied or deployed, including criminal justice, education, search engines, online marketplaces, emergency response, social media, medicine, and hiring.  }\fi In Figure \ref{fig:pie}, we report a subdivision of the surveyed datasets in different macrodomains. We mostly follow the area-category taxonomy by Scimago,\footnote{See the ``subject area'' and ``subject category'' drop down menus from \url{https://www.scimagojr.com/journalrank.php}, accessed on March 15, 2022} departing from it where appropriate. For example, we consider computer vision and linguistics macrodomains of their own, for the purposes of algorithmic fairness, as much fair ML work has been published in both disciplines. \ifconf\else{Below we present a description of each macrodomain and its main subdomains, summarized in detail in Table \ref{tab:domainwise}.}\fi

\textbf{Computer Science}. Datasets from this macrodomain are very well represented, comprising \emph{information systems, social media, library and information sciences, computer networks, and signal processing}. \emph{Information systems} heavily feature datasets on search engines for various items such as text, images, worker profiles, and real estate, retrieved in response to queries issued by users (Occupations in Google Images, Scientist+Painter, Zillow Searches, Barcelona Room Rental, Burst, TaskRabbit, Online Freelance Marketplaces, Bing US Queries, Symptoms in Queries). Other datasets represent problems of item recommendation, covering products, businesses, and movies (Amazon Recommendations, Amazon Reviews, Google Local, MovieLens,  FilmTrust). 
The remaining datasets in this subdomain represent knowledge bases (Freebase15k-237, Wikidata) and automated screening systems (CVs from Singapore, Pymetrics Bias Group). Datasets from \emph{social media} that are not focused on links and relationships between people are also considered part of computer science in this survey. These resources are often focused on text, powering tools and analyses of hate speech and toxicity (Civil Comments, Twitter Abusive Behavior, Twitter Offensive Language, Twitter Hate Speech Detection, Twitter Online Harassment), dialect (TwitterAAE), and political leaning (Twitter Presidential Politics). Twitter is by far the most represented platform, while datasets from Facebook (German Political Posts), Steeemit (Steemit), Instagram (Instagram Photos), Reddit (RtGender, Reddit Comments), Fitocracy (RtGender), and YouTube (YouTube Dialect Accuracy) are also present. Datasets from \emph{library and information sciences} are mainly focused on academic collaboration networks (Cora Papers, CiteSeer Papers, PubMed Diabetes Papers, ArnetMiner Citation Network, 4area, Academic Collaboration Networks), except for a dataset about peer review of scholarly manuscripts (Paper-Reviewer Matching).

\textbf{Social Sciences}. Datasets from social sciences are also plentiful, spanning \emph{law, education, social networks, demography, social work, political science, transportation, sociology} and \emph{urban studies}. \emph{Law} datasets are mostly focused on recidivism (Crowd Judgement, COMPAS, Recidivism of Felons on Probation, State Court Processing Statistics, Los Angeles City Attorney’s Office Records) and crime prediction (Strategic Subject List, Philadelphia Crime Incidents, Stop, Question and Frisk, Real-Time Crime Forecasting Challenge, Dallas Police Incidents, Communities and Crime), with a granularity spanning the range from individuals to communities. In the area of \emph{education} we find datasets that encode application processes (Nursery, IIT-JEE), student performance (Student, Law School, UniGe, ILEA, US Student Performance, Indian Student Performance, EdGap, Berkeley Students), including attempts at automated grading (Automated Student Assessment Prize), and placement information after school (Campus Recruitment). Some datasets on student performance support studies of differences across schools and educational systems, for which they report useful features (Law School, ILEA, EdGap), while the remaining datasets are more focused on differences in the individual condition and outcome for students, typically within the same institution. Datasets about \emph{social networks} mostly concern online social networks (Facebook Ego-networks, Facebook Large Network, Pokec Social Network, Rice Facebook Network, Twitch Social Networks, University Facebook Networks), except for High School Contact and Friendship Network, also featuring offline relations. \emph{Demography} datasets comprise census data from different countries (Dutch Census, Indian Census, National Longitudinal Survey of Youth, Section 203 determinations, US Census Data (1990)).  Datasets from \emph{social work} cover complex personal and social problems, including child maltreatment prevention (Allegheny Child Welfare), emergency response (Harvey Rescue), and drug abuse prevention (Homeless Youths' Social Networks, DrugNet). Resources from \emph{political science} describe registered voters (North Carolina Voters), electoral precincts (MGGG States), polling (2016 US Presidential Poll), and sortition (Climate Assembly UK). \emph{Transportation} data summarizes trips and fares from taxis (NYC Taxi Trips, Shanghai Taxi Trajectories), ride-hailing (Chicago Ridesharing, Ride-hailing App), and bike sharing services (Seoul Bike Sharing), along with public transport coverage (Equitable School Access in Chicago). \emph{Sociology} resources summarize online (Libimseti) and offline dating (Columbia University Speed Dating). Finally, we assign SafeGraph Research Release to \emph{urban studies}.

\textbf{Computer Vision}. This is an area of early success for artificial intelligence, where fairness typically concerns learned representations and equality of performance across classes. The surveyed articles feature several popular datasets on image classification (ImageNet, MNIST, Fashion MNIST, CIFAR), visual question answering (Visual Question Answering), segmentation and captioning (MS-COCO, Open Images Dataset). 
We find over ten face analysis datasets (Labeled Faces in the Wild, UTK Face, Adience, FairFace, IJB-A, CelebA, Pilot Parliaments Benchmark, MS-Celeb-1M, Diversity in Faces, Multi-task Facial Landmark, Racial Faces in the Wild, BUPT Faces), including one from experimental psychology (FACES), for which fairness is most often intended as the robustness of classifiers across different subpopulations, without much regard for downstream benefits or harms to these populations. Synthetic images are popular to study the relationship between fairness and disentangled representations (dSprites, Cars3D, shapes3D). Similar studies can be conducted on datasets with spurious correlations between subjects and backgrounds (Waterbirds, Benchmarking Attribution Methods) or gender and occupation (Athletes and health professionals). Finally, the Image Embedding Association Test dataset is a fairness benchmark to study biases in image embeddings across religion, gender, age, race, sexual orientation, disability, skin tone, and weight. It is worth noting that this significant proportion of computer vision datasets is not an artifact of including CVPR in the list of candidate conferences, which contributed just five additional datasets (Multi-task Facial Landmark, Office31, Racial Faces in the Wild, BUPT Faces, Visual Question Answering).

\textbf{Health}. This macrodomain, comprising medicine, psychology and pharmacology displays a notable diversity of subdomains interested by fairness concerns. Specialties represented in the surveyed datasets are mostly medical, including \emph{public health} (Antelope Valley Networks, Willingness-to-Pay for Vaccine, Kidney Matching, Kidney Exchange Program), \emph{cardiology} (Heart Disease, Arrhythmia, Framingham), \emph{endocrinology} (Diabetes 130-US Hospitals, Pima Indians Diabetes Dataset), \emph{health policy} (Heritage Health, MEPS-HC). Specialties such as \emph{radiology} (National Lung Screening Trial, MIMIC-CXR-JPG, CheXpert) and \emph{dermatology} (SIIM-ISIC Melanoma Classification, HAM10000) feature several image datasets for their strong connections with medical imaging. Other specialties include \emph{critical care medicine} (MIMIC-III), \emph{neurology} (Epileptic Seizures), \emph{pediatrics} (Infant Health and Development Program), \emph{sleep medicine} (Apnea), \emph{nephrology} (Renal Failure), \emph{pharmacology} (Warfarin) and \emph{psychology} (Drug Consumption, FACES). These datasets are often extracted from care data of multiple medical centers to study problems of automated diagnosis. Resources derived from longitudinal studies, including Framingham and Infant Health and Development Program are also present. Works of algorithmic fairness in this domain are typically concerned with obtaining models with similar performance for patients across race and sex.

\textbf{Linguistics}. In addition to the textual resources we already described, such as the ones derived from social media, several datasets employed in algorithmic fairness literature can be assigned to the domain of linguistics and Natural Language Processing (NLP). There are many examples of resources curated to be fairness benchmarks for different tasks, including machine translation (Bias in Translation Templates), sentiment analysis (Equity Evaluation Corpus), coreference resolution (Winogender, Winobias, GAP Coreference), named entity recognition (In-Situ), language models (BOLD) and word embeddings (WEAT). Other datasets have been considered for their size and importance for pretraining text representations (Wikipedia dumps, One billion word benchmark, BookCorpus, WebText) or their utility as NLP benchmarks (GLUE, Business Entity Resolution). Speech recognition resources have also been considered (TIMIT).

\textbf{Economics and Business}. This macrodomain comprises datasets from \emph{economics}, \emph{finance}, \emph{marketing}, and \emph{management information systems}. \emph{Economics} datasets mostly consist of census data focused on wealth (Adult, US Family Income, Poverty in Colombia, Costarica Household Survey) and other resources which summarize employment (ANPE), tariffs (US Harmonized Tariff Schedules), insurance (Italian Car Insurance), and division of goods (Spliddit Divide Goods). \emph{Finance} resources feature data on microcredit and peer-to-peer lending (Mobile Money Loans, Kiva, Prosper Loans Network), mortgages (HMDA), loans (German Credit, Credit Elasticities), credit scoring (FICO) and default prediction (Credit Card Default). \emph{Marketing} datasets describe marketing campaigns (Bank Marketing), customer data (Wholesale) and advertising bids (Yahoo! A1 Search Marketing). Finally, datasets from \emph{management information systems} summarize information about automated hiring (CVs from Singapore, Pymetrics Bias Group) and employee retention (IBM HR Analytics).

\ifconf\else{
\textbf{Miscellaneous}. This macrodomain contains several datasets originating from the \emph{news} domain (Yow news, Guardian Articles, Latin Newspapers, Adressa, Reuters 50 50, New York Times Annotated Corpus, TREC Robust04). Other resources include datasets on food (Sushi), sports (Fantasy Football, FIFA 20 Players, Olympic Athletes)
, and toy datasets (Toy Dataset 1--4).

\textbf{Arts and Humanities}. In this area we mostly find \emph{literature} datasets, which contain text from literary works (Shakespeare, Curatr British Library Digital Corpus, Victorian Era Authorship Attribution, Nominees Corpus, Riddle of Literary Quality), which are typically studied with NLP tools. Other datasets in this domain include domain-specific information systems about books (Goodreads Reviews), \emph{movies} (MovieLens) and \emph{music} (Last.fm, Million Song Dataset, Million Playlist Dataset). 

\textbf{Natural Sciences}. This domain is represented with three datasets from \emph{biology} (iNaturalist), \emph{biochemestry} (PP-Pathways) and \emph{plant science}, with the classic Iris dataset.

As a whole, many of these datasets encode fundamental human activities where algorithms and ADM systems have been studied and deployed. Alertness and attention to equity seems especially important in specific domains, including social sciences, computer science, medicine, and economics. Here the potential for impact may result in large benefits, but also great harm, particularly for vulnerable populations and minorities, more likely to be neglected during the design, training, and testing of an ADM. After concentrating on domains, in the next section we analyze the variety of tasks studied in works of algorithmic fairness and supported by these datasets.

\multipagetable{

\toprule
Domain & Sample datasets \\
\midrule
Computer Science &  \\
\hspace{0.4cm} social media &  \\
\hspace{0.8cm} toxicity and hate speech & Civil Comments, Wikipedia Toxic Comments, Twitter offensive language \\
\hspace{0.8cm} political leaning & Twitter Presidential Politics \\
\hspace{0.8cm} dialect & TwitterAAE \\
\hspace{0.4cm} library and information sciences & \\
\hspace{0.8cm} collaboration networks & Paper-Reviewer Matching, 4area, ArnetMiner Citation Network \\
\hspace{0.8cm} peer review & Paper-Reviewer Matching \\
\hspace{0.4cm} information systems &  \\
\hspace{0.8cm} search engines & Online Freelance Marketplaces, Bing US Queries, Symptoms in Queries \\
\hspace{0.8cm} recommender systems & Amazon Recommendations, Amazon Reviews, MovieLens \\
\hspace{0.8cm} knowledge bases & Freebase15k-237, Wikidata \\
\hspace{0.4cm} computer networks & KDD Cup 99 \\
\hspace{0.4cm} pattern recognition & Internet Ads \\
\hspace{0.4cm} signal processing & Vehicle \\

Social Sciences &  \\
\hspace{0.4cm} urban studies & SafeGraph Research Release \\
\hspace{0.4cm} social networks & University Facebook Networks, Pokec Social Network, Rice Facebook Network \\
\hspace{0.4cm} demography & US Census Data (1990), Dutch Census, National Longitudinal Survey of Youth \\
\hspace{0.4cm} sociology & Columbia University Speed Dating, Libimseti \\
\hspace{0.4cm} law &  \\
\hspace{0.8cm} recidivism prediction &  COMPAS, Recidivism of Felons on Probation, State Court Processing Statistics \\
\hspace{0.8cm} crime prediction &   Communities and Crime, Stop, Question and Frisk, Strategic Subject List \\
\hspace{0.4cm} political science & \\
\hspace{0.8cm} registered voters & North Carolina Voters\\
\hspace{0.8cm} electoral precincts & MGGG States \\
\hspace{0.8cm} polling & 2016 US Presidential Poll \\
\hspace{0.8cm} sortition & Climate Assembly UK  \\
\hspace{0.4cm} education &  \\
\hspace{0.8cm} application processes &  Nursery, IIT-JEE  \\
\hspace{0.8cm} student performance &  Student, Law School, UniGe  \\
\hspace{0.8cm} post-education placement &  Campus Recruitment  \\
\hspace{0.4cm} social work &  \\
\hspace{0.8cm} child maltreatment prevention &  Allegheny Child Welfare  \\
\hspace{0.8cm} emergency response &  Harvey Rescue  \\
\hspace{0.8cm} drug abuse prevention &  Homeless Youths' Social Networks, DrugNet  \\
\hspace{0.4cm} transportation & \\
\hspace{0.8cm} taxi trips & NYC Taxi Trips, Shanghai Taxi Trajectories \\
\hspace{0.8cm} ride hailing & Chicago Ridesharing, Ride-hailing App \\
\hspace{0.8cm} bike sharing & Seoul Bike Sharing \\
\hspace{0.8cm} public transport & Equitable School Access in Chicago \\
Computer Vision &  \\
\hspace{0.4cm} general purpose & ImageNet, MNIST, CIFAR \\
\hspace{0.4cm} face analysis &  CelebA, Pilot Parliaments Benchmar, FairFace \\
\hspace{0.4cm} synthetic &  dSprites, Cars3D, shapes3D \\
Health &  \\
\hspace{0.4cm} sleep medicine & Apnea \\
\hspace{0.4cm} critical care medicine & MIMIC-III \\
\hspace{0.4cm} public health & Kidney Exchange Program, Willingness-to-Pay for Vaccine, Kidney Matching \\
\hspace{0.4cm} cardiology & Arrhythmia, Heart Disease, Framingham \\
\hspace{0.4cm} neurology & Epileptic Seizures \\
\hspace{0.4cm} pediatrics & Infant Health and Development Program (IHDP) \\
\hspace{0.4cm} dermatology & HAM10000, SIIM-ISIC Melanoma Classification \\
\hspace{0.4cm} medicine & Stanford Medicine Research Data Repository \\
\hspace{0.4cm} pharmacology & Warfarin \\
\hspace{0.4cm} endocrinology & Diabetes 130-US Hospitals, Pima Indians Diabetes Dataset (PIDD) \\
\hspace{0.4cm} nephrology & Renal Failure \\
\hspace{0.4cm} radiology & CheXpert, MIMIC-CXR-JPG, National Lung Screening Trial (NLST) \\
\hspace{0.4cm} health policy & Heritage Health, MEPS-HC \\
\hspace{0.4cm} applied psychology & Drug Consumption \\
\hspace{0.4cm} experimental psychology & FACES \\
Economics and Business &  \\
\hspace{0.4cm} economics &  \\
\hspace{0.8cm} census & Adult, US Family Income, Poverty in Colombia \\
\hspace{0.8cm} employment & ANPE \\
\hspace{0.8cm} tariffs & US Harmonized Tariff Schedule \\
\hspace{0.8cm} insurance & Italian Car Insurance \\
\hspace{0.8cm} division of goods & Spliddit Divide Goods \\
\hspace{0.4cm} finance &  \\
\hspace{0.8cm} peer-to-peer lending & Mobile Money Loans, Kiva, Prosper Loans Network \\
\hspace{0.8cm} mortgages & HMDA \\
\hspace{0.8cm} credit scoring & FICO \\
\hspace{0.8cm} other credit & German Credit, Credit Card Default, Credit Elasticities \\
\hspace{0.4cm} marketing &  \\
\hspace{0.8cm} marketing campaigns & Bank Marketing\\
\hspace{0.8cm} advertising bids & Yahoo! A1 Search Marketing, Wholesale \\
\hspace{0.4cm} management information systems & \\
\hspace{0.8cm} automated hiring & Pymetrics Bias Group, CVs from Singapore \\
\hspace{0.8cm} employee retention & IBM HR Analytics \\
Linguistics &  \\
\hspace{0.4cm} general purpose & Wikipedia dumps, One billion word benchmark, BookCorpus \\
\hspace{0.4cm} fairness benchmarks & Bias in Translation Templates, Equity Evaluation Corpus, Winogender \\
Arts and Humanities &  \\
\hspace{0.4cm} music & Million Playlist Dataset (MPD), Million Song Dataset (MSD), Last.fm \\
\hspace{0.4cm} literature & Goodreads Reviews, Riddle of Literary Quality, Nominees Corpus \\
\hspace{0.4cm} movies & MovieLens, FilmTrust \\
Natural Sciences &  \\
\hspace{0.4cm} biology & iNaturalist Datasets \\
\hspace{0.4cm} biochemestry & PP-Pathways\\
\hspace{0.4cm} plant science & Iris \\
Miscellaneous &  \\
\hspace{0.4cm} news & TREC Robust04, New York Times Annotated Corpus, Reuters 50 50 \\
\hspace{0.4cm} sports & Fantasy Football, FIFA 20 Players, Olympic Athletes \\
\hspace{0.4cm} food & Sushi \\
    \bottomrule
    \caption{\novel{A selection of datasets through the lens of the domain taxonomy.}}
    \label{tab:domainwise}
}}\fi

\subsection{Task and setting}
\label{sec:tasks}

\ifconf\else{ Researchers and practitioners are showing an increasing interest in algorithmic fairness, proposing solutions for many different \emph{tasks}, including fair classification, regression, and ranking. At the same time, the academic community is developing an improved understanding of important challenges that run across different tasks in the algorithmic fairness space \citep{chouldechova2020snapshot}, also thanks to practitioner surveys \citep{holstein2019:improving} and studies of specific legal challenges \citep{andrus2021what}. To exemplify, the presence of noise corrupting labels for sensitive attributes represents a challenge that may apply across different tasks, including fair classification, regression, and ranking. We refer to these challenges as \emph{settings}, describing them in the second part of this section. While our work focuses on fair ML datasets, it is cognizant of the wide variety of tasks tackled in the algorithmic fairness literature, which are captured in a specific field of our data briefs. }\fi In this section, we provide an overview of common tasks and settings studied on these datasets, showing their variety and diversity. \ifconf\else{}\fi  Table \ref{tab:taskwise} summarizes the tasks and settings, listing, for each, the three most used datasets. When describing tasks and settings, we explicitly highlight datasets that are particularly relevant, even when outside of the top three. \ifconf\else{}\fi

\begin{table*}[h] 
    \caption{\novel{Most used datasets by algorithmic fairness task and setting.}}
    \label{tab:taskwise}
    \tasktable{
    \toprule
    \multicolumn{1}{c}{Task} & \multicolumn{1}{c}{Datasets} \\
    \midrule
    Fair classification & Adult; COMPAS; German Credit \\
    Fair regression & Communities and Crime; Law School; Student \\
    Fair ranking & MovieLens; German Credit; Kiva \\
    Fair matching & NYC Taxi Trips; Libimseti; Columbia University Speed Dating \\
    Fair risk assessment & COMPAS; Allegheny Child Welfare; Infant Health and Development Program (IHDP) \\
    Fair representation learning & Adult; COMPAS; dSprites \\
    Fair clustering & Adult; Bank Marketing; Diabetes 130-US Hospitals \\
    Fair anomaly detection & Adult; MNIST; Credit Card Default \\
    Fair districting & MGGG States \\
    Fair task assignment & Crowd Judgement; COMPAS \\
    Fair spatio-temporal process learning & Real-Time Crime Forecasting Challenge; Dallas Police Incidents; Harvey Rescue \\
    Fair graph diffusion/augmentation & University Facebook Networks; Antelope Valley Networks; Rice Facebook Network \\
    Fair resource allocation/subset selection & ML Fairness Gym; US Federal Judges; Climate Assembly UK \\
    Fair data summarization & Adult; Student; Credit Card Default \\
    Fair data generation & CelebA; MovieLens; shapes3D \\
    Fair graph mining & MovieLens; Freebase15k-237; PP-Pathways \\
    Fair pricing & Willingness-to-Pay for Vaccine; Credit Elasticities; Italian Car Insurance \\
    Fair advertising & Yahoo! A1 Search Marketing; North Carolina Voters; Instagram Photos \\
    Fair routing & Shanghai Taxi Trajectories \\
    Fair entity resolution & Winogender; Winobias; Business Entity Resolution \\
    Fair sentiment analysis & Popular Baby Names; Equity Evaluation Corpus (EEC); TwitterAAE \\
    Bias in word embeddings & Wikipedia dumps; Word Embedding Association Test (WEAT); Popular Baby Names \\
    Bias in language models & TwitterAAE; BOLD; GLUE \\
    Fair machine translation & Bias in Translation Templates \\
    Fair speech recognition & YouTube Dialect Accuracy; TIMIT \\
    \midrule
    \multicolumn{1}{c}{Setting} & \multicolumn{1}{c}{Datasets} \\
    \midrule
    Rich-subgroup fairness & Adult; COMPAS; Communities and Crime \\
    Fairness under unawareness & Adult; COMPAS; HMDA \\
    Limited-label fairness & Adult; German Credit; COMPAS \\
    Robust fairness & COMPAS; Adult; MEPS-HC \\
    Dynamical fairness & FICO; ML Fairness Gym; COMPAS \\
    Preference-based fairness & Adult; COMPAS; Toy Dataset 1 \\
    Multi-stage fairness & Adult; Heritage Health; Twitter Offensive Language \\
    Fair few-shot learning & Communities and Crime; Toy Dataset 1; Mobile Money Loans \\
    Fair private learning & UTK Face; CheXpert; FairFace \\
    Fair federated learning & Vehicle; Sentiment140; Shakespeare \\
    Fair incremental learning & ImageNet; CIFAR \\
    Fair active learning & Adult; German Credit; Heart Disease \\
    Fair selective classification & CheXpert; CelebA; Civil Comments \\
    \bottomrule
    }
\end{table*}

\subsubsection{Task} 
\ifconf\else{}\fi
\textbf{Fair classification} \citep{calders2010three,dwork2012fairness} is  the most common task by far. \ifconf\else{ Typically, it involves equalizing some measure of interest across subpopulations, such as the recall, precision, or accuracy for different racial groups. On the other hand, individually fair classification focuses on the idea that similar individuals (low distance in the covariate space) should be treated similarly (low distance in the outcome space), often formalized as a Lipschitz condition. }\fi Unsurprisingly, the most common datasets for fair classification are the most popular ones overall (\autoref{sec:popular}), i.e., Adult, COMPAS, and German Credit.

\textbf{Fair regression} \citep{berk2017convex} concentrates on models that predict a real-valued target, requiring the average loss to be balanced across groups. \ifconf\else{ Individual fairness in this context may require losses to be as uniform as possible across all individuals. }\fi Fair regression is a less popular task, often studied on the Communities and Crime dataset, where the task is predicting the rate of violent crimes in different communities.

\textbf{Fair ranking} \citep{yang2017measuring} requires ordering candidate items based on their relevance to a current need. Fairness concerns both the people producing the items that are being ranked \ifconf\else{ (e.g. artists) }\fi and those consuming the items\ifconf\else{  (users of a music streaming platform). It is typically studied in applications of recommendation (MovieLens, Last.fm, Million Song Dataset, Amazon Recommendations, Adressa) and search engines (Yahoo! c14B Learning to Rank, Microsoft Learning to Rank, TREC Robust04).}\fi

\textbf{Fair matching} \citep{kobren2019paper} \ifconf\else{is similar to ranking as they are both tasks defined on two-sided markets. This task, however, is focused on highlighting and matching pairs of items on both sides of the market, without emphasis on the ranking component. }\fi Datasets for this task are from diverse domains, including dating (Libimseti, Columbia University Speed Dating), transportation (NYC Taxi Trips, Ride-hailing App), and organ donation (Kidney Matching, Kidney Exchange Program).

\textbf{Fair risk assessment} \citep{coston2020counterfactual} studies algorithms that score instances in a dataset according to a predefined type of risk. \ifconf\else{ Relevant domains include healthcare and criminal justice. Key differences with respect to classification are an emphasis on real-valued scores rather than labels, and awareness that the risk assessment process can lead to interventions impacting the target variable. For this reason, fairness concerns are often defined in a counterfactual fashion. }\fi The most popular dataset for this task is COMPAS, followed by datasets from medicine (IHDP, Stanford Medicine Research Data Repository), social work (Allegheny Child Welfare), Economics (ANPE) and Education (EdGap).

\textbf{Fair representation learning} \citep{creager2019flexibly} concerns the study of features learnt by models as intermediate representations for inference tasks. \ifconf\else{ A popular line of work in this space, called \emph{disentaglement}, aims to learn representations where a single factor of import corresponds to a single feature. Ideally, this approach should select representations where sensitive attributes cannot be used as proxies for target variables. }\fi Cars3D and dSprites are popular datasets for this task, consisting of synthetic images depicting controlled shape types under a controlled set of rotations. Post-processing approaches are also applicable to obtain fair representations from biased ones via debiasing.

\textbf{Fair clustering} \citep{chierichetti2017fair} is an unsupervised task concerned with the division of a sample into homogenous groups. Fairness may be intended as an equitable representation of protected subpopulations in each cluster, or in terms of average distance from the cluster center. While Adult is the most common dataset, other resources often used for this task include Bank Marketing, Diabetes 130-US Hospitals, Credit Card Default and US Census Data (1990).

\textbf{Fair anomaly detection} \citep{zhang2021towards}, also called \textbf{outlier detection} \citep{davidson2020framework}, is aimed at identifying surprising or anomalous points in a dataset. Fairness requirements involve equalizing key measures (e.g. acceptance rate, recall, distribution of anomaly scores) across populations of interest. This problem is particularly relevant for \ifconf\else{ members of }\fi minority groups, who, in the absence of specific attention to dataset inclusivity, are less likely to fit the norm in the feature space.

\ifconf\else{ 
\textbf{Fair districting} \citep{schutzman2020:to} 
is the division of a territory into electoral districts for political elections. Fairness notions brought forth in this space are either outcome-based, requiring that seats earned by a party roughly match their share of the popular vote, or procedure-based, ignoring outcomes 
and requiring that counties or municipalities are split as little as possible. MGGG States is a reference resource for this task, providing precinct-level aggregated
information about demographics and political leaning of voters in US districts. 
}\fi

\textbf{Fair task assignment} and \textbf{truth discovery} \citep{goel2019crowdsourcing,li2020towards} are different subproblems in the same area, focused on the subdivision of work and the aggregation of answers in crowdsourcing. Fairness may be intended concerning errors in the aggregated answer, requiring error rates to be balanced across groups, or in terms of the work load imposed to workers. A dataset suitable for this task is Crowd Judgement, containing crowd-sourced recidivism predictions.

\ifconf\else{ 
\textbf{Fair spatio-temporal process learning} \citep{shang2020listwise} focuses on the estimation of models for processes which evolve in time and space. Surveyed applications include crime forecasting (Real-Time Crime Forecasting Challenge, Dallas Police Incidents) and disaster relief (Harvey Rescue), with fairness requirements focused on equalization of performance across different neighbourhoods and special attention to their racial composition.
}\fi

\textbf{Fair graph diffusion} \citep{farnad2020unifying} models and optimizes the propagation of information and influence over networks, and its probability of reaching individuals of different sensitive groups.
Applications include obesity prevention (Antelope Valley Networks) and drug-use prevention (Homeless Youths' Social Networks). 
\textbf{Fair graph augmentation} \citep{ramachandran2021gaea} is a similar task, defined on graphs which model access to resources based on existing infrastructure (e.g. transportation), which can be augmented under a budget to increase equity. This task has been proposed to improve school access (Equitable School Access in Chicago) and information availability in social networks (Facebook100). 

\textbf{Fair resource allocation/subset selection} \citep{babaioff2019fair,huang2020towards} can be formalized as a classification problem with constraints on the number of positives. Fairness requirements are similar to those of classification. Subset selection may be employed to choose a group of people from a wider set for a given task (US Federal Judges, Climate Assembly UK). Resource allocation concerns the division of goods (Spliddit Divide Goods) and resources (ML Fairness Gym, German Credit). 

\textbf{Fair data summarization} \citep{celis2018fair} refers to equity in data reduction. It may involve finding a small subset representative of a larger dataset (strongly linked to subset selection) or selecting the most important features (dimensionality reduction). Approaches for this task have been applied to select a subset of images (Scientist+Painter) or customers (Bank Marketing) that represent the underlying population across sensitive groups.

\ifconf\else{ 
\textbf{Fair data generation} \citep{xu2018fairgan} deals with generating ``fair'' data points and labels, which can be used as training or test sets. Approaches in this space may be used to ensure an equitable representation of protected categories in data generation processes learnt from biased datasets (CelebA, IBM HR Analytics), and to evaluate biases in existing classifiers (MS-Celeb-1M). Data generation may also be limited to synthesizing artificial sensitive attributes \citep{burke2018synthetic}.
}\fi

\textbf{Fair graph mining} \citep{kang2020inform} focuses on representations and prediction on graph structures. Fairness is defined as a lack of bias in representations or with respect to a final inference task defined on the graph. Fair graph mining approaches have been applied to knowledge bases (Freebase15k-237, Wikidata), collaboration networks (CiteSeer Paper, Academic Collaboration Networks) and social network datasets (Facebook Large Network, Twitch Social Networks).

\textbf{Fair pricing} \citep{kallus2021fairness} concerns learning and deploying an optimal pricing policy for revenue while maintaining equity of access to services and consumer welfare across groups. Employed datasets are from the economics (Credit Elasticities, Italian Car Insurance), transportation (Chicago Ridesharing), and public health domains (Willingness-to-Pay for Vaccine).

\textbf{Fair advertising} \citep{celis2019toward} is also concerned with access to goods and services. It comprises both bidding strategies and auction mechanisms which may be modified to reduce discrimination with respect to the gender or race composition of the audience that sees an ad. One publicly available dataset for this subtask is Yahoo! A1 Search Marketing.

\ifconf\else{ 
\textbf{Fair routing} \citep{qian2015scram} is the task of suggesting an optimal path from a starting location to a destination. For this task, experimentation has been carried out on a semi-synthetic traffic dataset (Shanghai Taxi Trajectories). The proposed fairness measure requires equalizing the driving cost per customer across all drivers.

\textbf{Fair entity resolution} \citep{cotter2019training} is a task focused on deciding whether multiple records refer to the same entity, which is useful, for instance, for the construction and maintenance of knowledge bases. Business Entity Resolution is a proprietary dataset for fair entity resolution, where constraints of performance equality across chain and non-chain businesses can be tested. Winogender and Winobias are publicly available datasets developed to study gender biases in pronoun resolution.  

\textbf{Fair sentiment analysis} \citep{kiritchenko2018examining} is a well-established instance of fair classification, where text snippets are typically classified as positive, negative, or neutral depending on the sentiment they express. Fairness is intended with respect to the entities mentioned in the text (e.g. men and women). The central idea is that the estimated sentiment for a sentence should not change if female entities (e.g. ``her'', ``woman'', ``Mary'') are substituted with their male counterparts (``him'', ``man'', ``James''). The Equity Evaluation Corpus is a benchmark developed to assess gender and race bias in sentiment analysis models.

\textbf{Bias in Word Embeddings (WEs)} \citep{bolukbasi2016man} is the study of undesired semantics and stereotypes captured by vectorial representations of words. WEs are typically trained on large text corpora (Wikipedia dumps) and audited for associations between gendered words (or other words connected to sensitive attributes) and stereotypical or harmful concepts, such as the ones encoded in WEAT.

\textbf{Bias in Language Models (LMs)} \citep{bordia2019identifying} is, quite similarly, the study of biases in LMs, which are flexible models of human language based on contextualized word representations, which can be employed in a variety of linguistics and NLP tasks. LMs are trained on large text corpora from which they may learn spurious correlations and stereotypes. The BOLD dataset is an evaluation benchmark for LMs, based on prompts that mention different socio-demographic groups. LMs complete these prompts into full sentences, which can be tested along different dimensions (sentiment, regard, toxicity, emotion and gender polarity).

\textbf{Fair Machine Translation (MT)} \citep{stanovsky2019evaluating} concerns automatic translation of text from a source language into a target one. MT systems can exhibit gender biases, such as a tendency to translate gender-neutral pronouns from the source language into gendered pronouns of the target language in accordance with gender stereotypes. For example, a ``nurse'' mentioned in a gender-neutral context in the source sentence may be rendered with feminine grammar in the target language. Bias in Translation Templates is a set of short templates to test such biases.

\textbf{Fair speech recognition} \citep{tatman2017:gender} requires accurate annotation of spoken language into text across different demographics. YouTube Dialect Accuracy is a dataset developed to audit the accuracy of YouTube's automatic captions across two genders and five dialects of English. Similarly, TIMIT is a classic speech recognition dataset annotated with American English dialect and gender of speaker.
}\fi

\subsubsection{Setting}
\ifconf\else{ 
As noted at the beginning of this section, there are several \emph{settings} (or challenges) that run across different tasks described above. Some of these settings are specific to fair ML, such as ensuring fairness across an exponential number of groups, or in the presence of noisy labels for sensitive attributes. Other settings are connected with common ML challenges, including few-shot and privacy-preserving learning. Below we describe common settings encountered in the surveyed articles. Most of these settings are tested on fairness datasets which are popular overall, i.e. Adult, COMPAS, and German Credit. We highlight situations where this is not the case, potentially due to a given challenge arising naturally in some other dataset.}\fi

\textbf{Rich-subgroup fairness} \citep{kearns2018preventing} is a setting where fairness properties are required to hold not only for a limited number of protected groups, but across an exponentially large number of subpopulations. This line of work represents an attempt to bridge the normative reasoning underlying individual and group fairness. 

\textbf{Fairness under unawareness} is a general expression to indicate problems where sensitive attributes are missing \citep{chen2019fairness}, encrypted \citep{kilbertus2018blind} or corrupted by noise \citep{lamy2019noisetolerant}. \ifconf\else{ These problems respond to real-world challenges related to the confidential nature of protected attributes, that individuals may wish to hide, encrypt, or obfuscate. }\fi This setting is most commonly studied on highly popular fairness dataset (Adult, COMPAS), moderately popular ones (Law School and Credit Card Default), and a dataset about home mortgage applications in the US (HMDA). 

\textbf{Limited-label fairness} comprises settings with limited information on the target variable, including situations where labelled instances are few \citep{ji2020can}, noisy \citep{wang2021fair}, or only available in aggregate form \citep{sabato2020bounding}. 

\textbf{Robust fairness} problems arise under perturbations to the training set \citep{huang2019stable}, adversarial attacks \citep{nanda2021fairness} and dataset shift \citep{singh2021fairness}. This line of research is often connected with work in robust machine learning, extending the stability requirements beyond accuracy-related metrics to fairness-related ones. 

\textbf{Dynamical fairness} \citep{liu2018delayed,damour2020fairness} entails repeated decisions in changing environments, potentially affected by the very algorithm that is being studied. Works in this space study the co-evolution of algorithms and populations on which they act over time. \ifconf\else{ For example, an algorithm that achieves equality of acceptance rates across protected groups in a static setting may generate further incentives for the next generation of individuals from historically disadvantaged groups. }\fi Popular resources for this setting are FICO and the ML Fairness GYM.

\textbf{Preference-based fairness} \citep{zafar2017from} denotes work informed\ifconf\else{, explicitly or implicitly, }\fi by the preferences of stakeholders. For data subjects this is related to notions of envy-freeness and loss aversion \citep{ali2019loss}; for policy-makers it permits an indication of how to trade-off different fairness measures \citep{zhang2020joint} or direct demonstrations of fair outcomes \citep{galhotra2021learning}.

\textbf{Multi-stage fairness} \citep{madras2018predict} refers to settings where several decision makers coexist in a compound decision-making process. Decision makers, both humans and algorithmic, may act with different levels of coordination. A fundamental question in this setting is how to ensure fairness under composition of different decision mechanisms. 

\textbf{Fair few-shot learning} \citep{zhao2020fair} aims at developing fair ML solutions in the presence of a small amount of data samples. The problem is closely related to, and possibly solved by, \textbf{fair transfer learning} \citep{coston2019fair}\ifconf\else{, where the goal is to exploit the knowledge gained on a problem to solve a different but related one. }\fi Datasets where this setting arises naturally are Communities and Crime, where one may restrict the training set to a subset of US states, and Mobile Money Loans, which consists of data from different African countries.

\textbf{Fair private learning} \citep{bagdasaryan2019differential,jagielski2019differentially} studies the interplay between privacy-preserving mechanisms and fairness constraints. \ifconf\else{ Works in this space consider the equity of machine learning models designed to avoid leakage of information about individuals in the training set. }\fi Common domains for datasets employed in this setting are face analysis (UTK Face, FairFace, Diversity in Face) and medicine (CheXpert, SIIM-ISIC Melanoma Classification, MIMIC-CXR-JPG).

Additional settings that are less common include \textbf{fair federated learning} \citep{li2020fair}, where algorithms are trained across multiple decentralized devices, \textbf{fair incremental learning} \citep{zhao2020maintaining}, where novel classes may be added to the learning problem over time, \textbf{fair active learning} \citep{noriegacampero2019active}, allowing for the acquisition of novel information during inference, and \textbf{fair selective classification} \citep{jones2021selective}, where predictions are issued only if model confidence is above a certain threshold.

\ifconf\else{ Overall, we found a variety of tasks defined on fairness datasets, ranging from generic, such as \emph{fair classification}, to narrow and specifically defined on certain datasets, such as \emph{fair districting} on MGGG States and \emph{fair truth discovery} on Crowd Judgement. Orthogonally to this dimension, many settings or challenges may arise to complicate these tasks, including noisy labels, system dynamics, and privacy concerns. Quite clearly, algorithmic fairness research has been expanding in both directions, by studying a variety of tasks under diverse and challenging settings. In the next section, we analyze the roles played in scholarly works by the surveyed datasets. }\fi



\ifconf\else{

\subsection{Role}
\label{sec:roles}

The datasets used in algorithmic fairness research can play different roles. For example, some may be used to train novel algorithms, while others are suited to test existing algorithms from a specific point of view. Chapter 7 of \citet{barocas2019fair}, describes six different roles of datasets in machine learning. We adopt their framework to analyse fair ML datasets, adding to the taxonomy two roles that are specific to fairness research. 

\emph{A source of real data}. While synthetic datasets and simulations may be suited to demonstrate specific properties of a novel method, the usefulness of an algorithm is typically established on data from the real world. More than a sign of immediate applicability to important challenges, good performance on real-world sources of data signals that the researchers did not make up the data to suit the algorithm. This is likely the most common role for fairness datasets, especially common for the ones hosted on the UCI ML repository, including Adult, German Credit, Communities and Crime, Diabetes 130-US Hospitals, Bank Marketing, Credit Card Default, US Census Data (1990). These resources owe their popularity in fair ML research to being a product of human processes and to encoding protected attributes. Quite simply, they are sources of real human data.

\emph{A catalyst of domain-specific progress}. Datasets can spur algorithmic insight and bring about domain-specific progress. Civil Comments is a great example of this role, powering the Jigsaw Unintended Bias in Toxicity Classification challenge. The challenge responds to a specific need in the space of automated moderation against toxic comments in online discussion. Early attempts at toxicity detection resulted in models which associate mentions of frequently attacked identities (e.g. gay) with toxicity, due to spurious correlations in training sets. The dataset and associated challenge tackle this issue by providing toxicity ratings for comments, along with labels encoding whether members of a certain group are mentioned, favouring measurement of undesired bias. Many other datasets can play a similar role, including, Winogender, Winobias and the Equity Evaluation Corpus. In a broader sense, COMPAS and the accompanying study \citep{angwin2016machine} have been an important catalyst, not for a specific task, but for fairness research overall. 

\emph{A way to numerically track progress on a problem}. This role is common for machine learning benchmarks that also provide human performance baselines. Algorithmic methods approaching or surpassing these baselines are often considered a sign that the task is ``solved'' and that harder benchmarks are required \citep{barocas2019fair}. Algorithmic fairness is a complicated, context-dependent, contested construct whose correct measurement is continuously debated. Due to this reason, we are unaware of any dataset with a similar role in the algorithmic fairness literature. 

\emph{A resource to compare models}. Practitioners interested in solving a specific problem may take a large set of algorithms and test them on a group of datasets that are representative of their problem, in order to select the most promising ones. For well-established ML challenges, there are often leaderboards providing a concise comparison between algorithms for a given task, which may be used for model selection. This setting is rare in the fairness literature, also due to inherent difficulties in establishing a single measure of interest in the field. One notable exception is represented by \citet{friedler2019comparative}, who employed a suite of four datasets (Adult, COMPAS, German Credit, Ricci) to compare the performance of four different approaches to fair classification.

\emph{A source of pre-training data}. Flexible, general-purpose models are often pre-trained to encode useful representations, which are later fine-tuned for specific tasks in the same domain. For example, large text corpora are often employed to train language models and word embeddings which are later specialized to support a variety of downstream NLP applications. Wikipedia dumps, for instance, are often used to train word embeddings and investigate their biases \citep{brunet2019understanding,liang2020artificial,papakyriakopulos2020bias}. Several algorithmic fairness works aim to study and mitigate undesirable biases in learnt representations. Corpora like Wikipedia dumps are used to obtain representations via realistic pretraining procedures that mimic common machine learning practice as closely as possible.

\emph{A source of training data}. Models for a specific task are typically learnt from training sets that encode relations between features and target variable in a representative fashion. One example from the fairness literature is Large Movie Review, used to train sentiment analysis models, later audited for fairness \citep{liang2020artificial}. For fairness audits, one alternative would be resorting to publicly available models, but sometimes a close control on the training corpus and procedure is necessary. Indeed, it is interesting to study issues of model fairness in relation to biases present in the respective training corpora, which can help explain the causes of bias \citep{brunet2019understanding}. Some works measure biases in internal model representations before and after fine-tuning on a training set, and regard the difference as a measure of bias in the training set. \citet{babaeianjelodar2020quantifying} employ this approach to measure biases in RtGender, Civil Comments, and datasets from GLUE.

\emph{A representative summary of a service}. Much important work in the fairness literature is focused on measuring fairness and harms in the real world. This line of work includes audits of products and services, which rely on datasets extracted from the application of interest. Datasets created for this purpose include Amazon Recommendations, Pymetrics Bias Group, Occupations in Google Images, Zillow Searches, Online Freelance Marketplaces, Bing US Queries, YouTube Dialect Accuracy. Several other datasets were originally created for this purpose and later repurposed in the fairness literature as sources of real data, including Stop Question and Frisk, HMDA, Law School, and COMPAS.

\emph{An important source of data}. Some datasets acquire a pivotal role in research and industry, to the point of being considered a de-facto standard for a given purpose. This status warrants closer scrutiny of the dataset, through which researchers aim to uncover potential biases and problematic aspects that may impact models and insights derived from the dataset. ImageNet, for instance, is a dataset with millions of images across thousands of categories. Since its release in 2011, this resource has been used to train, benchmark, and compare hundreds of computer vision models. Given its status in machine learning research, ImageNet has been the subject of two quantitative investigations analyzing its biases and other problematic aspects in the person subtree, uncovering issues of representation \citep{yang2020towards} and non-consensuality \citep{prabhu2020large}. A different data bias audit was carried out on SafeGraph Research Release. SafeGraph data captures mobility patterns in the US, with data from nearly 50 million mobile devices obtained and maintained by Safegraph, a private data company. Their recent academic release has become a fundamental resource for pandemic research, to the point of being used by the Centers for Disease Control and Prevention to measure the effectiveness of social distancing measures \citep{moreland2020timing}. To evaluate its representativeness for the overall US population, \citet{coston2021leveraging} have studied selection biases in this dataset. 

In algorithmic fairness research, datasets play similar roles to the ones they play in machine learning according to \citet{barocas2019fair}, including training, catalyzing attention, and signalling awareness of common data practices. 
One notable exception is that fairness datasets are not used to track algorithmic progress on a problem over time, likely due to the fact that there is no consensus on a single measure to be reported. On the other hand, two roles peculiar to fairness research are summarizing a service or product that is being audited, and representing an important resource whose biases and ethical aspects are particularly worthy of attention. We note that these roles are not mutually exclusive and that datasets can play multiple roles. COMPAS, for example, was originally curated to perform an audit of pretrial risk assessment tools and was later used extensively in fair ML research as a source of real human data, becoming, overall, a catalyst for fairness research and debate.

In sum, existing fairness datasets originate from a variety of domains, support diverse tasks, and play different roles in the algorithmic fairness literature. We hope our work will contribute to establishing principled data practices in the field, to guide an optimal usage of these resources. In the next section we continue our discussion on the key features of these datasets with a change of perspective, asking which lessons can be learnt from existing resources for the curation of novel ones.}\fi

\section{Best Practices for Dataset Curation}
\label{sec:best_practices}

\begin{figure*}
  \includegraphics[width=\customsize\textwidth]{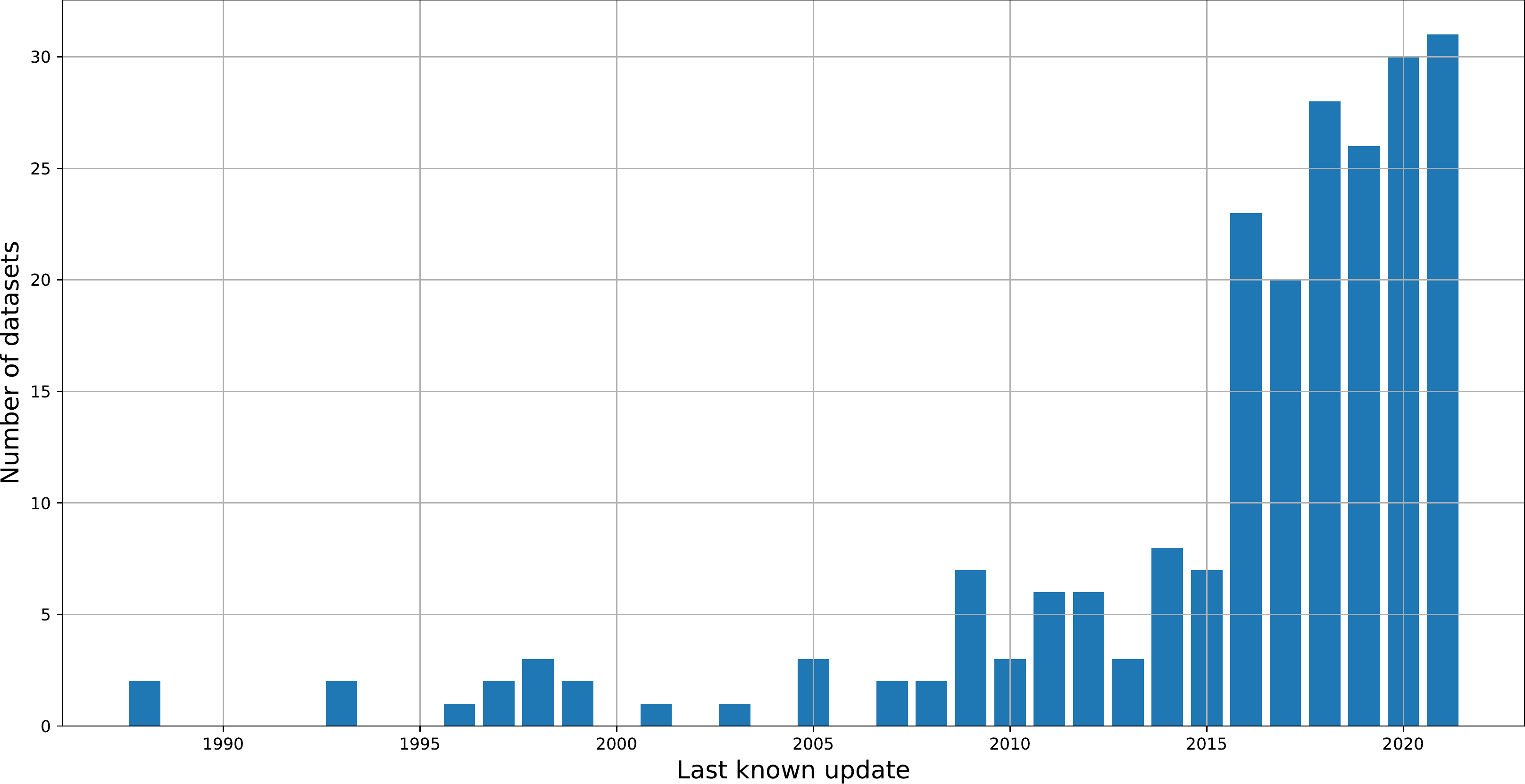}
\caption{Most datasets employed in algorithmic fairness were created or updated after 2015, with a clear growth in recent years.}
\label{fig:bar_year}
\end{figure*}

In this section, we analyze the surveyed datasets from different perspectives, typical of critical data studies, human-computer interaction, and computer-supported cooperative work. In particular, we discuss concerns of re-identification (\autoref{sec:reidentification}), consent (\autoref{sec:consent}), inclusivity (\autoref{sec:inclusivity}), sensitive attribute labeling (\autoref{sec:sensitive_attribute}) and transparency (\autoref{sec:transparency}). We describe a range of approaches and consideration to these topics, ranging from negligent to conscientious. Our aim is to make these concerns and related desiderata more visible and concrete, to help inform responsible curation of novel fairness resources, whose number has been increasing in recent years (Figure \ref{fig:bar_year}). 

\subsection{Re-identification}
\label{sec:reidentification}

\textbf{Motivation.} Data re-identification (or de-anonymization) is a practice through which instances in a dataset, theoretically representing people in an anonymized fashion, are successfully mapped back to the respective individuals. Their identity is thus discovered and associated with the information encoded in the dataset features. Examples of external re-identification attacks include de-anonymization of movie ratings from the Netflix prize dataset \citep{narayanan2008robust}, identification of profiles based on social media group membership \citep{wondracek2010practical}, and identification of people depicted in verifiably pornographic categories of ImageNet \citep{prabhu2020large}. These analyses were carried out as ``attacks'' by external teams for demonstrative purposes, but dataset curators and stakeholders may undertake similar efforts internally \citep{mckenna2019:history_drb}.

There are multiple harms connected to data re-identification, especially the ones featured in algorithmic fairness research, due to their social significance. Depending on the domain and breadth of information provided by a dataset, malicious actors may acquire information about mobility patterns, consumer habits, political leaning, psychological traits, and medical conditions of individuals, just to name a few. The potential for misuse is tremendous, including phishing attacks, blackmail, threat, and manipulation \citep{kroger2021data}. Face recognition datasets are especially prone to successful re-identification as, by definition, they contain information strongly connected with a person's identity. The problem also extends to general purpose computer vision datasets. In a recent dataset audit, \citet{prabhu2020large} found images of beach voyeurism and other non-consensual depictions in ImageNet, and were able to identify the victims using reverse image search engines, highlighting downstream risks of blackmail and other forms of abuse.

\textbf{Disparate consideration.} In this work, we find that fairness datasets are proofed against re-identification with a full range of measures and care. Perhaps surprisingly, some datasets allow for straightforward re-identification of individuals, providing their full names.
We do not discuss these resources here to avoid amplifying the harms discussed above.
Other datasets afford plausible re-identification, providing social media handles and aliases, such as Twitter Abusive Behavior, Sentiment140, Facebook Large Network, and Google Local. Columbia University Speed Dating may also fall in this category due to a restricted population from which the sample is drawn, and provision of age, field of study and ZIP code where participants grew up in addition. In contrast, many datasets come with strong guarantees against de-anonymization, which is especially typical of health data, such as MIMIC-III and Heritage Health \citep{el2012deidentification}. Indeed, health is a domain where a culture of patient record confidentiality is widely established and there is a strong attention to harm avoidance. Also datasets describing scholarly works and academic collaboration networks (Academic Collaboration Networks, PubMed Diabetes Papers, Cora, CiteSeer) are typically de-identified, with numerical IDs substituting names. This is possibly a sign of attention to anonymization from curators when the data represents potential colleagues. As a consequence, researchers are protected from related harms, but posterior annotation of sensitive attributes similarly to \citet{biega2019overview} becomes difficult or impossible. One notable exception is ArnetMiner Citation Network, derived from an online platform which is especially focused on data mining from academic social networks and profiling of researchers.

\textbf{Mitigating factors.} A wide range of factors, summarized in Table \ref{tab:reid}. may help to reduce the risk of re-identification. A first set of approaches concerns the distribution of data artefacts. Some datasets are simply kept private, minimizing risks in this regard. These include UniGe, US Student Performance, Apnea, Symptoms in Queries and Pymetrics Bias Group, the last two being proprietary datasets that are not disclosed to preserve intellectual property. Twitter Online Harrassment is available upon request to protect the identities of Twitter users that were included. Another interesting approach are mixed release strategies: NLSY has some publicly available data, while access to further information that may favour re-identification (e.g. ZIP code and census tract) is restricted. For crawl-based datasets, it is possible to keep a resource private while providing code to recreate it (Bias in Bios). While this may alleviate some concerns, it will not deter motivated actors. As a post-hoc remedy, proactive removal of problematic instances is also a possibility, as shown by recent work on ImageNet \citep{yang2020towards}.

\begin{table*}[h]
  \caption{\novel{Mitigating factors against re-identification.}}
  \label{tab:reid}
  \begin{tabular}{p{4.5cm}p{6.5cm}}
    \toprule
    \multicolumn{1}{c}{Mitigating factor} & \multicolumn{1}{c}{Example datasets}\\
    \midrule
    Controlled distribution & \\
    \hspace{0.5 cm} Private dataset & UniGe, Pymetrics Bias Group \\
    \hspace{0.5 cm} Availability upon request & Twitter Online Harrassment \\
    \hspace{0.5 cm} Mixed release strategy & NLSY \\
    \hspace{0.5 cm} Code-based reconstruction & Bias in Bios \\
    Data perturbation & \\
    \hspace{0.5 cm} Obfuscation &  Yahoo! c14B Learn to Rank, Microsoft Learning to Rank \\
    \hspace{0.5 cm} Top-coding &  Adult \\
    \hspace{0.5 cm} Blurring &  Chicago Ridesharing \\
    \hspace{0.5 cm} Targeted scrubbing &  ASAP \\
    \hspace{0.5 cm} Aggregation &  FICO \\
    Synthesis & \\
    \hspace{0.5 cm} Synthetic data &  Toy Dataset 1--4 \\
    \hspace{0.5 cm} Semi-synthetic data &  Antelope Valley Networks, Kidney Matching  \\
    \hspace{0.5 cm} Hypothetical profiles &  Italian Car Insurance  \\
    Age & German Credit \\
  \bottomrule
\end{tabular}
\end{table*}

Another family of approaches is based on redaction, aggregation, and injection of noise. Obfuscation typically involves the distribution of proprietary company data at a level of abstraction which maintains utility to a company while hindering reconstruction of the underlying human-readable data, which also makes re-identification highly unlikely (Yahoo! c14B Learn to Rank, Microsoft Learning to Rank). Noise injection can take many forms, such as top-coding (Adult), i.e., saturation of certain variables, and blurring (Chicago Ridesharing), i.e., disclosure at coarse granularity. Targeted scrubbing of identifiable information is also rather common, with ad-hoc techniques applied in different domains. For example, the curators of ASAP, a dataset featuring student essays, removed personally identifying information from the essays using named entity recognition and several heuristics. Finally, aggregation of data into subpopulations of interest also supports the anonymity of the underlying individuals (FICO). 

So far we have covered datasets that feature human data derived from real-world processes. Toy datasets, on the other hand, are perfectly safe from this point of view, however their social relevance is inevitably low. In this work we survey four popular ones, taken from \citet{zafar2017fairness,donini2018empirical,lipton2018does,singh2019policy}. Semi-synthetic datasets aim for the best of both worlds by generating artificial data from models that emulate the key characteristics of the underlying processes, as is the case with Antelope Valley Networks, Kidney Matching, and the generative adversarial network trained by  \citet{mcduff2019characterizing} on 
MS-Celeb-1M. Data synthesis may also be applied to augment datasets with artificial sensitive attributes in a principled fashion (MovieLens -- \citep{burke2018synthetic}). Finally, resources designed to externally probe services, algorithms, and platforms, to estimate the direct effect of a feature of interest (e.g.\ gender, race), may rely on hypothetical profiles \citep{bertrand2004emily,fabris2021algorithmic}. This approach can support evaluations of \emph{fairness through unawareness} \citep{grgichlaca2016case}, of which Italian Car Insurance is an example. 

One last important factor is the \emph{age} of a dataset. Re-identification of old information about individuals requires matching with auxiliary resources from the same period, which are less likely to be maintained than comparable resources from recent years. Moreover, even if successful, the consequences of re-identification are likely mitigated by dataset age, as old information about individuals is less likely to support harm against them. The German Credit dataset, for example, represents loan applicants from 1973--1975, whose re-identification and subsequent harm appears less likely than re-identification for more recent datasets in the same domain.

\textbf{Anonymization vs social relevance.} Utility and privacy are typically considered conflicting objectives for a dataset \citep{wieringa2021data}. If we define social relevance as the breadth and depth of societally useful insights that can be derived from a dataset, a similar conflict with privacy becomes clear. Old datasets hardly afford any insight that is actionable and relevant to current applications. Insight derived from synthetic datasets is inevitably questionable. Noise injection increases uncertainty and reduces the precision of claims. Obfuscation hinders subsequent annotation of sensitive attributes. Conservative release strategies increase friction and deter from obtaining and analyzing the data. The most socially relevant fairness datasets typically feature confidential information (e.g. criminal history and financial situation) in conjunction with sensitive attributes of individuals (e.g. race and sex). For these reasons, the social impact afforded by a dataset and the safety against re-identification of included individuals are potentially conflicting objectives that require careful balancing. In the next section we discuss informed consent, another important aspect for the privacy of data subjects.

\subsection{Consent}
\label{sec:consent}

\textbf{Motivation.} In the context of data, \emph{informed consent} is an agreement between a data processor and a subject, aimed at allowing collection and use of personal information while guaranteeing some control to the subject. It is emphasized in Article 7 and Recitals (42) and (43) of the General Data Protection Regulation \citep{eu2016gdpr}, requiring it to be freely given, specific, informed, and unambiguous. \citet{paullada2020data} note that in the absence of individual control on personal information, anyone with access to the data can process it with little oversight, possibly against the interest and well-being of data subjects.  Consent is thus an important tool in a healthy data ecosystem that favours development, trust, and dignity.

\textbf{Negative examples.} A separate framework, often conflated with consent, is copyright. Licenses such as Creative Commons discipline how academic and creative works can be shared and built upon, with proper credit attribution. According to the Creative Commons organization, however, their licenses are not suited to protect privacy and cover research ethics \citep{merkley2019use}. In computer vision, and especially in face recognition, consent and copyright are often considered and discussed jointly, and Creative Commons licenses are frequently taken as an all-inclusive permit encompassing intellectual property, consent, and ethics \citep{prabhu2020large}. \citet{merler2019diversity}, for example, mention privacy and copyright concerns in the construction of Diversity in Faces. These concerns are apparently jointly solved by obtaining images from YFCC-100M, due to the fact that ``a large portion of the photos have Creative Commons license''. Indeed lack of consent is a widespread and far-reaching problem in face recognition datasets \citep{keyes2019government}. \citet{prabhu2020large} find several examples of non-consensual images in large scale computer vision datasets. A particularly egregious example covered in this survey is MS-Celeb-1M, released in 2016 as the largest publicly available training set for face recognition in the world \citep{guo2016msceleb1m}. As suggested by its name, the dataset should feature only celebrities, ``to enable our training, testing, and re-distributing under certain licenses'' \citep{guo2016msceleb1m}. However, the dataset was later found to feature several people who are in no way celebrities, and must simply maintain an online presence. The dataset was retracted for this reason \citep{murgia2019microsoft}.

\textbf{Positive examples.} FACES, an experimental psychology dataset on emotion-related stimuli, represents a positive exception in the face analysis domain. Due its small cardinality, it was possible to obtain informed consent from every participant. One domain where informed consent doctrine has been well-established for decades is medicine; fairness datasets from this space are typically sensitive to the topic. Experiments such as randomized controlled trials always require consent elicitation and often discuss the process in the respective articles. Infant Health and Development Program (IHDP), for instance, is a dataset used to study fair risk assessment. It was collected through the IHDP program, carried out between 1985 and 1988 in the US to evaluate the effectiveness of comprehensive early intervention in reducing developmental and health problems in low birth weight premature infants. \citet{brooks1992effects} clearly state that ``of the  1302 infants who met enrollment criteria, 274 (21\%) had  parents who refused consent and 43 were withdrawn before entry into the assigned group''. Longitudinal studies require trust and continued participation. They typically produce insights and data thanks to participants who have read and signed an informed consent form. Examples of such datasets include Framingham, stemming from a study on cardiovascular disease, and the National Longitudinal Survey of Youth, following the lives of representative samples of US citizens, focusing on their labor market activities and other significant life events. Field studies and derived datasets (DrugNet, Homeless Youths' Social Networks) are also attentive to informed consent.

\textbf{The FRIES framework.} According to the Consentful Tech Project,\footnote{\url{https://www.consentfultech.io/}} consent should be \emph{\underline{F}reely given}, \emph{\underline{R}eversible}, \emph{\underline{I}nformed}, \emph{\underline{E}nthusiastic}, and \emph{\underline{S}pecific} (FRIES). Below we expand on these points and discuss some fairness datasets through the FRIES lens. Pokec Social Network summarizes the networks of Pokec users, a popular social network in Slovakia and Czech Republic. Due to default privacy settings being predefined as public, a wealth of information for each profile was collected by curators, including information on demographics, politics, education, marital status, and children \citep{takac2012data}. While privacy settings are a useful tool to control personal data, default public settings are arguably misleading and do not amount to \emph{freely given} consent. In the presence of more conservative predefined settings, a user can explicitly choose to publicly share their information. This may be interpreted as consent to share one's information here and now with other users; more loose interpretations favouring data collection and distribution are also possible, but they seem rather lacking in \emph{specificity}. It is far from clear that choosing public profile settings entails consent to become part of a study and a publicly available dataset for years to come.

This stands in contrast with Framingham and other datasets derived from medical studies, where consent may be provided or refused with fine granularity \citep{levy2010consent}. In this regard, let us consider a consent form from a recent Framingham exam \citep{framingham2021consent}. The form comes with five different consent boxes which cover participation in examination, use of resulting data, participation in genetic studies, sharing of data with external entities, and notification of findings to subject. Before the consent boxes, a well-structured document informs participants on the reasons for this study, clarifies that they can choose to drop out without penalties at any point, provides a point of contact, explains what will happen in the study and what are the risks to the subject. Some examples of accessible language and open explanations include the following:
\begin{itemize}
    \item ``You have the right to refuse to allow your data and samples to be used or shared for further research. Please check the appropriate box in the selection below.''
    \item ``There is a potential risk that your genetic information could be used to your disadvantage. For example, if genetic research findings suggest a serious health problem, that could be used to make it harder for you to get or keep a job or insurance.''
    \item ``However, we cannot guarantee total privacy. [\ldots] Once information is given to outside parties, we cannot promise that it will be kept private.''
\end{itemize}

Moreover, the consent form is accessible from a website that promises to deliver a Spanish version, showing attention to linguistic minorities. Overall, this approach seems geared towards trust and truly informed consent.

In some cases, consent is made unapplicable by necessity. Allegheny Child Welfare, for instance, stems from an initiative by the Allegheny County’s Department of Human Services to develop assistive tools to support child maltreatment hotline screening decisions. Individuals who resort to this service are in a situation of need and emergency that makes \emph{enthusiastic} consent highly unlikely. Similar considerations arise in any situations where data subjects are in a state of need and can only access a service by providing their data. A clear example is Harvey Rescue, the result of crowdsourced efforts to connect rescue parties with people requesting help in the Houston area. Moreover, the provision of data is mandatory in some cases, such as the US census, which conflicts with meaningful, let alone enthusiastic, consent.

Finally, consent should be \emph{reversible}, giving individuals a chance to revoke it and be removed from a dataset. This is an active area of research, studying specific tools for consent management \citep{albanese2020dynamic} and approaches for retroactive removal of an instance from a model's training set  \citep{ginart2019making}. Unfortunately, even when discontinued or redacted, some datasets remain available through backchannels and derivatives. MS-Celeb-1M is, again, a negative example in this regard. The dataset was removed by Microsoft after widespread criticism and claims of privacy infringement. Despite this fact, it remains available via academic torrents \citep{peng2021mitigating}. Moreover, MS-Celeb-1M was used as a source of images for several datasets derived from it, including the BUPT Faces and Racial Faces in the Wild datasets covered in this survey. This fact demonstrates that harms related to data artefacts are not simply remedied via retirement or redaction. Ethical considerations about consent and potential harms to people must be more than an afterthought and need to enter the discussion during design. 

\subsection{Inclusivity}
\label{sec:inclusivity}

\textbf{Motivation.} Issues of representation, inclusion and diversity are central to the fair ML community. Due to historical biases stemming from structural inequalities, some populations and their perspectives are underrepresented in certain domains and in related data artefacts \citep{jo2020lessons}. For example, the person subtree of ImageNet contains images that skew toward male, young and light skin individuals \citep{yang2020towards}. Female entities were found to be underrepresented in popular datasets for coreference resolution \citep{zhao2018gender}. Even datasets that match natural group proportions may support the development of biased tools with low accuracy for minorities. 

Recent works have demonstrated the disparate performance of tools on sensitive subpopulations in domains such as health care \citep{obermeyer2019dissecting}, speech recognition \citep{tatman2017:gender}, and computer vision \citep{Buolamwini2018gender}. Inclusivity and diversity are often considered a primary solution in this regard, both in training sets, which support the development of better models, and test sets, capable of flagging such issues.

\textbf{Positive examples.} Ideally, inclusivity should begin with a clear definition of data collection objectives \citep{jo2020lessons}. Indeed, we find that diversity and representation are strong points of datasets that were created to assess biases in services, products and algorithms (BOLD, HMDA, FICO, Law School, Scientist+Painter, CVs from Singapore, YouTube Dialect Accuracy, Pilot Parliaments Benchmark), which were designed and curated with special attention to sensitive groups. We also find instances of ex-post remedies to issues of diversity. As an example, the curators of ImageNet proposed a demographic balancing solution based on a web interface that removes the images of overrepresented categories \citep{yang2020towards}. A natural alternative is the collection of novel instances, a solution adopted for Framingham. This dataset stems from a study of key factors that contribute to cardiovascular disease, with participants recruited in Framingham, Massachusetts over multiple decades. Recent cohorts were especially designed to reflect a greater racial and ethnic diversity in the town \citep{tsao2015cohort}.

\textbf{Negative examples.} Among the datasets we surveyed, we highlight one whose low inclusivity is rather obvious. WebText is a 40 GB text dataset that supported training of the GPT-2 language model \citep{radford2019language}. The authors crawled every document reachable from outbound Reddit links that collected at least 3 \emph{karma}. While this was considered a useful heuristic to achieve size and quality, it ended up skewing this resource towards content appreciated by Reddit users, who are predominantly male, young, and enjoy good internet access. This should act as reminder that size does not guarantee diversity \citep{bender2021:dangers}, and that sampling biases are almost inevitable.

\textbf{Inclusivity is nuanced.} While inclusivity surely requires an attention to subpopulations, a more precise definition may depend on context and application. Based on the task at hand, an ideal sample may feature all subpopulations with equal presence, or proportionally to their share in the overall population. Let us call these the \emph{equal} and \emph{proportional} approach to diversity. The equal approach is typical  of datasets that are meant to be evaluation benchmarks (Pilot Parliaments Benchmark, Winobias) and allow for statistically significant statements on performance differences across groups. On the other hand, the proportional approach is rather common in datasets collected by census offices, such as US Census Data (1990), and in resources aimed precisely at studying issues of representation in services and products (Occupations in Google Images).

Open-ended collection of data is ideal to ensure that various cultures are represented in the manner in which they would like to be seen \citep{jo2020lessons}. Unfortunately, we found no instance of datasets where sensitive labels were self-reported according to open-ended responses. On the contrary, individuals with non-conforming gender identities were excluded from some datasets and analyses. Bing US Queries is a proprietary dataset used to study differential user satisfaction with the Bing search engine across different demographic groups. It consists of a subset of Bing users who provided their gender at registration according to a binary categorization, which misrepresents or simply excludes non-binary users from the subset. Moreover, a dataset may be inclusive and encode gender in a non-binary gender fashion (Climate Assembly UK), but, if used in conjunction with an auxiliary dataset where gender has binary encoding, a common solution is removing instances whose gender is neither female nor male \citep{flanigan2020neutralizing}.

\textbf{Inclusivity does not guarantee benefits.} To avoid downstream harms, inclusion by itself is insufficient. The context in which people and sensitive groups are represented should always be taken into account. Despite its overall skew towards male subjects, ImageNet has a high female-to-male ratio in classes such as \texttt{bra}, \texttt{bikini} and
\texttt{maillot}, which often feature images that are voyeuristic, pornographic, and non-consensual \citep{prabhu2020large}. Similarly, in MS-COCO, a famous dataset for object recognition, there is roughly a 1:3 female-to-male ratio, increasing to 0.95 for images of kitchens \citep{hendricks2018women}. This sort of representation is unlikely to benefit women in any way and, on the contrary, may contribute to reinforce stereotypes and support harmful biases.

Another clear (but often ignored) disconnect between the inclusion of a group and benefits to it is represented by the task at hand and, more in general, by possible uses afforded by a dataset. In this regard, we find many datasets from the face recognition domain, which are presented as resources geared towards inclusion (Diversity in Faces, BUPT Faces, UTK Face, FairFace, Racial Faces in the Wild). Attention to subpopulations in this context is still called ``diversity'' (Diversity in Faces, FairFace, Racial Faces in the Wild) or ``social awareness'' (BUPT Faces), but is driven by business imperatives and goals of robustness for a technology that can very easily be employed for surveillance purposes, and become detrimental to vulnerable populations included in datasets. In a similar vein, the FACES dataset has been used to measure age bias in emotion detection, a task whose applications and benefits for individuals remain dubious.

Overall, attention to subpopulations is an upside of many datasets we surveyed. However, inclusion, representation, and diversity can be defined in different ways according to the problem at hand. Individuals would rather be included on their own terms, and decide whether and how they should be represented. \novel{The problems of diversity and robustness have some clear commonalities, as the former can be seen as a means towards the latter, but it seems advisable to maintain a clear separation between the two, and to avoid equating either one with fairness. Algorithmic fairness will not be ``solved'' by simply collecting more data, or granting equal performance across different groups identified by a given sensitive attribute.}


\subsection{Sensitive Attribute Labelling}
\label{sec:sensitive_attribute}

\textbf{Motivation.} Datasets are often taken as factual information that supports objective computation and pattern extraction. The etymology of the word ``data'', meaning ``given'', is rather revealing in this sense. On the contrary, research in human-computer interaction, computer-supported cooperative work, and critical data studies argues that this belief is superficial, limited and potentially harmful \citep{muller2019how,crawford2021excavating}.

Data is, quite simply, a human-influenced entity \citep{miceli2021documenting}, determined by a chain of discretionary decisions on measurement, sampling and categorization, which shape how and by whom data will be collected and annotated, according to which taxonomy and based on which guidelines.  Data science professionals, often more cognizant of the context surrounding data than theoretical researchers, report significant awareness of how curation and annotation choices influence their data and its relation with the underlying phenomena \citep{muller2019how}. In an interview, a senior text classification researcher responsible for ground truth annotation shows consciousness of their own influence on datasets by stating ``I am the ground truth.'' \citep{muller2019how}.

Sensitive attributes, such as race and gender, are no exception in this regard. Inconsistencies in racial annotation are rather common within the same system \citep{lum2020impact} and, even more so, across different systems \citep{scheuerman2020how,khan2021one}. External annotation (either human or algorithmic) is essentially based on co-occurrence of specific traits with membership in a group, thus running the risk of encoding and reinforcing stereotypes. Self-reported labels overcome this issue, although they are still based on an imposed taxonomy, unless provided in an open-ended fashion. In this section, we discuss the practices through which sensitive attributes are annotated in datasets used in algorithmic fairness research, which are summarized in Table \ref{tab:sensitive_attributes}. 

\begin{table*}[h]
  \caption{\novel{Approaches to demographic data procurement.}}
  \label{tab:sensitive_attributes}
  \begin{tabular}{p{4.5cm}p{6.5cm}}
    \toprule
    \multicolumn{1}{c}{Approach} & \multicolumn{1}{c}{Example datasets}\\
    \midrule
    Self-reported labels & Bing US Queries, MovieLens, Libimset, Adult, HMDA, Law School, Sushi, Willingness-to-Pay for Vaccine\\
    Expert labels & Pilot Parliaments Benchmark \\
    Non-expert labels & CelebFaces Attributes, Diversity in Faces, FairFace, Occupations in Google Images \\
    ML algorithm & Racial Faces in the Wild, Instagram Photos, BUPT Faces, UTK Face \\
    ML algorithm $+$ annotators & FairFace, Open Images Dataset \\
    Rule- / knowledge-based algorithm & RtGender, Bias in Bios, Demographics on Twitter, TwitterAAE\\
  \bottomrule
\end{tabular}
\end{table*}

\textbf{Procurement of sensitive attributes.} Self-reported labels for sensitive attributes are typical of datasets that represent users of a service, who may report their demographics during registration (Bing US Queries, MovieLens, Libimseti), or that were gathered through surveys (HMDA, Adult, Law School, Sushi, Willingness-to-Pay for Vaccine). These are all resources for which collection of protected attributes was envisioned at design, potentially as an optional step. However, when sensitive attributes are not available, their annotation may be possible through different mechanisms.

A common approach is having sensitive attributes labelled by non-experts, often workers hired on crowdsourcing platforms. CelebFaces Attributes Dataset (CelebA) features images of celebrities from the CelebFaces dataset, augmented with annotations of landmark location and categorical attributes, including gender, skin tone and age, which were annotated by a ``professional labeling company'' \citep{liu2015deep}. In a similar fashion, Diversity in Faces consists of images labeled with gender and age by workers hired through the Figure Eight crowd-sourcing platform, while the creators of FairFace hired workers on Amazon Mechanical Turk to annotate gender, race, and age in a public image dataset. This practice also raises concerns of fair compensation of labour, which are not discussed in this work. 

Some creators employ algorithms to obtain sensitive labels. Face datasets curators often resort to the Face++ API (Racial Faces in the Wild, Instagram Photos, BUPT Faces) or other algorithms (UTK Face, FairFace). In essence labeling is classifying, hence measuring and reporting accuracy for this procedure would be in order, but rarely happens. Creators occasionally note that automated labels were validated (FairFace) or substantially enhanced (Open Images Dataset) by human annotators, and very seldom report inter-annotator agreement (Occupations in Google Images).

Other examples of external labels include the geographic origin of candidates in resumes (CVs from Singapore), political leaning of US Twitter profiles (Twitter Political Searches),  English dialect of tweets (TwitterAAE), and gender of subjects featured in image search results for professions (Occupations in Google Images). Annotation may also rely on external knowledge bases such as Wikipedia,\footnote{\url{https://en.wikipedia.org/wiki/Category:American_female_tennis_players}} as is the case with RtGender. In situations where text written by individuals is available, rule-based approaches exploiting gendered nouns (``woman'') or pronouns (``she'') are also applicable (Bias in Bios, Demographics on Twitter).

Some datasets may simply have no sensitive attribute. These are often used in works of individual fairness, but may occasionally support studies of group fairness. For example, dSprites is a synthetic computer vision dataset where regular covariates may play the role of sensitive variables \citep{locatello2019fairness}. Alternatively, datasets can be augmented with simulated demographics, as done by \citet{madnadi2017:building} who randomly assigned a native language to test-takers in ASAP, or through the technique of \citet{burke2018synthetic}, which they demonstrate on MovieLens. 

\textbf{Face datasets.} Posterior annotation is especially common in computer vision datasets. The Pilot Parliaments Benchmark, for instance, was devised as a testbed for face analysis algorithms. It consists of images of parliamentary representatives from three African and three European countries, that were labelled by a  surgical dermatologist with the Fitzpatrick skin type of the subjects \citep{fitzpatrick1988validity}. This is a dermatological scale for skin color, which can be retrieved from people's appearance. On the contrary, annotations of race or ethnicity from a photo are simplistic at best, and it should be clear that they actually capture \emph{perceived race} from the perspective of the annotator (FairFace, BUPT Faces). Careful nomenclature is an important first step to improve the transparency of a dataset and make the underlying context more visible.\footnote{In this article, we typically discuss sensitive attributes following the naming convention in the accompanying documentation of a dataset, avoiding a critical terminology discussion .}

Similarly to \citet{scheuerman2020how}, we find that documentation accompanying face recognition datasets hardly ever describes how specific taxonomies for gender and race were chosen, conveying a false impression of objectivity. A description of the annotation process is typically present, but minimal. For Multi-task Facial Landmark, for instance, we only know that ``The ground truths of the related tasks are labeled manually'' \citep{zhang2014facial}.


\textbf{Annotation trade-offs.} It is worth re-emphasizing that sensitive label assignment is a classification task that rests on assumptions. Annotation of race and gender in images, for example, is based on the idea that they can be accurately ascertained from pictures, which is an oversimplification of these constructs. The envisioned classes (e.g. binary gender) are another subjective choice stemming from the point of view of dataset curators and may reflect narrow or outdated conceptions and potentially harm the data subjects. In this regard a quote from the curators of MS-Celeb-1M, who do not annotate race, but consider it for their sampling strategy, is particularly striking: ``We cover all the major races in the world (Caucasian, Mongoloid, and Negroid)'' \citep{guo2016msceleb1m}. For these reasons, external annotation of sensitive attributes is controversial and inevitably influenced by dataset curators.

On the other hand, external annotation may be the only way to test specific biases. Occupations in Google Images, for instance, is an image dataset collected to study gender and skin tone diversity in image search results for various professions. 
The creators hired workers on  Amazon Mechanical Turk to label the gender (male, female) and Fitzpatrick skin tone (Type 1--6) of the primary person in each image. The Pilot Parliaments Benchmark was also annotated externally to obtain a benchmark for the evaluation of face analysis technology, with a balanced representation of gender and skin type. Different purposes can motivate data collection and annotation of sensitive attributes. Purposes and aims should be documented clearly, while also reflecting on other uses and potential for misuse of a dataset \citep{gebru2018datasheets}. Dataset curators may use documentation to discuss these aspects and specify limitations for the intended use of a resource \citep{peng2021mitigating}. In the next section we focus on documentation and why it represents a key component of data curation.


\subsection{Transparency}
\label{sec:transparency}

\textbf{Motivation.} Transparent and accurate documentation is a fundamental part of data quality. Its absence may lead to serious issues, including lack of reproducibility, concerns of scientific validity, ethical problems, and harms \citep{barocas2019fair}. Clear documentation can shine a light on inevitable choices made by dataset creators and on the context surrounding the data. In the absence of this information, the curation mechanism mediating reality and data is hidden; the data becomes one with its context, to the point that interpretation of numerical results can be misleading and overarching \citep{bao2021COMPASlicated}. 

\novel{
The ``ground truth'' labels (typically indicated with the letter $y$), which are the target of prediction tasks in some datasets, such as indications of recidivism in COMPAS, are especially sensitive in this regard. Indeed, not only accuracy and related quality metrics, but also measures of algorithmic fairness such as sufficiency and separation \citep{barocas2019fair} are based on $y$ labels and the ability of ML algorithms to replicate them, implicitly granting them a special status of truthfulness. In reality, however, these labels may be biased and incorrect due to multiple causes, including, very frequently, a disconnect between what we aim to measure in an ideal construct space (e.g., offense in the case of COMPAS) and what we can actually measure in the observed space (e.g., arrest) \citep{friedler2021impossibility}. Fair ML algorithms (measures) can only partly overcome (catch) these biases, and actually run the risk of further reifying them. Proper documentation does not solve this issue, but equips practitioners and researchers with the necessary awareness to handle these biases.
}

More broadly, good documentation should discuss and explain features, providing context about who collected and annotated the data, how, and for which purpose \citep{gebru2018datasheets,denton2020bringing}. This provides dataset users with information they can leverage to select appropriate datasets for their tasks and avoid unintentional misuse \citep{gebru2018datasheets}. Other actors, such as reviewers, may also access the official documentation of a dataset to ensure that it is employed in compliance with its stated purpose, guidelines, and terms of use \citep{peng2021mitigating}. Overall data documentation plays a fundamental role in increasing transparency and accountability \citep{hutchinson2021towards}, favouring responsible and reflexive data curation \citep{miceli2021documenting,jo2020lessons}, and a correct utilization of these resources \citep{paullada2020data}.

\textbf{Positive examples.} In this survey, we find examples of excellent documentation in datasets related to studies and experiments, including CheXpert, Framingham, and NLSY. Indeed, datasets curated by medical institutions and census offices are often well-documented. The ideal source of good documentation are descriptor articles published in conjunction with a dataset (e.g. MIMIC-III), typically offering stronger guarantees than web pages in terms of quality and permanence. Official websites hosting and distributing datasets are also important to collect updates, errata, and additional information that may not be available at the time of release. The Million Song Dataset and Goodreads Reviews, for instance, are available on websites which contain a useful overview of the respective dataset, a list of updates, code samples, pointers to documentation, and contacts for further questions. 

\textbf{Negative examples.} On the other hand, some datasets are opaque and poorly documented. Among publicly available ones, Arrhythmia is distributed with a description of the features but no context about the purposes, actors, and subjects involved in the data collection. Similarly, the whole curation process and composition of Multi-task Facial Landmark is described in a short paragraph, explaining it consists of 10,000 outdoor face images from the web that were labelled manually with gender. Most face datasets suffer from opaque documentation, especially concerning the choice of sensitive labels and their annotation. For semi-synthetic resources, proper documentation is especially important, to let users understand the broader applicability and implications of numerical analyses performed on a dataset. IBM HR Analytics is a resource about employee attrition, which the hosting website describes as containing ``fictional data'', without any additional information. Nonetheless, this data was plausibly generated in a principled fashion and (even partial) disclosure of the underlying data generation mechanism would benefit dataset users. 

\textbf{Retrospective documentation.} Good documentation may also be produced retrospectively \citep{bandy2021addressing,garbin2021structured}. German Credit is an interesting example of a dataset that was poorly documented for decades, until the recent publication of a report correcting severe coding mistakes \citep{gromping2019:sg}. As mentioned in Section \ref{sec:popular_german}, from the old documentation it seemed possible to retrieve the sex of data subjects from a feature jointly encoding sex and marital status. The \emph{dataset archaeology} work by \citet{gromping2019:sg} shows that this is not the case, which has particular relevance for many algorithmic fairness works using this dataset with sex as a protected feature, as this feature is simply not available. Numerical results obtained in this setting may be an artefact of the wrong coding with which the dataset has been, and still is, officially distributed in the \citet{hofmann1994:sg}. Until the report and the new redacted dataset \citep{gromping2019:sg2} become well-known, the old version will remain prevalent and more mistakes will be made. In other words, while the documentation debt for this particular dataset has been retrospectively addressed (\emph{opacity}), many algorithmic fairness works published after the report continue to use the German Credit dataset with sex as a protected attribute \citep{he2020geometric,yang2020fairness,baharlouei2020renyi,lohaus2020too,martinez2020minimax,wang2021fair}. This is an issue of documentation \emph{sparsity}, where the right information exists but does not reach interested parties, including researchers and reviewers.

Documentation is a fundamental part of data curation, with most responsibility resting on creators. However, dataset users can also play a role in mitigating the documentation debt by proactively looking for information about the resources they plan to use. Brief summaries discussing and motivating the chosen datasets can be included in scholarly articles, at least in supplementary materials when conflicting with page limitations. Indeed, documentation debt is a problem for the whole research community, which can be addressed collectively with retrospective contributions and clarifications. We argue that it is also up to individual researchers to seek contextual information for situating the data they want to use.





\novel{
\section{Broader Relevance to the Community}
\label{sec:relevance}

Along with the analyses presented in this work, through the lens of tasks supported, domains spanned, and roles played by algorithmic fairness datasets, we are releasing the underlying data briefs, as a further contribution for the research community.  Data briefs are a short documentation format providing essential information on datasets used in fairness research. Data briefs are composed of ten fields, detailed in appendix \ref{sec:databriefs}, derived from shared vocabularies such as Data Catalog Vocabulary (DCAT); to be compliant with the FAIR data principles \citep{wilkinson2016fair}, we also defined a schema called \texttt{fdo} to model the relationships between the terms and to make the links to external vocabularies explicit. We leverage \texttt{fdo} to format the data briefs as a Resource Description Framework (RDF) \citep{miller1998introduction}, and to make them available as linked open data, thus supporting data reuse, interoperability, and interpretability.\footnote{Schema publicly available at \url{https://fairnessdatasets.dei.unipd.it/schema/}; RDF publicly available at \url{https://zenodo.org/record/6518370\#.YnOSKFTMJhF}.} Our final aim is to release, update, and maintain a  web app, which can be queried by researchers and practitioners to find the most relevant datasets, according to their specific needs.\footnote{This resource will be released at \url{https://fairnessdatasets.dei.unipd.it/} \label{fn:web_app}} We envision several benefits for the algorithmic fairness and data studies communities, such as:

\begin{itemize}
    \item Informing the choice of datasets for experimental evaluations of fair ML methods, including domain-oriented and task-oriented search.
    \item Directing studies of data bias, and other quantitative and qualitative analyses, including retrospective documentation efforts, towards popular (or otherwise important) resources.
    \item Identifying areas and sub-problems that are understudied in the algorithmic fairness literature.
    \item Supporting multi-dataset studies, focused on resources united by a common characteristic, such as encoding a given sensitive attribute \citep{scheuerman2020how},  concerning computer vision \citep{fabbrizzi2021survey}, or being popular in the fairness literature \citep{lequysurvey2022}.
\end{itemize}
}








\section{Conclusions and Recommendations}
\label{sec:conclusion}
Algorithmic fairness is a young research area, undergoing a fast expansion, with diverse contributions in terms of methodology and applications. Progress in the field hinges on different resources, including, very prominently, datasets. In this work, we have surveyed hundreds of datasets used in the fair ML and algorithmic equity literature to help the research community reduce its documentation debt, improve the utilization of existing datasets, and the curation of novel ones. 

With respect to existing resources, we have shown that the most popular datasets in the fairness literature (Adult, COMPAS, and German Credit) have limited merits beyond originating from human processes and encoding protected attributes. On the other hand, several negative aspects call into question their current status of general-purpose fairness benchmarks, including contrived prediction tasks, noisy data, severe coding mistakes, limitations in encoding sensitive attributes, and age. 

We have documented over two hundred datasets to provide viable alternatives, annotating their domain, the tasks they support, and discussing the roles they play in works of algorithmic fairness. We have shown that the processes generating the data belong to many different domains, including, for instance, criminal justice, education, search engines, online marketplaces, emergency response, social media, medicine, hiring, and finance. At the same time, we have described a variety of tasks studied on these resources, ranging from generic, such as \emph{fair  classification}, to  narrow such as \emph{fair districting} and \emph{fair truth discovery}. Overall, such diversity of domains and tasks provides a glimpse into the variety of human activities and applications that can be impacted by automated decision making, and that can benefit from algorithmic fairness research. Tasks and domain annotations are made available in our data briefs to facilitate the work of researchers and practitioners interested in the study of algorithmic fairness applied to specific domains or tasks. By assembling sparse information on hundreds of datasets into a single document, we aim to provide a useful reference to support both domain-oriented and task-oriented dataset search. 

At the same time, we have analyzed issues connected to re-identification, consent, inclusivity, labeling, and transparency running across these datasets. By describing a range of approaches and attentiveness to these topics, we aim to make them more visible and concrete. On one hand, this may prove valuable to inform post-hoc data interventions aimed at mitigating potential harms caused by existing datasets. On the other hand, as novel datasets are increasingly curated, published, and adopted in fairness research, it is important to motivate these concerns, make them  tangible, and distill existing approaches into best practices, which we summarize below, for future endeavours of data curation. Our recommendations complement (and do not replace) a growing body of work studying key aspects in the life cycle of datasets \citep{gebru2018datasheets,jo2020lessons,prabhu2020large,crawford2021excavating,peng2021mitigating,hutchinson2021towards}.

Social relevance of data, intended as the breadth and depth of societally useful insights afforded by datasets, is a central requirement in fairness research. Unfortunately, this may conflict with user privacy, favouring re-identification or leaving consideration of consent in the background. Consent should be considered during the initial design of a dataset, in accordance with existing frameworks, such as the FRIES framework outlined in the Consentful Tech project. Moreover, different strategies are available to alleviate concerns of re-identification, including noise injection, conservative release, and (semi)synthetic data generation.  Algorithmic fairness is motivated by aims of justice and harm avoidance for people, which should be extended to data subjects. 

Inclusivity is also important for social relevance, as it allows for a wider representation, and supports analyses that take into account important groups. However, inclusivity is insufficient in itself. 
Possible uses afforded by a dataset should always be considered, evaluating costs and benefits for the data subjects and the wider population. In the absence of these considerations, acritical inclusivity runs the risk of simply supporting system robustness across sensitive attributes, such as race and gender, rebranded as fairness.

Sensitive attributes are a key ingredient to measure inclusion and increase the social relevance of a dataset. Although often impractical, it is typically preferable for sensitive attributes to be self-reported by data subjects. Externally assigned labels and taxonomies can harm individuals by erasing their needs and points of view. 
Sensitive attribute labelling is thus a shortcut whose advantages and disadvantages should be carefully weighted and, if chosen, it should be properly documented. Possible approaches based on human labour include expert and non-expert annotation, while automated approaches range from simple rule-based systems to complex and opaque algorithms. To label is to classify, hence measuring and reporting per-group accuracy is in order. Some labeling endeavours are more sensible than others: while skin tone can arguably be retrieved from pictures, annotations of race from an image actually capture \emph{perceived race} from the perspective of the annotator. Rigorous nomenclature favours better understanding and clarifies the subjectivity of certain labels.

Reliable documentation shines a light on inevitable choices made by dataset creators and on the context surrounding the data. This provides dataset users with information they can leverage to select appropriate datasets for their tasks and avoid unintentional misuse.  Datasets for which some curation choices are poorly documented may appear more objective at first sight. However, it should be clear that objective data and turbid data are very different things. Proper documentation increases transparency, trust, and understanding. At a minimum, it should include the purpose of a data artifact, a description of the sample, the features and related annotation procedures, along with an explicit discussion of the associated task, if any. It should also clarify who was involved in the different stages of the data development procedure, with special attention to annotation. Data documentation also supports reviewers and readers of academic research in assessing whether a dataset was selected with good reason and utilized in compliance with creators' guidelines.

Understanding and budgeting for all these aspects during early design phases, rather than after collection or release, can be invaluable for data subjects, data users, and society. While possible remedies exist, data is an extremely fluid asset allowing for easy reproduction and derivatives of all sorts; remedies applied to a dataset do not necessarily benefit its derivates. In this work, we have targeted the collective documentation debt of the algorithmic fairness community, resulting from the opacity surrounding certain resources and the sparsity of existing documentation. We have mainly targeted sparsity in a centralized documentation effort; as a result, we have found and described a range of weaknesses and best practices that can be adopted to reduce opacity and mitigate concerns of privacy and inclusion. Similarly to other types of data interventions, useful documentation can be produced after release, but, as shown in this work, the documentation debt may propagate nonetheless. In a mature research community, curators, users, and reviewers can all contribute to cultivating a data documentation culture and keep the overall documentation debt in check.

\begin{acknowledgements}
\ack{
The authors would like to thank the following researchers and dataset creators for the useful feedback on the data briefs: Alain Barrat, Luc Behaghel, Asia Biega, Marko Bohanec, Chris Burgess, Robin Burke, Alejandro Noriega Campero, Margarida Carvalho, Abhijnan Chakraborty, Robert Cheetham, Won Ik Cho, Paulo Cortez, Thomas Davidson, Maria De-Arteaga, Lucas Dixon, Danijela Djordjević, Michele Donini, Marco Duarte, Natalie Ebner, Elaine Fehrman, H. Altay Guvenir, Moritz Hardt, Irina Higgins, Yu Hen Hu, Rachel Huddart, Lalana Kagal, Dean Karlan, Vijay Keswani, Been Kim, Hyunjik Kim, Jiwon Kim, Svetlana Kiritchenko, Pang Wei Koh, Joseph A. Konstan, Varun Kumar, Jeremy Andrew Irvin, Jamie N. Larson, Jure Leskovec, Jonathan Levy, Andrea Lodi, Oisin Mac Aodha, Loic Matthey, Julian McAuley, Brendan McMahan, Sergio Moro, Luca Oneto, Orestis Papakyriakopoulos, Stephen Robert Pfohl, Christopher G. Potts, Mike Redmond, Kit Rodolfa, Ben Roshan, Veronica Rotemberg, Rachel Rudinger, Sivan Sabato, Kate Saenko, Mark D. Shermis, Daniel Slunge, David Solans, Luca Soldaini, Efstathios Stamatatos, Ryan Steed, Rachael Tatman, Schrasing Tong, Alan Tsang, Sathishkumar V E, Andreas van Cranenburgh, Lucy Vasserman, Roland Vollgraf, Alex Wang, Zeerak Waseem, Kellie Webster, Bryan Wilder, Nick Wilson, I-Cheng Yeh, Elad Yom-Tov, Neil Yorke-Smith, Michal Zabovsky, Yukun Zhu.
}
\end{acknowledgements}

\bibliographystyle{spbasic}      
\bibliography{biblio}   

\clearpage
\appendix
\begin{appendices}
\section{Data briefs}
\label{sec:databriefs}


Data briefs were drafted by the first author and reviewed by the remaining authors. 
For over 95\% of the surveyed datasets, we identified at least one contact involved in the data curation process or familiar with the dataset, who received a preliminary version of the respective data brief and a request for corrections and additions. Data briefs are meant as short documentation providing essential information on datasets used in fairness research. \novel{Data briefs are composed of ten fields derived from shared vocabularies such as Data Catalog Vocabulary (DCAT)\footnote{\url{http://www.w3.org/ns/dcat}, with namespace \texttt{dct}}; to be compliant with the FAIR data principles \citep{wilkinson2016fair}, we also defined a schema (with namespace \texttt{fdo}) to model the relationships between the terms, to make the links to external vocabularies explicit, and map the data briefs to a machine-readable RDF graph.\footnote{Schema publicly available at \url{https://fairnessdatasets.dei.unipd.it/schema/}; RDF graph publicly available at \url{https://zenodo.org/record/6518370\#.YnOSKFTMJhF}. To favour consultation and dynamical querying of the data briefs, we are working to release a web app at \url{https://fairnessdatasets.dei.unipd.it}.} The \texttt{fdo} schema has been defined by reusing, as much as possible, existing terminology from established vocabularies. In the following we detail the fields of the data briefs and present their correspondence to DCAT and \texttt{fdo} properties:}

\begin{description}
    \item[Description.] This is a free-text field reporting (1) the aim/purpose of a data artifact (i.e., why it was developed/collected), as stated by curators or inferred from context; (2) a high-level description of the available features; (3) the labeling procedure for annotated attributes, with special attention to sensitive ones, if any; (4) the envisioned ML task, if any. \novel{Corresponds to \texttt{dct:description} in DCAT.}
    \item[Affiliation of creators.] Typically derived from reports, articles, or official web pages presenting a dataset. Datasets can be derivatives of other datasets (e.g., Adult). We typically refer to the final resource while providing the prior context where appropriate. \novel{In DCAT vocabulary, it is the affiliation of a \texttt{dct:publisher} (for published resources) or a \texttt{dct:creator}.}
    \item[Domain.] The main field where the data is used (e.g., computer vision for ImageNet) or the field studying the processes and phenomena that produced the dataset (e.g., radiology for CheXpert). Corresponds to \texttt{fdo:Domain} in the \texttt{fdo} schema.
    \item[Tasks in fairness literature.] An indication of the task performed on the dataset in each surveyed article that uses the current resource. \novel{Corresponds to \texttt{fdo:Task}.}
    \item[Data spec.] The main format of the data. The envisioned categories are text, image, time-series, tabular data, and pairs. The latter denotes a special type of tabular data where rows and columns correspond to entities and cells to a relation between them, such as relevance for query-document pairs, ratings for user-item pairs, co-authorship relation for author-author pairs. A ``mixture'' category was added for resources with multimodal data. \novel{Corresponds to \texttt{dct:type} in DCAT.}
    \item[Sample size.] Dataset cardinality. Corresponds to \texttt{fdo:sampleSize} in \texttt{fdo}.
    \item [Year.] Last known update to the dataset. For resources whose collection and curation are ongoing (e.g., Framingham) we write ``present''. \novel{Corresponds to \texttt{dct:modified}.}
    \item[Sensitive features.] Sensitive attributes in the dataset. These are typically explicitly annotated, but may include implicit ones, such as textual references to people and their demographics in text datasets. References to gender, for instance, can easily be retrieved from English-language text corpora based on intrinsically gendered words, such as she, man, aunt. \novel{Corresponds to \texttt{fdo:sensitiveFeature}}.
    \item[Link.] A link to the website where the resource can be downloaded or requested. \novel{Corresponds to \texttt{dcat:landingPage}}.
    \item[Further information.] Reference to works and web pages describing the dataset. 
\end{description}

Following the algorithmic fairness literature, we define sensitive features as encoding membership to groups that are salient for society and have some special protection based on the law, including race, ethnicity, sex, gender, and age. We may occasionally stretch this definition and report features considered sensitive in some works, such as political leaning or education, so long as they reflect essential divisions in society. We also report domain-specific attributes considered sensitive in a given context, such as language for Section 203 determinations or brand ownership for Amazon Recommendations. We follow the language of the available documentation for the names and values of sensitive features, including distinctions between race and ethnicity. For datasets that report geographical information at any granularity (GPS coordinates, neighbourhoods, countries) we report ``geography'' among the sensitive attributes. If an article considers features to be sensitive in an arbitrary fashion (e.g., sepal width in the Iris dataset), we do not report it in the respective field.

For the dataset domain, we follow the area-category taxonomy defined by Scimago,\footnote{\url{https://www.scimagojr.com/journalrank.php}} with the addition of ``news'', ``social media'', ``social networks'', ``sports'' and ``food''. Table \ref{tab:domainwise} contains a summary of the surveyed datasets through this domain-based taxonomy.  Tasks in the fairness literature were labeled via open coding. The final taxonomy is detailed in Section \ref{sec:tasks}. We distinguish between works that are more focused on evaluation rather than a proposal of novel solutions by writing, e.g. ``fair ranking evaluation'' instead of ``fair ranking''. We use ``evaluation'' as a broad term for works focusing on analyses of algorithms, products, platforms, or datasets and their properties from multiple fairness and accuracy perspectives. With some abuse of nomenclature, we also use this label for works that focus on properties of fairness metrics \citep{pleiss2017fairness}. Unless otherwise specified, ``fairness evaluation'' is about fair classification, which is the most common task. Exploratory approaches focused on discovering biases that are not fully specified ex-ante are indicated with the label ``bias discovery''. 

\ifconf\else{}\fi 

\subsection{2010 Frequently Occurring Surnames}
\begin{itemize}
    \item \textbf{Description}: this dataset reports all surnames occurring 100 or more times in the 2010 US Census, broken down by race (White, Black, Asian and Pacific Islander (API), American Indian and Alaskan Native only (AIAN), multiracial, or Hispanic).
    \item \textbf{Affiliation of creators}: US Census Bureau.
    \item \textbf{Domain}: linguistics.
    \item \textbf{Tasks in fairness literature}: fair subset selection under unawareness \citep{mehrotra2021mitigating}.
    \item \textbf{Data spec}: tabular data.
    \item \textbf{Sample size}: $\sim200$K surnames.
    \item \textbf{Year}: 2016.
    \item \textbf{Sensitive features}: race.
    \item \textbf{Link}: \url{https://www.census.gov/topics/population/genealogy/data/2010_surnames.html}
    \item \textbf{Further info}: \url{https://www2.census.gov/topics/genealogy/2010surnames/surnames.pdf}
\end{itemize}

\subsection{2016 US Presidential Poll}
\begin{itemize}
    \item \textbf{Description}: this dataset was collected and maintained by FiveThirtyEight, a website specialized in opinion poll analysis. This resource was developed with the goal of providing an aggregated estimate based on multiple polls, weighting each input according to sample size, recency, and historical accuracy of the polling organization. For each poll, the dataset provides the period of data collection, its sample size, the pollster conducting it, their rating, and a url linking to the source data.
    \item \textbf{Affiliation of creators}: FiveThirtyEight.
    \item \textbf{Domain}: political science.
    \item \textbf{Tasks in fairness literature}: limited-label fairness evaluation \citep{sabato2020bounding}.
    \item \textbf{Data spec}: tabular data.
    \item \textbf{Sample size}: $\sim13$K poll results.
    \item \textbf{Year}: 2016.
    \item \textbf{Sensitive features}: geography.
    \item \textbf{Link}: \url{http://projects.fivethirtyeight.com/general-model/president_general_polls_2016.csv}
    \item \textbf{Further info}: \url{https://projects.fivethirtyeight.com/2016-election-forecast/}
\end{itemize}

\subsection{4area}
\begin{itemize}
    \item \textbf{Description}: this dataset was extracted from DBLP to study the problem of topic modeling on documents connected by links in a graph structure. The creators extracted from DBLP articles published at 20 major conferences from four related areas, i.e., database, data mining, machine learning, and information retrieval. Each author is associated with four continuous variables based on the fraction of research papers published in these areas. The associated task is the prediction of these attributes.
    \item \textbf{Affiliation of creators}: University of Illinois at Urbana-Champaign.
    \item \textbf{Domain}: library and information sciences.
    \item \textbf{Tasks in fairness literature}: fair clustering \citep{harb2020kfc}.
    \item \textbf{Data spec}: author-author pairs.
    \item \textbf{Sample size}: $\sim30$K nodes (authors) connected by $\sim200$K edges (co-author relations).
    \item \textbf{Year}: 2009.
    \item \textbf{Sensitive features}: author.
    \item \textbf{Link}: not available
    \item \textbf{Further info}: \citet{sun2009itopic}
\end{itemize}

\subsection{Academic Collaboration Networks}
\begin{itemize}
    \item \textbf{Description}: these dataset represent two collaboration networks from the preprint server arXiv, covering scientific papers submitted to the astrophysics (AstroPh) and condensed matter (CondMat) physics categories. Each node in the network is an author, with links indicating co-authorship of one or more articles. Nodes are indicated with ids, hence information about the researchers in the graph is not immediately available. These datasets were developed to study the evolution of graphs over time.
    \item \textbf{Affiliation of creators}: Carnegie Mellon University; Cornell University.
    \item \textbf{Domain}: library and information sciences.
    \item \textbf{Tasks in fairness literature}: fair graph mining \citep{kang2020inform}.
    \item \textbf{Data spec}: author-author pairs.
    \item \textbf{Sample size}: $\sim$19K nodes (authors) connected by $\sim200$K edges (indications of co-authorship) (AstroPh). $\sim$23K nodes connected by $\sim93$K edges (CondMat).
    \item \textbf{Year}: 2009.
    \item \textbf{Sensitive features}: none.
    \item \textbf{Link}: \url{http://snap.stanford.edu/data/ca-AstroPh.html} (AstroPh) and \url{http://snap.stanford.edu/data/ca-CondMat.html} (CondMat)
    \item \textbf{Further info}: \citet{leskovec2007graph}
\end{itemize}

\subsection{Adience}
\begin{itemize}
    \item \textbf{Description}: this resource was developed to favour the study of automated age and gender identification from images of faces. Photos were sourced from Flickr albums, among the ones automatically uploaded from iPhone and made available under Creative Commons license. All images were manually labeled for age, gender and identity ``using both the images themselves and any available contextual information''. These annotations are fundamental for the tasks associated with this dataset, i.e. age and gender estimation. One author of \citet{Buolamwini2018gender} labeled each image in Adience with Fitzpatrick skin type.
    \item \textbf{Affiliation of creators}: Adience; Open University of Israel.
    \item \textbf{Domain}: computer vision.
    \item \textbf{Tasks in fairness literature}: data bias evaluation \citep{Buolamwini2018gender},
    robust fairness evaluation \citep{nanda2021fairness}.
    \item \textbf{Data spec}: image.
    \item \textbf{Sample size}: $\sim30$K images of $\sim2$K subjects.
    \item \textbf{Year}: 2014.
    \item \textbf{Sensitive features}: age, gender, skin type.
    \item \textbf{Link}: \url{https://talhassner.github.io/home/projects/Adience/Adience-data.html}
    \item \textbf{Further info}: \citet{eidinger2014:age,Buolamwini2018gender}
\end{itemize}

\subsection{Adressa}
\begin{itemize}
    \item \textbf{Description}: this dataset was curated as part of the RecTech project on recommendation technology owned by  Adresseavisen (shortened to Adressa) a large Norwegian newspaper. It summarizes one week of traffic to the newspaper website by both subscribers and non-subscribers, during February 2017. The dataset describes reading events, i.e. a reader accessing an article, providing access timestamps and user information inferred from their IP. Specific information about the articles is also available, including author, keywords, body, and mentioned entities. The dataset curators also worked on an extended version of the dataset (Adressa 20M), ten times larger than the one described here.
    \item \textbf{Affiliation of creators}: Norwegian University of Science and Technology; Adresseavisen.
    \item \textbf{Domain}: news, information systems.
    \item \textbf{Tasks in fairness literature}: fair ranking \citep{chakraborty2019equality}.
    \item \textbf{Data spec}: user-article pairs.
    \item \textbf{Sample size}: $\sim3$M ratings by $\sim15$M readers over $\sim1K$ articles.
    \item \textbf{Year}: 2018.
    \item \textbf{Sensitive features}: geography.
    \item \textbf{Link}: \url{http://reclab.idi.ntnu.no/dataset/}
    \item \textbf{Further info}: \citep{gulla2017adressa}
\end{itemize}

\subsection{Adult}
\label{sec:adult_db}
\begin{itemize}
    \item \textbf{Description}: this dataset was created as a resource to benchmark the performance of machine learning algorithms on socially relevant data. Each instance is a person who responded to the March 1994 US Current Population Survey, represented along demographic and socio-economic dimensions, with features describing their profession, education, age, sex, race, personal and financial condition. The dataset was extracted from the census database, preprocessed, and donated to UCI Machine Learning Repository in 1996 by Ronny Kohavi and Barry Becker. A binary variable encoding whether respondents' income is above \$50,000 was chosen as the target of the prediction task associated with this resource. See Appendix \ref{sec:adult} for extensive documentation.
    \item \textbf{Affiliation of creators}: Silicon Graphics Inc.
    \item \textbf{Domain}: economics.
    \item \textbf{Tasks in fairness literature}: fairness evaluation \citep{sharma2020:cc,cardoso2019framework,oneto2019taking,friedler2019comparative,chen2018why,lipton2018does,pleiss2017fairness,diciccio2020evaluating,speicher2018unified,feldman2015certifying,maity2021statistical,kim2020fact,liu2019implicit,williamson2019fairness,vonkugelgen2021fairness,ngong2020towards,jabbari2020empirical,huan2020fairness,zliobaite2015relation,islam2021can,segal2021fairness}, 
    fair classification \citep{he2020geometric,sharma2020data,goel2018nondiscriminatory,raff2018fair,zhang2018mitigating,hu2020fair,celis2019classification,yang2020fairness,cho2020fair,savani2020intraprocessing,wu2019cpfairness,donini2018empirical,quadrianto2017recycling,calmon2017optimized,xu2020algorithmic,zhang2017achieving,yurochkin2021sensei,vargo2021individually,chuang2021fair,roh2021fairbatch,yurochkin2020training,baharlouei2020renyi,lohaus2020too,martinez2020minimax,roh2020frtrain,celis2020data,cotter2019training,gordaliza2019obtaining,wang2019repairing,agarwal2018reductions,creager2021exchanging,delobelle2020ethical,ogura2020convex,feldman2015certifying,zafar2017fairness,fish2015fair,raff2018gradient,mhasawade2021causal,perrone2021fair,shah2021rawslian,sharma2021fairn}, 
    fair clustering \citep{abbasi2021fair,ghadiri2021socially,harb2020kfc,ahmadian2020fair,huang2019coresets,bera2019fair,chierichetti2017fair,brubach2020pairwise,mahabadi2020individual,backurs2019scalable,mary2019fairnessaware,wang2020augmented,berk2017convex,beutel2017data}, 
    fair clustering under unawareness \citep{esmaeili2020probabilistic},
    fair active classification \citep{noriegacampero2019active,bakker2019dadi,bakker2021beyond}, 
    fair preference-based classification \citep{ali2019loss,ustun2019fairness,mukherjee2020two}, 
    fair classification under unawareness \citep{lahoti2020fairness,wang2020robust,mozannar2020fair,kilbertus2018blind}, 
    fair anomaly detection \citep{zhang2021towards,shekhar2021fairod},
    fairness evaluation under unawareness \citep{awasthi2021evaluating},
    robust fairness evaluation \citep{black2021leaveoneout},
    data bias evaluation \citep{beretta2021detecting},
    rich-subgroup fairness evaluation \citep{kearns2019empirical,chouldechova2017fairer},
    fair representation learning \citep{ruoss2020learning,zhao2019inherent,zhao2020conditional,louizos2016variational,quadrianto2019discovering,madras2018learning},
    fair multi-stage classification \citep{hu2020fairmultiple,goel2020importance},
    robust fair classification \citep{mandal2020ensuring,huang2019stable,rezaei2021robust},
    dynamical fair classification \citep{zhang2019group},
    fair ranking evaluation \citep{kallus2019fairness},
    fair data summarization \citep{chiplunkar2020how,jones2020fair,kleindessner2019fair,celis2018fair,halabi2020fairness,belitz2021automating},
    fair regression \citep{agarwal2019fair},
    limited-label fair classification \citep{chzhen2019leveraging,wang2021fair,choi2020group},
    limited-label fairness evaluation \citep{ji2020can}, preference-based fair clustering \citep{galhotra2021learning}.
    \item \textbf{Data spec}: tabular data.
    \item \textbf{Sample size}: $\sim50$K instances.
    \item \textbf{Year}: 1996.
    \item \textbf{Sensitive features}: age, sex, race.
    \item \textbf{Link}: \url{https://archive.ics.uci.edu/ml/datasets/adult}
    \item \textbf{Further info}: \citet{kohavi1996scaling,kohavi1994adult_uci,usdeptcomm1995current,hardt2021facing,mckenna2019:history,mckenna2019:history_drb}
\end{itemize}

\subsection{Allegheny Child Welfare}
\begin{itemize}
    \item \textbf{Description}: this dataset stems from an initiative by the Allegheny County’s Department of Human Services to develop assistive tools to support child maltreatment hotline screening decisions. Referrals received by Allegheny County via a hotline between September 2008 and April 2016 were assembled into a dataset. To obtain a relevant history and follow-up time for each referral, a subset of samples spanning the period from April 2010 to April 2014 is considered. Each data point pertains to a referral for suspected child abuse or neglect and contains a wealth of information from the integrated data management systems of Allegheny County. This data includes cross-sector administrative information for individuals associated with a report of child abuse or neglect, including data from child protective services, mental health services, drug, and alcohol services. The target to be estimated by risk models is future child harm, as measured e.g. by re-referrals, which complements the role of the screening staff who are focused on the information currently available about the referral.
    \item \textbf{Affiliation of creators}: Allegheny County Department of Human Services; Auckland University of Technology; University of Southern California; University of Auckland; University of California.
    \item \textbf{Domain}: social work.
    \item \textbf{Tasks in fairness literature}: fairness evaluation of risk assessment \citep{coston2020counterfactual},
    fair risk assessment \citep{mishler2021fairness}.
    \item \textbf{Data spec}: tabular data.
    \item \textbf{Sample size}: $\sim80$K calls.
    \item \textbf{Year}: 2019.
    \item \textbf{Sensitive features}: age, race, gender of child.
    \item \textbf{Link}: not available
    \item \textbf{Further info}: \citet{vaithianathan2017developing}
\end{itemize}

\subsection{Amazon Recommendations}
\begin{itemize}
    \item \textbf{Description}: this dataset was crawled to study anti-competitive behaviour on Amazon, and the extent to which Amazon’s private label products are recommended on the platform. Considering the categories \emph{backpack} and \emph{battery}, where Amazon is known to have a strong private label presence, the creators gathered a set of organic and sponsored recommendations from Amazon.in, exploiting snowball sampling. Metadata for each product was also collected, including user rating, number of reviews, brand, seller.
    \item \textbf{Affiliation of creators}: Indian Institute of Technology; Max Planck Institute for Software Systems.
    \item \textbf{Domain}: information systems.
    \item \textbf{Tasks in fairness literature}: fair ranking evaluation \citep{dash2021when}.
    \item \textbf{Data spec}: item-recommendation pairs.
    \item \textbf{Sample size}: $\sim1$M recommendations associated with $\sim20$K items.
    \item \textbf{Year}: 2021.
    \item \textbf{Sensitive features}: brand ownership.
    \item \textbf{Link}: not available
    \item \textbf{Further info}: \citet{dash2021when}
\end{itemize}

\subsection{Amazon Reviews}
\begin{itemize}
    \item \textbf{Description}: this is large-scale dataset of over ten million products and respective reviews on Amazon, spanning more than two decades. It was created to study the problem of image-based recommendation and its dynamics. Rich metadata are available for both products and reviews. Reviews consist of ratings, text, reviewer name, and review ID, while products include title, price, image, and sales rank of product.
    \item \textbf{Affiliation of creators}: University of California, San Diego.
    \item \textbf{Domain}: information systems.
    \item \textbf{Tasks in fairness literature}: fair ranking \citep{patro2019incremental}.
    \item \textbf{Data spec}: user-product pairs (reviews).
    \item \textbf{Sample size}: $\sim200$M reviews of products.
    \item \textbf{Year}: 2018.
    \item \textbf{Sensitive features}: none.
    \item \textbf{Link}: \url{https://nijianmo.github.io/amazon/index.html}
    \item \textbf{Further info}: \citet{mcauley2015imagebased,he2016ups}
\end{itemize}

\subsection{ANPE}
\begin{itemize}
    \item \textbf{Description}: this dataset  represents a large randomized controlled trial, assigning job seekers in France to a program run by the Public employment agency (ANPE), or to a program outsourced to private providers by the Unemployment insurance organization (Unédic). The data involves 400 public employment branches and over 200,000 job-seekers. Data about job seekers includes their demographics, their placement program and the subsequent duration of unemployment spells.
    \item \textbf{Affiliation of creators}: Paris School of Economics; Institute of Labor Economics; CREST; ANPE; Unédic; Direction de l’Animation de la Recherche et des Études Statistiques.
    \item \textbf{Domain}: economics.
    \item \textbf{Tasks in fairness literature}: fairness evaluation of risk assessment \citep{kallus2019assessing}.
    \item \textbf{Data spec}: tabular data.
    \item \textbf{Sample size}: $\sim200$K job seekers.
    \item \textbf{Year}: 2012.
    \item \textbf{Sensitive features}: age, gender, nationality.
    \item \textbf{Link}: \url{https://www.openicpsr.org/openicpsr/project/113904/version/V1/view?path=/openicpsr/113904/fcr:versions/V1/Archive&type=folder}
    \item \textbf{Further info}: \citet{behaghel2014private}
\end{itemize}

\subsection{Antelope Valley Networks}
\begin{itemize}
    \item \textbf{Description}: this a set of synthetic datasets generated to study the problem of influence maximization for obesity prevention. Samples of agents are generated to emulate the demographic and obesity distribution across regions in the Antelope Valley in California, exploiting data from the US Census, the Los Angeles County Department of Public Health, and Los Angeles Times Mapping L.A. project. Each agent in the network has a geographic region, gender, ethnicity, age, and connections to other agents, which are more frequent for agents with similar attributes. Agents are also assigned a weight status, which may change based on interactions with other agents in their ego-network, emulating social learning.
    \item \textbf{Affiliation of creators}: National University of Singapore; National University of Southern California.
    \item \textbf{Domain}: public health.
    \item \textbf{Tasks in fairness literature}: fair graph diffusion \citep{farnad2020unifying}.
    \item \textbf{Data spec}: agent-agent pairs.
    \item \textbf{Sample size}: $\sim20$ synthetic networks, containing $\sim500$ individuals each.
    \item \textbf{Year}: 2019.
    \item \textbf{Sensitive features}: ethnicity, gender, age, geography.
    \item \textbf{Link}: \url{https://github.com/bwilder0/fair_influmax_code_release}
    \item \textbf{Further info}: \citet{wilder2018optimizing,tsang2019group}
\end{itemize}

\subsection{Apnea}
\begin{itemize}
    \item \textbf{Description}: this dataset results from a sleep medicine study focused on establishing important factors for the automated diagnosis of Obstructive Sleep Apnea (OSA). The task associated with this dataset is the prediction of medical condition (OSA/no OSA) from available patient features, which include demographics, medical history, and symptoms.
    \item \textbf{Affiliation of creators}: Massachusetts Institute of Technology; Massachusetts General Hospital; Harvard Medical School.
    \item \textbf{Domain}: sleep medicine.
    \item \textbf{Tasks in fairness literature}: fair preference-based classification \citep{ustun2019fairness}.
    \item \textbf{Data spec}: mixture (time series and tabular data).
    \item \textbf{Sample size}: $\sim2$K patients.
    \item \textbf{Year}: 2016.
    \item \textbf{Sensitive features}: age, sex.
    \item \textbf{Link}: not available
    \item \textbf{Further info}: \citet{ustun2016clinical}
\end{itemize}

\subsection{ArnetMiner Citation Network}
\begin{itemize}
    \item \textbf{Description}: this dataset is one of the many resources made available by the ArnetMiner online service. The ArnetMiner system was developed for the extraction and mining of data from academic social networks, with a focus on profiling of researchers. The DBLP Citation Network is extracted from academic resources, such as DBLP, ACM and MAG (Microsoft Academic Graph). The dataset captures the relationships between scientific articles and their authors in a connected graph structure. It can be used for tasks such as community discovery, topic modeling, centrality and influence analysis. In its latest versions, the dataset comprises over 20 fields, including paper title, keywords, abstract, venue, year, along with authors, and their affiliations. The ArnetMiner project was partially funded by the Chinese National High-tech R\&D Program, the National Science Foundation of China, IBM China Research Lab, the Chinese Young Faculty Research Funding program and Minnesota China Collaborative Research Program.
    \item \textbf{Affiliation of creators}: Tsinghua University; IBM.
    \item \textbf{Domain}: library and information sciences.
    \item \textbf{Tasks in fairness literature}: fair graph mining \citep{buyl2020debayes}.
    \item \textbf{Data spec}: article-article pairs.
    \item \textbf{Sample size}: $\sim5$M papers connected by $\sim50$M citations.
    \item \textbf{Year}: 2021.
    \item \textbf{Sensitive features}: author.
    \item \textbf{Link}: \url{http://www.arnetminer.org/citation}
    \item \textbf{Further info}: \citet{tang2008arnetminer}; \url{https://www.aminer.org/}
\end{itemize}

\subsection{Arrhythmia}
\begin{itemize}
    \item \textbf{Description}: data provenance for this set of patient records seems uncertain. The first work referencing this dataset dates to 1997 and details a machine learning approach for the diagnosis of arrhythmia, which presumably motivated its collection. Each data point describes a different patient; features include demographics, weight and height and clinical measurements from ECG signals, along with the diagnosis of a cardiologist into 16 different classes of arrhythmia (including none), which represents the target variable.
    \item \textbf{Affiliation of creators}: Bilkent University; Baskent University.
    \item \textbf{Domain}: cardiology.
    \item \textbf{Tasks in fairness literature}: fair classification \citep{donini2018empirical,mary2019fairnessaware}, 
    robust fair classification \citep{rezaei2021robust},
    limited-label fair classification \citep{chzhen2019leveraging}.
    \item \textbf{Data spec}: tabular data.
    \item \textbf{Sample size}: $\sim500$ patients.
    \item \textbf{Year}: 1997.
    \item \textbf{Sensitive features}: age, sex.
    \item \textbf{Link}: \url{https://archive.ics.uci.edu/ml/datasets/arrhythmia}
    \item \textbf{Further info}: \citet{guvenir1997supervised}
\end{itemize}

\subsection{Athletes and health professionals}
\begin{itemize}
    \item \textbf{Description}: the datasets were developed to study the effects of bias in image classification. The health professional dataset (doctors and nurses) contains race and gender as sensitive features and the athlete dataset (basketball and volleyball players) contains gender and jersey color as sensitive features. Each subgroup, separated by combinations of sensitive features, is roughly balanced at ~200 images. The collected data was manually examined by the curators to remove stylized images and images containing both females and males.
    \item \textbf{Affiliation of creators}: Massachusetts Institute of Technology.
    \item \textbf{Domain}: computer vision.
    \item \textbf{Tasks in fairness literature}: bias discovery \citep{tong2020investigating}.
    \item \textbf{Data spec}: image.
    \item \textbf{Sample size}: $\sim800$ images of athletes and $\sim500$ images of health professionals.
    \item \textbf{Year}: 2020.
    \item \textbf{Sensitive features}: Gender (both), race (health professionals), jersey color (athletes).
    \item \textbf{Link}: \url{https://github.com/ghayat2/Datasets}
    \item \textbf{Further info}: \citet{tong2020investigating}
\end{itemize}

\subsection{Automated Student Assessment Prize (ASAP)}
\begin{itemize}
    \item \textbf{Description}: this dataset was collected to evaluate the feasibility of automated  essay scoring. It consists of a collection of essays by US students in grade levels 7--10, rated by at least two human raters. The dataset comes with a predefined training/validation/test split and powers the Hewlett Foundation Automated Essay Scoring competition on Kaggle. The curators tried to remove personally identifying information from the essays using Named Entity Recognizer (NER) and several heuristics.
    \item \textbf{Affiliation of creators}: University of Akron; The Common Pool; OpenEd Solutions.
    \item \textbf{Domain}: education.
    \item \textbf{Tasks in fairness literature}: fair regression evaluation \citep{madnadi2017:building}.
    \item \textbf{Data spec}: text.
    \item \textbf{Sample size}: $\sim20$K student essays.
    \item \textbf{Year}: 2012.
    \item \textbf{Sensitive features}: none.
    \item \textbf{Link}: \url{https://www.kaggle.com/c/asap-aes/data/}
    \item \textbf{Further info}: \citet{shermis2014stateoftheart}
\end{itemize}

\subsection{Bank Marketing}
\begin{itemize}
    \item \textbf{Description}: often simply called \emph{Bank} dataset in the fairness literature, this resource was produced to support a study of success factors in telemarketing of long-term deposits within a Portuguese bank, with data collected over the period 2008--2010. Each data point represents a telemarketing phone call and includes client-specific features (e.g. job, education), features about the marketing phone call (e.g. day of the week and duration) and meaningful environmental features (e.g. euribor). The classification target is a binary variable indicating client subscription to a term deposit.
    \item \textbf{Affiliation of creators}: Instituto Universitário de Lisboa (ISCTE-IUL), ISTAR, Lisboa; University of Minho.
    \item \textbf{Domain}: marketing.
    \item \textbf{Tasks in fairness literature}: fair classification \citep{savani2020intraprocessing,baharlouei2020renyi,zafar2017fairness,diana2021minimax,shah2021rawslian},
    fair clustering \citep{abbasi2021fair,harb2020kfc,ahmadian2020fair,huang2019coresets,bera2019fair,mahabadi2020individual,backurs2019scalable,chierichetti2017fair},
    fair data summarization \citep{halabi2020fairness},
    fair classification under unawareness \citep{kilbertus2018blind},
    fairness evaluation \citep{lipton2018does,islam2021can},
    limited-label fairness evaluation \citep{ji2020can}, preference-based fair clustering \citep{galhotra2021learning}.
    \item \textbf{Data spec}: tabular data.
    \item \textbf{Sample size}: $\sim40$K phone contacts.
    \item \textbf{Year}: 2012.
    \item \textbf{Sensitive features}: age.
    \item \textbf{Link}: \url{https://archive.ics.uci.edu/ml/datasets/Bank+Marketing}
    \item \textbf{Further info}: \citet{moro2014datadriven}
\end{itemize}

\subsection{Barcelona Room Rental}
\begin{itemize}
    \item \textbf{Description}: this dataset summarizes the operations of a room rental platform in Barcelona over 30 months, from January 2017 through June 2019. It contains information about over $60,000$ users, divided into those seeking (seeker) and those listing (lister) a room. The data consists of lister-seeker pairs, such that a seeker is recommended for a room and lister. Recommendations are provided by a set of different recommender systems (recsys). For each pair, the data reports the rank in which each seeker was listed, the recsys providing the recommendation, and the post-recommendation interaction, if any, along with demographic information on both users. Textual indications of ``gay-friendliness'' in user profiles is treated as a sensitive feature (among others), as sexual orientation was previously found to be a discriminating factor in access to housing.
    \item \textbf{Affiliation of creators}: University Pompeu Fabra; Eurecat; Institute for Political Economy and Governance; ISI Foundation.
    \item \textbf{Domain}: information systems.
    \item \textbf{Tasks in fairness literature}: fair ranking evaluation \citep{solans2021comparing}.
    \item \textbf{Data spec}: lister-seeker pairs.
    \item \textbf{Sample size}: $\sim4$M pairs.
    \item \textbf{Year}: 2021.
    \item \textbf{Sensitive features}: gender, age, spoken language, ``gay-friendliness''.
    \item \textbf{Link}: not available
    \item \textbf{Further info}: \citet{solans2021comparing}
\end{itemize}

\subsection{Benchmarking Attribution Methods (BAM)}
\begin{itemize}
    \item \textbf{Description}: this dataset was developed to evaluate different explainability methods in computer vision. It was constructed by pasting object pixels from MS-COCO \citep{lin2014microsoft} into scene images from MiniPlaces \citep{zhou2018:places}. Objects are rescaled to a variable proportion between one third and one half of the scene images onto which they are pasted. Both scene images and object images belong to ten different classes, for a total of 100 possible combinations. Scene images were chosen between the ones that do not contain the objects from the ten MS-COCO classes. 
    This dataset enables users to freely control how each object is correlated with scenes, from which ground truth explanations can be formed. The creators also propose a few quantitative metrics to evaluate interpretability methods by either contrasting different inputs in the same dataset or contrasting two models with the same input.
    \item \textbf{Affiliation of creators}: Google.
    \item \textbf{Domain}: computer vision.
    \item \textbf{Tasks in fairness literature}: fair representation learning \citep{david2020debiasing}.
    \item \textbf{Data spec}: image.
    \item \textbf{Sample size}: $\sim100$K images over 10 object classes and 10 image classes.
    \item \textbf{Year}: 2020.
    \item \textbf{Sensitive features}: none.
    \item \textbf{Link}: \url{https://github.com/google-research-datasets/bam}
    \item \textbf{Further info}: \citet{yang2019benchmarking}
\end{itemize}

\subsection{Berkeley Students}
\begin{itemize}
    \item \textbf{Description}: this dataset holds anonymized student records at UC Berkeley from Spring 2012 through Fall 2019. It consistst of enrollment information on a per-semester basis for tens of thousands of students. For each enrollment, student course scores are provided, along with student demographic information, including gender, race, entry status and parental income. The dataset supports evaluations of equity in educational outcome as well as grade predictions for academic support interventions. It is maintained by the University's Enterprise Data and Analytics unit.
    \item \textbf{Affiliation of creators}: University of California, Berkeley.
    \item \textbf{Domain}: education.
    \item \textbf{Tasks in fairness literature}: fair classification \citep{jiang2021towards}.
    \item \textbf{Data spec}: tabular data.
    \item \textbf{Sample size}: $\sim2$M enrollments across $\sim80$K students.
    \item \textbf{Year}: 2021.
    \item \textbf{Sensitive features}: gender, race.
    \item \textbf{Link}: not available
    \item \textbf{Further info}: \citet{jiang2021towards}
\end{itemize}

\subsection{Bias in Bios}
\begin{itemize}
    \item \textbf{Description}: this dataset was developed as a large-scale study of gender bias in occupation classification. It consists of online biographies of professionals scraped from the Common Crawl. Biographies are detected in crawls when they match the regular expression ``$<$name$>$ is a(n) $<$title$>$'', with $<$title$>$ being one of twenty-eight common occupations. The gender of each person in the dataset is identified via the third person gendered pronoun, typically used in professional biographies. The envisioned task mirrors that of a job search automated system in a two-sided labor marketplace, i.e. automated occupation classification. The dataset curators provide python code to recreate the dataset from old Common Crawls. 
    \item \textbf{Affiliation of creators}: Carnegie Mellon University; University of Massachusetts Lowell; Microsoft; LinkedIn.
    \item \textbf{Domain}: linguistics, information systems.
    \item \textbf{Tasks in fairness literature}: fairness evaluation \citep{dearteaga2019bias}, fair classification \citep{yurochkin2021sensei}.
    \item \textbf{Data spec}: text.
    \item \textbf{Sample size}: $\sim400$K biographies.
    \item \textbf{Year}: 2018.
    \item \textbf{Sensitive features}: gender.
    \item \textbf{Link}: \url{https://github.com/Microsoft/biosbias}
    \item \textbf{Further info}: \citet{dearteaga2019bias}
\end{itemize}

\subsection{Bias in Translation Templates}
\begin{itemize}
    \item \textbf{Description}: this resource was developed to study the problem of gender biases in machine translation. It consists of a set of short templates of the form \texttt{One thing about the man/woman, [he/she] is [a \#\#]}, where \texttt{[he/she]} can be a gender-neutral or gender-specific pronoun, and \texttt{[a \#\#]} refers to a profession or conveys sentiment. Templates are built so that the part before the comma acts as a gender-specific clue, and the part after the comma contains information about gender and sentiment/profession. Accurate translations should correctly match the grammatical gender before and after the comma, in every word where it is required by the target language. The curators identify a set of languages to which this template is easily applicable, namely German, Korean, Portuguese, and Tagalog, which are chosen for their different properties with respect to grammatical gender. Depending on which language pair is being considered for translation, the curators identify a set of criteria for the evaluation of translation quality, with special emphasis on the correctness of grammatical gender.
    \item \textbf{Affiliation of creators}: Seoul National University.
    \item \textbf{Domain}: linguistics.
    \item \textbf{Tasks in fairness literature}: bias evaluation of machine translation \citep{cho2021towards}.
    \item \textbf{Data spec}: text.
    \item \textbf{Sample size}: $\sim1$K templates.
    \item \textbf{Year}: 2021.
    \item \textbf{Sensitive features}: gender.
    \item \textbf{Link}: \url{https://github.com/nolongerprejudice/tgbi-x}
    \item \textbf{Further info}: \citet{cho2021towards}
\end{itemize}

\subsection{Bing US Queries}
\begin{itemize}
    \item \textbf{Description}: this dataset was created to investigate differential user satisfaction with the Bing search engine across different demographic groups. The authors selected log data of a random subset of Bing’s desktop and laptop users from the English-speaking US market over a two week period. The data was preprocessed by cleaning spam and bot queries, and it was enriched with user demographics, namely age (bucketed) and gender (binary), which were self-reported by users during account registration and automatically validated by the dataset curators. Moreover, queries were labeled with topic information. Finally, four different signals were extracted from search logs, namely graded utility, reformulation rate, page click count, and successful click count.
    \item \textbf{Affiliation of creators}: Microsoft.
    \item \textbf{Domain}: information systems.
    \item \textbf{Tasks in fairness literature}: fair ranking evaluation \citep{mehrotra2017auditing}.
    \item \textbf{Data spec}: query-result pairs.
    \item \textbf{Sample size}: $\sim30$M (non-unique) queries issued by $\sim4$M distinct users.
    \item \textbf{Year}: 2017.
    \item \textbf{Sensitive features}: age, gender.
    \item \textbf{Link}: not available
    \item \textbf{Further info}: \citet{mehrotra2017auditing}
\end{itemize}

\subsection{BOLD}
\begin{itemize}
    \item \textbf{Description}: this resource is a benchmark to measure biases of language models with respect to sensitive demographic attributes. The creators identified six attributes (e.g. race, profession) and values of said attribute (e.g. African American, flight nurse) for which they gather prompts from English Language Wikipedia, either from pages about the group (e.g. ``A flight nurse is a registered'') or people representing it (e.g. ``Over the years, Isaac Hayes was able''). Prompts are fed to different language models, whose outputs are automatically labelled for sentiment, regard, toxicity, emotion and gender polarity. These labels are also validated by human annotators hired on Amazon Mechanical Turk.
    \item \textbf{Affiliation of creators}: Amazon; University of California, Santa Barbara.
    \item \textbf{Domain}: linguistics.
    \item \textbf{Tasks in fairness literature}: bias evaluation in language models \citep{dhamala2021bold}.
    \item \textbf{Data spec}: text.
    \item \textbf{Sample size}: $\sim20$K prompts.
    \item \textbf{Year}: 2021.
    \item \textbf{Sensitive features}: gender, race, religion, profession, political leaning.
    \item \textbf{Link}: \url{https://github.com/amazon-research/bold}
    \item \textbf{Further info}: \citet{dhamala2021bold}
\end{itemize}

\subsection{BookCorpus}
\begin{itemize}
    \item \textbf{Description}: this dataset was developed for the problem of learning general representations of text useful for different downstream tasks. It consist of text from 11,038 books from the web by unpublished authors available on \url{https://www.smashwords.com/} in 2015. The BookCorpus contains thousands of duplicate books (only 7,185 are unique) and many contain copyright restrictions. The GPT \citep{radford2018improving} and BERT \citep{devlin2019BERT} language models were trained on this dataset.
    \item \textbf{Affiliation of creators}: University of Toronto; Massachusetts Institute of Technology.
    \item \textbf{Domain}: linguistics.
    \item \textbf{Tasks in fairness literature}: data bias evaluation \citep{tan2019assessing}.
    \item \textbf{Data spec}: text.
    \item \textbf{Sample size}: $\sim$ 1B words in $\sim$74M sentences from $\sim$11K books.
    \item \textbf{Year}: unknown.
    \item \textbf{Sensitive features}: textual references to people and their demographics.
    \item \textbf{Link}: not available
    \item \textbf{Further info}: \citet{zhu2015aligning,bandy2021addressing}
\end{itemize}

\subsection{BUPT Faces}
\label{sec:bupt}
\begin{itemize}
    \item \textbf{Description}: this resource consists of two datasets, developed as a large scale collection, suitable for training face verification algorithms operating on diverse populations. The underlying data collection procedure mirrors the one from RFW (\autoref{sec:rfw}), including sourcing from MS-Celeb-1M and automated annotation of so-called \emph{race} into one of four categories:  Caucasian, Indian, Asian and African. For categories where not enough images were readily available, the authors resort to the FreeBase celebrity list, downloading images of people from Google and cleaning them "both automatically and manually". The remaining images were obtained from MS-Celeb-1M (\autoref{sec:msceleb}), on which the BUPT Faces datasets are heavily based.
    \item \textbf{Affiliation of creators}: Beijing University of Posts and Telecommunications.
    \item \textbf{Domain}: computer vision.
    \item \textbf{Tasks in fairness literature}: fair reinforcement learning \citep{wang2020mitigating}, fair classification \citep{xu2021consistent}, fair representation learning \citep{gong2021mitigating}.
    \item \textbf{Data spec}: image.
    \item \textbf{Sample size}: $\sim2$M images of $\sim40$K celebrities (BUPT-Globalface); $\sim1$M images of $\sim30$K celebrities (BUPT-Balancedface).
    \item \textbf{Year}: 2019.
    \item \textbf{Sensitive features}: race.
    \item \textbf{Link}: \url{http://www.whdeng.cn/RFW/Trainingdataste.html}
    \item \textbf{Further info}: \citet{wang2020mitigating}
\end{itemize}

\subsection{Burst}
\begin{itemize}
    \item \textbf{Description}: Burst is a free provider of stock photography powered by Shopify. This dataset features a subset of Burst images used as a resource to test algorithms for fair image retrieval and ranking, aimed at providing, in response to a query, a collection of photos that is balanced across demographics. Images come with human-curated tags annotated internally by the Burst team.
    \item \textbf{Affiliation of creators}: Shopify.
    \item \textbf{Domain}: information systems.
    \item \textbf{Tasks in fairness literature}: fair ranking \citep{karako2018using}.
    \item \textbf{Data spec}: image.
    \item \textbf{Sample size}: $\sim3$K images.
    \item \textbf{Year}: present.
    \item \textbf{Sensitive features}: gender.
    \item \textbf{Link}: not available
    \item \textbf{Further info}: \citet{karako2018using}; \url{https://burst.shopify.com/}
\end{itemize}

\subsection{Business Entity Resolution}
\begin{itemize}
    \item \textbf{Description}: A proprietary Google dataset, where the task is to predict whether a pair of business descriptions describe the same real business.
    \item \textbf{Affiliation of creators}: Google.
    \item \textbf{Domain}: linguistics.
    \item \textbf{Tasks in fairness literature}: fair entity resolution \citep{cotter2019training}.
    \item \textbf{Data spec}: text.
    \item \textbf{Sample size}: $\sim$15K samples.
    \item \textbf{Year}: 2019.
    \item \textbf{Sensitive features}: geography, business size.
    \item \textbf{Link}: not available
    \item \textbf{Further info}: \citet{cotter2019training}
\end{itemize}

\subsection{Campus Recruitment}
\begin{itemize}
    \item \textbf{Description}: this dataset was published to Kaggle in 2020 by Ben Roshan, who was then enrolled in an MBA in Business Analytics at Jain University Bangalore. The provenance of this dataset is not clear. It was provided by a Jain University professor as a class resource to study and experiment with data analysis. It encodes information about students at an Indian institution, including their degree, their performance in school and placement information at the end of school, including salary.
    \item \textbf{Affiliation of creators}: Jain University Bangalore.
    \item \textbf{Domain}: education.
    \item \textbf{Tasks in fairness literature}: fair data generation \citep{liu2021rawlsnet}.
    \item \textbf{Data spec}: tabular data.
    \item \textbf{Sample size}: $\sim200$ students.
    \item \textbf{Year}: 2020.
    \item \textbf{Sensitive features}: gender.
    \item \textbf{Link}: \url{https://www.kaggle.com/datasets/benroshan/factors-affecting-campus-placement}
    \item \textbf{Further info}: 
\end{itemize}

\subsection{Cars3D}
\begin{itemize}
    \item \textbf{Description}: this dataset consists of CAD-generated models of 199 cars rendered from from 24 rotation angles. Originally devised for visual analogy making, it is also used for more general research on learning disentangled representation.
    \item \textbf{Affiliation of creators}: University of Michigan.
    \item \textbf{Domain}: computer vision.
    \item \textbf{Tasks in fairness literature}: fair representation learning \citep{locatello2019fairness}.
    \item \textbf{Data spec}: image.
    \item \textbf{Sample size}: $\sim5$K images.
    \item \textbf{Year}: 2020.
    \item \textbf{Sensitive features}: none.
    \item \textbf{Link}: \url{https://github.com/google-research/disentanglement_lib/tree/master/disentanglement_lib/data/ground_truth}
    \item \textbf{Further info}: \citet{reed2015deep}
\end{itemize}

\subsection{CelebA}
\begin{itemize}
    \item \textbf{Description}: CelebFaces Attributes Dataset (CelebA) features images of celebrities from the CelebFaces dataset, augmented with annotations of landmark location and binary attributes. The attributes, ranging from highly subjective features (e.g. attractive, big nose) and potentially offensive (e.g. double chin) to more objective ones (e.g. black hair) were annotated by a ``professional labeling company''.
    \item \textbf{Affiliation of creators}: Chinese University of Hong Kong.
    \item \textbf{Domain}: computer vision.
    \item \textbf{Tasks in fairness literature}: fair classification \citep{savani2020intraprocessing,kim2019multiaccuracy,chuang2021fair,lohaus2020too,creager2019flexibly,jung2021fair}, 
    fair anomaly detection \citep{zhang2021towards}, 
    bias discovery \citep{amini2019uncovering}
    fair anomaly detection \citep{zhang2021towards},
    fairness evaluation of private classification \citep{cheng2021can},
    fairness evaluation of selective classification \citep{jones2021selective},
    fairness evaluation \citep{wang2020towards,segal2021fairness},
    fair representation learning \citep{quadrianto2019discovering},
    fair data summarization \citep{chiplunkar2020how},
    fair data generation \citep{choi2020fair,ramaswamy2021fair}.
    \item \textbf{Data spec}: image.
    \item \textbf{Sample size}: $\sim200$K face images of over $\sim10$K unique individuals.
    \item \textbf{Year}: 2015.
    \item \textbf{Sensitive features}: gender, age, skin tone.
    \item \textbf{Link}: \url{http://mmlab.ie.cuhk.edu.hk/projects/CelebA.html}
    \item \textbf{Further info}: \citet{liu2015deep}
\end{itemize}

\subsection{CheXpert}
\begin{itemize}
    \item \textbf{Description}: this dataset consists of chest X-ray images from patients that have been treated at the Stanford Hospital between October 2002 and July 2017. Each radiograph, either frontal or lateral, is annotated for the presence of 14 observations related to medical conditions. Most annotations were automatically extracted from free text radiology reports and validated against a set of 1,000 held-out reports, manually reviewed by a radiologist. For a subset of the X-ray images, high-quality labels are provided by a group of 3 radiologists. The task associated with this dataset is the automated diagnosis of medical conditions from radiographs.
    \item \textbf{Affiliation of creators}: Stanford University.
    \item \textbf{Domain}: radiology.
    \item \textbf{Tasks in fairness literature}: fairness evaluation of selective classification \citep{jones2021selective}, fairness evaluation of private classification \citep{cheng2021can}.
    \item \textbf{Data spec}: image.
    \item \textbf{Sample size}: $\sim200$K chest radiographs from $60$K patients.
    \item \textbf{Year}: 2019.
    \item \textbf{Sensitive features}: sex, age (of patient).
    \item \textbf{Link}: \url{https://stanfordmlgroup.github.io/competitions/chexpert/}
    \item \textbf{Further info}: \citet{irvin2019chexpert,garbin2021structured}
\end{itemize}

\subsection{Chicago Ridesharing}
\begin{itemize}
    \item \textbf{Description}: this resource describes all trips reported by ridesharing companies to the City of Chicago, starting November 2018. It is the result of an ongoing transparency effort, following the introduction of a city-wide ordinance requiring the disclosure of trips and and fares on part of transportation network providers. For each trip, this dataset reports geographical information (pickup and dropoff), duration and cost. To avoid individual re-identification, the granularity of times and locations is reduced to the nearest 15-minutes interval and census tract. Moreover, for rare combinations of census tract an interval, location data is provided at coarser granularity (community area).
    \item \textbf{Affiliation of creators}: City of Chicago.
    \item \textbf{Domain}: transportation.
    \item \textbf{Tasks in fairness literature}: fair pricing evaluation \citep{pandey2021disparate}.
    \item \textbf{Data spec}: tabular data.
    \item \textbf{Sample size}: $\sim200$M trips.
    \item \textbf{Year}: present.
    \item \textbf{Sensitive features}: geography.
    \item \textbf{Link}: \url{https://data.cityofchicago.org/Transportation/Transportation-Network-Providers-Trips/m6dm-c72p}
    \item \textbf{Further info}: \url{http://dev.cityofchicago.org/open\%20data/data\%20portal/2020/04/28/tnp-trips-2019-additional.html}; \url{http://dev.cityofchicago.org/open\%20data/data\%20portal/2019/04/12/tnp-taxi-privacy.html}
\end{itemize}

\subsection{CIFAR}
\begin{itemize}
    \item \textbf{Description}: CIFAR-10 and CIFAR-100 are a labelled subset of the 80 million tiny images database. CIFAR consists of 32x32 colour images that students were paid to annotate. The project, aimed at advancing the effectiveness of supervised learning techniques in computer vision, was funded by the the Canadian Institute for Advanced Research, after which the dataset is named.
    \item \textbf{Affiliation of creators}: University of Toronto.
    \item \textbf{Domain}: computer vision.
    \item \textbf{Tasks in fairness literature}: fair classification \citep{wang2020towards,jung2021fair}, 
    fair incremental learning \citep{zhao2020maintaining},
    robust fairness evaluation \citep{nanda2021fairness}.
    \item \textbf{Data spec}: image.
    \item \textbf{Sample size}: $\sim6K$ images x 10 classes (CIFAR-10) or $600$ images x 100 classes (CIFAR-100).
    \item \textbf{Year}: 2009.
    \item \textbf{Sensitive features}: none.
    \item \textbf{Link}: \url{https://www.cs.toronto.edu/~kriz/cifar.html}
    \item \textbf{Further info}: \citet{krizhevsky2009learning}
    \item \textbf{Variants}: CIFAR-10S \citep{wang2020towards} is a modified version specifically aimed at studying biases in image classification across an artificial sensitive attribute (color/grayscale).
\end{itemize}

\subsection{CiteSeer Papers}
\begin{itemize}
    \item \textbf{Description}: this dataset was created to study the problem of link-based classification of connected entities. The creators extracted a network of papers from CiteSeer, belonging to one of six categories: Agents, Artificial Intelligence, Database, Human Computer Interaction, Machine Learning and Information Retrieval. Each article is associated with a bag-of-word representation, and the associated task is classification into one of six topics.
    \item \textbf{Affiliation of creators}: University of Maryland.
    \item \textbf{Domain}: library and information sciences.
    \item \textbf{Tasks in fairness literature}: fair graph mining \citep{li2021on}.
    \item \textbf{Data spec}: paper-paper pairs.
    \item \textbf{Sample size}: $\sim3$K articles connected by $\sim5$K citations.
    \item \textbf{Year}: 2016.
    \item \textbf{Sensitive features}: none.
    \item \textbf{Link}: \url{http://networkrepository.com/citeseer.php}
    \item \textbf{Further info}: \citet{lu2003:link}
\end{itemize}

\subsection{Civil Comments}
\begin{itemize}
    \item \textbf{Description}: this dataset derives from an archive of the Civil Comments platform, a browser plugin for independent news sites, whose users peer-reviewed each other's comments with civility ratings. When the plugin shut down, they decided to make comments and metadata available, including the crowd-sourced toxicity ratings. A subset of this dataset was later annotated with a variety of sensitive attributes, capturing whether members of a certain group are mentioned in comments. This dataset powers the Jigsaw Unintended Bias in Toxicity Classification challenge.
    \item \textbf{Affiliation of creators}: Jigsaw; Civil Comments.
    \item \textbf{Domain}: social media.
    \item \textbf{Tasks in fairness literature}: fair toxicity classification \citep{adragna2020fairness,yurochkin2021sensei,chuang2021fair},
    fairness evaluation of selective classification \citep{jones2021selective},
    fair robust toxicity classification \citep{adragna2020fairness},
    fairness evaluation of toxicity classification \citep{hutchinson2020unintented}, fairness evaluation \citep{babaeianjelodar2020quantifying}.
    \item \textbf{Data spec}: text.
    \item \textbf{Sample size}: $\sim2$M comments.
    \item \textbf{Year}: 2019.
    \item \textbf{Sensitive features}: race/ethnicity, gender, sexual orientation, religion, disability.
    \item \textbf{Link}: \url{https://www.kaggle.com/c/jigsaw-unintended-bias-in-toxicity-classification}
    \item \textbf{Further info}: \citet{borkan2019nuanced}
\end{itemize}

\subsection{Climate Assembly UK}
\begin{itemize}
    \item \textbf{Description}: this resource was curated to study the problem of subset selection for \emph{sortition}, a political system where decisions are taken by a subset of the whole voting population selected at random. The data describes participants to Climate Assembly UK, a panel organized by the Sortition Foundation in 2020. With the goal of understanding public opinion on how the UK can meet greenhouse gas emission targets. The panel consisted of 110 UK residents selected from a pool of 1,715 who responded to an invitation from the Sortition Foundation reaching $\sim60K$ citizens. Features for each subject in the pool describe their demographics and climate concern level.
    \item \textbf{Affiliation of creators}: Carnegie Mellon University; Harvard University; Sortition Foundation.
    \item \textbf{Domain}: political science.
    \item \textbf{Tasks in fairness literature}: fair subset selection \citep{flanigan2020neutralizing}.
    \item \textbf{Data spec}: tabular data.
    \item \textbf{Sample size}: $\sim2$K pool participants.
    \item \textbf{Year}: 2020.
    \item \textbf{Sensitive features}: gender, age, education, urban/rural, geography, ethnicity.
    \item \textbf{Link}: not available
    \item \textbf{Further info}: \citet{flanigan2020neutralizing}; \url{https://www.climateassembly.uk/}
\end{itemize}

\subsection{Columbia University Speed Dating}
\begin{itemize}
    \item \textbf{Description}: this dataset is a result of a speed dating experiment aimed at understanding preferences in mate selection in men and women. Subjects were recruited from students at Columbia University. Fourteen rounds were conducted with different proportions of male and female subjects, over the period 2002--2004, with participants meeting each potential mate for four minutes and rating them thereafter on six attributes. They also provide an overall evaluation of each potential mate and a binary decision indicating interest in meeting again. Before an event, each participant filled in a survey disclosing their preferences, expectations, and demographics. The inference task associated with this dataset is optimal recommendation in symmetrical two-sided markets.
    \item \textbf{Affiliation of creators}: Columbia University; Harvard University; Stanford University.
    \item \textbf{Domain}: sociology.
    \item \textbf{Tasks in fairness literature}: fair matching \citep{zheng2018fairness}, preference-based fair ranking \citep{paraschakis2020matchmaking}.
    \item \textbf{Data spec}: person-person pairs.
    \item \textbf{Sample size}: $\sim10$K dating records involving $\sim400$ people.
    \item \textbf{Year}: 2016.
    \item \textbf{Sensitive features}: gender, age, race, geography.
    \item \textbf{Link}: \url{https://data.world/annavmontoya/speed-dating-experiment}
    \item \textbf{Further info}: \citet{fisman2006gender}
\end{itemize}

\subsection{Communities and Crime}
\begin{itemize}
    \item \textbf{Description}: this dataset was curated to develop a software tool supporting the work of US police departments. It was especially aimed at identifying similar precincts to exchange best practices and share experiences among departments. The creators were supported by the police departments of Camden (NJ) and Philadelphia (PA). The factors included in the dataset were the ones deemed most important to define similarity of communities from the perspective of law enforcement; they were chosen with the help of law enforcement officials from partner institutions and academics of criminal justice, geography and public policy. The dataset includes socio-economic factors (aggregate data on age, income, immigration, and racial composition) obtained from the 1990 US census, along with information about policing (e.g. number of police cars available) based on the 1990 Law Enforcement Management and Administrative Statistics survey, and crime data derived from the 1995 FBI Uniform Crime Reports. In its released version on UCI, the task associated with the dataset is predicting the total number of violent crimes per 100K population in each community. The most referenced version of this dataset was preprocessed with a normalization step; after receiving multiple requests, the creators also published an unnormalized version.
    \item \textbf{Affiliation of creators}: La Salle University; Rutgers University.
    \item \textbf{Domain}: law.
    \item \textbf{Tasks in fairness literature}: fair classification \citep{yang2020fairness,Sharifi2019average,heidari2018fairness,lohaus2020too,cotter2019training,creager2019flexibly,cotter2018training},
    fair regression evaluation \citep{heidari2019moral}, 
    fair few-shot learning  \citep{slack2020fairness,slack2019fair},
    rich-subgroup fairness evaluation \citep{kearns2019empirical},
    rich-subgroup fair classification \citep{kearns2018preventing},
    fair regression \citep{chzhen2020fair,chzhen2020fairwassertein,romano2020achieving,agarwal2019fair,mary2019fairnessaware,komiyama2018nonconvex,ogura2020convex,berk2017convex,diana2021minimax},
    fair representation learning \citep{ruoss2020learning},
    robust fair classification \citep{mandal2020ensuring},
    fair private classification \citep{jagielski2019differentially},
    fairness evaluation of transfer learning \citep{lan2017discriminatory}, preference-based fair clustering \citep{galhotra2021learning}.
    \item \textbf{Data spec}: tabular data.
    \item \textbf{Sample size}: $\sim2K$ communities.
    \item \textbf{Year}: 2009.
    \item \textbf{Sensitive features}: race, geography.
    \item \textbf{Link}: \url{https://archive.ics.uci.edu/ml/datasets/communities+and+crime} and \url{http://archive.ics.uci.edu/ml/datasets/communities+and+crime+unnormalized}
    \item \textbf{Further info}: \citet{redmond2002datadriven}
\end{itemize}

\subsection{COMPAS}
\label{sec:compas_db}
\begin{itemize}
    \item \textbf{Description}: this dataset was created for an external audit of racial biases in the Correctional Offender Management Profiling for Alternative Sanctions (COMPAS) risk assessment tool developed by Northpointe (now Equivant), which estimates the likelihood of a 
    defendant becoming a recidivist. Instances represent defendants scored by COMPAS in Broward County, Florida, between 2013--2014, reporting their demographics, criminal record, custody and COMPAS scores. Defendants' public criminal records were obtained from the Broward County Clerk’s Office website matching them based on date of birth, first and last names. The dataset was augmented with jail records and COMPAS scores provided by the Broward County Sheriff’s Office. Finally, public incarceration records were downloaded from the Florida Department of Corrections website. Instances are associated with two target variables (is\_recid and is\_violent\_recid), indicating whether defendants were booked in jail for a criminal offense (potentially violent) that occurred after their COMPAS screening but within two years. See Appendix \ref{sec:compas} for extensive documentation.
    \item \textbf{Affiliation of creators}: ProPublica.
    \item \textbf{Domain}: law.
    \item \textbf{Tasks in fairness literature}: fair classification \citep{he2020geometric,sharma2020data,goel2018nondiscriminatory,oneto2019taking,celis2019classification,canetti2019from,cho2020fair,savani2020intraprocessing,donini2018empirical,heidari2018fairness,russell2017when,quadrianto2017recycling,calmon2017optimized,diciccio2020evaluating,xu2020algorithmic,vargo2021individually,roh2021fairbatch,maity2021statistical,lohaus2020too,roh2020frtrain,celis2020data,cotter2019training,mary2019fairnessaware,wang2019repairing,delobelle2020ethical,ogura2020convex,lum2016statistical,zafar2017fairnessbeyond,berk2017convex,wadsworth2018achieving,diana2021minimax,perrone2021fair,ali2021accounting}, 
    fairness evaluation \citep{cardoso2019framework,mcnamara2019equalized,kasy2021fairness,taskesen2021statistical,friedler2019comparative,wick2019unlocking,zhang2018equality,pleiss2017fairness,chaibubneto2020causal,speicher2018unified,corbettdavies2017algorithmic,liu2019implicit,agarwal2018reductions,ngong2020towards,jabbari2020empirical,chouldechova2017fair,grgichlaca2016case,islam2021can},
    fair risk assessment \citep{coston2020counterfactual,mishler2021fairness,nabi2019learning}, 
    fair \emph{task assignment} \citep{goel2019crowdsourcing}, 
    fair classification under unawareness \citep{lahoti2020fairness,lamy2019noisetolerant,chzhen2019leveraging,kilbertus2018blind},
    data bias evaluation \citep{beretta2021detecting},
    fair representation learning \citep{ruoss2020learning,zhao2020conditional,bower2018debiasing},
    robust fair classification \citep{mandal2020ensuring,rezaei2021robust,biswas2021ensuring},
    dynamical fairness evaluation \citep{zhang2020how},
    fair reinforcement learning \citep{metevier2019offline},
    fair ranking evaluation \citep{kallus2019fairness,yang2017measuring},
    fair multi-stage classification \citep{madras2018predict},
    dynamical fair classification \citep{valera2018enhancing},
    preference-based fair classification \citep{zafar2017from,ustun2019fairness},
    fair regression \citep{komiyama2018nonconvex},
    fair multi-stage classification \citep{goel2020importance},
    limited-label fair classification \citep{chzhen2019leveraging,wang2021fair,choi2020group},
    robust fairness evaluation \citep{slack2020fairness,slack2019fairness},
    rich subgroup fairness evaluation \citep{chouldechova2017fairer,zhang2017identifying}.
    \item \textbf{Data spec}: tabular data.
    \item \textbf{Sample size}: $\sim12$K defendants.
    \item \textbf{Year}: 2016.
    \item \textbf{Sensitive features}: sex, age, race.
    \item \textbf{Link}: \url{https://github.com/propublica/compas-analysis}
    \item \textbf{Further info}: \citet{angwin2016machine, larson2016how}
\end{itemize}

\subsection{Cora Papers}
\begin{itemize}
    \item \textbf{Description}: this resource was produced within the wider development effort for \emph{Cora}, an Internet portal for computer science research papers available in the early 2000s. The portal supported keyword search, topical categorization of articles, and citation mapping. This dataset consists of articles and citation links between them. It contains bag-of-word representations for the text of each article, and the associated task is classification into one of seven topics.
    \item \textbf{Affiliation of creators}: Just Research Carnegie Mellon University; Massachusetts Institute of Technology; Univeristy of Maryland;  Lawrence Livermore National Laboratory.
    \item \textbf{Domain}: library and information sciences.
    \item \textbf{Tasks in fairness literature}: .
    \item \textbf{Data spec}: article-article pairs.
    \item \textbf{Sample size}: $\sim3$K articles connected by $\sim5$K citations.
    \item \textbf{Year}: 2019.
    \item \textbf{Sensitive features}: none.
    \item \textbf{Link}: \url{https://relational.fit.cvut.cz/dataset/CORA}
    \item \textbf{Further info}: \citet{mccallum2000automating,sen2008collective}
\end{itemize}

\subsection{Costarica Household Survey}
\begin{itemize}
    \item \textbf{Description}: this data comes from the national household survey of Costa Rica, performed by the national institute of statistics and census (Instituto Nacional de Estadística y Censos). The survey is aimed at measuring the socio-economical situation in the country and informing public policy. The data collection procedure is specially designed to allow for precise conclusions with respect to six different regions of the country and about differences in urban vs rural areas; stratification along these variables is deemed suitable. The 2018 survey contains a special section on the crimes suffered by respondents.
    \item \textbf{Affiliation of creators}: Instituto Nacional de Estadística y Censos.
    \item \textbf{Domain}: economics.
    \item \textbf{Tasks in fairness literature}: fair classification \citep{noriegacampero2020algorithmic}.
    \item \textbf{Data spec}: tabular data.
    \item \textbf{Sample size}: $\sim13$K households.
    \item \textbf{Year}: 2018.
    \item \textbf{Sensitive features}: sex, age, birthplace, disability, geography, family size.
    \item \textbf{Link}: \url{https://www.inec.cr/encuestas/encuesta-nacional-de-hogares}
    \item \textbf{Further info}: \url{https://www.inec.cr/sites/default/files/documetos-biblioteca-virtual/enaho-2018.pdf}
\end{itemize}

\subsection{Credit Card Default}
\begin{itemize}
    \item \textbf{Description}: this dataset was built to investigate automated mechanisms for credit card default prediction following a wave of defaults in Taiwan connected to patters of card over-issuing and over-usage. The dataset contains payment history of customers of an important Taiwanese bank, from April to October 2005. Demographics, marital status, and education of customers are also provided, along with the amount of credit and a binary variable encoding default on payment, which is the target variable of the associated task.
    \item \textbf{Affiliation of creators}: Chung-Hua University; Thompson Rivers University.
    \item \textbf{Domain}: finance.
    \item \textbf{Tasks in fairness literature}: fair classification \citep{cho2020fair,berk2017convex}, 
    fair clustering \citep{harb2020kfc,ghadiri2021socially,harb2020kfc,bera2019fair},
    fair clustering under unawareness \citep{esmaeili2020probabilistic},
    fair classification under unawareness \citep{wang2020robust},
    fair data summarization \citep{Tantipongpipat2019multicriteria,samadi2018price},
    fairness evaluation \citep{lipton2018does}, fair anomaly detection \citep{shekhar2021fairod}.
    \item \textbf{Data spec}: tabular data.
    \item \textbf{Sample size}: $\sim30$K credit card holders.
    \item \textbf{Year}: 2016.
    \item \textbf{Sensitive features}: gender, age.
    \item \textbf{Link}: \url{https://archive.ics.uci.edu/ml/datasets/default+of+credit+card+clients}
    \item \textbf{Further info}: \citet{yeh2009comparisons}
\end{itemize}

\subsection{Credit Elasticities}
\begin{itemize}
    \item \textbf{Description}: this dataset stems from a randomized trial conducted by a consumer lender in South Africa to study loan price elasticity. Prior customers were contacted by mail with limited-time loan offers at variable and randomized interest rates. The aim of the study was understanding the relationship between interest rate and customer acceptance rates, along with the benefits for the lender. Customers who accepted and received formal approval, filled in a short survey with factors of interest for the study, including demographics, education, and prior borrowing history.
    \item \textbf{Affiliation of creators}: Yale University; Dartmouth College.
    \item \textbf{Domain}: finance.
    \item \textbf{Tasks in fairness literature}: fair pricing evaluation \citep{kallus2021fairness}.
    \item \textbf{Data spec}: tabular data.
    \item \textbf{Sample size}: $\sim50$K clients.
    \item \textbf{Year}: 2008.
    \item \textbf{Sensitive features}: gender, age, geography.
    \item \textbf{Link}: \url{http://doi.org/10.3886/E113240V1}
    \item \textbf{Further info}: \citet{karlan2008credit}
\end{itemize}

\subsection{Crowd Judgement}
\begin{itemize}
    \item \textbf{Description}: this dataset was assembled to compare the performance of the COMPAS recidivism risk prediction system against that of non-expert human assessors \citep{dressel2018accuracy}. A subset of 1,000 defendants were selected from the COMPAS dataset. Crowd-sourced assessors were recruited through Amazon Mechanical Turk. They were presented with a summary of each defendant, including demographics and previous criminal history, and asked to predict whether they would recidivate within 2 years of their most recent crime. These judgements, assembled via plain majority voting, ended up exhibiting accuracy and fairness levels comparable to that displayed by the COMPAS system. While this dataset was assembled for an experiment, it was later used to study the problem of fairness in crowdsourced judgements.
    \item \textbf{Affiliation of creators}: Dartmouth College.
    \item \textbf{Domain}: law.
    \item \textbf{Tasks in fairness literature}: fair \emph{truth discovery} \citep{li2020towards}, fair \emph{task assignment} \citep{li2020towards,goel2019crowdsourcing} 
    \item \textbf{Data spec}: judge-defendant pair.
    \item \textbf{Sample size}: $\sim1$K defendants from COMPAS and $\sim400$ crowd-sourced labellers. Each defendant is judged by 20 different labellers.
    \item \textbf{Year}: 2018.
    \item \textbf{Sensitive features}: sex, age and race of defendants and crowd-sourced judges.
    \item \textbf{Link}: \url{https://farid.berkeley.edu/downloads/publications/scienceadvances17/}
    \item \textbf{Further info}: \citep{dressel2018accuracy}
    \item \textbf{Variants}: a similar dataset was collected by \citet{wang2019empirical}.
\end{itemize}

\subsection{Curatr British Library Digital Corpus}
\begin{itemize}
    \item \textbf{Description}: this dataset is a subset of English language digital texts from the British Library focused on volumes of 19th-century fiction, obtained through the Curatr platform. It was selected for the well-researched presence of stereotypical and binary concepts of gender in this literary production. The goal of the creators was studying gender biases in large text corpora and their relationship with biases in word embeddings trained on those corpora.
    \item \textbf{Affiliation of creators}: University College Dublin.
    \item \textbf{Domain}: literature.
    \item \textbf{Tasks in fairness literature}: data bias evaluation \citep{leavy2020mitigating}.
    \item \textbf{Data spec}: text.
    \item \textbf{Sample size}: $\sim20$K  books.
    \item \textbf{Year}: 2020.
    \item \textbf{Sensitive features}: textual references to people and their demographics.
    \item \textbf{Link}: \url{http://curatr.ucd.ie/}
    \item \textbf{Further info}: \citet{leavy2019curatr}
\end{itemize}

\subsection{CVs from Singapore}
\begin{itemize}
    \item \textbf{Description}: this dataset was developed to test demographic biases in resume filtering. In particular, the authors studied nationality bias in automated resume filtering in Singapore, across the three major ethnic groups of the city state: Chinese, Malaysian and Indian. The dataset consists of 135 resumes (45 per ethnic group) used for application to finance jobs in Singapore, collected by Jai Janyani. The dataset only includes resumes for which the origin of the candidates can be reliably inferred to be either Chinese, Malaysian, or Indian from education and initial employment. The dataset also comprises 9 finance job postings from China, Malaysia, and India (3 per country). All job-resume pairs are rated for relevance/suitability by three annotators.
    \item \textbf{Affiliation of creators}: University of Maryland.
    \item \textbf{Domain}: information systems, management information systems.
    \item \textbf{Tasks in fairness literature}: fair ranking \citep{deshpande2020mitigating}.
    \item \textbf{Data spec}: text.
    \item \textbf{Sample size}: $\sim100$ resumes.
    \item \textbf{Year}: 2020.
    \item \textbf{Sensitive features}: ethnic group.
    \item \textbf{Link}: not available
    \item \textbf{Further info}: \citet{deshpande2020mitigating}
\end{itemize}

\subsection{Dallas Police Incidents}
\begin{itemize}
    \item \textbf{Description}: this dataset is due to the Dallas OpenData initiative\footnote{\url{https://www.dallasopendata.com/}} and ``reflects crimes as reported to the Dallas Police Department'' beginning June 1, 2014. Each incident comes with rich spatio-temporal data, information about the victim, the officers involved and the type of crime. A subset of the dataset is available on Kaggle\footnote{\url{https://www.kaggle.com/carrie1/dallaspolicereportedincidents}}.
    \item \textbf{Affiliation of creators}: Dallas Police Department.
    \item \textbf{Domain}: law.
    \item \textbf{Tasks in fairness literature}: fair spatio-temporal process learning \citep{shang2020listwise}.
    \item \textbf{Data spec}: tabular.
    \item \textbf{Sample size}: $\sim800$K incidents.
    \item \textbf{Year}: present.
    \item \textbf{Sensitive features}: age, race, and gender (of victim), geography.
    \item \textbf{Link}: \url{https://www.dallasopendata.com/Public-Safety/Police-Incidents/qv6i-rri7}
    \item \textbf{Further info}: 
\end{itemize}

\subsection{Demographics on Twitter}
\begin{itemize}
    \item \textbf{Description}: this dataset was developed to test demographic classifiers on Twitter data. In particular, the tasks associated with this resource are the automatic inference of gender, age, location and political orientation of users. The true values for these attributes, which act as a ground truth for learning algorithms, were inferred from tweets and user bios, such as the ones containing the regexp "I'm a $<$gendered noun$>$", with gendered nouns including mother, woman, father, man.
    \item \textbf{Affiliation of creators}: Massachusetts Institute of Technology.
    \item \textbf{Domain}: social media.
    \item \textbf{Tasks in fairness literature}: fairness evaluation of sentiment analysis \citep{shen2018darling}.
    \item \textbf{Data spec}: mixture.
    \item \textbf{Sample size}: $\sim80$K profiles.
    \item \textbf{Year}: 2017.
    \item \textbf{Sensitive features}: gender, age, political orientation, geography.
    \item \textbf{Link}: not available
    \item \textbf{Further info}: \citet{vijayaraghavan2017twitter}
\end{itemize}

\subsection{Diabetes 130-US Hospitals}
\begin{itemize}
    \item \textbf{Description}: this dataset contains 10 years of care data from 130 US hospitals extracted from Health Facts, a clinical database associated with a multi-institution data collection program. The dataset was extracted to study the association between the measurement of HbA1c (glycated hemoglobin) in human bloodstream and early hospital readmission, and was donated to UCI in 2014. The dataset includes patient demographics, in-hospital procedures, and diagnoses, along with information about subsequent readmissions.
    \item \textbf{Affiliation of creators}: Virginia Commonwealth University; University of Cordoba; Polish Academy of Sciences.
    \item \textbf{Domain}: endocrinology.
    \item \textbf{Tasks in fairness literature}: fair clustering \citep{chierichetti2017fair,bera2019fair,backurs2019scalable,mahabadi2020individual,huang2019coresets,bera2019fair}.
    \item \textbf{Data spec}: tabular data.
    \item \textbf{Sample size}: $\sim100$K patients.
    \item \textbf{Year}: 2014.
    \item \textbf{Sensitive features}: age, race, gender.
    \item \textbf{Link}: \url{https://archive.ics.uci.edu/ml/datasets/diabetes+130-us+hospitals+for+years+1999-2008}
    \item \textbf{Further info}: \citet{strack2014impact}
\end{itemize}

\subsection{Diversity in Faces (DiF)}
\begin{itemize}
    \item \textbf{Description}: this large dataset was created to favour the development and evaluation of robust face analysis algorithms across diverse demographics and domain-specific features, such as craniofacial distances and facial contrast). One million images of people's faces from Flickr were labelled, mostly automatically, according to 10 different coding schemes, comprising, e.g., cranio-facial measurements, pose, and demographics. Age and gender were inferred both automatically and by human workers. Statistics about the diversity of this dataset along these coded measures are available in the accompanying report.
    \item \textbf{Affiliation of creators}: IBM.
    \item \textbf{Domain}: computer vision.
    \item \textbf{Tasks in fairness literature}: fair representation learning \citep{quadrianto2019discovering}, 
    fairness evaluation of private classification \citep{bagdasaryan2019differential}.
    \item \textbf{Data spec}: image.
    \item \textbf{Sample size}: $\sim1$M images.
    \item \textbf{Year}: 2019.
    \item \textbf{Sensitive features}: skin color, age, and gender.
    \item \textbf{Link}: \url{https://www.ibm.com/blogs/research/2019/01/diversity-in-faces/}
    \item \textbf{Further info}: \citet{merler2019diversity}
\end{itemize}

\subsection{Drug Consumption}
\begin{itemize}
    \item \textbf{Description}: this dataset was collected by Elaine Fehrman between March 2011 and March 2012 after receiving approval from relevant ethics boards from the University of Leicester. The goal of this dataset is to seek patterns connecting an individual’s risk of drug consumption with demographics and psychometric measurements of the Big Five personality traits (NEO-FFI-R), impulsivity (BIS-11), and sensation seeking (ImpSS). The study employed an online survey tool from Survey Gizmo to recruit participants world-wide; over 93\% of the final usable sample reported living in an English-speaking country. Target variables summarize the consumption of 18 psychoactive substances on an ordinal scale ranging from never using the drug to using it over a decade ago, or in the last decade, year, month, week, or day. The 18 substances considered in the study are classified as central nervous system depressants, stimulants, or hallucinogens and comprise the following: alcohol, amphetamines, amyl nitrite, benzodiazepines, cannabis, chocolate, cocaine, caffeine, crack, ecstasy, heroin, ketamine, legal highs, LSD, methadone, magic mushrooms, nicotine, and Volatile Substance Abuse (VSA), along with one fictitious drug (Semeron) introduced to identify over-claimers. A version of the dataset donated to the UCI Machine Learning Repository is associated with 18 prediction tasks, i.e. one per substance.
    \item \textbf{Affiliation of creators}: Rampton Hospital; Nottinghamshire Healthcare NHS Foundation Trust; University of Leicester; University of Nottingham; University of Salahaddin.
    \item \textbf{Domain}: applied psychology.
    \item \textbf{Tasks in fairness literature}: fair classification \citep{donini2018empirical,mary2019fairnessaware},
    evaluation of data bias \citep{beretta2021detecting},
    limited-label fair classification \citep{chzhen2019leveraging},
    robust fair classification \citep{rezaei2021robust}.
    \item \textbf{Data spec}: tabular data.
    \item \textbf{Sample size}: $\sim2$K respondents.
    \item \textbf{Year}: 2016.
    \item \textbf{Sensitive features}: age, gender, ethnicity, geography.
    \item \textbf{Link}: \url{https://archive.ics.uci.edu/ml/datasets/Drug+consumption+\%28quantified\%29}
    \item \textbf{Further info}: \citet{fehrman2017five, fehrman2019personality}
\end{itemize}

\subsection{DrugNet}
\begin{itemize}
    \item \textbf{Description}: this dataset was collected to study drug consumption patterns in connection with social ties and behaviour of drug users. This work puts particular emphasis on situations at risk of disease transmission and to assess the opportunity for prevention via recruitment of peer educators to demonstrate, disseminate and support HIV prevention practices among their connections. Participants were recruited in Hartford neighbourhoods of high drug-use activity, mostly via street outreach and recruitment by early participants. Eligibility criteria included being at least 18 years old, using an illicit drug, and signing an informed consent form. Each participant provided data about their drug use, most common sites of usage, HIV risk practices associated with drug use and sexual behavior, and social ties deemed important by the respondent and their demographics.
    \item \textbf{Affiliation of creators}: Institute for Community Research of Hartford; Hispanic Health Council, Hartford; Boston College.
    \item \textbf{Domain}: social work, social networks.
    \item \textbf{Tasks in fairness literature}: fair graph clustering \citep{kleindessner2019guarantees}.
    \item \textbf{Data spec}: person-person pairs.
    \item \textbf{Sample size}: $\sim300$ people.
    \item \textbf{Year}: 2016.
    \item \textbf{Sensitive features}: ethnicity, sex, age.
    \item \textbf{Link}: \url{https://sites.google.com/site/ucinetsoftware/datasets/covert-networks/drugnet}
    \item \textbf{Further info}: \citet{weeks2002social}
\end{itemize}

\subsection{dSprites}
\begin{itemize}
    \item \textbf{Description}: this dataset was assembled by researchers affiliated with Google DeepMind as an artificial benchmark for unsupervised methods aimed at learning disentangled data representations. Each image in the dataset consists of a black-and-white sprite with variable shape, scale, orientation and position. Together these are the \emph{generative factors} underlying each image. Ideally, systems trained on this data should learn disentangled representations, such that latent image representations are clearly associated with changes in a single generative factor.
    \item \textbf{Affiliation of creators}: Google.
    \item \textbf{Domain}: computer vision.
    \item \textbf{Tasks in fairness literature}: fair representation learning \citep{locatello2019fairness,creager2019flexibly}.
    \item \textbf{Data spec}: image.
    \item \textbf{Sample size}: $\sim700$K images.
    \item \textbf{Year}: 2017.
    \item \textbf{Sensitive features}: none.
    \item \textbf{Link}: \url{https://github.com/deepmind/dsprites-dataset}
    \item \textbf{Further info}: \citet{Higgins2017betaVAELB}
\end{itemize}

\subsection{Dutch Census}
\begin{itemize}
    \item \textbf{Description}: this dataset was derived from the 2001 census carried out by the Dutch Central Bureau for Statistics to gather data about family composition, economic activities, levels of education, and occupation of Dutch citizens and foreigners from various countries of origin. A version of the dataset commonly employed in the fairness research literature has been preprocessed and made available online. The associated task is the classification of individuals into high-income and low-income professions.
    \item \textbf{Affiliation of creators}: Bournemouth University; TU Eindhoven.
    \item \textbf{Domain}: demography.
    \item \textbf{Tasks in fairness literature}: fair classification \citep{agarwal2018reductions,xu2020algorithmic,zhang2017achieving,lohaus2020too}, 
    fairness evaluation \citep{cardoso2019framework}.
    \item \textbf{Data spec}: tabular data.
    \item \textbf{Sample size}: $\sim60$K respondents.
    \item \textbf{Year}: 2001.
    \item \textbf{Sensitive features}: sex, age, citizenship.
    \item \textbf{Link}: \url{https://sites.google.com/site/conditionaldiscrimination/}
    \item \textbf{Further info}: \citet{zliobaite2011handling}; \url{https://microdata.worldbank.org/index.php/catalog/2102/data-dictionary/F2?file_name=NLD2001-P-H}; \url{https://www.cbs.nl/nl-nl/publicatie/2004/31/the-dutch-virtual-census-of-2001}
\end{itemize}

\subsection{EdGap}
\begin{itemize}
    \item \textbf{Description}: this dataset focuses on education performance in different US counties, with a focus on inequality of opportunity and its connection to socioeconomic factors. Along with average SAT and ACT test scores by county, this dataset reports socioeconomic data from the American Community Survey by the Bureau of Census, including  household income, unemployment, adult educational attainment, and family structure. Importantly, some states require all students to take ACT or SAT tests, while others do not. As a result, average test scores are inherently higher in states that do not require all students to test, and they are not directly comparable to average scores in states where testing is mandatory.
    \item \textbf{Affiliation of creators}: Memphis Teacher Residency.
    \item \textbf{Domain}: education.
    \item \textbf{Tasks in fairness literature}: fair risk assessment \citep{he2020inherent}.
    \item \textbf{Data spec}: tabular data.
    \item \textbf{Sample size}: $\sim2$K counties.
    \item \textbf{Year}: 2019.
    \item \textbf{Sensitive features}: geography.
    \item \textbf{Link}: \url{https://www.edgap.org/}
    \item \textbf{Further info}: 
\end{itemize}

\subsection{Epileptic Seizures}
\begin{itemize}
    \item \textbf{Description}: this dataset was curated to study electroencephalographic (EEG) time series in relation to epilepsy. The dataset consists of EEG recordings from healthy volunteers with eyes closed and eyes open, and from epilepsy patients during seizure-free intervals and during epileptic seizures. Volunteers and patients are recorded for 23.6-sec. A version of this dataset, used in fairness research, was donated to UCI Machine Learning Repository by researchers affiliated with Rochester Institute of Technology in 2017, with a classification task based on the patients' condition and state at the time of recording. The data was later removed from UCI at the original curators' request.
    \item \textbf{Affiliation of creators}: University of Bonn.
    \item \textbf{Domain}: neurology.
    \item \textbf{Tasks in fairness literature}: robust fairness evaluation \citep{black2021leaveoneout}.
    \item \textbf{Data spec}: time series.
    \item \textbf{Sample size}: $\sim500$ individuals, each summarized by $\sim4$K-points time series.
    \item \textbf{Year}: 2017.
    \item \textbf{Sensitive features}: none.
    \item \textbf{Link}: \url{https://archive.ics.uci.edu/ml/datasets/Epileptic+Seizure+Recognition}; \url{http://epileptologie-bonn.de/cms/upload/workgroup/lehnertz/eegdata.html}
    \item \textbf{Further info}: \citet{andrzejak2001indications}
\end{itemize}

\subsection{Equitable School Access in Chicago}
\begin{itemize}
    \item \textbf{Description}: this resource was assembled from disparate sources to evaluate school access in Chicago for different race groups. A transportation network was inferred from data on public bus lines available on the Chicago Transit Authority website. Data on school location and quality evaluation was obtained from the Chicago Public School data portal. Finally, demographic information on race representation in different tracts was retrieved from the 2010 US census.
    \item \textbf{Affiliation of creators}: Salesforce.
    \item \textbf{Domain}: transportation.
    \item \textbf{Tasks in fairness literature}: fair graph augmentation \citep{ramachandran2021gaea}.
    \item \textbf{Data spec}: location-location pairs.
    \item \textbf{Sample size}: $\sim2K$ nodes (locations), connected by $\sim8$K edges (bus lines).
    \item \textbf{Year}: 2020.
    \item \textbf{Sensitive features}: race.
    \item \textbf{Link}: \url{https://github.com/salesforce/GAEA}
    \item \textbf{Further info}: \citet{ramachandran2021gaea}
\end{itemize}

\subsection{Equity Evaluation Corpus (EEC)}
\begin{itemize}
    \item \textbf{Description}: this dataset was compiled to audit sentiment analysis systems for gender and race bias. It is based on 11 short sentence templates; 7 templates include emotion words, while the remaining 4 do not. Moreover, each sentence includes one gender- or race-associated word, such as names predominantly associated with African American or European American people. Gender-related words consist of names, nouns, and pronouns.
    \item \textbf{Affiliation of creators}: National Research Council Canada.
    \item \textbf{Domain}: linguistics.
    \item \textbf{Tasks in fairness literature}: fair sentiment analysis evaluation \citep{liang2020artificial}.
    \item \textbf{Data spec}: text.
    \item \textbf{Sample size}: $\sim9$K sentences.
    \item \textbf{Year}: 2018.
    \item \textbf{Sensitive features}: race, gender.
    \item \textbf{Link}: \url{https://saifmohammad.com/WebPages/Biases-SA.html}
    \item \textbf{Further info}: \citet{kiritchenko2018examining}
\end{itemize}

\subsection{Facebook Ego-networks}
\begin{itemize}
    \item \textbf{Description}: this dataset was collected to study the problem of  identifying users’ social circles, i.e. categorizing links between nodes in a social network. The data represents ten ego-networks whose central user was asked to fill in a survey and manually identify the circles to which their friends belonged. Features from each profile, including education, work and location are anonymized.
    \item \textbf{Affiliation of creators}: Stanford University.
    \item \textbf{Domain}: social networks.
    \item \textbf{Tasks in fairness literature}: fair graph mining \citep{li2021on}.
    \item \textbf{Data spec}: user-user pairs.
    \item \textbf{Sample size}: $\sim4$K people connected by $\sim90$K friend relations.
    \item \textbf{Year}: 2012.
    \item \textbf{Sensitive features}: geography, gender.
    \item \textbf{Link}: \url{https://snap.stanford.edu/data/egonets-Facebook.html}
    \item \textbf{Further info}: \citet{leskovec2012learning}
\end{itemize}

\subsection{Facebook Large Network}
\begin{itemize}
    \item \textbf{Description}: this dataset was developed to study the effectiveness of node embeddings for learning tasks defined on graphs. The dataset concentrates on verified Facebook pages of politicians, governmental organizations, television shows, and companies, represented as nodes, while edges represent mutual likes. In addition, each page comes with node embeddings which are extracted from the textual description of each page. The original task on this dataset is page category classification.
    \item \textbf{Affiliation of creators}: University of Edinburgh.
    \item \textbf{Domain}: social networks.
    \item \textbf{Tasks in fairness literature}: fair graph mining evaluation \citep{kang2020inform}.
    \item \textbf{Data spec}: page-page pairs.
    \item \textbf{Sample size}: $\sim$20K nodes (pages) connected by $\sim200$K edges (mutual likes).
    \item \textbf{Year}: 2019.
    \item \textbf{Sensitive features}: none.
    \item \textbf{Link}: \url{http://snap.stanford.edu/data/facebook-large-page-page-network.html}
    \item \textbf{Further info}: \citet{rozemberczki2021multi}
\end{itemize}

\subsection{FACES}
\begin{itemize}
    \item \textbf{Description}: this resource contains images of Caucasian individuals of variable age and gender under six predefined facial expressions (neutrality, sadness, disgust, fear, anger, and happiness). This dataset is described as a database of emotion-related stimuli for scientific research. Subjects were hired through a model agency in Berlin, and suitably informed about the purpose of the photo-shooting session, thereafter signing an informed consent document. Each model reported their own age and gender. The necessary facial expressions were carefully explained with the help of a manual, with attention to the position of muscles. Photographs were obtained and post-processed in a standardized fashion, and later validated by raters of different ages with respect to the perceived expression and age of subjects. At a later stage, images were also annotated for attractiveness and distinctiveness. Currently, a small subset of the images is publicly available, while the full dataset is available after registration.
    \item \textbf{Affiliation of creators}: Max Planck Institute for Human Development.
    \item \textbf{Domain}: computer vision, experimental psychology.
    \item \textbf{Tasks in fairness literature}: fairness evaluation \citep{kim2021age}.
    \item \textbf{Data spec}: image.
    \item \textbf{Sample size}: $\sim2$K images of $\sim200$ people.
    \item \textbf{Year}: 2010.
    \item \textbf{Sensitive features}: age, gender.
    \item \textbf{Link}: \url{https://faces.mpib-berlin.mpg.de/imeji/}
    \item \textbf{Further info}: \citet{ebner2010faces}
\end{itemize}

\subsection{FairFace}
\begin{itemize}
    \item \textbf{Description}: this dataset was developed as a balanced resource for face analysis with diverse race, gender and age composition. The associated task is race, gender and age classification. Starting from a large public image dataset (Yahoo YFCC100M), the authors sampled images incrementally to ensure diversity with respect to race, for which they considered seven categories: White, Black, Indian, East Asian, Southeast Asian, Middle East, and Latino. Sensitive attributes were annotated by workers on Amazon Mechanical Turk, and also through a model based on these annotations. Faces with low agreement between model and annotators were manually re-verified by the dataset curators. This dataset was annotated automatically with a binary Fitzpatrick skin tone label \citep{cheng2021can}.
    \item \textbf{Affiliation of creators}: University of California, Los Angeles.
    \item \textbf{Domain}: computer vision.
    \item \textbf{Tasks in fairness literature}: fairness evaluation of private classification \citep{cheng2021can}.
    \item \textbf{Data spec}: image.
    \item \textbf{Sample size}: $\sim100$K images.
    \item \textbf{Year}: 2019.
    \item \textbf{Sensitive features}: race, age, gender, skin tone.
    \item \textbf{Link}: \url{https://github.com/joojs/fairface}
    \item \textbf{Further info}: \citet{karkkainen2019fairface}
\end{itemize}

\subsection{Fantasy Football}
\begin{itemize}
    \item \textbf{Description}: this resource was curated to study the problem of fair ranking aggregation. The creators collected rankings of National Football League players from the top 25 experts on the popular fantasy sports website FantasyPros. The data covers 16 weeks during the 2019 football season. Players are assigned to different sensitive groups based on the conference of their team (American Football Conference or National Football Conference). The data available online concentrates on wide receivers.
    \item \textbf{Affiliation of creators}: Worcester Polytechnic Institute.
    \item \textbf{Domain}: sports.
    \item \textbf{Tasks in fairness literature}: fair ranking evaluation \citep{kuhlman2021measuring}.
    \item \textbf{Data spec}: player-expert pairs.
    \item \textbf{Sample size}: $\sim50$ players, ranked by 25 experts (on a weekly basis), over 16 weeks.
    \item \textbf{Year}: 2020.
    \item \textbf{Sensitive features}: football conference.
    \item \textbf{Link}: \url{https://arcgit.wpi.edu/cakuhlman/VLDB2020/tree/master/charts/data}
    \item \textbf{Further info}: \citet{kuhlman2020rank}
\end{itemize}

\subsection{Fashion MNIST}
\begin{itemize}
    \item \textbf{Description}: this dataset is based on product assortement from the Zalando website. It contains gray-scale resized versions of thumbnail images of unique clothing products, labeled by in-house fashion experts according to their category, including e.g. trousers, coat and shirt. The envisioned task is object classification. The dataset, sharing the same size and structure as MNIST, was developed to provide a harder and more representative task, and to replace MNIST as a popular computer vision benchmark.
    \item \textbf{Affiliation of creators}: Zalando.
    \item \textbf{Domain}: computer vision.
    \item \textbf{Tasks in fairness literature}: robust fairness evaluation \citep{black2021leaveoneout}.
    \item \textbf{Data spec}: image.
    \item \textbf{Sample size}: $\sim70$K images across 10 product categories.
    \item \textbf{Year}: 2017.
    \item \textbf{Sensitive features}: none.
    \item \textbf{Link}: \url{https://github.com/zalandoresearch/fashion-mnist}
    \item \textbf{Further info}: \citet{xiao2017fashionmnist}
\end{itemize}

\subsection{FICO}
\begin{itemize}
    \item \textbf{Description}: based on a sample of 301,536 TransUnion TransRisk scores from 2003, this dataset was created to study the problem of adjusting predictors for compliance with the equality of opportunity fairness metric. The TransUnion data was preprocessed and aggregated to summarize the CDF of risk scores by race (Non-Hispanic white, Black, Hispanic, Asian). The original data comes from a 2007 report to the US Congress on credit scoring and its effects on the availability and affordability of credit carried out by a dedicated Federal Reserve working group. The collection, creation, processing, and aggregation was carried out by the working group; the data was later scraped by the creators, who made it available without any modification.
    \item \textbf{Affiliation of creators}: Google; University of Texas at Austin; Toyota Technological Institute at Chicago.
    \item \textbf{Domain}: finance.
    \item \textbf{Tasks in fairness literature}: fairness evaluation \citep{hardt2016equality}, 
    dynamical fair classification \citep{liu2020disparate}, 
    dynamical fairness evaluation \citep{zhang2020how,liu2018delayed,creager2020causal},
    fair resource allocation \citep{goelz2019paradoxes}.
    \item \textbf{Data spec}: tabular data.
    \item \textbf{Sample size}: N/As. CDFs are provided over risk scores which are normalized (0-100\%)  and quantized with step 0.5\%.
    \item \textbf{Year}: 2016.
    \item \textbf{Sensitive features}: race.
    \item \textbf{Link}: \url{https://github.com/fairmlbook/fairmlbook.github.io/tree/master/code/creditscore/data}
    \item \textbf{Further info}: \citet{usfr2007report,hardt2016equality,barocas2019fair}
\end{itemize}

\subsection{FIFA 20 Players}
\begin{itemize}
    \item \textbf{Description}: this dataset was scraped by Stefano Leone and made available on Kaggle. It includes the players' data for the Career Mode from FIFA 15 to FIFA 20, a popular football game. Several tasks are envisioned for this dataset, including a historical comparison of players.
    \item \textbf{Affiliation of creators}: unknown.
    \item \textbf{Domain}: sports.
    \item \textbf{Tasks in fairness literature}: fairness evaluation under unawareness \citep{awasthi2021evaluating}.
    \item \textbf{Data spec}: tabular data.
    \item \textbf{Sample size}: $\sim20$K players.
    \item \textbf{Year}: 2019.
    \item \textbf{Sensitive features}: geography.
    \item \textbf{Link}: \url{https://www.kaggle.com/stefanoleone992/fifa-20-complete-player-dataset}
    \item \textbf{Further info}: 
\end{itemize}

\subsection{FilmTrust}
\begin{itemize}
    \item \textbf{Description}: this dataset was crawled from the entire FilmTrust website, a movie recommendation service with a social network component. The dataset comprises user-movie ratings on a 5-star scale and user-user indications of trust about movie taste. This resource can be used to train and evaluate recommender systems.
    \item \textbf{Affiliation of creators}: Northeastern University; Nanyang Technological University; American University of Beirut; University of Cambridge.
    \item \textbf{Domain}: information systems, movies.
    \item \textbf{Tasks in fairness literature}: fair ranking \citep{liu2018personalizing}.
    \item \textbf{Data spec}: user-movie pairs and user-user pairs.
    \item \textbf{Sample size}: $\sim40$K ratings by $\sim2$K users over $\sim2$K movies.
    \item \textbf{Year}: 2011.
    \item \textbf{Sensitive features}: none.
    \item \textbf{Link}: \url{https://guoguibing.github.io/librec/datasets.html}
    \item \textbf{Further info}: \citet{guo2016novel}
\end{itemize}

\subsection{Framingham}
\begin{itemize}
    \item \textbf{Description}: the Framingham Heart Study began in 1948 under the direction of the National Heart, Lung, and Blood Institute (NHLBI), with the goal of identifying key factors that contribute to cardiovascular disease, given a mounting epidemic of cardiovascular disease whose etiology was mostly unknown at the time. Six different cohorts have been recruited over the years among citizens of Framingham, Massachusetts, without symptoms of cardiovascular disease. After the original cohort, two more were enrolled from the children and grandchildren of the first one. Additional cohorts were also started to reflect the increased racial and ethnic diversity in the town of Framingham. Participants in the study report on their habits (e.g. physical activity, smoking) and undergo regular physical examination and laboratory tests.
    \item \textbf{Affiliation of creators}: National Heart, Lung, and Blood Institute (NHLBI); Boston University.
    \item \textbf{Domain}: cardiology.
    \item \textbf{Tasks in fairness literature}: fair ranking evaluation \citep{kallus2019fairness}.
    \item \textbf{Data spec}: mixture.
    \item \textbf{Sample size}: $\sim15$K respondents.
    \item \textbf{Year}: present.
    \item \textbf{Sensitive features}: age, sex, race.
    \item \textbf{Link}: \url{https://framinghamheartstudy.org/}
    \item \textbf{Further info}: \citet{kannel1979diabetes, tsao2015cohort}
\end{itemize}

\subsection{Freebase15k-237}
\begin{itemize}
    \item \textbf{Description}: Freebase was a collaborative knowledge base which allowed its community members to fill in structured data about diverse entities and relations between them. This database was developed from a prior Freebase dataset \citep{bordes2013:translating}, pruning it from redundant relations and augmenting it with textual relationships from the ClueWeb12 corpus. The creators of this dataset worked on the joint optimization of entity knowledge base and representations of the entities' textual relations, with the goal of providing representations of entities suited for knowledge base completion.
    \item \textbf{Affiliation of creators}: Microsoft; Stanford University.
    \item \textbf{Domain}: information systems.
    \item \textbf{Tasks in fairness literature}: fair graph mining \citep{bose2019compositional}, fairness evaluation in graph mining \citep{fisher2020measuring}.
    \item \textbf{Data spec}: entity-relation-entity triples.
    \item \textbf{Sample size}: $\sim15$K entities connected by $170$K edges (relations).
    \item \textbf{Year}: 2016.
    \item \textbf{Sensitive features}: demographics of people featured in entities and their relations.
    \item \textbf{Link}: \url{https://www.microsoft.com/en-us/download/details.aspx?id=52312}
    \item \textbf{Further info}: \citet{toutanova2015representing}
\end{itemize}

\subsection{GAP Coreference}
\begin{itemize}
    \item \textbf{Description}: this resource was developed as a gender-balanced coreference resolution dataset, useful for auditing gender-dependent differences in the accuracy of existing pronoun resolution algorithms and for training new algorithms that are less gender-biased. The dataset consists of thousands of ambiguous pronoun-name pairs in sentences extracted from Wikipedia. Several measures are taken to avoid the success of naïve heuristics and to favour diversity. Most notably, while the initial (automated) stage of the data collection pipeline extracts contexts with a female:male ratio of 1:9, feminine pronouns are oversampled to achieve a 1:1 ratio. Each example is presented to and annotated  for coreference by three in-house workers.
    \item \textbf{Affiliation of creators}: Google.
    \item \textbf{Domain}: linguistics.
    \item \textbf{Tasks in fairness literature}: data bias evaluation \citep{kocijan2020gap}.
    \item \textbf{Data spec}: text.
    \item \textbf{Sample size}: $\sim9$K sentences.
    \item \textbf{Year}: 2018.
    \item \textbf{Sensitive features}: gender.
    \item \textbf{Link}: \url{https://github.com/google-research-datasets/gap-coreference}
    \item \textbf{Further info}: \citet{webster2018mind}
\end{itemize}

\subsection{German Credit}
\label{sec:german_db}
\begin{itemize}
    \item \textbf{Description}: the German Credit dataset was created to study the problem of automated credit decisions at a regional Bank in southern Germany. Instances represent loan applicants from 1973 to 1975, who were deemed creditworthy and were granted a loan, bringing about a natural selection bias. The data summarizes their financial situation, credit history and personal situation, including housing and number of liable people. A binary variable encoding whether each loan recipient punctually payed every installment is the target of a classification task. Among covariates, marital status and sex are jointly encoded in a single variable. Many documentation mistakes are present in the UCI entry associated with this resource \citep{hofmann1994:sg}. Due to one of these mistakes, users of this dataset are led to believe that the variable sex can be retrieved from the joint marital\_status-sex variable, however this is false. A revised version with correct variable encodings, called South German Credit, was donated to \citet{gromping2019:sg2} with an accompanying report \citep{gromping2019:sg}. See Appendix \ref{sec:german} for extensive documentation.
    \item \textbf{Affiliation of creators}: Hypo Bank (OP/EDV-VP); Universität Hamburg; Strathclyde University (German Credit); Beuth University of Applied Sciences Berlin (South German Credit).
    \item \textbf{Domain}: finance.
    \item \textbf{Tasks in fairness literature}: fair classification \citep{he2020geometric,sharma2020data,raff2018fair,celis2019classification,yang2020fairness,donini2018empirical,vargo2021individually,baharlouei2020renyi,lohaus2020too,martinez2020minimax,mary2019fairnessaware,delobelle2020ethical,raff2018gradient,perrone2021fair,sharma2021fairn},
    fairness evaluation \citep{friedler2019comparative,feldman2015certifying},
    fair active resource allocation \citep{cai2020:fa}, 
    preference-based fair classification \citep{zhang2020joint}, 
    fair active classification \citep{noriegacampero2019active}, 
    fair classification under unawareness \citep{kilbertus2018blind},
    robust fairness evaluation \citep{black2021leaveoneout},
    fair representation learning \citep{ruoss2020learning,louizos2016variational},
    fair reinforcement learning \citep{metevier2019offline},
    fair ranking evaluation \citep{kallus2019fairness,wu2018discrimination,yang2017measuring},
    fair ranking \citep{singh2019policy,bower2021individually},
    fair multi-stage classification \citep{goel2020importance},
    limited-label fair classification \citep{chzhen2019leveraging,wang2021fair,choi2020group},
    limited-label fairness evaluation \citep{ji2020can}.
    \item \textbf{Data spec}: tabular data.
    \item \textbf{Sample size}: $\sim1$K.
    \item \textbf{Year}: 1994 (German Credit); 2020 (South German Credit).
    \item \textbf{Sensitive features}: age, geography.
    \item \textbf{Link}: \url{https://archive.ics.uci.edu/ml/datasets/statlog+(german+credit+data)} (German Credit); \url{https://archive.ics.uci.edu/ml/datasets/South+German+Credit+\%28UPDATE\%29} (South German Credit)
    \item \textbf{Further info}: \citet{gromping2019:sg}
\end{itemize}

\subsection{German Political Posts}
\begin{itemize}
    \item \textbf{Description}: this dataset was used as a training set for German word embeddings, with the goal of investigating biases in word representations. The authors used the Facebook and Twitter APIs to collect posts and comments from the social media channels of six main political parties in Germany (CDU/CSU, SPD, Bundnis90/Die Grünen, FDP, Die Linke, AfD). Facebook posts are from the period 2015--2018, while tweets were collected between January and October 2018. Overall, the dataset consists of millions of posts, for a total of half a billion tokens. A subset of the Facebook comments (100,000) were labeled by human annotators based on whether they contain sexist content, with four sub-labels indicating sexist comments, sexist buzzwords, gender-related compliments, statements against gender equality and assignment of gender stereotypical roles to people.
    \item \textbf{Affiliation of creators}: Technical University of Munich.
    \item \textbf{Domain}: social media.
    \item \textbf{Tasks in fairness literature}: bias evaluation in WEs \citep{papakyriakopulos2020bias}.
    \item \textbf{Data spec}: text.
    \item \textbf{Sample size}: $\sim20$M posts comments and tweets.
    \item \textbf{Year}: 2020.
    \item \textbf{Sensitive features}: textual references to people and their demographics.
    \item \textbf{Link}: not available
    \item \textbf{Further info}: \citet{papakyriakopulos2020bias}
\end{itemize}

\subsection{GLUE}
\label{sec:glue}
\begin{itemize}
    \item \textbf{Description}: this benchmark was assembled to reliably evaluate the progress of natural language processing models. It consists of multiple datasets and associated tasks from the natural language processing domain, including paraphrase detection, textual entailment, sentiment analysis and question answering. Given the quick progress registered by language models on GLUE, a similar benchmark called SuperGLUE was subsequently released comprising more challenging and diverse tasks \citep{wang2019:superglue}.
    \item \textbf{Affiliation of creators}: New York University; University of Washington; DeepMind.
    \item \textbf{Domain}: linguistics.
    \item \textbf{Tasks in fairness literature}: fairness evaluation \citep{babaeianjelodar2020quantifying,rudinger2017:social}, 
    bias evaluation in language models \citep{cheng2021fairfil},
    fairness evaluation of selective classification \citep{jones2021selective}.
    \item \textbf{Data spec}: text.
    \item \textbf{Sample size}: $\sim 100 - 400$K samples. Datasets have variable sizes spanning three orders of magnitude.
    \item \textbf{Year}: 2018.
    \item \textbf{Sensitive features}: none.
    \item \textbf{Link}: \url{https://gluebenchmark.com/}
    \item \textbf{Further info}: \citet{wang2018glue}
\end{itemize}

\subsection{Goodreads Reviews}
\begin{itemize}
    \item \textbf{Description}: there are several versions of this dataset, corresponding to different crawls. Here we refer to the most well documented one by \citet{wan2018item}. This resource consists of anonymized reviews collected from public user \emph{book shelves}. Rich metadata is available for books and reviews, including. authors, country code, publisher, userid, rating, timestamp, and text. A few medium-size subsamples focused on specific book genres are available. The task typically associated with this resource is book recommendation.
    \item \textbf{Affiliation of creators}: University of California, San Diego.
    \item \textbf{Domain}: literature, information systems.
    \item \textbf{Tasks in fairness literature}: fair ranking evaluation \citep{raj2020comparing}, fairness evaluation \citep{chen2018why}.
    \item \textbf{Data spec}: user-book pairs.
    \item \textbf{Sample size}: $\sim200$M records from $\sim900$K users over $\sim2$M books.
    \item \textbf{Year}: 2019.
    \item \textbf{Sensitive features}: author.
    \item \textbf{Link}: \url{https://sites.google.com/eng.ucsd.edu/ucsdbookgraph/}
    \item \textbf{Further info}: \citet{wan2018item}
\end{itemize}

\subsection{Google Local}
\begin{itemize}
    \item \textbf{Description}: this dataset contains reviews and ratings from millions of users on local businesses from five different continents. Businesses are labelled with nearly 50 thousand categories. This resource was collected as a real world example of interactions between users and ratable items, with the goal of testing novel recommendation approaches. The dataset comprises data that is specific to users (e.g. places lived), businesses (e.g. GPS coordinates), and reviews (e.g. timestamps).
    \item \textbf{Affiliation of creators}: University of California, San Diego.
    \item \textbf{Domain}: information systems.
    \item \textbf{Tasks in fairness literature}: fair ranking \citep{patro2019incremental}.
    \item \textbf{Data spec}: user-business pairs.
    \item \textbf{Sample size}: $\sim10$M reviews and ratings from $\sim5$M users on $\sim3$M local businesses.
    \item \textbf{Year}: 2018.
    \item \textbf{Sensitive features}: geography.
    \item \textbf{Link}: \url{https://cseweb.ucsd.edu/~jmcauley/datasets.html#google_local}
    \item \textbf{Further info}: \citet{he2017translationbased}
\end{itemize}

\subsection{Greek Websites}
\begin{itemize}
    \item \textbf{Description}: this dataset was created to demonstrate the \emph{bias goggles} tools, which enables users to explore diverse bias aspects connected with popular Greek web domains. The dataset is a subset of the Greek web, crawled from Greek websites that cover politics and sports, represent big industries, or are generally popular. Starting from a seed of hundreds of websites, crawlers followed the links up to depth 7, avoiding popular sites such as Facebook and Twitter. The final dataset has a graph structure, comprising pages and links between them.
    \item \textbf{Affiliation of creators}: FORTH-ICS, University of Crete.
    \item \textbf{Domain}: .
    \item \textbf{Tasks in fairness literature}: bias discovery\citep{konstantakis2020bias}.
    \item \textbf{Data spec}: page-page pairs.
    \item \textbf{Sample size}: $\sim900$k pages from $\sim 90$k domains.
    \item \textbf{Year}: 2020.
    \item \textbf{Sensitive features}: none.
    \item \textbf{Link}: \url{https://pangaia.ics.forth.gr/bias-goggles/about.html#Dataset}
    \item \textbf{Further info}: \citet{konstantakis2020bias}
\end{itemize}

\subsection{Guardian Articles}
\begin{itemize}
    \item \textbf{Description}: this dataset consists of articles from \emph{The Guardian}, retrieved from The Guardian Open Platform API. In particular, the authors crawled every article that appeared on the website between 2009 and 2018. They created this dataset to demonstrate a framework for the identification of gender biases in training data for machine learning.
    \item \textbf{Affiliation of creators}: University College Dublin.
    \item \textbf{Domain}: news.
    \item \textbf{Tasks in fairness literature}: data bias evaluation \citep{leavy2020mitigating}.
    \item \textbf{Data spec}: text.
    \item \textbf{Sample size}: unknown.
    \item \textbf{Year}: 2020.
    \item \textbf{Sensitive features}: textual references to people and their demographics.
    \item \textbf{Link}: not available
    \item \textbf{Further info}: \citet{leavy2020mitigating}
\end{itemize}

\subsection{HAM10000}
\begin{itemize}
    \item \textbf{Description}: the dataset comprises 10,015 dermatoscopic images collected over a period of 20 years the Department of Dermatology at the Medical University of Vienna, Austria and the skin cancer practice of Cliff Rosendahl in Queensland, Australia. Images were acquired and stored through different modalities; each image depicts a lesion and comes with metadata detailing the region of skin lesion, patient demographics, and diagnosis, which is the target variable. The dataset was employed for the lesion disease classification of the ISIC 2018 challenge.
    \item \textbf{Affiliation of creators}: Medical University of Vienna; University of Queensland.
    \item \textbf{Domain}: dermatology.
    \item \textbf{Tasks in fairness literature}: fair classification \citep{martinez2020minimax}.
    \item \textbf{Data spec}: image.
    \item \textbf{Sample size}: $\sim$10K images.
    \item \textbf{Year}: 2018.
    \item \textbf{Sensitive features}: age, sex.
    \item \textbf{Link}: \url{https://doi.org/10.7910/DVN/DBW86T}
    \item \textbf{Further info}: \citet{tschandl2018ham10000}
\end{itemize}

\subsection{Harvey Rescue}
\begin{itemize}
    \item \textbf{Description}: this dataset is the result of crowdsourced efforts to connect rescue parties with people requesting help in the Houston area, mostly due to the flooding caused by Hurricane Harvey. Most requests are from August 28, 2017, and were sent via social media; they are timestamped and associated with the location of the people seeking help.
    \item \textbf{Affiliation of creators}: Harvey Relief Handiworks; Harvey Relief Coalition.
    \item \textbf{Domain}: social work.
    \item \textbf{Tasks in fairness literature}: fair spatio-temporal process learning \citep{shang2020listwise}.
    \item \textbf{Data spec}: tabular data.
    \item \textbf{Sample size}: $\sim$1K help requests.
    \item \textbf{Year}: 2017.
    \item \textbf{Sensitive features}: geography.
    \item \textbf{Link}: not available
    \item \textbf{Further info}: \url{http://harveyrelief.handiworks.co/}
\end{itemize}

\subsection{Heart Disease}
\begin{itemize}
    \item \textbf{Description}: this dataset is a collection of medical data from separate groups of patients referred for cardiac catheterisation and coronary angiography at 5 different medical centers, namely the Cleveland Clinic (data from 1981--1984), the Hungarian Institute of Cardiology in Budapest (1983--1987), the Long Beach Veterans Administration Medical Center (1984--1987) and the University Hospitals of Basel and Zurich (1985). The binary target variable in this dataset encodes a diagnosis of Coronary artery disease. Covariates relate to patient demographics, exercise data (e.g. maximum heart rate) and routine test data (e.g. resting blood pressure). Overall, 76 covariates are available but 14 are recommended. Names and social security numbers of the patients were initially available, but have been removed from the publicly available dataset.
    \item \textbf{Affiliation of creators}: Veterans Administration Medical Center, Long Beach; Hungarian Institute of Cardiology, Budapest; University Hospital, Zurich; University Hospital, Basel; Studer Corporation; Stanford University.
    \item \textbf{Domain}: cardiology.
    \item \textbf{Tasks in fairness literature}: fairness evaluation \citep{pleiss2017fairness}, fair active classification \citep{noriegacampero2019active}.
    \item \textbf{Data spec}: tabular data.
    \item \textbf{Sample size}: $\sim1$K patients.
    \item \textbf{Year}: 1988.
    \item \textbf{Sensitive features}: age, sex.
    \item \textbf{Link}: \url{https://archive.ics.uci.edu/ml/datasets/heart+disease}
    \item \textbf{Further info}: \citet{detrano1989international}
\end{itemize}

\subsection{Heritage Health}
\begin{itemize}
    \item \textbf{Description}: this dataset was developed as part of the Heritage Health Prize competition with the goal of reducing the cost of health care by decreasing the number of avoidable hospitalizations. The competition requires predicting the number of days a patient will spend in hospital during the 12 months following a cutoff date. The dataset features basic demographic information about patients, along with data about prior hospitalizations (e.g. length of stay and diagnosis), laboratory tests and prescriptions.
    \item \textbf{Affiliation of creators}: CHEO Research Institute, Inc; University of Ottawa; University of Maryland; Privacy Analytics, Inc; Kaggle; Heritage Provider Network.
    \item \textbf{Domain}: health policy.
    \item \textbf{Tasks in fairness literature}: fair multi-stage classification \citep{madras2018predict}, 
    fair representation learning \citep{louizos2016variational}, 
    fair classification \citep{raff2018fair,raff2018gradient},
    fair transfer learning \citep{madras2018learning}, fairness evaluation \citep{islam2021can}.
    \item \textbf{Data spec}: tabular data.
    \item \textbf{Sample size}: $\sim150$K patients.
    \item \textbf{Year}: 2011.
    \item \textbf{Sensitive features}: age, sex.
    \item \textbf{Link}: \url{https://www.kaggle.com/c/hhp/data}
    \item \textbf{Further info}: \citet{el2012deidentification}
\end{itemize}

\subsection{High School Contact and Friendship Network}
\begin{itemize}
    \item \textbf{Description}: this dataset was developed to compare and contrast different methods commonly employed to measure human interaction and build the underlying social network. Data corresponds to interactions and friendship relations between students of a French high school in Marseilles. The authors consider four different methods of network data collection, namely face-to-face contacts measured by two concurrent methods (sensors and diaries), self-reported friendship surveys, and Facebook links.
    \item \textbf{Affiliation of creators}: Aix Marseille Université; Université de Toulon; Centre national de la recherche scientifique; ISI Foundation.
    \item \textbf{Domain}: social networks.
    \item \textbf{Tasks in fairness literature}: fair graph clustering \citep{kleindessner2019guarantees}.
    \item \textbf{Data spec}: student-student pairs.
    \item \textbf{Sample size}: $\sim300$ students.
    \item \textbf{Year}: 2015.
    \item \textbf{Sensitive features}: gender.
    \item \textbf{Link}: \url{http://www.sociopatterns.org/datasets/high-school-contact-and-friendship-networks/}
    \item \textbf{Further info}: \citet{mastrandrea2015contact}
\end{itemize}

\subsection{HMDA}
\begin{itemize}
    \item \textbf{Description}: The Home Mortgage Disclosure Act (HMDA) is a US federal law from 1975 mandating that financial institutions maintain and disclose information about mortgages to the public. Companies submit a Loan Application Register (LAR) to the Federal Financial Institutions Examination Council FFIEC who maintain and disclose the data. The LAR format is subject to changes, such as the one which happened in 2017. From 2018 onward, entries to the LAR comprise information about the financial institution (e.g.\ geography, id), the applicants (e.g.\ demographics, income), the house (e.g.\ value, construction method), the mortgage conditions (type, interest rate, amount) and the outcome. Ethnicity, race, and sex of applicants are self-reported.
    \item \textbf{Affiliation of creators}: Federal Financial Institutions Examination Council.
    \item \textbf{Domain}: finance.
    \item \textbf{Tasks in fairness literature}: fairness evaluation under unawareness \citep{chen2019fairness,kallus2020assessing}.
    \item \textbf{Data spec}: tabular data.
    \item \textbf{Sample size}: $\sim200$M records.
    \item \textbf{Year}: present.
    \item \textbf{Sensitive features}: sex, geography, race, ethnicity.
    \item \textbf{Link}: \url{https://ffiec.cfpb.gov/data-browser/}
    \item \textbf{Further info}: \url{https://ffiec.cfpb.gov/}; \url{https://www.consumerfinance.gov/data-research/hmda/}
\end{itemize}

\subsection{Homeless Youths' Social Networks}
\begin{itemize}
    \item \textbf{Description}: this dataset was collected to study methamphetamine use norms among homeless youth in association with their social networks. A sample of homeless youth aged 13--25 years was recruited between 2011---2012 from two drop-in centers in California. After obtaining informed consent/assent, participants filled in a survey and answered questions from  an interview. The survey included questions on demographics, migratory status, educational status and housing. To reconstruct the social network between them, each participant provided information for up to 50 people with whom they had interacted during the previous 30 days.
    \item \textbf{Affiliation of creators}: University of Denver; University of Southern California.
    \item \textbf{Domain}: social work.
    \item \textbf{Tasks in fairness literature}: fair graph diffusion \citep{rahmattalabi2019exploring}.
    \item \textbf{Data spec}: person-person pairs.
    \item \textbf{Sample size}: $\sim300$ youth.
    \item \textbf{Year}: 2015.
    \item \textbf{Sensitive features}: age, gender, sexual orientation, race and ethnicity.
    \item \textbf{Link}: not available
    \item \textbf{Further info}: \citet{barman2016sociometric}
\end{itemize}

\subsection{IBM HR Analytics}
\begin{itemize}
    \item \textbf{Description}: based on the information available on Kaggle, this is a fictional dataset created by IBM data scientists. It describes employees along dimensions that may be relevant for attrition, the target variable encoding employee departure. Available covariates include information on employee background (education, number of prior companies), work satisfaction (recent promotions, environment and job satisfaction) and seniority (years at the company, years in current role, job level).
    \item \textbf{Affiliation of creators}: IBM.
    \item \textbf{Domain}: information systems, management information systems.
    \item \textbf{Tasks in fairness literature}: fair data generation \citep{liu2021rawlsnet}.
    \item \textbf{Data spec}: tabular data.
    \item \textbf{Sample size}: $\sim1K$ employees.
    \item \textbf{Year}: 2019.
    \item \textbf{Sensitive features}: gender.
    \item \textbf{Link}: \url{https://www.kaggle.com/datasets/pavansubhasht/ibm-hr-analytics-attrition-dataset}
    \item \textbf{Further info}: \url{https://github.com/IBM/employee-attrition-aif360}
\end{itemize}

\subsection{IIT-JEE}
\begin{itemize}
    \item \textbf{Description}: this dataset was released in response to a Right to Information application filed in June 2009, and contains country-wide results for the Joint Entrance Exam (EET) to Indian Institutes of Technology (IITs), a group of prestigious engineering schools in India. The dataset contains the marks obtained by every candidate who took the test in 2009, divided according to the specific Math, Physics, and Chemistry sections of the test. Demographics such as ZIP code, gender, and birth categories (ethnic categories relating to the caste system) are also included.
    \item \textbf{Affiliation of creators}: Indian Institute of Technology, Kharagpur.
    \item \textbf{Domain}: education.
    \item \textbf{Tasks in fairness literature}: fair ranking \citep{celis2020interventions}.
    \item \textbf{Data spec}: tabular data.
    \item \textbf{Sample size}: $\sim400$K students.
    \item \textbf{Year}: 2009.
    \item \textbf{Sensitive features}: gender, birth category.
    \item \textbf{Link}: not available
    \item \textbf{Further info}: \citet{celis2020interventions}
\end{itemize}

\subsection{IJB-A}
\begin{itemize}
    \item \textbf{Description}: the IARPA Janus Benchmark A (IJB-A) dataset was proposed as a face recognition benchmark with wide geographic representation and pose variation for subjects. It consists of \emph{in-the-wild} images and videos of 500 subjects, obtained through internet searches over Creative Commons licensed content. The subjects were manually specified by the creators of the dataset to ensure broad geographic representation. The tasks associated with the dataset are face identification and verification. The dataset curators also collected the subjects' skin color and gender, through an unspecified annotation procedure. Similar protected attributes (gender and Fitzpatrick skin type) were labelled by one author of \citet{Buolamwini2018gender}.
    \item \textbf{Affiliation of creators}: Noblis; National Institute of Standards and Technology (NIST); Intelligence Advanced Research Projects Activity (IARPA); Michigan State University.
    \item \textbf{Domain}: computer vision.
    \item \textbf{Tasks in fairness literature}: data bias evaluation \citep{Buolamwini2018gender}.
    \item \textbf{Data spec}: image.
    \item \textbf{Sample size}: $\sim6$K images of $\sim500$ subjects.
    \item \textbf{Year}: 2015.
    \item \textbf{Sensitive features}: gender, skin color.
    \item \textbf{Link}: \url{https://www.nist.gov/itl/iad/image-group/ijb-dataset-request-form}
    \item \textbf{Further info}: \citet{klare2015pushing}
\end{itemize}

\subsection{ILEA}
\begin{itemize}
    \item \textbf{Description}: this dataset was created by the Inner London Education Authority (ILEA) considering data from 140 British schools. It comprises the results of public examinations taken by students of age 16 over the period 1985--1987. These values are used as a measurement of school effectiveness, with emphasis on quality of education and equality of opportunity for students of different backgrounds and ethnicities. Student-level records report their sex and ethnicity, while school-level factors include the percentage of students eligible for free meals and the percentage of girls in each institute.
    \item \textbf{Affiliation of creators}: Inner London Education Authority (ILEA).
    \item \textbf{Domain}: education.
    \item \textbf{Tasks in fairness literature}: fair representation learning \citep{oneto2019learning,oneto2020exploiting}.
    \item \textbf{Data spec}: unknown.
    \item \textbf{Sample size}: $\sim30$K students from $140$ secondary schools.
    \item \textbf{Year}: unknown.
    \item \textbf{Sensitive features}: age, sex, ethnicity.
    \item \textbf{Link}: not available
    \item \textbf{Further info}: \citep{nuttall1989differential,goldstein1991multilevel}
\end{itemize}

\subsection{Image Embedding Association Test (iEAT)}
\begin{itemize}
    \item \textbf{Description}: the Image Embedding Association Test (iEAT) is a resource for quantifying biased associations between representations of social concepts and attributes in images. It mimics seminal work on biases in WEs \citep{caliksan2017semantics}, following the Implicit Association Test (IAT) from social psychology \citep{greenwald1998measuring}. The curators identified several combinations of target concepts (e.g. young) and attributes (e.g. pleasant), testing similarities between representations of these concepts learnt  by unsupervised computer vision models. For each attribute/concept they obtained a set of images from the IAT, the CIFAR-100 dataset or Google Image Search, which act as the source of images and the associated sensitive attribute labels.
    \item \textbf{Affiliation of creators}: Carnegie Mellon University; George Washington University.
    \item \textbf{Domain}: computer vision.
    \item \textbf{Tasks in fairness literature}: fairness evaluation of learnt representations \citep{steed2021image}.
    \item \textbf{Data spec}: image.
    \item \textbf{Sample size}: $\sim200$ image for 15 iEATs.
    \item \textbf{Year}: 2021.
    \item \textbf{Sensitive features}: religion, gender, age, race, sexual orientation, disability, skin tone, weight.
    \item \textbf{Link}: \url{https://github.com/ryansteed/ieat/tree/master/data}
    \item \textbf{Further info}: \citet{steed2021image}
\end{itemize}

\subsection{ImageNet}
\begin{itemize}
    \item \textbf{Description}: Imagenet is one of the most influential machine learning dataset of the 2010s. Much important work on computer vision, including early breakthroughs in deep learning has been sparked by ImageNet Large Scale Visual Recognition Challenge (ILSVRC), a competition held yearly from 2010 to 2017. The most used portion of ImageNet is indeed the data powering the classification task in ILSVRC 2012, featuring 1,000 classes, over 100 of which represent different dog breeds.     Recently, several problematic biases were found in the \texttt{person} subtree of ImageNet, tracing their causes and proposing approaches to remove them \citep{prabhu2020large,yang2020towards,crawford2021excavating}.
    \item \textbf{Affiliation of creators}: Princeton University.
    \item \textbf{Domain}: computer vision.
    \item \textbf{Tasks in fairness literature}: fair classification \citep{dwork2018decoupled}, 
    bias discovery \citep{amini2019uncovering},
    data bias evaluation \citep{yang2020towards},
    fair incremental learning \citep{zhao2020maintaining},
    fairness evaluation \citep{dwork2017decoupled}.
    \item \textbf{Data spec}: image.
    \item \textbf{Sample size}: $\sim14$M images depicting  $\sim20$K categories (synsets).
    \item \textbf{Year}: 2021.
    \item \textbf{Sensitive features}: people's gender and other sensitive annotations may be present in synsets from the person subtree.
    \item \textbf{Link}: \url{https://image-net.org/}
    \item \textbf{Further info}: \citet{deng2009imagenet,barocas2019fair,prabhu2020large,yang2020towards,crawford2021excavating}
\end{itemize}

\subsection{In-Situ}
\begin{itemize}
    \item \textbf{Description}: this dataset was curated to measure biases in named entity recognition algorithms, based on gender, race and religion of people represented by entities. The authors exploit census data to build a list of 123 names typical of men and women of different race and religion. Next, they extract 289 sentences mentioning people from the  CoNLL 2003 NER test data \citep{tjong-kim-sang2003:introduction}, itself derived from Reuters 1990s news stories. Finally, they substitute the unigram person entity from the CoNLL 2003 shared task with each of names obtained previously as specific to a demographic group.
    \item \textbf{Affiliation of creators}: Twitter.
    \item \textbf{Domain}: linguistics.
    \item \textbf{Tasks in fairness literature}: fairness evaluation in entity recognition \citep{mishra2020assessing}.
    \item \textbf{Data spec}: text.
    \item \textbf{Sample size}: $\sim50$K sentences.
    \item \textbf{Year}: 2020.
    \item \textbf{Sensitive features}: gender, race and religion.
    \item \textbf{Link}: \url{https://github.com/napsternxg/NER_bias}
    \item \textbf{Further info}: \citet{mishra2020assessing}
\end{itemize}

\subsection{iNaturalist Datasets}
\begin{itemize}
    \item \textbf{Description}: these datasets were curated as challenging real-world benchmarks for large-scale fine-grained visual classification and feature visually similar classes with large class imbalance. They consist of images of plants and animals from iNaturalist, a social network where nature enthusiasts share information and observations about biodiversity. There are four different releases of the dataset: 2017, 2018, 2019, and 2021. A subset of the images are also annotated with bounding boxes and have additional metadata such as where and when the images were captured.
    \item \textbf{Affiliation of creators}: California Institute of Technology; University of Edinburgh; Google; Cornell University; iNaturalist.
    \item \textbf{Domain}: biology.
    \item \textbf{Tasks in fairness literature}: fairness evaluation of private classification \citep{bagdasaryan2019differential}.
    \item \textbf{Data spec}: image.
    \item \textbf{Sample size}: $\sim3$M images from $\sim10$K different species of plants and animals.
    \item \textbf{Year}: 2021.
    \item \textbf{Sensitive features}: none.
    \item \textbf{Link}: \url{https://github.com/visipedia/inat_comp}
    \item \textbf{Further info}: \cite{vanhorn2018inaturalist,van2021benchmarking}
\end{itemize}

\subsection{Indian Census}
\begin{itemize}
    \item \textbf{Description}: very little information seems to be available on this dataset. It represents a count of residents of 35 Indian states, repeated every ten years between 1951 and 2001.
    \item \textbf{Affiliation of creators}: Office of the Registrar General of India.
    \item \textbf{Domain}: demography.
    \item \textbf{Tasks in fairness literature}: fairness evaluation of private resource allocation \citep{pujol2020fair}.
    \item \textbf{Data spec}: tabular data.
    \item \textbf{Sample size}: $\sim30$ state.
    \item \textbf{Year}: unknown.
    \item \textbf{Sensitive features}: geography.
    \item \textbf{Link}: \url{https://www.indiabudget.gov.in/budget_archive/es2006-07/chapt2007/tab97.pdf}
    \item \textbf{Further info}: 
\end{itemize}

\subsection{Indian Student Performance}
\begin{itemize}
    \item \textbf{Description}: this dataset was curated to support educational data mining algorithms. The creators collected data from three colleges of Assam, India (Duliajan College, Doomdooma College, and Digboi College). Each data point represents a student, summarizing information on their demographics (gender, caste), family (occupation and qualification of parents), and school fruition (study hours, attendance, home-to-school travel). Among the latter there are four variables summarizing student performance in different classes and examinations, which represent the response variable of a prediction task.
    \item \textbf{Affiliation of creators}: Dibrugarh University; Sana’a University; Abdelmalek Essaâdi University.
    \item \textbf{Domain}: education.
    \item \textbf{Tasks in fairness literature}: fair data summarization \citep{belitz2021automating}.
    \item \textbf{Data spec}: tabular data.
    \item \textbf{Sample size}: $\sim300$ students.
    \item \textbf{Year}: 2018.
    \item \textbf{Sensitive features}: gender, caste, geography.
    \item \textbf{Link}: \url{https://archive.ics.uci.edu/ml/datasets/Student+Academics+Performance}
    \item \textbf{Further info}: \citet{hussain2018educational}
\end{itemize}

\subsection{Infant Health and Development Program (IHDP)}
\begin{itemize}
    \item \textbf{Description}: this dataset is the result of the IHDP program carried out between 1985 and 1988 in the US. A longitudinal randomized trial was conducted to evaluate the effectiveness of comprehensive early intervention in reducing developmental and health problems in low birth weight premature infants. Families in the experimental group received an intervention based on an educational program delivered through home visits, a daily center-based program and a parent supporting group. Children in the study were assessed across multiple cognitive, behavioral, and health dimensions longitudinally in four phases at ages 3, 5, 8, and 18. The dataset also contains information on household composition, source of health care, parents' demographics and employment.
    \item \textbf{Affiliation of creators}: unknown.
    \item \textbf{Domain}: pediatrics.
    \item \textbf{Tasks in fairness literature}: fair risk assessment \citep{madras2019fairness,yi2019fair}.
    \item \textbf{Data spec}: mixture.
    \item \textbf{Sample size}: $\sim1$K infants.
    \item \textbf{Year}: 1993.
    \item \textbf{Sensitive features}: race and ethnicity (of parents), age (maternal), gender (of infant).
    \item \textbf{Link}: \url{https://www.icpsr.umich.edu/web/HMCA/studies/9795}
    \item \textbf{Further info}: \citet{brooks1992effects}
\end{itemize}

\subsection{Instagram Photos}
\begin{itemize}
    \item \textbf{Description}: this dataset was crawled from Instagram to explore trade-offs between fairness and revenue in platforms that serve ads to their users. The authors crawled metadata from photos (location and tags) and users (names), using Kevin Systrom as a seed user and cascading into profiles that like or comment photos. The curators concentrated on cities with enough geotagged data, namely New York and Los Angeles. Moreover, they labeled the users with gender and race. Gender was labeled via US social security data, using the proportion of babies with a given name registered with either gender. Gender was only assigned to users with a first name for which there were both at least 50 births and 95\% of recorded births were one gender. Race were labeled using the Face++ API on a subset of photos. Photos were not downloaded, rather they were fed to Face++ via their publicly available URL. Finally, the ground truth labels were validated by two research assistants. To emulate a location-based advertisement model, the creators devised a task aimed at predicting what topics a user will be interested in, given their locations from previous check-ins.
    \item \textbf{Affiliation of creators}: Columbia University.
    \item \textbf{Domain}: social media.
    \item \textbf{Tasks in fairness literature}: fair advertising \citep{riederer2017price}.
    \item \textbf{Data spec}: unknown.
    \item \textbf{Sample size}: $\sim1$M photos from $\sim40$K users.
    \item \textbf{Year}: 2017.
    \item \textbf{Sensitive features}: race, gender, geography.
    \item \textbf{Link}: not available
    \item \textbf{Further info}: \citet{riederer2017price}
\end{itemize}

\subsection{Internet Ads}
\begin{itemize}
    \item \textbf{Description}: this dataset was assembled to study the problem of automated advertisement removal in browsers. It consists of images crawled from randomly generated urls, manually classified as ad/no-ad. Image encodings are derived from raw html, thus containing no information about pixel values, but rather encoding width, height, anchor text and image source. The associated task is classifying each image encoding as an ad or a no-ad image.
    \item \textbf{Affiliation of creators}: University College Dublin.
    \item \textbf{Domain}: pattern recognition.
    \item \textbf{Tasks in fairness literature}: fair anomaly detection \citep{shekhar2021fairod}.
    \item \textbf{Data spec}: tabular data.
    \item \textbf{Sample size}: $\sim3$K image encodings.
    \item \textbf{Year}: 1998.
    \item \textbf{Sensitive features}: none.
    \item \textbf{Link}: \url{https://archive.ics.uci.edu/ml/datasets/internet+advertisements}
    \item \textbf{Further info}: \citet{kushmerick1999learning}
\end{itemize}

\subsection{Iris}
\begin{itemize}
    \item \textbf{Description}: the most popular dataset on the UCI Machine Learning Repository was created by E. Anderson and popularized by R.A. Fisher in the pattern recognition community in the 1930s. The measurements in this collection represent the length and width of sepal and petals of different Iris flowers, collected to evaluate the morphological variation of different Iris species. The typical learning task associated with this dataset is labelling the species based on the available measurements.
    \item \textbf{Affiliation of creators}: Missouri Botanical Garden; Washington University.
    \item \textbf{Domain}: plant science.
    \item \textbf{Tasks in fairness literature}: fair clustering \citep{chen2019proportionally,abbasi2021fair}.
    \item \textbf{Data spec}: tabular data.
    \item \textbf{Sample size}: $\sim100$ samples from three species of Iris.
    \item \textbf{Year}: 1988.
    \item \textbf{Sensitive features}: none.
    \item \textbf{Link}: \url{https://archive.ics.uci.edu/ml/datasets/iris}
    \item \textbf{Further info}: \citep{anderson1936species,fisher1936use}
\end{itemize}

\subsection{Italian Car Insurance}
\begin{itemize}
    \item \textbf{Description}: this resource was curated to study discriminatory practices in the Italian car insurance market. More specifically, the data was collected to estimate the direct effect of gender and birthplace on yearly quoted premiums. It was collected in 2020 from a popular Italian car insurance comparison website, where the curators tried different hypothetical driver profiles and collected the quotes provided by nine companies. Along with gender and birthplace, additional driver features include age, city of residence, insured vehicle, mileage, and a summary of claim history.
    \item \textbf{Affiliation of creators}: University of Padua; Carnegie Mellon University; University of Udine.
    \item \textbf{Domain}: economics.
    \item \textbf{Tasks in fairness literature}: fair pricing evaluation \citep{fabris2021algorithmic}.
    \item \textbf{Data spec}: tabular data.
    \item \textbf{Sample size}: $\sim2$K driver profiles.
    \item \textbf{Year}: 2021.
    \item \textbf{Sensitive features}: gender, birthplace.
    \item \textbf{Link}: not available
    \item \textbf{Further info}: \citet{fabris2021algorithmic}
\end{itemize}

\subsection{KDD Cup 99}
\begin{itemize}
    \item \textbf{Description}: this dataset was developed for a data mining competition on cybersecurity, focused on building an automated network intrusion detector based on TCP dump data. The task is predicting whether a connection is legitimate and inoffensive or symptomatic of an attack, such as denial-of-service or user-to-root; tens of attack classes have been simulated and annotated within this dataset. The available features include basic TCP/IP information, network traffic and contextual features, such as number of failed login attempts.
    \item \textbf{Affiliation of creators}: Massachusetts Institute of Technology.
    \item \textbf{Domain}: computer networks.
    \item \textbf{Tasks in fairness literature}: fair clustering \citep{chen2019proportionally}.
    \item \textbf{Data spec}: tabular data.
    \item \textbf{Sample size}: $\sim7$M connections.
    \item \textbf{Year}: 1999.
    \item \textbf{Sensitive features}: none.
    \item \textbf{Link}: \url{http://kdd.ics.uci.edu/databases/kddcup99/kddcup99.html}
    \item \textbf{Further info}: \citet{tavallaee2009detailed}
\end{itemize}

\subsection{Kidney Exchange Program}
\begin{itemize}
    \item \textbf{Description}: this dataset is based on data of the Canadian Kidney Paired Donation Program (KPD) to study strategic behavior among entities controlling part of the incompatible patient-donor pairs. Based on data from the Canadian Blood Services on the KPD and census, these instances were generated. The random instance generator is available upon request. The instances are weighted graphs. The incompatible patient-donor pairs represent the vertices of the graph, an arc means that the donor of a vertex is compatible with the patient of another vertex, and weights represent the benefit of the donation. Compatibility is encoded based on true blood type distribution and risk of transplant rejection.
    \item \textbf{Affiliation of creators}: Université de Montréal; Polytechnique de Montréal.
    \item \textbf{Domain}: public health.
    \item \textbf{Tasks in fairness literature}: fair matching evaluation \citep{farnadi2019enhancing}.
    \item \textbf{Data spec}: patient-donor pairs.
    \item \textbf{Sample size}: 180.
    \item \textbf{Year}: 2020.
    \item \textbf{Sensitive features}: blood type, geography.
    \item \textbf{Link}: \url{https://github.com/mxmmargarida/KEG}
    \item \textbf{Further info}: \citet{carvalho2019game}
\end{itemize}

\subsection{Kidney Matching}
\begin{itemize}
    \item \textbf{Description}: this dataset was created via a simulator based on real data provided by the Organ and Tissue Authority of Australia. The data was validated against additional information from the Australian Bureau of Statistics, the Public and Research sets, and Wikipedia. The simulator models the probability distribution over the Blood Type and State of donors and patients, along with the quality of a donated organ (summarized by Kidney Donor Patient Index) and of a patient (quantified by the Expected Post-Transplant Survival). The envisioned task for this data is optimal matching of organs and patients.
    \item \textbf{Affiliation of creators}: unknown.
    \item \textbf{Domain}: public health.
    \item \textbf{Tasks in fairness literature}: fairness matching evaluation \citep{mattei2018fairness}.
    \item \textbf{Data spec}: tabular data.
    \item \textbf{Sample size}: unknown.
    \item \textbf{Year}: 2018.
    \item \textbf{Sensitive features}: age, geography, blood type.
    \item \textbf{Link}: not available
    \item \textbf{Further info}: \citet{mattei2018axiomatic}
\end{itemize}

\subsection{Kiva}
\begin{itemize}
    \item \textbf{Description}: this dataset was obtained from \url{kiva.org}, a non-profit organization allowing low-income entrepreneurs and students to borrow money through loan crowdfunding. The data summarizes all transactions occurred in 2017. Transactions are typically between 25\$ to 50\$ and range from 5\$ to 10,000\$. Features include information about the loan, such as its purpose, sector and amount, and data specific to the borrower and their demographics. Women are prevalent in this dataset, probably due to the priorities of partner organizations and the easier access to capital enjoyed by men in many countries.
    \item \textbf{Affiliation of creators}: Kiva; DePaul University.
    \item \textbf{Domain}: finance.
    \item \textbf{Tasks in fairness literature}: fair ranking \citep{burke2018balanced,liu2018personalizing,sonboli2020and}, bias discovery \citep{sonboli2019localized}.
    \item \textbf{Data spec}: tabular data.
    \item \textbf{Sample size}: $\sim1$M transactions involving $\sim100$K loans and $\sim200$K users.
    \item \textbf{Year}: 2018.
    \item \textbf{Sensitive features}: gender, geography, activity.
    \item \textbf{Link}: not available
    \item \textbf{Further info}: \citet{sonboli2019localized}
\end{itemize}

\subsection{Labeled Faces in the Wild (LFW)}
\begin{itemize}
    \item \textbf{Description}: LFW is a public benchmark for face verification, maintained by researchers affiliated with the University of Massachusetts. It was built to measure the progress of face verification systems in unconstrained settings  (e.g. variable pose, illumination, resolution). The dataset consists of images of people who appeared in the news, labelled with the name of the respective individual. According to perception of human coders who were later asked to annotate this dataset, images mostly skew white, male and below 60.
    \item \textbf{Affiliation of creators}: University of Massachussets, Amherst; Stony Brook University.
    \item \textbf{Domain}: computer vision.
    \item \textbf{Tasks in fairness literature}: fair data summarization \citep{samadi2018price}, 
    fair clustering \citep{ghadiri2021socially}, 
    robust fairness evaluation \citep{black2021leaveoneout}, fairness evaluation \citep{segal2021fairness}.
    \item \textbf{Data spec}: image.
    \item \textbf{Sample size}: $\sim13$K face images of $\sim6$K individuals.
    \item \textbf{Year}: 2007.
    \item \textbf{Sensitive features}: gender, age, race.
    \item \textbf{Link}: \url{http://vis-www.cs.umass.edu/lfw/}
    \item \textbf{Further info}: \citet{huang2007labeled,han2014age,gebru2018datasheets}
\end{itemize}

\subsection{Large Movie Review}
\begin{itemize}
    \item \textbf{Description}: a set of reviews from IMDB, collected, filtered and preprocessed by researchers affiliated with Stanford University. Polarity judgements are balanced in terms of positive and negative reviews and automatically inferred from star-based ratings, so that 7 or more is positive, while 4 or less is considered negative. The dataset was collected to provide a large benchmark for sentiment analysis algorithms.
    \item \textbf{Affiliation of creators}: Stanford University.
    \item \textbf{Domain}: linguistics.
    \item \textbf{Tasks in fairness literature}: fair sentiment analysis evaluation \citep{liang2020artificial}.
    \item \textbf{Data spec}: text.
    \item \textbf{Sample size}: $\sim50$K reviews.
    \item \textbf{Year}: 2011.
    \item \textbf{Sensitive features}: textual references to people and their demographics.
    \item \textbf{Link}: \url{https://ai.stanford.edu/~amaas/data/sentiment/}
    \item \textbf{Further info}: \citet{maas2011learning}
\end{itemize}

\subsection{Last.fm}
\begin{itemize}
    \item \textbf{Description}: the Last.fm datasets were collected via the Last.fm API with the purpose of studying music consumption, discovery and recommendation on the web. Two datasets are provided: LFM1K, comprising timestamped listening habits of a limited user sample ($\sim$1K) at song granularity, and LFM360K, containing the top 50 most played artists of a wider user population ($\sim$360K).
    \item \textbf{Affiliation of creators}: Barcelona Music and Audio Technologies; Universitat Pompeu Fabra.
    \item \textbf{Domain}: music, information systems.
    \item \textbf{Tasks in fairness literature}: fair ranking evaluation \citep{ekstrand2018all}.
    \item \textbf{Data spec}: user-song pairs (LFM1K); user-artist pairs (LFM360K).
    \item \textbf{Sample size}: $\sim$19M timestamped records of $\sim$1K users playing songs from $\sim$170K artists (LFM1K); $\sim20$M play counts (user-artist pairs) for $\sim$400K users over $\sim$300K artists (LFM360K).
    \item \textbf{Year}: 2010.
    \item \textbf{Sensitive features}: user age, gender, geography; artist.
    \item \textbf{Link}: \url{http://ocelma.net/MusicRecommendationDataset/}
    \item \textbf{Further info}: \citet{Celma:Springer2010}
\end{itemize}

\subsection{Latin Newspapers}
\begin{itemize}
    \item \textbf{Description}: this dataset was built to study gender bias in language models and their connection with the corpora they have been trained on. It was built crawling articles from the websites of three newspapers from Chile, Peru, and Mexico. More detailed information about this resource seems to be missing.
    \item \textbf{Affiliation of creators}: Capital One.
    \item \textbf{Domain}: news.
    \item \textbf{Tasks in fairness literature}: data bias evaluation \citep{florez2019unintended}.
    \item \textbf{Data spec}: text.
    \item \textbf{Sample size}: $\sim60$K articles.
    \item \textbf{Year}: 2019.
    \item \textbf{Sensitive features}: textual references to people and their demographics.
    \item \textbf{Link}: not available
    \item \textbf{Further info}: \citet{florez2019unintended}
\end{itemize}

\subsection{Law School}
\begin{itemize}
    \item \textbf{Description}: This dataset was collected to study performance in law school and bar examination of minority examinees in connection with affirmative action programs established after 1967 and subsequent anecdotal reports suggesting low bar passage rates for black examinees. Students, law schools, and state boards of bar examiners contributed to this dataset. The study tracks students who entered law school in fall 1991 through three or more years of law school and up to five administrations of the bar examination. Variables include demographics of candidates (e.g. age, race, sex), their academic performance (undergraduate GPA, law school admission test, and GPA), personal condition (e.g. financial responsibility for others during law school)  along with information about law schools and bar exams (e.g. geographical area where it was taken). The associated task in machine learning is prediction of passage of the bar exam.
    \item \textbf{Affiliation of creators}: Law School Admission Council (LSAC).
    \item \textbf{Domain}: education.
    \item \textbf{Tasks in fairness literature}: fair classification \citep{yang2020fairness,cho2020fair,russell2017when,agarwal2018reductions,berk2017convex},
    rich-subgroup fairness evaluation \citep{kearns2019empirical}, 
    fair classification under unawareness \citep{lahoti2020fairness,lamy2019noisetolerant}, 
    fairness evaluation \citep{black2020fliptest,kusner2017counterfactual},
    fair regression \citep{chzhen2020fair,chzhen2020fairwassertein,agarwal2019fair,komiyama2018nonconvex},
    fair representation learning \citep{ruoss2020learning},
    robust fair classification \citep{mandal2020ensuring},
    limited-label fair classification \citep{wang2021fair}.
    \item \textbf{Data spec}: tabular data.
    \item \textbf{Sample size}: $\sim20$K examinees.
    \item \textbf{Year}: 1998.
    \item \textbf{Sensitive features}: sex, race, age.
    \item \textbf{Link}: not available
    \item \textbf{Further info}: \citet{wightman1998lsac}
\end{itemize}

\subsection{Libimseti}
\begin{itemize}
    \item \textbf{Description}: this dataset was collected to explore the effectiveness of recommendations in online dating services based on collaborative filtering. It was collected in collaboration with employees of the dating platform libimseti.cz, one of the largest Czech dating websites at the time. The data consists of anonymous ratings provided by (and to) users of the web service on a 10-point scale.
    \item \textbf{Affiliation of creators}: Charles University in Prague; Libimseti.
    \item \textbf{Domain}: sociology, information systems.
    \item \textbf{Tasks in fairness literature}: fair matching \citep{tziavelis2019equitable}.
    \item \textbf{Data spec}: user-user pairs.
    \item \textbf{Sample size}: $\sim$10M ratings over $\sim$200K users.
    \item \textbf{Year}: 2007.
    \item \textbf{Sensitive features}: gender.
    \item \textbf{Link}: \url{http://colfi.wz.cz/}
    \item \textbf{Further info}: \citet{brozovsky2006recommender,brozovsky2007recommender}
\end{itemize}

\subsection{Los Angeles City Attorney’s Office Records}
\begin{itemize}
    \item \textbf{Description}: this dataset was extracted from the Los Angeles City Attorney’s case management system. It consists of a collection of records aimed at powering data-driven approaches to decision making and resource allocation for misdemeanour recidivism reduction via individually tailored social service interventions. Focusing on cases handled by the office between 1995--2017, the data includes information about jail bookings, charges, court appearances, outcomes, and demographics.
    \item \textbf{Affiliation of creators}: Los Angeles City Attorney’s Office; University of Chicago.
    \item \textbf{Domain}: law.
    \item \textbf{Tasks in fairness literature}: fair classification \citep{rodolfa2020case}.
    \item \textbf{Data spec}: tabular data.
    \item \textbf{Sample size}: $\sim1$M unique individuals associated with $\sim2$M cases.
    \item \textbf{Year}: 2020.
    \item \textbf{Sensitive features}: race, ethnicity.
    \item \textbf{Link}: not available
    \item \textbf{Further info}: \citep{rodolfa2020case}
\end{itemize}

\subsection{MEPS-HC}
\begin{itemize}
    \item \textbf{Description}: the Medical Expenditure Panel Survey (MEPS) data is collected by the US Department of Health and Human Services, to survey healthcare spending and utilization by US citizens.  Overall, this is a set of large-scale surveys of families and individuals, their employers, and medical providers (e.g. doctors, hospitals, pharmacies). The Household Component (HC) focuses on households and individuals, who provide information about their demographics, medical conditions and expenses, health insurance coverage, and access to care. Individuals included in a panel undergo five rounds of interviews over two years. Healthcare expenditure is often regarded as a target variable in machine learning applications, where it has been used as a proxy for healthcare utilization, with the goal of identifying patients in need.
    \item \textbf{Affiliation of creators}: Agency for Healthcare Research and Quality.
    \item \textbf{Domain}: health policy.
    \item \textbf{Tasks in fairness literature}: fair transfer learning \citep{coston2019fair}, 
    fair regression \citep{romano2020achieving}, 
    fairness evaluation \citep{singh2019understanding}, robust fair classification \citep{biswas2021ensuring}, fair classification \citep{sharma2021fairn}.
    \item \textbf{Data spec}: tabular data.
    \item \textbf{Sample size}: $\sim30$K, variable on a yearly basis.
    \item \textbf{Year}: present.
    \item \textbf{Sensitive features}: gender, ethnicity, age.
    \item \textbf{Link}: \url{https://meps.ahrq.gov/mepsweb/data_stats/download_data_files.jsp}
    \item \textbf{Further info}: \url{https://www.ahrq.gov/data/meps.html}
\end{itemize}

\subsection{MGGG States}
\begin{itemize}
    \item \textbf{Description}: developed by the Metric Geometry and Gerrymandering Group\footnote{\url{https://mggg.org/}}, this dataset contains precinct-level aggregated information about demographics and political leaning of voters in each district. The data hinges on several distinct sources of data, including GIS mapping files from the US Census. Bureau\footnote{\url{https://www.census.gov/geographies/mapping-files.html}}, demographic data from IPUMS\footnote{\url{https://www.nhgis.org/}} and election data from MIT Election and Data Science \footnote{\url{https://electionlab.mit.edu/}}. Source and precise data format vary by state.
    \item \textbf{Affiliation of creators}: Tufts University.
    \item \textbf{Domain}: political science.
    \item \textbf{Tasks in fairness literature}: fair districting for electoral precincts \citep{schutzman2020:to}.
    \item \textbf{Data spec}: mixture.
    \item \textbf{Sample size}: variable number of precincts (thousands) per state.
    \item \textbf{Year}: 2021.
    \item \textbf{Sensitive features}: race, political affiliation (representation in different precincts).
    \item \textbf{Link}: \url{https://github.com/mggg-states}
    \item \textbf{Further info}: \url{https://mggg.org/}
\end{itemize}

\subsection{Microsoft Learning to Rank}
\begin{itemize}
    \item \textbf{Description}: this dataset was released to spur advances in learning to rank algorithms, capable of producing a list of documents in response to a text query, ranked according to their relevance for the query. The dataset contains relevance judgements for query-document pairs, obtained ``from a retired labeling set'' of the Bing search engine. Over 100 numerical features are provided for each query-document pair, summarizing the salient lexical properties of the pair and the quality of the webpage, including its page rank.
    \item \textbf{Affiliation of creators}: Microsoft.
    \item \textbf{Domain}: information systems.
    \item \textbf{Tasks in fairness literature}: fair ranking \citep{bower2021individually}.
    \item \textbf{Data spec}: query document pairs.
    \item \textbf{Sample size}: $\sim30$K queries.
    \item \textbf{Year}: 2013.
    \item \textbf{Sensitive features}: none.
    \item \textbf{Link}: \url{https://www.microsoft.com/en-us/research/project/mslr/}
    \item \textbf{Further info}: \citep{qin2013introducing}
\end{itemize}

\subsection{Million Playlist Dataset (MPD)}
\begin{itemize}
    \item \textbf{Description}: this dataset powered the 2018 RecSys Challenge on automatic playlist continuation. It consists of a sample of public Spotify playlists created by US Spotify users between 2010--2017. Each playlist consists of a title, track list and additional metadata. For each track, MPD provides the title, artist, album, duration and Spotify pointers. User data is anonymized. The dataset was augmented with record label information crawled from the web \citep{knees2019towards}.
    \item \textbf{Affiliation of creators}: Spotify; Johannes Kepler University; University of Massachusetts.
    \item \textbf{Domain}: music, information systems.
    \item \textbf{Tasks in fairness literature}: data bias evaluation \citep{knees2019towards}.
    \item \textbf{Data spec}: tabular data.
    \item \textbf{Sample size}: $\sim1$M playlists containing $\sim2$M unique tracks by $\sim300$K artists.
    \item \textbf{Year}: 2018.
    \item \textbf{Sensitive features}: artist, record label.
    \item \textbf{Link}: \url{https://www.aicrowd.com/challenges/spotify-million-playlist-dataset-challenge}
    \item \textbf{Further info}: \citet{chen2018recsys}
\end{itemize}

\subsection{Million Song Dataset (MSD)}
\begin{itemize}
    \item \textbf{Description}: this dataset was created as a large-scale benchmark for algorithms in the musical domain. Song data was acquired through The Echo Nest API, capturing a wide array of information about the song (duration, loudness, key, tempo, etc.) and the artist (name, id, location, etc.). In total the dataset creators retrieved one million songs, and for each song 55 fields are provided as metadata. This dataset also powers the Million Song Dataset Challenge, integrating the MSD with implicit feedback from taste profiles gather from an undisclosed set of applications.
    \item \textbf{Affiliation of creators}: Columbia University; The Echo Nest.
    \item \textbf{Domain}: music, information systems.
    \item \textbf{Tasks in fairness literature}: dynamical evaluation of fair ranking \citep{ferraro2019artist}.
    \item \textbf{Data spec}: user-song pairs.
    \item \textbf{Sample size}: $\sim50$M play counts over $\sim1$M users and $\sim400$K songs.
    \item \textbf{Year}: 2012.
    \item \textbf{Sensitive features}: artist; geography.
    \item \textbf{Link}: \url{http://millionsongdataset.com/}; \url{https://www.kaggle.com/c/msdchallenge}
    \item \textbf{Further info}: \citet{bertinmahieux2011million,mcfee2012milllion}
\end{itemize}

\subsection{MIMIC-CXR-JPG}
\begin{itemize}
    \item \textbf{Description}: this dataset was curated to encourage research in medical computer vision. It consists of  chest x-rays sourced from the Beth Israel Deaconess Medical Center between 2011--2016. Each image is tagged with one or more of fourteen labels, derived from the corresponding free-text radiology reports via natural language processing tools. A subset of 687 report-label pairs have been validated by a board of certified radiologists with 8 years of experience.
    \item \textbf{Affiliation of creators}: Massachusetts Institute of Technology; Beth Israel Deaconess Medical Center; Stanford University; Harvard Medical School; National Library of Medicine.
    \item \textbf{Domain}: radiology.
    \item \textbf{Tasks in fairness literature}: fairness evaluation of private classification \citep{cheng2021can}.
    \item \textbf{Data spec}: images.
    \item \textbf{Sample size}: $\sim400$K images of $\sim70$K patients.
    \item \textbf{Year}: 2019.
    \item \textbf{Sensitive features}: sex.
    \item \textbf{Link}: \url{https://physionet.org/content/mimic-cxr-jpg/2.0.0/}
    \item \textbf{Further info}: \citep{johnson2019mimic}
\end{itemize}

\subsection{MIMIC-III}
\begin{itemize}
    \item \textbf{Description}: this dataset was extracted from a database of patients admitted to critical care units at the Beth Israel Deaconess Medical Center in Boston (MA), following the widespread adoption of digital health records in US hospitals. Data comprises vital signs, medications, laboratory measurements, notes and observations by care providers, fluid balance, procedure codes, diagnostic codes, imaging reports, length of stay, survival data, and demographics. The dataset spans over a decade of intensive care unit stays for adult and neonatal patients.
    \item \textbf{Affiliation of creators}: Massachusetts Institute of Technology; Beth Israel Deaconess Medical Center; A*STAR.
    \item \textbf{Domain}: critical care medicine.
    \item \textbf{Tasks in fairness literature}: fair classification \citep{martinez2020minimax}, 
    fairness evaluation \citep{chen2018why,zhang2020hurtful},
    robust fair classification \citep{singh2021fairness}.
    \item \textbf{Data spec}: mixture.
    \item \textbf{Sample size}: $\sim$ 60K patients.
    \item \textbf{Year}: 2016.
    \item \textbf{Sensitive features}: age, ethnicity, gender.
    \item \textbf{Link}: \url{https://mimic.mit.edu/}
    \item \textbf{Further info}: \citet{johnson2016mimiciii}
\end{itemize}

\subsection{ML Fairness Gym}
\begin{itemize}
    \item \textbf{Description}: this resource was developed to study the long-term behaviour and emergent properties of fair ML systems. It is an extension of OpenAI Gym \citep{brockman2016openai}, simulating the actions of agents within environments as Markov Decision Processes. As of 2021, four environments have been released. (1) \emph{Lending} emulates the decisions of a bank, based on perceived credit-worthiness of individuals, which is distributed according to an artificial sensitive feature. (2) \emph{Attention allocation} concentrates on agents tasked with monitoring sites for incidents. (3) \emph{College admission} relates to sequential game theory, where agents represent colleges and environments contain students capable of strategically manipulating their features at different costs, for instance through preparation courses. (4) \emph{Infectious disease} models the problem of vaccine allocation and its long-term consequences on people in different demographic groups.
    \item \textbf{Affiliation of creators}: Google.
    \item \textbf{Domain}: N/A.
    \item \textbf{Tasks in fairness literature}: dynamical fair resource allocation \citep{atwood2019fair,damour2020fairness}, dynamical fair classification \citep{damour2020fairness}.
    \item \textbf{Data spec}: time series.
    \item \textbf{Sample size}: variable.
    \item \textbf{Year}: 2020.
    \item \textbf{Sensitive features}: synthetic.
    \item \textbf{Link}: \url{https://github.com/google/ml-fairness-gym}
    \item \textbf{Further info}: \citet{damour2020fairness}
\end{itemize}

\subsection{MNIST}
\begin{itemize}
    \item \textbf{Description}: one of the most famous resources in computer vision, this dataset was created from an earlier database released by the National Institute of Standards and Technology (NIST). It consists of hand-written digits collected among high-school students and Census Bureau employees, which have to be correctly labelled by image processing systems. Several augmentations have also been used in the fairness literature, discussed at the end of this section.
    \item \textbf{Affiliation of creators}: AT\&T Labs.
    \item \textbf{Domain}: computer vision.
    \item \textbf{Tasks in fairness literature}: fair clustering \citep{harpeled2019near,li2020deep}, 
    fair anomaly detection \citep{zhang2021towards},
    fair classification \citep{creager2021exchanging}, fairness evaluation \citep{segal2021fairness}.
    \item \textbf{Data spec}: image.
    \item \textbf{Sample size}: $\sim70$K images across 10 digits.
    \item \textbf{Year}: 1998.
    \item \textbf{Sensitive features}: none.
    \item \textbf{Link}: \url{http://yann.lecun.com/exdb/mnist/}
    \item \textbf{Further info}: \citet{lecun1998gradientbased,barocas2019fair}
    \item \textbf{Variants}: \begin{itemize}
        \item MNIST-USPS \citep{li2020deep}: merge with USPS dataset of handwritten digits \citep{hull1994:database}.
        \item Color-reverse MNIST \citep{li2020deep} or MNIST-Invert \citep{zhang2021towards}: images from MNIST, reversed via $p=255-p$ for each pixel $p$.
        \item Color MNIST \citep{arjovsky2020invariant}: images from MNIST colored red or green based on class label.
        \item C-MNIST: images from MNIST, such that both digits and background are colored.
    \end{itemize}.
\end{itemize}

\subsection{Mobile Money Loans}
\begin{itemize}
    \item \textbf{Description}: this dataset captures the ongoing collaboration between some banks and mobile network operators in East Africa. Phone data, including mobile money transactions, is used as ``soft'' financial data to create a credit score. Mobile money (bank-less) transactions represent a low-barrier tool for the financial inclusion of the poor and are fairly popular in some African countries.
    \item \textbf{Affiliation of creators}: unknown.
    \item \textbf{Domain}: finance.
    \item \textbf{Tasks in fairness literature}: fair transfer learning \citep{coston2019fair}.
    \item \textbf{Data spec}: tabular data.
    \item \textbf{Sample size}: $\sim200$K people.
    \item \textbf{Year}: unknown.
    \item \textbf{Sensitive features}: age, gender.
    \item \textbf{Link}: not available
    \item \textbf{Further info}: \citet{speakman2018:three}
\end{itemize}

\subsection{MovieLens}
\begin{itemize}
    \item \textbf{Description}: first released in 1998, MovieLens datasets represent user ratings from the movie recommender platform run by the GroupLens research group from the University of Minnesota. While different datasets have been released by GroupLens, in this section we concentrate on MovieLens 1M, the one predominantly used in fairness research. User-system interactions take the form of a quadruple (UserID, MovieID, Rating, Timestamp), with ratings expressed on a 1-5 star scale. The dataset also reports user demographics such as age and gender, which is voluntarily provided by the users.
    \item \textbf{Affiliation of creators}: University of Minnesota.
    \item \textbf{Domain}: information systems, movies.
    \item \textbf{Tasks in fairness literature}: fair ranking  \citep{burke2018balanced,sonboli2020and,dickens2020hyperfair,farnadi2018fairnessaware,liu2018personalizing}, 
    fair ranking evaluation \citep{ekstrand2018all,yao2017beyond,yao2017new},
    fair data summarization \citep{halabi2020fairness},
    fair representation learning \citep{oneto2020exploiting,oneto2019learning},
    fair graph mining \citep{buyl2020debayes,bose2019compositional},
    fair data generation \citep{burke2018synthetic}.
    \item \textbf{Data spec}: user-movie pairs.
    \item \textbf{Sample size}: $\sim1$M reviews by $\sim6$K users over $\sim4$K movies.
    \item \textbf{Year}: 2003.
    \item \textbf{Sensitive features}: gender, age.
    \item \textbf{Link}: \url{https://grouplens.org/datasets/movielens/1m/}
    \item \textbf{Further info}: \citet{harper2015movielens}
\end{itemize}

\subsection{MS-Celeb-1M}
\label{sec:msceleb}
\begin{itemize}
    \item \textbf{Description}: this dataset was created as a large scale public benchmark for face recognition. The creators cover a wide range of countries and emphasizes diversity echoing outdated notions of race: ``We cover all the major races in the world (Caucasian, Mongoloid, and Negroid)'' \citep{guo2016msceleb1m}.  While (in theory) containing only images of celebrities, the dataset was found to feature people who simply must maintain an online presence, and was retracted for this reason. Despite termination of the hosting website, the dataset is still searched for, available and used to build new fairness datasets, such as RFW (\autoref{sec:rfw}) and BUPT Faces (\autoref{sec:bupt}). The dataset was recently augmented with gender and nationality data automatically inferred from biographies of people \citep{mcduff2019characterizing}. From nationality, a race-related attribute was also annotated on a subset of 20,000 images.
    \item \textbf{Affiliation of creators}: Microsoft.
    \item \textbf{Domain}: computer vision.
    \item \textbf{Tasks in fairness literature}: fairness evaluation through artificial data generation \citep{mcduff2019characterizing}.
    \item \textbf{Data spec}: image.
    \item \textbf{Sample size}: $\sim10$M images representing $\sim100$K people.
    \item \textbf{Year}: 2016.
    \item \textbf{Sensitive features}: gender, race, geography.
    \item \textbf{Link}: not available
    \item \textbf{Further info}: \citet{guo2016msceleb1m,mcduff2019characterizing,murgia2019microsoft}
\end{itemize}

\subsection{MS-COCO}
\begin{itemize}
    \item \textbf{Description}: this dataset was created with the goal of improving the state of the art in object recognition. The dataset consists of over 300,000 labeled images collected from Flickr.  Each image was annotated based on whether it contains one or more of the 91 object types proposed by the authors. Segmentations are also provided to indicate the region where objects are located in each image. Finally, five human-generated captions are provided for each image. Annotation, segmentation and captioning were performed by human annotators hired on Amazon Mechanical Turk. A subset of the images depicting people have been  augmented with gender labels ``man'' and ``woman'' based on whether captions mention one word but not the other \citep{zhao2017men,hendricks2018women}.
    \item \textbf{Affiliation of creators}: Cornell University;  Toyota Technological Institute; Facebook; Microsoft; Brown University;  California Institute of Technology; University of California at Irvine.
    \item \textbf{Domain}: computer vision.
    \item \textbf{Tasks in fairness literature}: fair representation learning \citep{david2020debiasing}, fair classification \citep{hendricks2018women}.
    \item \textbf{Data spec}: image.
    \item \textbf{Sample size}: $\sim300$K images.
    \item \textbf{Year}: 2014.
    \item \textbf{Sensitive features}: gender.
    \item \textbf{Link}: \url{https://cocodataset.org/}
    \item \textbf{Further info}: \citet{lin2014microsoft}
\end{itemize}

\subsection{Multi-task Facial Landmark (MTFL)}
\begin{itemize}
    \item \textbf{Description}: this dataset was developed to evaluate the effectiveness of multi-task learning in problems of facial landmark detection. The dataset builds upon an existing collection of outdoor face images sourced from the web already labelled with bounding boxes and landmarks \citep{yi2013deep}, by annotating whether subjects are smiling or wearing glasses, along with their gender and pose. These annotations, whose provenance is not documented, allow researchers to define additional classification tasks for their multi-task learning pipeline.
    \item \textbf{Affiliation of creators}: The Chinese University of Hong Kong.
    \item \textbf{Domain}: computer vision.
    \item \textbf{Tasks in fairness literature}: fair clustering \citep{li2020deep}.
    \item \textbf{Data spec}: image.
    \item \textbf{Sample size}: $\sim10$K images.
    \item \textbf{Year}: 2014.
    \item \textbf{Sensitive features}: gender.
    \item \textbf{Link}: \url{http://mmlab.ie.cuhk.edu.hk/projects/TCDCN.html}
    \item \textbf{Further info}: \citet{zhang2014facial,zhang2015learning}
\end{itemize}

\subsection{National Longitudinal Survey of Youth}
\begin{itemize}
    \item \textbf{Description}: the National Longitudinal Surveys from the US Bureau of Labor Statistics follow the lives of representative samples of US citizens, focusing on their labor market activities and other significant life events. Subjects periodically provide responses to questions about their education, employment, housing, income, health, and more. Two different cohorts were started in 1979 (NLSY79) and (NLSY97), which have been associated with machine learning tasks of income prediction and GPA prediction respectively.
    \item \textbf{Affiliation of creators}: US Bureau of Labor Statistics.
    \item \textbf{Domain}: demography.
    \item \textbf{Tasks in fairness literature}: fair regression \citep{komiyama2018nonconvex,chzhen2020fairwassertein,chzhen2020fair}.
    \item \textbf{Data spec}: tabular data.
    \item \textbf{Sample size}: $\sim10$K respondents (NLSY79); $\sim9$K respondents (NLSY97).
    \item \textbf{Year}: present.
    \item \textbf{Sensitive features}: age, race, sex.
    \item \textbf{Link}: \url{https://www.bls.gov/nls/nlsy79.htm} (NLSY79); \url{https://www.bls.gov/nls/nlsy97.htm} (NLSY97)
    \item \textbf{Further info}: 
\end{itemize}

\subsection{National Lung Screening Trial (NLST)}
\begin{itemize}
    \item \textbf{Description}: the NLST was a randomized controlled trial aimed at understanding whether imaging through low-dose helical computed tomography reduces lung cancer mortality relative to chest radiography. Participants were recruited at 33 screening centers across the US, among subjects deemed at risk of lung cancer based on age and smoking history, and were made aware of the trial. A breadth of features about participants is available, including demographics, disease history, smoking history, family history of lung cancer, type, and results of screening exams.
    \item \textbf{Affiliation of creators}: National Cancer Institute's Division of Cancer Prevention, Division of Cancer Treatment and Diagnosis.
    \item \textbf{Domain}: radiology.
    \item \textbf{Tasks in fairness literature}: fair preference-based classification \citep{ustun2019fairness}.
    \item \textbf{Data spec}: image.
    \item \textbf{Sample size}: $\sim50$K participants.
    \item \textbf{Year}: 2020.
    \item \textbf{Sensitive features}: age, ethnicity, race, sex.
    \item \textbf{Link}: \url{https://cdas.cancer.gov/nlst/}
    \item \textbf{Further info}: \citet{national2011national}; \url{https://www.cancer.gov/types/lung/research/nlst}
\end{itemize}

\subsection{New York Times Annotated Corpus}
\begin{itemize}
    \item \textbf{Description}: this corpus contains nearly two million articles published in The New York Times over the period 1987--2007. For some articles, annotations by library scientists are available, including topics, mentioned entities, and summaries. The data is provided in News Industry Text Format (NITF).
    \item \textbf{Affiliation of creators}: The New York Times.
    \item \textbf{Domain}: news.
    \item \textbf{Tasks in fairness literature}: bias evaluation in WEs \citep{brunet2019understanding}.
    \item \textbf{Data spec}: text.
    \item \textbf{Sample size}: $\sim2$M articles.
    \item \textbf{Year}: 2008.
    \item \textbf{Sensitive features}: textual references to people and their demographics.
    \item \textbf{Link}: \url{https://catalog.ldc.upenn.edu/LDC2008T19}
    \item \textbf{Further info}: 
\end{itemize}

\subsection{Nominees Corpus}
\begin{itemize}
    \item \textbf{Description}: this corpus was curated to study gender-related differences in literary production, with attention to perception of quality. It consists of fifty Dutch-language fiction novels nominated for either the AKO Literatuurprijs(shortlist) or the Libris Literatuur Prijs (longlist) in the period 2007--2012. The corpus was curated to control for nominee gender and country of origin. Word counts, LIWC counts, and metadata for this dataset are available at \url{http://dx.doi.org/10.17632/tmp32v54ss.2}.
    \item \textbf{Affiliation of creators}: University of Amsterdam.
    \item \textbf{Domain}: literature.
    \item \textbf{Tasks in fairness literature}: fairness evaluation \citep{koolen2017:stereotypes}.
    \item \textbf{Data spec}: text.
    \item \textbf{Sample size}: $\sim50$ novels.
    \item \textbf{Year}: 2017.
    \item \textbf{Sensitive features}: gender, geography (of author).
    \item \textbf{Link}: not available
    \item \textbf{Further info}: \citet{koolen2017:stereotypes,koolen2018reading}
\end{itemize}

\subsection{North Carolina Voters}
\begin{itemize}
    \item \textbf{Description}: US voter data is collected, curated, and maintained for multiple reasons. Data about voters in North Carolina is collected publicly as part of voter registration requirements and also privately. Private companies curating these datasets sell voter data as part of products, which include outreach lists and analytics. These datasets include voters' full names, address, demographics, and party affiliation.
    \item \textbf{Affiliation of creators}: North Carolina State Board of Elections.
    \item \textbf{Domain}: political science.
    \item \textbf{Tasks in fairness literature}: data bias evaluation \citep{coston2021leveraging}, fair clustering \citep{abbasi2021fair}, fairness evaluation of advertisement \citep{speicher2018potential}.
    \item \textbf{Data spec}: tabular data.
    \item \textbf{Sample size}: $\sim8$M voters.
    \item \textbf{Year}: present.
    \item \textbf{Sensitive features}: race, ethnicity, age, geography.
    \item \textbf{Link}: \url{https://www.ncsbe.gov/results-data/voter-registration-data}
    \item \textbf{Further info}: 
    \item \textbf{Variants}: a privately curated version of this dataset is maintained by L2.\footnote{\url{https://l2-data.com/states/north-carolina/}}.
\end{itemize}

\subsection{Nursery}
\begin{itemize}
    \item \textbf{Description}: this dataset encodes applications for a nursery school in Ljubljana, Slovenia. To favour transparent and objective decision-making, a computer-based decision support system was developed for the selection and ranking of applications. The target variable reported is thus the output of an expert systems based on a set of rules, taking as an input information about the family, including housing, occupation and financial status, included in the dataset. The variables were reportedly constructed in a careful manner, taking into account laws that were in force at that time and following advice given by leading experts in that field. However, the variables also appear to be coded rather subjectively. For example, the variable \emph{social condition} admits as a value \emph{Slightly problematic}, allegedly reserved for ``When education ability of parents is low (unequal, inconsistent education, exaggerated pretentiousness or indulgence, neurotic reactions of parents), or there are improper relations in family (easier forms of parental personality disturbances, privileged or ignored children, conflicts in the family)''. Given that the true map between inputs and outputs is known, this resource is mostly useful to evaluate methods of structure discovery.
    \item \textbf{Affiliation of creators}: University of Maribor; Jožef Stefan Institute; University of Ljubljana; Center for Public Enterprises in Developing Countries.
    \item \textbf{Domain}: education.
    \item \textbf{Tasks in fairness literature}: fair classification \citep{romano2020achieving}.
    \item \textbf{Data spec}: tabular data.
    \item \textbf{Sample size}: $\sim10$K combinations of input data (hypothetical applicants).
    \item \textbf{Year}: 1997.
    \item \textbf{Sensitive features}: family wealth.
    \item \textbf{Link}: \url{https://archive.ics.uci.edu/ml/datasets/nursery}
    \item \textbf{Further info}: \citet{olave1989application}
\end{itemize}

\subsection{NYC Taxi Trips}
\begin{itemize}
    \item \textbf{Description}: this dataset was collected through a Freedom of Information Law request from the NYC Taxi and Limousine Commission. Data points represent New York taxi trips over 4 years (2010--2013), complete with spatio-temporal data, trip duration, number of passengers, and cost. Reportedly, the dataset contains a large number of errors, including misreported trip distance, duration, and GPS coordinates. Overall, these errors account for 7\% of all trips in the dataset.
    \item \textbf{Affiliation of creators}: University of Illinois.
    \item \textbf{Domain}: transportation.
    \item \textbf{Tasks in fairness literature}: fair matching \citep{lesmana2019balancing,nanda2020:bt}.
    \item \textbf{Data spec}: tabular data.
    \item \textbf{Sample size}: $\sim700$M taxi trips.
    \item \textbf{Year}: 2016.
    \item \textbf{Sensitive features}: none.
    \item \textbf{Link}: \url{https://experts.illinois.edu/en/datasets/new-york-city-taxi-trip-data-2010-2013-2}
    \item \textbf{Further info}: \url{https://bit.ly/3yrT8jt}
    \item \textbf{Variants}: a similar, smaller dataset was obtained by Chris Whong from the NYC Taxi and Limousine Commission under the Freedom of Information Law.\footnote{\url{http://www.andresmh.com/nyctaxitrips/}}.
\end{itemize}

\subsection{Occupations in Google Images}
\begin{itemize}
    \item \textbf{Description}: this dataset was collected to study gender and skin tone diversity in image search results for jobs, and its relation with gender and race conentration in different professions. The dataset consists of the top 100 results for 96 occupations from  Google Image Search, collected in December 2019. The creators hired workers on  Amazon Mechanical Turk to label the gender (male, female) and Fitzpatrick skin tone (Type 1--6) of the primary person in each image, adding ``Not applicable'' and ``Cannot determine'' as possible options. Three labels were collected for each image, to which the majority label was assigned where possible.
    \item \textbf{Affiliation of creators}: Yale Universiy.
    \item \textbf{Domain}: information systems.
    \item \textbf{Tasks in fairness literature}: fair subset selection under unawareness \citep{mehrotra2021mitigating}.
    \item \textbf{Data spec}: image.
    \item \textbf{Sample size}: $\sim10$K images of $\sim$100 occupations.
    \item \textbf{Year}: 2019.
    \item \textbf{Sensitive features}: gender, skin tone (inferred).
    \item \textbf{Link}: \url{https://drive.google.com/drive/u/0/folders/1j9I5ESc-7NRCZ-zSD0C6LHjeNp42RjkJ}
    \item \textbf{Further info}: \citet{celis2020implicit}
\end{itemize}

\subsection{Office31}
\begin{itemize}
    \item \textbf{Description}: this dataset was curated to support domain adaptation algorithms for computer vision systems. It features images of 31 different office tools (e.g. chair, keyboard, printer) from 3 different domains: listings on Amazon, high quality camera images, low quality webcam shots.
    \item \textbf{Affiliation of creators}: University of California, Berkeley.
    \item \textbf{Domain}: computer vision.
    \item \textbf{Tasks in fairness literature}: fair clustering \citep{li2020deep}.
    \item \textbf{Data spec}: image.
    \item \textbf{Sample size}: $\sim4$K images.
    \item \textbf{Year}: 2011.
    \item \textbf{Sensitive features}: none.
    \item \textbf{Link}: \url{https://paperswithcode.com/dataset/office-31}
    \item \textbf{Further info}: \citet{saenko2010adapting}
\end{itemize}

\subsection{Olympic Athletes}
\begin{itemize}
    \item \textbf{Description}: this is a historical sports-related dataset on the modern Olympic Games from their first edition in 1896 to the 2016 Rio Games. The dataset was consolidated by Randi H Griffin utilizing SportsReference as the primary source of information. For each athlete, the dataset comprises demographics, height, weight, competition, and medal.
    \item \textbf{Affiliation of creators}: unknown.
    \item \textbf{Domain}: sports.
    \item \textbf{Tasks in fairness literature}: fair clustering \citep{huang2019coresets}.
    \item \textbf{Data spec}: tabular data.
    \item \textbf{Sample size}: $\sim300$K athletes.
    \item \textbf{Year}: 2018.
    \item \textbf{Sensitive features}: sex, age.
    \item \textbf{Link}: \url{https://www.kaggle.com/heesoo37/120-years-of-olympic-history-athletes-and-results}
    \item \textbf{Further info}: \url{https://www.sports-reference.com/}
\end{itemize}

\subsection{Omniglot}
\begin{itemize}
    \item \textbf{Description}: this dataset was designed to study the problem of automatically learning basic visual concepts. It consists of handwritten characters from different alphabets drawn online via Amazon Mechanical Turk by 20 different people.
    \item \textbf{Affiliation of creators}: New York University; University of Toronto; Massachusetts Institute of Technology.
    \item \textbf{Domain}: computer vision.
    \item \textbf{Tasks in fairness literature}: fair few-shot learning \citep{li2020fair}.
    \item \textbf{Data spec}: image.
    \item \textbf{Sample size}: $\sim2$K images  from 50 different alphabets.
    \item \textbf{Year}: 2019.
    \item \textbf{Sensitive features}: none.
    \item \textbf{Link}: \url{https://github.com/brendenlake/omniglot}
    \item \textbf{Further info}: \citet{lake2015humanlevel}
\end{itemize}

\subsection{One billion word benchmark}
\begin{itemize}
    \item \textbf{Description}: this dataset was proposed in 2014 as a benchmark for language models. The authors sourced English textual data from the EMNLP 6th workshop on Statistical Machine Translation\footnote{\url{http://statmt.org/wmt11/training-monolingual.tgz}}, more specifically the Monolingual language model training data, comprising a news crawl from 2007--2011 and data from the European Parliament website. Preprocessing includes removal of duplicate sentences, rare words (appearing less than 3 times) and mapping out-of-vocabulary words to the \texttt{$<$UNK$>$} token. The ELMo contextualized WEs \citep{peters2018:deep} were trained on this benchmark.
    \item \textbf{Affiliation of creators}: Google; University of Edinburgh; Cantab Research Ltd.
    \item \textbf{Domain}: linguistics.
    \item \textbf{Tasks in fairness literature}: data bias evaluation \citep{tan2019assessing}.
    \item \textbf{Data spec}: text.
    \item \textbf{Sample size}: $\sim800$M words.
    \item \textbf{Year}: 2014.
    \item \textbf{Sensitive features}: textual references to people and their demographics.
    \item \textbf{Link}: \url{https://opensource.google/projects/lm-benchmark}
    \item \textbf{Further info}: \citet{chelba2014billion}
\end{itemize}

\subsection{Online Freelance Marketplaces}
\begin{itemize}
    \item \textbf{Description}: this dataset was created to audit racial and gender biases on TaskRabbit and Fiverr, two popular online freelancing marketplaces. The dataset was built by crawling workers' profiles from both websites, including metadata, activities, and past job reviews. Profiles were later annotated with perceived demographics (gender and race) by Amazon Mechanical Turk based on profile images. On TaskRabbit, the authors executed search queries for all task categories in the 10 largest cities where the service is available, logging workers' ranking in search results. On Fiverr,  they concentrated on 9 tasks of diverse nature. The total number of queries that were issued on each platform, resulting in as many search result pages, is not explicitly stated.
    \item \textbf{Affiliation of creators}: Northeastern University, GESIS Leibniz Institute for the Social Sciences, University of Koblenz-Landau, ETH Zürich.
    \item \textbf{Domain}: information systems.
    \item \textbf{Tasks in fairness literature}: fairness evaluation \citep{hannak2017bias}.
    \item \textbf{Data spec}: query-result pairs.
    \item \textbf{Sample size}: $\sim10$K workers (Fiverr); $\sim4$K (TaskRabbit).
    \item \textbf{Year}: 2017.
    \item \textbf{Sensitive features}: gender, race.
    \item \textbf{Link}: not available
    \item \textbf{Further info}: \citet{hannak2017bias}
\end{itemize}

\subsection{Open Images Dataset}
\begin{itemize}
    \item \textbf{Description}: this dataset was curated to improve and measure the performance of computer vision algorithms. Images with CC-BY license were downloaded from Flickr, and further filtered to remove near-duplicates, inappropriate content, and images appearing elsewhere in the internet. Different versions of this dataset were released, progressively adding a wealth of information on these images, including labels, bounding boxes, segmentation masks, visual relationships, and localized narratives. Bounding boxes relate to 600 classes, including ``person'', which admits ``girl'', ``boy'', ``woman'', and ``man'' as a subclass. 
    Image-level labels are generated automatically and verified by humans, resulting in annotations for a subset of present and absent classes (positive and negative image-level labels). Based on the positive image-level labels, spatial annotations are produced by human annotators: bounding boxes \citep{kuznetsova2020open}, visual relationships \citep{kuznetsova2020open}, and validation+test segmentations are drawn fully manually \citep{benenson2019large}; while segmentations in train are drawn using an interactive algorithm \citep{benenson2019large}. Further, independent of any other annotations, rich localized dense image captions are collected by asking humans to provide detailed free-form image descriptions while they hover the mouse over the regions they describe (Localized Narratives \citep{pont2020connecting}).
    \item \textbf{Affiliation of creators}: Google.
    \item \textbf{Domain}: computer vision.
    \item \textbf{Tasks in fairness literature}: data bias evaluation \citep{schumann2021step}, fairness evaluation \citep{aka2021measuring}.
    \item \textbf{Data spec}: image.
    \item \textbf{Sample size}: $\sim9$M images.
    \item \textbf{Year}: 2020.
    \item \textbf{Sensitive features}: gender, age.
    \item \textbf{Link}: \url{https://storage.googleapis.com/openimages/web/index.html}
    \item \textbf{Further info}: \citep{kuznetsova2020open,schumann2021step,benenson2019large,pont2020connecting}
\end{itemize}

\subsection{Paper-Reviewer Matching}
\begin{itemize}
    \item \textbf{Description}: this dataset summarizes the peer review assignment process of 3 different conferences, namely one edition of Medical Imaging and Deep Learning (MIDL) and two editions of the Conference on Computer Vision and Pattern Recognition (called CVPR and CVPR2018). The data, provided by OpenReview and the Computer Vision Foundation, consist of a matrix of paper-reviewer affinities, a set of coverage constraints to ensure each paper is properly reviewed, and a set of upper bound constraints to avoid imposing an excessive burden on reviewers.
    \item \textbf{Affiliation of creators}: unknown.
    \item \textbf{Domain}: library and information sciences.
    \item \textbf{Tasks in fairness literature}: fair matching \citep{kobren2019paper}.
    \item \textbf{Data spec}: paper-reviewer pairs.
    \item \textbf{Sample size}: $\sim200$ reviewers for $\sim100$ papers (MIDL); $\sim1$K reviewers for $\sim3$K papers (CVPR). $\sim3$K reviewers for $\sim5$K papers (CVPR2018).
    \item \textbf{Year}: 2019.
    \item \textbf{Sensitive features}: none.
    \item \textbf{Link}: not available
    \item \textbf{Further info}: \citet{kobren2019paper}
\end{itemize}

\subsection{Philadelphia Crime Incidents}
\begin{itemize}
    \item \textbf{Description}: this dataset is provided as part of OpenDataPhilly initiative. It summarizes hundreds of thousands of crime incidents handled by the Philadelphia Police Department over a period of ten years (2006--2016). The dataset comes with fine spatial and temporal granularity and has been used to monitor seasonal and historical trends and measure the effect of police strategies.
    \item \textbf{Affiliation of creators}: Philadelphia Police Department.
    \item \textbf{Domain}: law.
    \item \textbf{Tasks in fairness literature}: fair resource allocation \citep{elzayn2019fair}.
    \item \textbf{Data spec}: tabular data.
    \item \textbf{Sample size}: $\sim1$M crime incidents.
    \item \textbf{Year}: present.
    \item \textbf{Sensitive features}: geography.
    \item \textbf{Link}: \url{https://www.opendataphilly.org/dataset/crime-incidents}
    \item \textbf{Further info}: 
\end{itemize}

\subsection{Pilot Parliaments Benchmark (PPB)}
\begin{itemize}
    \item \textbf{Description}: this dataset was developed as a benchmark with a balanced representation of gender and skin type to evaluate the performance of face analysis technology. The dataset features images of parliamentary representatives from three African countries (Rwanda, Senegal, South Africa) and three European countries (Iceland, Finland, Sweden) to achieve a good balance between skin type and gender while reducing potential harms connected with lack of consent from the people involved. Three annotators provided gender and Fitzpatrick labels. A certified surgical dermatologist provided the definitive Fitzpatrick skin type labels. Gender was annotated based on name, gendered title, and photo appearance.
    \item \textbf{Affiliation of creators}: Massachusetts Institute of Technology; Microsoft.
    \item \textbf{Domain}: computer vision.
    \item \textbf{Tasks in fairness literature}: fair classification \citep{kim2019multiaccuracy,amini2019uncovering},
    fairness evaluation \citep{Buolamwini2018gender,raji2019actionable},
    bias discovery \citep{kim2019multiaccuracy,amini2019uncovering}.
    \item \textbf{Data spec}: image.
    \item \textbf{Sample size}: $\sim1$K images of $\sim1$K individuals.
    \item \textbf{Year}: 2018.
    \item \textbf{Sensitive features}: gender, skin type.
    \item \textbf{Link}: \url{http://gendershades.org/}
    \item \textbf{Further info}: \citet{Buolamwini2018gender}
\end{itemize}

\subsection{Pima Indians Diabetes Dataset (PIDD)}
\begin{itemize}
    \item \textbf{Description}: this resource owes its name to the respective entry on the UCI repository (now unavailable), and was derived from a medical study of Native Americans from the Gila River Community, often called Pima. The study was initiated in the 1960s by the National Institute of Diabetes and Digestive and Kidney Diseases and found a large prevalence of \emph{diabetes mellitus} in this population. The dataset commonly available nowadays represents a subset of the original study, focusing on women of age 21 or older. It reports whether they tested positive for diabetes, along with eight covariates that were found to be significant risk factors for this population. These include the number of pregnancies, skin thickness, and body mass index, based on which algorithms should predict the test results.
    \item \textbf{Affiliation of creators}: Logistics Management Institute; National Institute of Diabetes Digestive and Kidney Diseases; John Hopkins University.
    \item \textbf{Domain}: endocrinology.
    \item \textbf{Tasks in fairness literature}: fairness evaluation \citep{sharma2020:cc}, fair clustering \citep{chen2019proportionally}.
    \item \textbf{Data spec}: tabular data.
    \item \textbf{Sample size}: $\sim800$ subjects.
    \item \textbf{Year}: 2016.
    \item \textbf{Sensitive features}: age.
    \item \textbf{Link}: \url{https://www.kaggle.com/uciml/pima-indians-diabetes-database}
    \item \textbf{Further info}: \citet{smith1988using,radin2017digital}
\end{itemize}

\subsection{Pokec Social Network}
\begin{itemize}
    \item \textbf{Description}: this graph dataset summarizes the networks of Pokec users, a social network service popular in Slovakia and Czech Republic. Due to default privacy settings being predefined as public, a wealth of information for each profile was collected by curators including information on demographics, politics, education, marital status, and children wherever available. This resource was collected to perform data analysis in social networks.
    \item \textbf{Affiliation of creators}: University of Zilina.
    \item \textbf{Domain}: social networks.
    \item \textbf{Tasks in fairness literature}: fair data summarization \citep{halabi2020fairness}.
    \item \textbf{Data spec}: user-user pairs.
    \item \textbf{Sample size}: $\sim2$M nodes (profiles) connected by $\sim30$M edges (friendship relations).
    \item \textbf{Year}: 2013.
    \item \textbf{Sensitive features}: gender, geography, age.
    \item \textbf{Link}: \url{https://snap.stanford.edu/data/soc-pokec.html}
    \item \textbf{Further info}: \citet{takac2012data}
\end{itemize}

\subsection{Popular Baby Names}
\label{sec:names}
\begin{itemize}
    \item \textbf{Description}: this dataset summarizes birth registration in New York City, focusing on names sex and race of newborns, providing a reliable source of data to assess naming trends in New York. A similar nation-wide database is maintained by the US Social Security Administration.
    \item \textbf{Affiliation of creators}: City of New York, Department of Health and Mental Hygiene (NYC names); United States Social Security Administration (US names).
    \item \textbf{Domain}: linguistics.
    \item \textbf{Tasks in fairness literature}: fair sentiment analysis \citep{yurochkin2020training,mukherjee2020two}, bias discovery in WEs \citep{swinger2019what}.
    \item \textbf{Data spec}: tabular data.
    \item \textbf{Sample size}: $\sim3$K unique names (NYC names); $\sim30$K unique names (US names).
    \item \textbf{Year}: 2021.
    \item \textbf{Sensitive features}: sex, race.
    \item \textbf{Link}: \url{https://catalog.data.gov/dataset/popular-baby-names} (NYC names); \url{https://www.ssa.gov/oact/babynames/limits.html} (US names)
    \item \textbf{Further info}: 
\end{itemize}

\subsection{Poverty in Colombia}
\begin{itemize}
    \item \textbf{Description}: this dataset stems from an official survey of households performed yearly by the Colombian national statistics department (Departamento Administrativo Nacional de Estadística). The survey is aimed at soliciting information about employment, income, and demographics. The data serves as an input for studies on poverty in Colombia.
    \item \textbf{Affiliation of creators}: Departamento Administrativo Nacional de Estadística.
    \item \textbf{Domain}: economics.
    \item \textbf{Tasks in fairness literature}: fair classification \citep{noriegacampero2020algorithmic}.
    \item \textbf{Data spec}: tabular data.
    \item \textbf{Sample size}: unknown.
    \item \textbf{Year}: 2018.
    \item \textbf{Sensitive features}: age, sex, geography.
    \item \textbf{Link}: \url{https://www.dane.gov.co/index.php/estadisticas-por-tema/pobreza-y-condiciones-de-vida/pobreza-y-desigualdad/pobreza-monetaria-y-multidimensional-en-colombia-2018}
    \item \textbf{Further info}: \url{https://www.dane.gov.co/files/investigaciones/condiciones_vida/pobreza/2018/bt_pobreza_monetaria_18.pdf}
\end{itemize}

\subsection{PP-Pathways}
\begin{itemize}
    \item \textbf{Description}: this dataset represents a network of physical interactions between proteins that are experimentally documented in humans. The dataset was assembled to study the problem of automated discovery of the proteins (nodes) associated with a given disease. Starting from a few known disease-associated proteins and a a map of protein-protein interactions (edges), the task is to find the full list of proteins associated with said disease.
    \item \textbf{Affiliation of creators}: Stanford University; Chan Zuckerberg Biohub.
    \item \textbf{Domain}: biology.
    \item \textbf{Tasks in fairness literature}: fair graph mining \citep{kang2020inform}.
    \item \textbf{Data spec}: protein-protein pairs.
    \item \textbf{Sample size}: $\sim20$K proteins (nodes) linked by $\sim300$K physical interactions.
    \item \textbf{Year}: 2018.
    \item \textbf{Sensitive features}: none.
    \item \textbf{Link}: \url{http://snap.stanford.edu/biodata/datasets/10000/10000-PP-Pathways.html}
    \item \textbf{Further info}: \citet{agrawal2018large}
\end{itemize}

\subsection{Prosper Loans Network}
\begin{itemize}
    \item \textbf{Description}: this dataset represents transactions on the Prosper marketplace, a famous peer-to-peer lending service where US-based users can register as lenders or borrowers. This resource has a graph structure and covers the period 2005--2011. Loan records include user ids, timestamps, loan amount, and rate. The dataset was first associated with a study of arbitrage and its profitability in a peer-to-peer lending system.
    \item \textbf{Affiliation of creators}: Prosper; University College Dublin.
    \item \textbf{Domain}: finance.
    \item \textbf{Tasks in fairness literature}: fair classification \citep{li2020fairness}.
    \item \textbf{Data spec}: lender-borrower pairs.
    \item \textbf{Sample size}: $\sim3$M loan records involving $\sim100$K people.
    \item \textbf{Year}: 2015.
    \item \textbf{Sensitive features}: none.
    \item \textbf{Link}: \url{http://mlg.ucd.ie/datasets/prosper.html}
    \item \textbf{Further info}: \citet{redmond2013temporal}
\end{itemize}

\subsection{PubMed Diabetes Papers}
\begin{itemize}
    \item \textbf{Description}: this dataset was created to study the problem of classification of connected entities via active learning. The creators extracted a set of articles related to diabetes from PubMed, along with their citation network. The task associated with the dataset is inferring a label specifying the type of diabetes addressed in each publication. For this task, TF/IDF-weighted term frequencies of every article are available.
    \item \textbf{Affiliation of creators}: University of Maryland.
    \item \textbf{Domain}: library and information sciences.
    \item \textbf{Tasks in fairness literature}: fair graph mining \citep{li2021on}.
    \item \textbf{Data spec}: article-article pairs.
    \item \textbf{Sample size}: $\sim20$K articles connected by $\sim40$K citations.
    \item \textbf{Year}: 2020.
    \item \textbf{Sensitive features}: none.
    \item \textbf{Link}: \url{https://linqs.soe.ucsc.edu/data}
    \item \textbf{Further info}: \citet{namata2012query}
\end{itemize}

\subsection{Pymetrics Bias Group}
\begin{itemize}
    \item \textbf{Description}: Pymetrics is a company that offers a candidate screening tool to employers. Candidates play a core set of twelve games, derived from psychological studies. The resulting gamified psychological measurements are exploited to build predictive models for hiring, where positive examples are provided by high-performing employees from the employer. Pymetrics staff maintain a \emph{Pymetrics Bias Group} dataset for internal fairness audits by asking players to fill in an optional demographic survey after they complete the games.
    \item \textbf{Affiliation of creators}: Pymetrics.
    \item \textbf{Domain}: information systems, management information systems.
    \item \textbf{Tasks in fairness literature}: fairness evaluation \citep{wilson2021building}.
    \item \textbf{Data spec}: tabular data.
    \item \textbf{Sample size}: $\sim10$K users.
    \item \textbf{Year}: 2021.
    \item \textbf{Sensitive features}: gender, race.
    \item \textbf{Link}: not available
    \item \textbf{Further info}: \citet{wilson2021building}
\end{itemize}

\subsection{Race on Twitter}
\begin{itemize}
    \item \textbf{Description}: this dataset was collected to power applications of user-level race prediction on Twitter. Twitter users were hired through Qualtrics, were they filled in a survey providing their Twitter handle and demographics, including race, gender, age, education, and income. The dataset creators downloaded the most recent 3,200 tweets by the users who provided their handle. The data, allegedly released in an anonymized and aggregated format, appears to be unavailable.
    \item \textbf{Affiliation of creators}: University of Pennsylvania.
    \item \textbf{Domain}: social media.
    \item \textbf{Tasks in fairness literature}: fairness evaluation \citep{ballburack2021differential}.
    \item \textbf{Data spec}: text.
    \item \textbf{Sample size}: $\sim5$M tweets from $\sim4$K users.
    \item \textbf{Year}: 2018.
    \item \textbf{Sensitive features}: race, gender, age.
    \item \textbf{Link}: \url{http://www.preotiuc.ro/}
    \item \textbf{Further info}: \citet{preotiucpietro2018user}
\end{itemize}

\subsection{Racial Faces in the Wild (RFW)}
\label{sec:rfw}
\begin{itemize}
    \item \textbf{Description}: this dataset was developed as a benchmark for face verification algorithms operating on diverse populations. The dataset comprises 4 clusters of images extracted from MS-Celeb-1M (\autoref{sec:msceleb}), a dataset that was discontinued by Microsoft due to privacy violations. Clusters are of similar size and contain individuals labelled Caucasian, Asian, Indian and African. Half of the labels (Asian, Indian) are derived from the ``Nationality attribute of FreeBase celebrities''; the remaining half (Caucasian, African) is automatically estimated via the Face++ API. This attribute is referred to as ``race'' by the authors, who also assert ``carefully and manually'' cleaning every image. Clusters feature multiple images of each individual to allow for face verification applications.
    \item \textbf{Affiliation of creators}: Beijing University of Posts; Telecommunications and Canon Information Technology (Beijing).
    \item \textbf{Domain}: computer vision.
    \item \textbf{Tasks in fairness literature}: fair reinforcement learning \citep{wang2020mitigating}, fair representation learning \citep{gong2021mitigating}.
    \item \textbf{Data spec}: image.
    \item \textbf{Sample size}: $\sim50$K images of $\sim10$K individuals.
    \item \textbf{Year}: 2019.
    \item \textbf{Sensitive features}: race (inferrred).
    \item \textbf{Link}: \url{http://www.whdeng.cn/RFW/testing.html}
    \item \textbf{Further info}: \citet{wang2019racial}
\end{itemize}

\subsection{Real-Time Crime Forecasting Challenge}
\begin{itemize}
    \item \textbf{Description}: this dataset was assembled and released by the US National Institute of Justice in 2017 with the goal of advancing the state of automated crime forecasting. It consists of calls-for-service (CFS) records provided by the Portland Police Bureau for the period 2012--2017. Each CFS record contains spatio-temporal data and crime-related categories. The dataset was released as part of a challenge with a toal prize of 1,200,000\$.
    \item \textbf{Affiliation of creators}: National Institute of Justice.
    \item \textbf{Domain}: law.
    \item \textbf{Tasks in fairness literature}: fair spatio-temporal process learning \citep{shang2020listwise}.
    \item \textbf{Data spec}: tabular data.
    \item \textbf{Sample size}: $\sim$ 700K CFS records.
    \item \textbf{Year}: 2017.
    \item \textbf{Sensitive features}: geography.
    \item \textbf{Link}: \url{https://nij.ojp.gov/funding/real-time-crime-forecasting-challenge-posting#data}
    \item \textbf{Further info}: \citet{conduent2018:real}
\end{itemize}

\subsection{Recidivism of Felons on Probation}
\begin{itemize}
    \item \textbf{Description}: this dataset covers probation cases of persons who were sentenced in 1986 in 32 urban and suburban US jurisdictions. It was assembled to study the behaviour of individuals on probation and their compliance with court orders across states. Possible outcomes include successful discharge, new felony rearrest, and absconding. The information on probation cases was frequently obtained through manual reviews and transcription of probation files, mostly by college students. Variables include probationer's demographics, educational level, wage, history of convictions, disciplinary hearings and probation sentences. The final dataset consists of $\sim10$K probation cases ``representative of 79,043 probationers''.
    \item \textbf{Affiliation of creators}: US Department of Justice; National Association of Criminal Justice Planners.
    \item \textbf{Domain}: law.
    \item \textbf{Tasks in fairness literature}: limited-label fair classification \citep{wang2020augmented}.
    \item \textbf{Data spec}: tabular data.
    \item \textbf{Sample size}: $\sim10$K probation cases.
    \item \textbf{Year}: 2005.
    \item \textbf{Sensitive features}: sex, race, ethnicity, age.
    \item \textbf{Link}: \url{https://www.icpsr.umich.edu/web/NACJD/studies/9574}
    \item \textbf{Further info}: \url{https://bjs.ojp.gov/data-collection/recidivism-survey-felons-probation}
\end{itemize}

\subsection{Reddit Comments}
\begin{itemize}
    \item \textbf{Description}: this resource consists of Reddit comments and relative metadata, crawled and made available online for research purposes. While the available dumps cover the period 2006-2021, below the ``sample size'' field refers to comments from 2014 used in one surveyed work.
    \item \textbf{Affiliation of creators}: Pushshift data.
    \item \textbf{Domain}: social media, linguistics.
    \item \textbf{Tasks in fairness literature}: bias evaluation in language models \citep{guo2021detecting}.
    \item \textbf{Data spec}: text.
    \item \textbf{Sample size}: $\sim500$M comments.
    \item \textbf{Year}: 2021.
    \item \textbf{Sensitive features}: textual references to people and their demographics.
    \item \textbf{Link}: \url{https://files.pushshift.io/reddit/comments/}
    \item \textbf{Further info}: \citet{guo2021detecting}
\end{itemize}

\subsection{Renal Failure}
\begin{itemize}
    \item \textbf{Description}: the dataset was created to compare the performance of two different algorithms for automated renal failure risk assessment. Considering patients who received care at NYU Langone Medical Center, each entry encodes their health records, demographics, disease history, and lab results. The final version of the dataset has a cutoff date, considering only patients who did not have kidney failure by that time, and reporting, as a target ground truth, whether they proceeded to have kidney failure within the next year.
    \item \textbf{Affiliation of creators}: New York University; New York University Langone Medical Center.
    \item \textbf{Domain}: nephrology.
    \item \textbf{Tasks in fairness literature}: fairness evaluation \citep{williams2019quantification}.
    \item \textbf{Data spec}: tabular data.
    \item \textbf{Sample size}: $\sim2$M patients.
    \item \textbf{Year}: 2019.
    \item \textbf{Sensitive features}: age, gender, race.
    \item \textbf{Link}: not available
    \item \textbf{Further info}: \citet{williams2019quantification}
\end{itemize}

\subsection{Reuters 50 50}
\begin{itemize}
    \item \textbf{Description}: this dataset was extracted from the Reuters Corpus Volume 1 (RCV1), a large corpus of newswire stories, to study the problem of authorship attribution. The 50 most prolific authors were selected from RCV1, considering only texts labeled corporate/industrial. The dataset consists of short news stories from these authors, labelled with the name of the author.
    \item \textbf{Affiliation of creators}: University of the Aegean.
    \item \textbf{Domain}: news.
    \item \textbf{Tasks in fairness literature}: fair clustering \citep{harb2020kfc}.
    \item \textbf{Data spec}: text.
    \item \textbf{Sample size}: $\sim5$K articles.
    \item \textbf{Year}: 2011.
    \item \textbf{Sensitive features}: author, textual references to people and their demographics.
    \item \textbf{Link}: \url{http://archive.ics.uci.edu/ml/datasets/Reuter_50_50}
    \item \textbf{Further info}: \citet{houvardas2006n}
\end{itemize}

\subsection{Ricci}
\begin{itemize}
    \item \textbf{Description}: this dataset relates to the US supreme court labor case on discrimination \emph{Ricci vs DeStefano} (2009), connected with the disparate impact doctrine. It represents 118 firefighter promotion tests, providing the scores and race of each test taker. Eighteen firefighters from the New Haven Fire Department claimed ``reverse discrimination'' after the city refused to certify a promotion examination where they had obtained high scores. The reasons why city officials avoided certifying the examination included concerns of potential violation of the `four-fiths' rule, as, given the vacancies at the time, no black firefighter would be promoted. The dataset was published and popularized by Weiwen Miao for pedagogical use.
    \item \textbf{Affiliation of creators}: Haverford College.
    \item \textbf{Domain}: law.
    \item \textbf{Tasks in fairness literature}: fairness evaluation \citep{feldman2015certifying,friedler2019comparative},
    limited-label fairness evaluation \citep{ji2020can}.
    \item \textbf{Data spec}: tabular data.
    \item \textbf{Sample size}: $\sim100$ test takers.
    \item \textbf{Year}: 2018.
    \item \textbf{Sensitive features}: race.
    \item \textbf{Link}: \url{http://jse.amstat.org/jse_data_archive.htm}; \url{https://github.com/algofairness/fairness-comparison/tree/master/fairness/data/raw}
    \item \textbf{Further info}: \citet{gastwirth2009formal,miao2010did}
\end{itemize}

\subsection{Rice Facebook Network}
\begin{itemize}
    \item \textbf{Description}: this dataset repesents the Facebook sub-network of students and alumni of Rice University. It consists of a crawl of reachable profiles in the Rice Facebook network, augmented with academic information obtained from Rice University directories. This collection was created to study the problem of inferring unknown attributes in a social network based on the network graph and attributes that are available for a fraction of users.
    \item \textbf{Affiliation of creators}: MPI-SWS; Rice University; Northeastern University.
    \item \textbf{Domain}: social networks.
    \item \textbf{Tasks in fairness literature}: fair graph diffusion \citep{ali2019fairness}.
    \item \textbf{Data spec}: user-user pairs.
    \item \textbf{Sample size}: $\sim1$K profiles connected by $40$K edges.
    \item \textbf{Year}: 2010.
    \item \textbf{Sensitive features}: none.
    \item \textbf{Link}: not available
    \item \textbf{Further info}: \citet{mislove2010you}
\end{itemize}

\subsection{Riddle of Literary Quality}
\begin{itemize}
    \item \textbf{Description}: this text corpus was assembled to study the factors that correlate with the acceptance of a text as literary (or non-literary) and good (or bad). It consists of 401 Dutch-language novels published between 2007--2012. These works were selected for being bestsellers or often lent from libraries in the period 2009--2012. Due to copyright reasons, the data is not publicly available.
    \item \textbf{Affiliation of creators}: Huygens ING – KNAW; University of Amsterdam; Fryske Akademy.
    \item \textbf{Domain}: literature.
    \item \textbf{Tasks in fairness literature}: fairness evaluation \citep{koolen2017:stereotypes}.
    \item \textbf{Data spec}: text.
    \item \textbf{Sample size}: $\sim400$ novels.
    \item \textbf{Year}: 2017.
    \item \textbf{Sensitive features}: gender (of author).
    \item \textbf{Link}: not available
    \item \textbf{Further info}: \citet{koolen2017:stereotypes}; \url{https://literaryquality.huygens.knaw.nl/}
\end{itemize}

\subsection{Ride-hailing App}
\begin{itemize}
    \item \textbf{Description}: this dataset was gathered from a ride-hailing app operating in an undisclosed major Asian city. It summarizes spatio-temporal data about ride requests (jobs) and assignments to drivers during 29 consecutive days. The data tracks the position and status of taxis logging data every 30-90 seconds.
    \item \textbf{Affiliation of creators}: Max Planck Institute for Software Systems; Max Planck Institute for Informatics.
    \item \textbf{Domain}: transportation.
    \item \textbf{Tasks in fairness literature}: fair matching \citep{suhr2019twosided}.
    \item \textbf{Data spec}: driver-job pairs.
    \item \textbf{Sample size}: $\sim1$K drivers handling $\sim200$K job requests.
    \item \textbf{Year}: 2019.
    \item \textbf{Sensitive features}: geography.
    \item \textbf{Link}: not available
    \item \textbf{Further info}: \citet{suhr2019twosided}
\end{itemize}

\subsection{RtGender}
\begin{itemize}
    \item \textbf{Description}: this dataset captures differences in online commenting behaviour to posts and videos of female and male users. It was created by collecting posts and top-level comments from four platforms: Facebook, Reddit, Fitocracy, TED talks. For each of the four sources, the possibility to reliably report the gender of the poster or presenter shaped the data collection procedure. Authors of posts and videos were selected among users self-reporting their gender or public figures for which gender annotations were available. For instance, the authors created two Facebook-based datasets: one containing all posts and associated top-level comments for all 412 members of US parliament who have public Facebook pages, and a similar one for 105 American public figures (journalists, novelists, actors, actresses, etc.). The gender of these figures was derived based on their presence on Wikipedia category pages relevant for gender.\footnote{e.g. \url{https://en.wikipedia.org/wiki/Category:American_female_tennis_players}} The gender of commenters and a reliable ID to identify them across comments may be useful for some analyses. The authors report commenters' first names and a randomized ID, which should support these goals, while reducing chances of re-identification based on last name and Facebook ID.
    \item \textbf{Affiliation of creators}: Stanford University; University of Michigan; Carnegie Mellon University.
    \item \textbf{Domain}: social media, linguistics.
    \item \textbf{Tasks in fairness literature}: fairness evaluation \citep{babaeianjelodar2020quantifying}.
    \item \textbf{Data spec}: text.
    \item \textbf{Sample size}: $\sim2$M posts with $\sim25$M comments.
    \item \textbf{Year}: 2018.
    \item \textbf{Sensitive features}: gender.\footnote{Annotations for Facebook and TED come from Wikipedia and \citet{mirkin2015:motivating} respectively. Reddit and Fitocracy rely on self-reported labels.}.
    \item \textbf{Link}: \url{https://nlp.stanford.edu/robvoigt/rtgender/}
    \item \textbf{Further info}: \citep{voigt2018rtgender}
\end{itemize}

\subsection{SafeGraph Research Release}
\begin{itemize}
    \item \textbf{Description}: this dataset captures mobility patterns in the US and Canada. It is maintained by SafeGraph, a data company powering analytics about access to Points-of-Interest (POI) and mobility, including pandemic research. SafeGraph data is sourced from millions of mobile devices, whose users allow location tracking by some apps. The \emph{Research Release} dataset consists of aggregated estimates of hourly visit counts to over 6 million POI. Given the increasing importance of SafeGraph data, directly influencing not only private initiative but also public policy, audits of data representativeness are being carried out both internally \citep{squire2019measuring} and externally \citep{coston2021leveraging}.
    \item \textbf{Affiliation of creators}: Safegraph.
    \item \textbf{Domain}: urban studies.
    \item \textbf{Tasks in fairness literature}: data bias evaluation \citep{coston2021leveraging}.
    \item \textbf{Data spec}: mixture.
    \item \textbf{Sample size}: $\sim7$M POI.
    \item \textbf{Year}: present.
    \item \textbf{Sensitive features}: geography.
    \item \textbf{Link}: \url{https://www.safegraph.com/academics}
    \item \textbf{Further info}: \url{https://docs.safegraph.com/v4.0/docs}
\end{itemize}

\subsection{Scientist+Painter}
\begin{itemize}
    \item \textbf{Description}: this resource was crawled to study the problem of fair and diverse representation in subsets of instances selected from a large dataset, with a focus on gender concentration in professions. The dataset consists of approximately 800 images that equally represent male scientists, female scientists, male painters, and female painters. These images were gathered from Google image search, selecting the top 200 medium sized JPEG files that passed the strictest level of Safe Search filtering. Then, each image was processed to obtain sets of 128-dimensional SIFT descriptors. The descriptors are combined, subsampled and then clustered using k-means into 256 clusters.
    \item \textbf{Affiliation of creators}: École Polytechnique Fédérale de Lausanne (EPFL); Microsoft; University of California, Berkeley.
    \item \textbf{Domain}: information systems.
    \item \textbf{Tasks in fairness literature}: fair data summarization \citep{celis2016fair,celis2018fair}.
    \item \textbf{Data spec}: image.
    \item \textbf{Sample size}: $\sim800$ images.
    \item \textbf{Year}: 2016.
    \item \textbf{Sensitive features}: male/female.
    \item \textbf{Link}: \url{goo.gl/hNukfP}
    \item \textbf{Further info}: \citet{celis2016fair}
\end{itemize}

\subsection{Section 203 determinations}
\begin{itemize}
    \item \textbf{Description}: this dataset is created in support of the language minority provisions of the Voting Rights Act, Section 203. The data contains information about limited-English proficient voting population by jurisdiction, which is used to determine whether election materials must be printed in minority languages. For each combination of language protected by Section 203 and US jurisdiction, the dataset provides information about total population, population of voting age, US citizen population of voting age, combining this information with language spoken at home and overall English proficiency.
    \item \textbf{Affiliation of creators}: US Census Bureau.
    \item \textbf{Domain}: demography.
    \item \textbf{Tasks in fairness literature}: fairness evaluation of private resource allocation \citep{pujol2020fair}.
    \item \textbf{Data spec}: tabular data.
    \item \textbf{Sample size}: $\sim600$K combinations of jurisdictions and languages potentially spoken therein.
    \item \textbf{Year}: 2017.
    \item \textbf{Sensitive features}: geography, language.
    \item \textbf{Link}: \url{https://www.census.gov/data/datasets/2016/dec/rdo/section-203-determinations.html}
    \item \textbf{Further info}: \url{https://www.census.gov/programs-surveys/decennial-census/about/voting-rights/voting-rights-determination-file.2016.html}
\end{itemize}

\subsection{Sentiment140}
\begin{itemize}
    \item \textbf{Description}: this dataset was created to study the problem of sentiment analysis in social media, envisioning applications of product quality and brand reputation analysis via Twitter monitoring. The sentiment of tweets, retrieved via Twitter API, is automatically inferred based on the presence of emoticons conveying joy or sadness. This dataset is part of the LEAF benchmark for federated learning. In federated learning settings, devices correspond to accounts.
    \item \textbf{Affiliation of creators}: Stanford University.
    \item \textbf{Domain}: social media.
    \item \textbf{Tasks in fairness literature}: fair federated learning \citep{li2020fair}.
    \item \textbf{Data spec}: text.
    \item \textbf{Sample size}: $\sim2$M tweets by $\sim600$K accounts.
    \item \textbf{Year}: 2012.
    \item \textbf{Sensitive features}: textual references to people and their demographics.
    \item \textbf{Link}: \url{http://help.sentiment140.com/home}
    \item \textbf{Further info}: \citet{go2009twitter}
\end{itemize}

\subsection{Seoul Bike Sharing}
\begin{itemize}
    \item \textbf{Description}: this resource, summarizing hourly public rental history of \emph{Seoul Bikes}, was curated to study the problem of bike sharing demand prediction. The data was downloaded from the Seoul Public Data Park website of South Korea and spans one year of utilization (December 2017 to November 2018) of Seoul Bikes, a bike sharing system that started in 2015. This dataset consists of hourly information about weather (e.g. temperature, solar radiation, rainfall) and time (date, time, season, holiday), along with the number of bikes rented at each hour, which is the target of a prediction task.
    \item \textbf{Affiliation of creators}: Sunchon National University.
    \item \textbf{Domain}: transportation.
    \item \textbf{Tasks in fairness literature}: fair regression \citep{diana2021minimax}.
    \item \textbf{Data spec}: time series.
    \item \textbf{Sample size}: $\sim9$K hourly points.
    \item \textbf{Year}: 2020.
    \item \textbf{Sensitive features}: none.
    \item \textbf{Link}: \url{https://archive.ics.uci.edu/ml/datasets/Seoul+Bike+Sharing+Demand}
    \item \textbf{Further info}: \citet{sathishkumar2020rule,sathishkumar2020using}, \url{https://data.seoul.go.kr/index.do}
\end{itemize}

\subsection{Shakespeare}
\begin{itemize}
    \item \textbf{Description}: this dataset is available as part of the LEAF benchmark for federated learning \citep{caldas2018leaf}. It is built from ``The Complete Works of William Shakespeare'', where each speaking role represents a different device. The task envisioned for this dataset is next character prediction.
    \item \textbf{Affiliation of creators}: Google; Carnegie Mellon University; Determined AI.
    \item \textbf{Domain}: literature.
    \item \textbf{Tasks in fairness literature}: fair federated learning \citep{li2020fair}.
    \item \textbf{Data spec}: text.
    \item \textbf{Sample size}: $\sim4$M tokens over $\sim1$K speaking roles.
    \item \textbf{Year}: 2020.
    \item \textbf{Sensitive features}: textual references to people and their demographics.
    \item \textbf{Link}: 
    \url{https://www.tensorflow.org/federated/api_docs/python/tff/simulation/datasets/shakespeare}
    \item \textbf{Further info}: \citet{mcmahan2017communicationefficient,caldas2018leaf}
\end{itemize}

\subsection{Shanghai Taxi Trajectories}
\begin{itemize}
    \item \textbf{Description}: this semi-synthetic dataset represents the road network and traffic patterns of Shanghai. Trajectories were collected from thousands of taxis operating in Shanghai. Spatio-temporal traffic patterns were extracted from these trajectories and used to build the dataset.
    \item \textbf{Affiliation of creators}: Shanghai Jiao Tong University; CITI-INRIA Lab.
    \item \textbf{Domain}: transportation.
    \item \textbf{Tasks in fairness literature}: fair routing \citep{qian2015scram}.
    \item \textbf{Data spec}: unknown.
    \item \textbf{Sample size}: unknown.
    \item \textbf{Year}: 2015.
    \item \textbf{Sensitive features}: geography.
    \item \textbf{Link}: not available
    \item \textbf{Further info}: \citet{qian2015scram}
\end{itemize}

\subsection{shapes3D}
\begin{itemize}
    \item \textbf{Description}: this dataset is an artificial benchmark for unsupervised methods aimed at learning disentangled data representations. It consists of imagesof 3D shapes in a walled environment, with variable floor colour, wall colour, object colour, scale, shape and orientation.
    \item \textbf{Affiliation of creators}: DeepMind; Wayve.
    \item \textbf{Domain}: computer vision.
    \item \textbf{Tasks in fairness literature}: fair representation learning \citep{locatello2019fairness}, fair data generation \citep{choi2020fair}.
    \item \textbf{Data spec}: image.
    \item \textbf{Sample size}: $\sim500$K images.
    \item \textbf{Year}: 2018.
    \item \textbf{Sensitive features}: none.
    \item \textbf{Link}: \url{https://github.com/deepmind/3d-shapes}
    \item \textbf{Further info}: \citet{kim2018disentangling}
\end{itemize}

\subsection{SIIM-ISIC Melanoma Classification}
\begin{itemize}
    \item \textbf{Description}: this dataset was developed to advance the study of automated melanoma classification. The resource consists of dermoscopy images from six medical centers. Images in the dataset are tagged with a patient identifier, allowing lesions from the same patient to be mapped to one another. Images were queried from medical databases among patients with dermoscopy imaging from 1998 to 2019, ranging in quality from 307,200 to 24,000,000 pixels. A curated subset is employed for the 2020 ISIC Grand Challenge.\footnote{\url{https://www.kaggle.com/c/siim-isic-melanoma-classification}} This dataset was annotated automatically with a binary Fitzpatrick skin tone label \citep{cheng2021can}.
    \item \textbf{Affiliation of creators}: Memorial Sloan Kettering Cancer Center; University of Queensland; University of Athens; IBM; Universitat de Barcelona; Melanoma Institute Australia; Sydney Melanoma Diagnostic Center; Emory University; Medical University of Vienna; Mayo Clinic; SUNY Downstate Medical School; Stony brook Medical School; Rabin Medical Center; Weill Cornell Medical College.
    \item \textbf{Domain}: dermatology.
    \item \textbf{Tasks in fairness literature}: fairness evaluation of private classification \citep{cheng2021can}.
    \item \textbf{Data spec}: image.
    \item \textbf{Sample size}: $\sim30$K images of $\sim2$K patients.
    \item \textbf{Year}: 2020.
    \item \textbf{Sensitive features}: skin type.
    \item \textbf{Link}: url{https://doi.org/10.34970/2020-ds01}
    \item \textbf{Further info}: \citet{rotemberg2021patient}
\end{itemize}

\subsection{SmallNORB}
\begin{itemize}
    \item \textbf{Description}: this dataset was assembled by researchers affiliated with New York University as a benchmark for robust object recognition under variable pose and lighting conditions. It consists of images of 50 different toys  belonging to 5 categories (four-legged animals, human figures, airplanes, trucks, and cars) obtained by 2 different cameras.
    \item \textbf{Affiliation of creators}: New York University; NEC Labs America.
    \item \textbf{Domain}: computer vision.
    \item \textbf{Tasks in fairness literature}: fair representation learning \citep{locatello2019fairness}.
    \item \textbf{Data spec}: image.
    \item \textbf{Sample size}: $\sim100$K images.
    \item \textbf{Year}: 2005.
    \item \textbf{Sensitive features}: none.
    \item \textbf{Link}: \url{https://cs.nyu.edu/~ylclab/data/norb-v1.0-small/}
    \item \textbf{Further info}: \citet{lecun2004learning}
\end{itemize}

\subsection{Spliddit Divide Goods}
\begin{itemize}
    \item \textbf{Description}: this dataset summarizes instances of usage of the \emph{divide goods} feature of Spliddit, a not-for-profit academic endeavor providing easy access to fair division methods. A typical use case for the service is inheritance division. Participants express their preferences by dividing 1,000 points between the available goods. In response, the service provides suggestions that are meant to maximize the overall satisfaction of all stakeholders.
    \item \textbf{Affiliation of creators}: Spliddit.
    \item \textbf{Domain}: economics.
    \item \textbf{Tasks in fairness literature}: fair preferece-based resource allocation \citep{babaioff2019fair}.
    \item \textbf{Data spec}: tabular data.
    \item \textbf{Sample size}: $\sim1$K division instances.
    \item \textbf{Year}: 2016.
    \item \textbf{Sensitive features}: none.
    \item \textbf{Link}: not available
    \item \textbf{Further info}: \citet{caragiannis2016unreasonable}; \url{http://www.spliddit.org/apps/goods}
\end{itemize}

\subsection{Stanford Medicine Research Data Repository}
\begin{itemize}
    \item \textbf{Description}: this is a data lake/repository developed at Stanford University, supporting a number of data sources and access pipelines. The aim of the underlying project is favouring access to clinical data for research purposes through flexible and robust management of medical data. The data comes from Stanford Health Care, the Stanford Children’s Hospital, the University Healthcare Alliance and Packard Children's Health Alliance clinics.
    \item \textbf{Affiliation of creators}: Stanford University.
    \item \textbf{Domain}: medicine.
    \item \textbf{Tasks in fairness literature}: fair risk assessment \citep{pfohl2019creating}.
    \item \textbf{Data spec}: mixture.
    \item \textbf{Sample size}: $\sim$3M individuals.
    \item \textbf{Year}: present.
    \item \textbf{Sensitive features}: race, ethnicity, gender, age.
    \item \textbf{Link}: \url{https://starr.stanford.edu/}
    \item \textbf{Further info}: \citet{lowe2009stride,datta2020new}
\end{itemize}

\subsection{State Court Processing Statistics (SCPS)}
\begin{itemize}
    \item \textbf{Description}: this resource was curated as part of the SCPS program. The program tracked felony defendants from charging by the prosecutor until disposition of their cases for a maximum of 12 months (24 months for murder cases). The data represents felony cases filed in approximately 40 populous US counties in the period 1990-2009. Defendants are summarized by 106 variables summarizing demographics, arrest charges, criminal history, pretrial release and detention, adjudication, and sentencing.
    \item \textbf{Affiliation of creators}: US Department of Justice.
    \item \textbf{Domain}: law.
    \item \textbf{Tasks in fairness literature}: fairness evaluation of multi-stage classification \citep{green2019disparate}.
    \item \textbf{Data spec}: tabular data.
    \item \textbf{Sample size}: $\sim200$K defendants.
    \item \textbf{Year}: 2014.
    \item \textbf{Sensitive features}: gender, race, age, geography.
    \item \textbf{Link}: \url{https://www.icpsr.umich.edu/web/NACJD/studies/2038/datadocumentation}
    \item \textbf{Further info}: \url{https://bjs.ojp.gov/data-collection/state-court-processing-statistics-scps}
\end{itemize}

\subsection{Steemit}
\begin{itemize}
    \item \textbf{Description}: this resource was collected to test novel approaches for personalized content recommendation in social networks. It consists of two separate datasets summarizing interactions in the Spanish subnetwork and the English subsnetwork of Steemit, a blockchain-based social media website. The datasets summarize user-post interactions in a binary fashion, using comments as a proxy for positive engagement. The datasets cover a whole year of commenting activities over the period 2017--2018 and comprise the text of posts.
    \item \textbf{Affiliation of creators}: Hong Kong University of Science and Technology; WeBank.
    \item \textbf{Domain}: social media.
    \item \textbf{Tasks in fairness literature}: fairness evaluation \citep{xiao2019beyond}.
    \item \textbf{Data spec}: user-post pairs.
    \item \textbf{Sample size}: $\sim50$K users interacting over $\sim200$K posts.
    \item \textbf{Year}: 2019.
    \item \textbf{Sensitive features}: textual references to people and their demographics.
    \item \textbf{Link}: \url{https://github.com/HKUST-KnowComp/Social-Explorative-Attention-Networks}
    \item \textbf{Further info}: \citet{xiao2019beyond}
\end{itemize}

\subsection{Stop, Question and Frisk}
\begin{itemize}
    \item \textbf{Description}: Stop, Question and Frisk (SQF) is an expression that commonly refers to a New York City policing program under which officers can briefly detain, question, and search a citizen if the officer has a reasonable suspicion of criminal activity. Concerns about race-based disparities in this practice have been expressed multiple times, especially in connection with the subjective nature of ``reasonable suspicion'' and the fact that being in a ``high-crime area'' lawfully lowers the bar of want may constitute reasonable suspicion. The NYPD has a policy of keeping track of most stops, recording them in UF-250 forms which are maintained centrally and distributed by the NYPD. The form includes several information such as place and time of a stop, the duration of the stop and its outcome along with data on demographics and physical appearance of the suspect. Currently available data pertains to years 2003--2020.
    \item \textbf{Affiliation of creators}: New York Police Department.
    \item \textbf{Domain}: law.
    \item \textbf{Tasks in fairness literature}: preference-based fair classification \citep{zafar2017from}, 
    robust fair classification \citep{kallus2018residual}, 
    fair classification under unawareness \citep{kilbertus2018blind},
    fairness evaluation \citep{goel2017combatting}, fair classification \citep{ali2021accounting}.
    \item \textbf{Data spec}: tabular data.
    \item \textbf{Sample size}: $\sim1$M records.
    \item \textbf{Year}: 2021.
    \item \textbf{Sensitive features}: race, age, sex, geography.
    \item \textbf{Link}: \url{https://www1.nyc.gov/site/nypd/stats/reports-analysis/stopfrisk.page}
    \item \textbf{Further info}: \citet{gelman2007analysis, goel2016precinct}
\end{itemize}

\subsection{Strategic Subject List}
\begin{itemize}
    \item \textbf{Description}: this dataset was funded through a Bureau of Justice Assistance grant and leveraged by the Illinois Institute of Technology to develop the Chicago Police Department’s Strategic Subject Algorithm. The algorithm provides a risk score which reflects an individual’s probability of being involved in a shooting incident either as a victim or an offender. For each individual, the dataset provides information about the circumstances of their arrest, their demographics and criminal history. The dataset covers arrest data from the period 2012--2016; the associated program was discontinued in 2019.
    \item \textbf{Affiliation of creators}: Chicago Police Department; Illinois Institute of Technology.
    \item \textbf{Domain}: law.
    \item \textbf{Tasks in fairness literature}: fairness evaluation \citep{black2020fliptest}.
    \item \textbf{Data spec}: tabular data.
    \item \textbf{Sample size}: $\sim400$K individuals.
    \item \textbf{Year}: 2020.
    \item \textbf{Sensitive features}: ace, sex, age.
    \item \textbf{Link}: \url{https://data.cityofchicago.org/Public-Safety/Strategic-Subject-List-Historical/4aki-r3np}
    \item \textbf{Further info}: \citet{hollywood2019real}
\end{itemize}

\subsection{Student}
\begin{itemize}
    \item \textbf{Description}: the data was collected from two Portuguese public secondary schools in the Alentejo region, to investigate student achievement prediction and identify decisive factors in student success. The data tracks student performance in Mathematics and Portuguese through school year 2005-2006 and is complemented by demographic, socio-econonomical, and personal data obtained through a questionnaire. Numerical grades (20-point scale) collected by students over three terms are typically the target of the associated prediction task.
    \item \textbf{Affiliation of creators}: University of Minho.
    \item \textbf{Domain}: education.
    \item \textbf{Tasks in fairness literature}: fair regression \citep{chzhen2020fairwassertein,chzhen2020fair,heidari2019on},
    rich-subgroup fairness evaluation \citep{kearns2019empirical},
    fair data summarization \citep{jones2020fair,belitz2021automating}.
    \item \textbf{Data spec}: tabular data.
    \item \textbf{Sample size}: $\sim600$ students.
    \item \textbf{Year}: 2014.
    \item \textbf{Sensitive features}: sex, age.
    \item \textbf{Link}: \url{https://archive.ics.uci.edu/ml/datasets/student+performance}
    \item \textbf{Further info}: \citet{cortez2008using}
\end{itemize}

\subsection{Sushi}
\begin{itemize}
    \item \textbf{Description}: this dataset was sourced online via a commercial survey service to evaluate rank-based approaches to solicit preferences and provide recommendations. The dataset captures the preferences for different types of sushi held by people in different areas of Japan. These are encoded both as ratings in a 5-point scale and ordered lists of preferences, which recommenders should learn via collaborative filtering. Demographic data was also collected to study geographical preference patterns.
    \item \textbf{Affiliation of creators}: Japanese National Institute of Advanced Industrial Science and Technology (AIST).
    \item \textbf{Domain}: .
    \item \textbf{Tasks in fairness literature}: fair data summarization \citep{chiplunkar2020how}.
    \item \textbf{Data spec}: user-sushi pairs.
    \item \textbf{Sample size}: $\sim5$K respondents.
    \item \textbf{Year}: 2016.
    \item \textbf{Sensitive features}: gender, age, geography.
    \item \textbf{Link}: \url{https://www.kamishima.net/\%20sushi/}
    \item \textbf{Further info}: \citet{kamishima2003:nantonac}
\end{itemize}

\subsection{Symptoms in Queries}
\begin{itemize}
    \item \textbf{Description}: the purpose of this dataset is to study, using only aggregate statistics, the fairness and accuracy of a classifier that predicts whether an individual has a certain type of cancer based on their Bing search queries. The dataset does not include individual data points. It provides, for each US state, and for 18 types of cancer, the proportion of individuals who have this cancer in the state according to CDC 2019 data,\footnote{\url{https://gis.cdc.gov/Cancer/USCS/DataViz.html}} and the proportion of individuals who are predicted to have this cancer according to the classifier that was calculated using Bing queries.
    \item \textbf{Affiliation of creators}: Microsoft; Ben-Gurion University of the Negev.
    \item \textbf{Domain}: information systems, public health.
    \item \textbf{Tasks in fairness literature}: limited-label fairness evaluation \citep{sabato2020bounding}.
    \item \textbf{Data spec}: tabular data.
    \item \textbf{Sample size}: statistics for $\sim20$ cancer types across $\sim50$ US states.
    \item \textbf{Year}: 2020.
    \item \textbf{Sensitive features}: geography.
    \item \textbf{Link}: \url{https://github.com/sivansabato/bfa/blob/master/cancer_data.m}
    \item \textbf{Further info}: \citet{sabato2020bounding}
\end{itemize}

\subsection{TAPER Twitter Lists}
\begin{itemize}
    \item \textbf{Description}: this resource was collected to study the problem of personalized expert recommendation, leveraging Twitter lists where users labelled other users as relevant for (or expert in) a given topic. The creators started from a seed dataset of over 12 million geo-tagged Twitter lists, which they filtered to only keep US-based users in topics: news, music, technology, celebrities, sports, business, politics, food, fashion, art, science, education, marketing, movie, photography, and health. A subset of this dataset was annotated with user race (whites and non-whites) via Face++ \citep{zhu2018fairnessaware}.
    \item \textbf{Affiliation of creators}: Texas A\&M University.
    \item \textbf{Domain}: social media.
    \item \textbf{Tasks in fairness literature}: fair ranking \citep{zhu2018fairnessaware}.
    \item \textbf{Data spec}: user-topic pairs.
    \item \textbf{Sample size}: $\sim10$K Twitter lists featuring $\sim8$K list members.
    \item \textbf{Year}: 2016.
    \item \textbf{Sensitive features}: race.
    \item \textbf{Link}: not available
    \item \textbf{Further info}: \citet{ge2016taper}
\end{itemize}

\subsection{TaskRabbit}
\begin{itemize}
    \item \textbf{Description}: this resource was assembled to study the effectiveness of fair ranking approaches in improving outcomes for protected groups in online hiring. It consists of the top 10 results returned by the online freelance marketplace TaskRabbit for three queries: ``Shopping'', ``Event staffing'', and ``Moving Assistace''. The geographic location for a query was especially selected to yield a ranking with 3 female candidates among the top 10, with most of them appearing in the bottom 5, which may be a motivating condition for a fairness intervention. Candidates' gender was manually labelled by creators based on pronoun usage and profile pictures. For each profile, the authors extracted information on job suitability, including TaskRabbit relevance scores, number of completed tasks and positive reviews.
    \item \textbf{Affiliation of creators}: Technische Universität Berlin; Harvard University.
    \item \textbf{Domain}: information systems.
    \item \textbf{Tasks in fairness literature}: fair ranking evaluation \citep{suhr2021does}, multi-stage fairness evaluation \citep{suhr2021does}.
    \item \textbf{Data spec}: query-worker pairs.
    \item \textbf{Sample size}: 3 rankings (one per query) of $\sim10$ workers.
    \item \textbf{Year}: 2021.
    \item \textbf{Sensitive features}: gender.
    \item \textbf{Link}: not available
    \item \textbf{Further info}: \citet{suhr2021does}
\end{itemize}

\subsection{TIMIT}
\begin{itemize}
    \item \textbf{Description}: this resource was curated to power studies of phonetics and to evaulate systems of automated speech recognition. The dataset features speakers of different American English dialects, and includes time-aligned orthographic, phonetic and word transcriptions. Utterances are sampled at a 16kHz frequency.
    \item \textbf{Affiliation of creators}: University of Pennsylvania; National Institute of Standards and Technology; Massachusetts Institute of Technology; SRI International; Texas Instruments.
    \item \textbf{Domain}: linguistics.
    \item \textbf{Tasks in fairness literature}: fairness evaluation of speech recognition \citep{segal2021fairness}.
    \item \textbf{Data spec}: time series.
    \item \textbf{Sample size}: $\sim600$ speakers, each uttering $\sim10$ sentences.
    \item \textbf{Year}: 1993.
    \item \textbf{Sensitive features}: dialect, gender.
    \item \textbf{Link}: \url{https://catalog.ldc.upenn.edu/LDC93S1}
    \item \textbf{Further info}: \url{https://en.wikipedia.org/wiki/TIMIT}
\end{itemize}

\subsection{Toy Dataset 1}
\begin{itemize}
    \item \textbf{Description}: this dataset consists of $\sim4$K points generated as follows. Binary class labels $y$ are generated at random for each point. Next, two-dimensional features $x$ are assigned to each point, sampling from gaussian distributions whose mean and variance depend on $y$, so that $p(x|y=1) = \mathcal{N}([2; 2], [5, 1; 1, 5])$;  $p(x|y=-1) = \mathcal{N}([-2; -2], [10, 1; 1, 3])$. Finally, each point's sensitive attribute $z$ is sampled from a Bernoulli distribution so that $p(z = 1) = p(x'|y = 1)/(p(x'|y = 1) + p(x'|y = 1))$, where $x'$ is a rotated version of $x$: $x' = [\text{cos}(\phi),  -\text{sin}(\phi); \text{sin}(\phi), \text{cos}(\phi)]x$. Parameter $\phi$ controls the correlation between class label $y$ and sensitive attribute $z$.
    \item \textbf{Affiliation of creators}: Max Planck Institute for Software Systems.
    \item \textbf{Domain}: N/A.
    \item \textbf{Tasks in fairness literature}: fair classification \citep{zafar2017fairness,roh2020frtrain}, fair preference-based classification \citep{zafar2017from,ali2019loss},  fair few-shot learning  \citep{slack2019fair,slack2020fairness}, fair classification under unawareness \citep{kilbertus2018blind}.
    \item \textbf{Data spec}: tabular data.
    \item \textbf{Sample size}: $\sim4$K points.
    \item \textbf{Year}: 2017.
    \item \textbf{Sensitive features}: N/A.
    \item \textbf{Link}: \url{https://github.com/mbilalzafar/fair-classification/tree/master/disparate_impact/synthetic_data_demo}
    \item \textbf{Further info}: \citet{zafar2017fairness}
\end{itemize}

\subsection{Toy Dataset 2}
\begin{itemize}
    \item \textbf{Description}: this dataset contains synthetic relevance judgements over pairs of queries and documents that are biased against a minority group. For each query, there are 10 candidate documents, 8 from group $G_0$ and 2 from minority group $G_1$. Each document is associated with a feature vector $(x_1, x_2)$, with both components sampled uniformly at random from the interval $(0,3)$. The relevance of documents is set to $y=x_1+x_2$ and clipped between 0 and 5. Feature $x_2$ is then corrupted and replaced by zero for group $G_1$, leading to a biased representation between groups, such that any use of $x_2$ should lead to unfair rankings.
    \item \textbf{Affiliation of creators}: Cornell University.
    \item \textbf{Domain}: N/A.
    \item \textbf{Tasks in fairness literature}: fair ranking \citep{singh2019policy,bower2021individually}.
    \item \textbf{Data spec}: query-document pairs.
    \item \textbf{Sample size}: $\sim1$K relevance judgements overs  $\sim100$ queries with $\sim10$ candidate documents.
    \item \textbf{Year}: 2019.
    \item \textbf{Sensitive features}: N/A.
    \item \textbf{Link}: \url{https://github.com/ashudeep/Fair-PGRank}
    \item \textbf{Further info}: \citet{singh2019policy}
\end{itemize}

\subsection{Toy Dataset 3}
\begin{itemize}
    \item \textbf{Description}: this dataset was created to demonstrate undesirable properties of a family of fair classification approaches. Each instance in the dataset is associated with a sensitive attribute $z$, a target variable $y$ encoding employability, one feature that is important for the problem at hand and correlated with $z$ (work\_experience) and a second feature which is unimportant yet also correlated with $z$ (hair\_length). The data generating process is the following:
    
    \begin{align*}
     z_i &\sim \text{Bernoulli}(0.5) \nonumber \\
    \text{hair\_length}_i| z_i = 1 &\sim  35 \cdot \text{Beta}(2, 2) \nonumber \\
    \text{hair\_length}_i| z_i = 0 &\sim  35 \cdot \text{Beta}(2, 7) \nonumber \\
    \text{work\_exp}_i| z_i  &\sim \text{Poisson}(25 + 6z_i) - \text{Normal}(20,0.2) \nonumber \\
    y_i| \text{work\_exp}_i  &\sim 2 \cdot \text{Bernoulli}(p_i) - 1, \nonumber \\
    \text{where} p_i &= 1/ (1 + \text{exp}[-(-25.5 + 2.5\text{work\_exp})]) \nonumber \\
    \end{align*}.
    \item \textbf{Affiliation of creators}: Carnegie Mellon University; University of California, San Diego.
    \item \textbf{Domain}: N/A.
    \item \textbf{Tasks in fairness literature}: fairness evaluation \citep{lipton2018does,black2020fliptest}.
    \item \textbf{Data spec}: tabular data.
    \item \textbf{Sample size}: $\sim2$K points.
    \item \textbf{Year}: 2018.
    \item \textbf{Sensitive features}: N/A.
    \item \textbf{Link}: not available
    \item \textbf{Further info}: \citet{lipton2018does}
\end{itemize}

\subsection{Toy Dataset 4}
\begin{itemize}
    \item \textbf{Description}: in this toy example, features are generated according to four 2-dimensional isotropic Gaussian distributions with different mean $\mu$ and variance $\sigma^2$. Each of the four distributions corresponds to a different combination of binary label $y$ and protected attribute $s$ as follows: 
    (1) $s=a, y=+1: \mu=(-1,-1), \sigma^2=0.8$;
    (2) $s=a, y=-1: \mu=(1,1), \sigma^2=0.8$; 
    (3) $s=b; y=+1: \mu=(0.5,-0.5), \sigma^2=0.5$; 
    (4) $s=b, y=-1: \mu=(0.5,0.5), \sigma^2=0.5$.
    \item \textbf{Affiliation of creators}: Istituto Italiano di Tecnologia; University of Genoa; University of Waterloo; University College London.
    \item \textbf{Domain}: N/A.
    \item \textbf{Tasks in fairness literature}: fair classification \citep{donini2018empirical}, fairness evaluation \citep{williamson2019fairness}.
    \item \textbf{Data spec}: tabular data.
    \item \textbf{Sample size}: $\sim6$K points.
    \item \textbf{Year}: 2018.
    \item \textbf{Sensitive features}: N/A.
    \item \textbf{Link}: \url{https://github.com/jmikko/fair_ERM}
    \item \textbf{Further info}: \citet{donini2018empirical}
\end{itemize}

\subsection{TREC Robust04}
\begin{itemize}
    \item \textbf{Description}: this classic information retrieval collection is a set of topics, documents and relevance judgements collected as part of the Text REtrieval Conference (TREC) 2004 Robust Retrieval Track to catalyze research improving the consistency of information retrieval technology. Documents are taken from articles published during the 1990s in the Financial Times Limited, the Federal Register, the Foreign Broadcast Information Service, and the Los Angeles Times. Graded relevance (not relevant, relevant, highly relevant) was judged by human assessors for a subset of all possible topic-document combinations, which were selected as ``promising'' by the automated systems that entered the TREC initiative. The associated task is predicting the relevance of documents for various textual queries.
    \item \textbf{Affiliation of creators}: National Institute of Standards and Technology.
    \item \textbf{Domain}: news, information systems.
    \item \textbf{Tasks in fairness literature}: fair ranking evaluation \citep{gerritse2020effect}.
    \item \textbf{Data spec}: query-document pairs.
    \item \textbf{Sample size}: $\sim300$K relevance judgements over $\sim200$ queries and $\sim500$K documents.
    \item \textbf{Year}: 2005.
    \item \textbf{Sensitive features}: textual references to people and their demographics.
    \item \textbf{Link}: \url{https://trec.nist.gov/data/t13_robust.html}
    \item \textbf{Further info}: \citet{voorhees2005overview}
\end{itemize}

\subsection{Twitch Social Networks}
\begin{itemize}
    \item \textbf{Description}: this dataset was developed to study the effectiveness of node embeddings for learning tasks defined on graphs. This resource concentrates on Twitch content creators streaming in 6 different languages. The dataset has users as nodes, mutual friendships as edges, and node embedddings summarizing  games liked, location and streaming habits. The original task on this dataset is predicting whether a streamer uses explicit language.
    \item \textbf{Affiliation of creators}: University of Edinburgh.
    \item \textbf{Domain}: social networks.
    \item \textbf{Tasks in fairness literature}: fair graph mining \citep{kang2020inform}.
    \item \textbf{Data spec}: user-user pairs.
    \item \textbf{Sample size}: $\sim$30K nodes (users) connected by $\sim400$K edges (mutual friendship).
    \item \textbf{Year}: 2019.
    \item \textbf{Sensitive features}: none.
    \item \textbf{Link}: \url{http://snap.stanford.edu/data/twitch-social-networks.html}
    \item \textbf{Further info}: \citet{rozemberczki2021multi}
\end{itemize}

\subsection{Twitter Abusive Behavior}
\begin{itemize}
    \item \textbf{Description}: this dataset is the result of an eight-month crowdsourced study of various forms of abusive behavior on Twitter. The authors began by considering a wide variety of inappropriate speech categories, analyzing how they are used by amateur annotators hired on CrowdFlower. After two exploratory rounds, they merged some labels and eliminated others, converging to a final four-class categorization into (normal, spam, abusive, hateful), requiring five crowdsourced judgements per tweet. Tweets were sampled according to a boosted random sampling technique. A large part of the dataset is randomly sampled, with the addition of tweets that are likely to belong to one or more of the minority (non-normal) classes. The dataset is available as a table mapping tweet IDs to behavior category, making it possible to identify Twitter users in this dataset.
    \item \textbf{Affiliation of creators}: Aristotle University of Thessaloniki; Cyprus University of Technology; Telefonica; University of Alabama at Birmingham; University College London.
    \item \textbf{Domain}: social media.
    \item \textbf{Tasks in fairness literature}: fairness evaluation of harmful content detection \citep{ballburack2021differential}.
    \item \textbf{Data spec}: text.
    \item \textbf{Sample size}: $\sim100$K tweets.
    \item \textbf{Year}: 2018.
    \item \textbf{Sensitive features}: textual references to people and their demographics.
    \item \textbf{Link}: \url{https://github.com/ENCASEH2020/hatespeech-twitter}
    \item \textbf{Further info}: \citet{founta2018large}
\end{itemize}

\subsection{Twitter Hate Speech Detection}
\begin{itemize}
    \item \textbf{Description}: this dataset was developed to study the problem of automated hate speech detection. The creators used the Twitter API to search for tweets containing racist and sexist terms and hashtags. The annotation was carried out by the authors, with an external review by a 25-year-old woman studying gender studies. After identifying a list of eleven criteria to identify hate speech against a minority, each tweet was labelled as sexism, racism or none. The task associated with this resource is hate speech detection. The dataset is available as a table mapping tweet IDs to hate speech category, making it possible to identify Twitter users in this dataset.
    \item \textbf{Affiliation of creators}: University of Copenhagen.
    \item \textbf{Domain}: social media.
    \item \textbf{Tasks in fairness literature}: fairness evaluation \citep{ballburack2021differential}.
    \item \textbf{Data spec}: text.
    \item \textbf{Sample size}: $\sim20$K tweets.
    \item \textbf{Year}: 2016.
    \item \textbf{Sensitive features}: textual references to people and their demographics.
    \item \textbf{Link}: 
    \item \textbf{Further info}: \citet{waseem2016hateful}
\end{itemize}

\subsection{Twitter Offensive Language}
\begin{itemize}
    \item \textbf{Description}: this dataset was developed to study the problem of automated hate speech detection, and to distinguish between hate speech and other kinds of offensive language. The creators used the Twitter API to search for tweets containing terms from a hate speech lexicon compiled by \emph{Hatebase.org}. Workers on CrowdFlower annotated a random subset of these tweets as hate speech, offensive but not hate speech, or neither offensive nor hate speech. Workers were explicitly told that the mere presence of a slur word does not amount to hate speech. Three of more workers annotated each tweet.
    \item \textbf{Affiliation of creators}: Cornell University; Qatar Computing Research Institute.
    \item \textbf{Domain}: social media.
    \item \textbf{Tasks in fairness literature}: fairness evaluation \citep{ballburack2021differential}, fair multi-stage classification \citep{keswani2021towards}.
    \item \textbf{Data spec}: text.
    \item \textbf{Sample size}: $\sim20$K tweets.
    \item \textbf{Year}: 2017.
    \item \textbf{Sensitive features}: textual references to people and their demographics.
    \item \textbf{Link}: \url{https://github.com/t-davidson/hate-speech-and-offensive-language/tree/master/data}
    \item \textbf{Further info}: \citet{davidson2017automated}
\end{itemize}

\subsection{Twitter Online Harrassment}
\begin{itemize}
    \item \textbf{Description}: this dataset was developed as multidisciplinary resource to study online harrassment. The authors searched a stream of tweets for keywords likely to denote violent, offensive, threatening or hateful content based on race, gender, religion and sexual orientation. They developed coding guidelines to label a tweet as harrassing or non/harrassing and spent three weeks reviewing and refining it, annotating sample tweets as a group, and discussing the results. The curators are not publicly sharing the dataset due to Twitter terms of service restrictions and privacy concerns about individuals whose tweets are included; researchers can request access.
    \item \textbf{Affiliation of creators}: University of Maryland.
    \item \textbf{Domain}: social media.
    \item \textbf{Tasks in fairness literature}: fairness evaluation \citep{ballburack2021differential}.
    \item \textbf{Data spec}: text.
    \item \textbf{Sample size}: $\sim40$K tweets.
    \item \textbf{Year}: 2017.
    \item \textbf{Sensitive features}: textual references to people and their demographics.
    \item \textbf{Link}: not available
    \item \textbf{Further info}: \citet{golbeck2017large}
\end{itemize}

\subsection{Twitter Political Searches}
\begin{itemize}
    \item \textbf{Description}: this dataset was collected to study political biases in Twitter search results, due to political leaning of tweets and biases in the Twitter ranking algorithm. The authors identified 25 popular political queries in December 2015, and collected relevant tweets during a week in which two presidential debates occurred, via the Twitter streaming API. Tweets were annotated based on users' political leaning. Users' leaning was automatically inferred from their topics of interest, via a classifier trained on representative sets of democratic and republican users. Both the accuracy of classifiers and the validity of user leaning as a proxy for tweet leaning was validated by workers recruited on Amazon Mechanical Turk.
    \item \textbf{Affiliation of creators}: Max Planck Institute for Software Systems; University of Illinois at Urbana-Champaign; Indian Institute of Engineering Science and Technology, Shibpur; Adobe Research.
    \item \textbf{Domain}: social media.
    \item \textbf{Tasks in fairness literature}: social media.
    \item \textbf{Data spec}: query-result pairs.
    \item \textbf{Sample size}: $\sim30$K search results containing $\sim30$K distinct tweets from $\sim20$K users.
    \item \textbf{Year}: 2016.
    \item \textbf{Sensitive features}: political leaning.
    \item \textbf{Link}: not available
    \item \textbf{Further info}: \citet{kulshrestha2017quantifying}
\end{itemize}

\subsection{Twitter Presidential Politics}
\begin{itemize}
    \item \textbf{Description}: this dataset was created by collecting tweets, through the Twitter API, from 576 accounts linked to presidential candidates and members of congress, from the entire account history until December 2019. Out of all the accounts considered, 258 accounts were classified as Republican and 318 as Democratic. The dataset was collected to build a political bias subspace from word embeddings, which could be a flexible tool to quantitatively investigate political leaning in text-based media.
    \item \textbf{Affiliation of creators}: Clarkson University.
    \item \textbf{Domain}: social media.
    \item \textbf{Tasks in fairness literature}: bias audit \citep{gordon2020studying}.
    \item \textbf{Data spec}: text.
    \item \textbf{Sample size}: $\sim1$M tweets from $\sim500$ accounts.
    \item \textbf{Year}: 2020.
    \item \textbf{Sensitive features}: political leaning.
    \item \textbf{Link}: not available
    \item \textbf{Further info}: \citet{gordon2020studying}
\end{itemize}

\subsection{Twitter Trending Topics}
\begin{itemize}
    \item \textbf{Description}: this dataset was used to study the problem of fair recommendation. It comprises a random sample (1\%) of all tweets posted in the US between February and July 2017, obtained  through the Twitter Streaming API. This sample is paired with a collection of trending Twitter topics queried every 15-minutes through the Twitter REST API in July 2017. User interest in each topic was inferred using Twitter lists and follower-followee graphs. Finally, user demographics were also annotated to evaluate how user interest in different topics skews with respect to race, age, and gender. These attributes were obtained feeding user profile images to Face++.
    \item \textbf{Affiliation of creators}: Indian Institute of Technology Kharagpur; Max Planck Institute for Software Systems; Grenoble INP.
    \item \textbf{Domain}: social media.
    \item \textbf{Tasks in fairness literature}: fair ranking \citep{chakraborty2019equality}.
    \item \textbf{Data spec}: text.
    \item \textbf{Sample size}: $\sim200$M tweets by $\sim10$M users and $\sim10$K trending topics.
    \item \textbf{Year}: 2018.
    \item \textbf{Sensitive features}: race, age, and gender.
    \item \textbf{Link}: not available
    \item \textbf{Further info}: \citet{chakraborty2019equality}
\end{itemize}

\subsection{TwitterAAE}
\begin{itemize}
    \item \textbf{Description}: this resource was developed to study the use of dialect language on social media. The authors used Twitter APIs to collect public tweets sent on mobile phones from US users in 2013. They devise a distant supervision approach based on geolocation to annotate the probable language/dialect of the tweet, distinguishing between African American English (AAE) and Standard American English (SAE). To validate their approach, the creators studied the phonological and syntactic divergence of AAE tweets vs. SAE tweets, ensuring they align with linguistic phenomena that typically distinguish these variants of English.
    \item \textbf{Affiliation of creators}: University of Massachusetts Amherst.
    \item \textbf{Domain}: social media, linguistics.
    \item \textbf{Tasks in fairness literature}: fairness evaluation of sentiment analysis \citep{shen2018darling}, 
    fairness evaluation of private classification \citep{bagdasaryan2019differential}, fairness evaluation \citep{ballburack2021differential},
    robust fair language model \citep{hashimoto2018fairness},
    fairness evaluation of language identification \citep{blodgett2017racial}.
    \item \textbf{Data spec}: text.
    \item \textbf{Sample size}: $\sim8$M tweets.
    \item \textbf{Year}: 2016.
    \item \textbf{Sensitive features}: dialect (related to race).
    \item \textbf{Link}: \url{http://slanglab.cs.umass.edu/TwitterAAE/}
    \item \textbf{Further info}: \citet{blodgett2016demographic}
\end{itemize}

\subsection{US Harmonized Tariff Schedules (HTS)}
\begin{itemize}
    \item \textbf{Description}: this resource represents a comprehensive classification system for goods imported in the US, which defines the applicable tariffs. It defines a fine-grained categorization for goods, based e.g. on their material and shape. The chapter on apparel was explicitly criticized for its differential treatment of men's and women's clothing, effectively resulting in discriminatory tariffs for consumers.
    \item \textbf{Affiliation of creators}: US International Trade Commission.
    \item \textbf{Domain}: economics.
    \item \textbf{Tasks in fairness literature}: fairness evaluation \citep{luong2016classification}.
    \item \textbf{Data spec}: tabular data.
    \item \textbf{Sample size}: unknown.
    \item \textbf{Year}: present.
    \item \textbf{Sensitive features}: gender.
    \item \textbf{Link}: \url{https://hts.usitc.gov/current}
    \item \textbf{Further info}: \citet{barbaro2007apparel}
\end{itemize}

\subsection{UniGe}
\begin{itemize}
    \item \textbf{Description}: this dataset is connected with the \emph{DROP@UNIGE} project, aimed at studying the dynamics of university dropout, focusing on the University of Genoa as a case study. In ML fairness literature, the most common version of the dataset focuses on students who enrolled in 2017. Students are associated with attributes describing their ethnicity, gender, financial status, and prior school experience. The target variable encodes early academic success, as summarized by students' grades at the end of the first semester.
    \item \textbf{Affiliation of creators}: University of Genoa.
    \item \textbf{Domain}: education.
    \item \textbf{Tasks in fairness literature}: fair regression \citep{chzhen2020fairwassertein,chzhen2020fair}, fair representation learning \citep{oneto2019learning,oneto2020exploiting}.
    \item \textbf{Data spec}: tabular data.
    \item \textbf{Sample size}: $\sim5$K students.
    \item \textbf{Year}: unknown.
    \item \textbf{Sensitive features}: ethnicity, gender, financial status.
    \item \textbf{Link}: not availalbe
    \item \textbf{Further info}: \citet{oneto2017dropout}
\end{itemize}

\subsection{University Facebook Networks}
\begin{itemize}
    \item \textbf{Description}: a collection of 100 datasets shared with researchers in anonymized format by Adam D’Angelo of Facebook. The datasets used in the fairness literature consist of a 2005 snapshot from the Facebook network of the Universities of Oklahoma (Oklahoma97), North Carolina (UNC28), Caltech (Caltech36), Reed College (Reed98), and Michigan State (Michigan23), and links between them. User data comprises gender, class year, and anonymized data fields representing high school, major, and dormitory residences.
    \item \textbf{Affiliation of creators}: Facebook; University of North Carolina; Harvard University; University of Oxford.
    \item \textbf{Domain}: social networks.
    \item \textbf{Tasks in fairness literature}: fair graph mining \citep{li2021on}, fair graph augmentation \citep{ramachandran2021gaea}.
    \item \textbf{Data spec}: user-user pairs.
    \item \textbf{Sample size}: $\sim20$K people connected by $\sim1$M friend relations (Oklahoma97); $\sim20$K people connected by $\sim1$M friend relations (UNC28); $\sim30$K people connected by $\sim1$M friend relations (Michigan23); $\sim1$K people connected by $\sim20$K friend relations (Reed98); $\sim1$K people connected by $\sim20$K friend relations (Caltech36).
    \item \textbf{Year}: 2017.
    \item \textbf{Sensitive features}: gender.
    \item \textbf{Link}: \url{http://networkrepository.com/socfb-Oklahoma97.php} (Oklahoma97); \url{http://networkrepository.com/socfb-UNC28.php} (UNC28); \url{https://networkrepository.com/socfb-Michigan23.php} (Michigan23);
    \url{https://networkrepository.com/socfb-Reed98.php} (Reed98);
    \url{https://networkrepository.com/socfb-Caltech36.php} (Caltech36)
    \item \textbf{Further info}: \citet{red2011comparing}
\end{itemize}

\subsection{US Census Data (1990)}
\begin{itemize}
    \item \textbf{Description}: this resource is a one percent sample extracted from the 1990 US census data as a benchmark for clustering algorithms on large datasets. It contains a variety of features about different aspects of participants' lives, including demographics, wealth, and military service.
    \item \textbf{Affiliation of creators}: Microsoft.
    \item \textbf{Domain}: demography.
    \item \textbf{Tasks in fairness literature}: fair clustering \citep{backurs2019scalable,huang2019coresets,bera2019fair}, 
    fair clustering under unawareness \citep{esmaeili2020probabilistic},
    limited-label fairness evaluation \citep{sabato2020bounding}.
    \item \textbf{Data spec}: tabular data.
    \item \textbf{Sample size}: $\sim2$M respondents.
    \item \textbf{Year}: 1999.
    \item \textbf{Sensitive features}: age, sex.
    \item \textbf{Link}: \url{https://archive.ics.uci.edu/ml/datasets/US+Census+Data+(1990)}
    \item \textbf{Further info}: \citet{meek2002learning}
\end{itemize}

\subsection{US Family Income}
\begin{itemize}
    \item \textbf{Description}: this resource was compiled from the Current Population Survey (CPS) Annual Social and Economic (ASEC) Supplement. It contains income data for over 80,000 thousand US families, broken down by age and race (White, Black, Asian, and Hispanic).
    \item \textbf{Affiliation of creators}: US Bureau of Labor Statistics; US Census Bureau.
    \item \textbf{Domain}: economics.
    \item \textbf{Tasks in fairness literature}: fair subset selection under unawareness \citep{mehrotra2021mitigating}.
    \item \textbf{Data spec}: tabular data.
    \item \textbf{Sample size}: 4 races x 12 age categories x 41 income categories.
    \item \textbf{Year}: 2020.
    \item \textbf{Sensitive features}: age, race.
    \item \textbf{Link}: \url{https://www.census.gov/data/tables/time-series/demo/income-poverty/cps-finc/finc-02.html}
    \item \textbf{Further info}: \url{https://www2.census.gov/programs-surveys/cps/techdocs/cpsmar20.pdf}
\end{itemize}

\subsection{US Federal Judges}
\begin{itemize}
    \item \textbf{Description}: this dataset was extracted from \citet{epstein2013behavior} to study the problem of judicial subset selection from the point of view of justice, fairness and interpretability. Given the fact that in several judicial systems a subset of judges is selected from the whole judicial body to decide the outcome of appeals, the creators extract cases were three judges are required from \citet{epstein2013behavior}, covering the period 2000--2004. They emulate prior probabilities of affirmance/reversal for specific judges based on their past decisions. The task associated with this dataset is the optimal selection of a subset of judges, so that the procedure is interpretable, the subset contains at least one female (junior) judge and the decision of the subset coincides with the decision of the whole judicial body.
    \item \textbf{Affiliation of creators}: Yale University.
    \item \textbf{Domain}: law.
    \item \textbf{Tasks in fairness literature}: fair subset selection \citep{huang2020towards}.
    \item \textbf{Data spec}: judge-case pairs.
    \item \textbf{Sample size}: $\sim$300 judges selected for $\sim2$K cases.
    \item \textbf{Year}: 2020.
    \item \textbf{Sensitive features}: gender.
    \item \textbf{Link}: not available
    \item \textbf{Further info}: \citet{huang2020towards}
\end{itemize}

\subsection{US Student Performance}
\begin{itemize}
    \item \textbf{Description}: this resource represents students at an undisclosed US research university, spanning the Fall 2014 to Spring 2019 terms. The associated task is predicting student success based on university administrative records. Student features include demographics and academic information on prior achievement and standardized test scores.
    \item \textbf{Affiliation of creators}: Cornell University.
    \item \textbf{Domain}: education.
    \item \textbf{Tasks in fairness literature}: fairness evaluation \citep{lee2020evaluation}.
    \item \textbf{Data spec}: tabular data.
    \item \textbf{Sample size}: unknown.
    \item \textbf{Year}: 2020.
    \item \textbf{Sensitive features}: gender, racial-ethnic group.
    \item \textbf{Link}: not available
    \item \textbf{Further info}: \citet{lee2020evaluation}
\end{itemize}

\subsection{UTK Face}
\begin{itemize}
    \item \textbf{Description}: the dataset was developed as a diverse resource for face regression and progression (models of aging), where diversity is intended with respect to age, gender and race. The creators sourced part of the images from two existing datasets (Morph and CACD datasets). To increase the representation of some age groups, additional images were crawled from major search engines based on specific keywords (e.g., baby). Age, gender, and race were estimated through an algorithm and validated by a human annotator.
    \item \textbf{Affiliation of creators}: University of Tennessee.
    \item \textbf{Domain}: computer vision.
    \item \textbf{Tasks in fairness literature}: robust fairness evaluation \citep{nanda2021fairness}, 
    fairness evaluation of private classification \citep{bagdasaryan2019differential}, fairness evaluation \citep{segal2021fairness}, fair classification \citep{jung2021fair}.
    \item \textbf{Data spec}: image.
    \item \textbf{Sample size}: $\sim20$K face images.
    \item \textbf{Year}: 2017.
    \item \textbf{Sensitive features}: age, gender, race (inferred).
    \item \textbf{Link}: \url{https://susanqq.github.io/UTKFace/}
    \item \textbf{Further info}: \citet{zhifei2017age}
\end{itemize}

\subsection{Vehicle}
\begin{itemize}
    \item \textbf{Description}: this dataset comprises measurements from a distributed network of acoustic, seismic, and infrared sensors, as different types of military vehicles are driven in their proximity. This dataset was developed as part of a project supported by DARPA for the task of vehicle detection and type classification.
    \item \textbf{Affiliation of creators}: University of Wisconsin-Madison.
    \item \textbf{Domain}: signal processing.
    \item \textbf{Tasks in fairness literature}: fair federated learning \citep{li2020fair}.
    \item \textbf{Data spec}: time series.
    \item \textbf{Sample size}: unknown.
    \item \textbf{Year}: 2013.
    \item \textbf{Sensitive features}: none.
    \item \textbf{Link}: \url{http://www.ecs.umass.edu/mduarte/Software.html}
    \item \textbf{Further info}: \citet{duarte2004vehicle}
\end{itemize}

\subsection{Victorian Era Authorship Attribution}
\begin{itemize}
    \item \textbf{Description}: this resource was developed to benchmark different authorship attribution techniques. Querying the Gdelt database, the creators focus on English language authors from the 19th century with at least five books available. The corpus was split into text fragments of 1,000 words each. Only the most frequent 10,000 words were kept, while the remaining ones were removed.
    \item \textbf{Affiliation of creators}: Purdue University.
    \item \textbf{Domain}: literature.
    \item \textbf{Tasks in fairness literature}: fair clustering \citep{harb2020kfc}.
    \item \textbf{Data spec}: text.
    \item \textbf{Sample size}: $\sim100$K text fragments.
    \item \textbf{Year}: 2018.
    \item \textbf{Sensitive features}: textual references to people and their demographics.
    \item \textbf{Link}: \url{http://archive.ics.uci.edu/ml/datasets/Victorian+Era+Authorship+Attribution}
    \item \textbf{Further info}: \citet{gungor2018benchmarking}
\end{itemize}

\subsection{Visual Question Answering (VQA)}
\begin{itemize}
    \item \textbf{Description}: this dataset is curated as a benchmark for open-ended visual question answering. The collection features both real images from MS-COCO \citep{lin2014microsoft} and abstract scenes with human figures. Questions and answers were compiled by workers on Mechanical Turk who were instructed to formulate questions that require seeing the associated image for a correct answer.
    \item \textbf{Affiliation of creators}: Georgia Institute of Technology; Carnegie Mellon University; Army Research Lab;   Facebook AI Research.
    \item \textbf{Domain}: computer vision.
    \item \textbf{Tasks in fairness literature}: bias discovery \citep{manjunatha2019explicit}.
    \item \textbf{Data spec}: mixture (image, text).
    \item \textbf{Sample size}: $\sim1$M questions over $\sim300$K images.
    \item \textbf{Year}: 2017.
    \item \textbf{Sensitive features}: visual and textual references to gender.
    \item \textbf{Link}: \url{https://visualqa.org/}
    \item \textbf{Further info}: \citet{goyal2017making}
\end{itemize}

\subsection{Warfarin}
\begin{itemize}
    \item \textbf{Description}: this dataset was collected as part of a study about algorithmic estimation of optimal warfarin dosage as an oral anticoagulation treatment. The study was carried out by the International Warfarin Pharmacogenetics  Consortium, comprising 21 research groups from 9 countries and 4 continents. The dataset was co-curated by staff at the Pharmacogenomics Knowledge Base (PharmGKB) including, for thousands of patients at centers around the world, their demographics, comorbities, other medications and genetic factors, along with the steady-state dose of warfarin that led to stable levels of anticoagulation without adverse events.
    \item \textbf{Affiliation of creators}: PharmGKB; International Warfarin Pharmacogenetics Consortium.
    \item \textbf{Domain}: pharmacology.
    \item \textbf{Tasks in fairness literature}: fairness evaluation under unawareness \citep{kallus2020assessing}.
    \item \textbf{Data spec}: tabular data.
    \item \textbf{Sample size}: $\sim6K$ patients.
    \item \textbf{Year}: 2009.
    \item \textbf{Sensitive features}: sex, ethnicity, age.
    \item \textbf{Link}: \url{https://www.pharmgkb.org/downloads}
    \item \textbf{Further info}: \citet{international2009estimation}
\end{itemize}

\subsection{Waterbirds}
\begin{itemize}
    \item \textbf{Description}: this computer vision dataset consists of photos where subjects and backgrounds are carefully paired to induce spurious correlations. Subjects are birds, taken from the CUB dataset \citep{wah2011Caltechucsd}, divided into waterbirds and landbirds. Pixel-level segmentation  masks are exploited to cut out subjects and paste them onto land or water backgrounds from the Places dataset \citep{zhou2018:places}. While in the provided validation and test splits both landbirds and waterbirds appear with the same frequency on either background, the training split is imbalanced so that 95\% of all waterbirds are placed against a water background and 95\% of all landbirds are depicted against a land background.
    \item \textbf{Affiliation of creators}: Stanford University; Microsoft.
    \item \textbf{Domain}: computer vision.
    \item \textbf{Tasks in fairness literature}: fairness evaluation of selective classification \citep{jones2021selective}.
    \item \textbf{Data spec}: image.
    \item \textbf{Sample size}: $\sim10$K images.
    \item \textbf{Year}: 2021.
    \item \textbf{Sensitive features}: none.
    \item \textbf{Link}: \url{https://github.com/ejones313/worst-group-sc/tree/main/src/data}
    \item \textbf{Further info}: \citet{sagawa2020distributionally}
\end{itemize}

\subsection{WebText}
\begin{itemize}
    \item \textbf{Description}: this resource is a web scrape collected to train the GPT-2 language model. The authors considered all outbound links from Reddit which collected at least 3 \emph{karma}. This inclusion criterion signals that the link received some upvotes by redditors and is treated as a quality heuristic for the webpage. To extract text data from each link, a combination of Dragnet \citep{peters2013content} and Newspaper\footnote{\url{https://github.com/codelucas/newspaper}} extractors was exploited. The curators performed deduplication and removed all Wikipedia pages to reduce text overlap with Wikipedia-based datasets.
    \item \textbf{Affiliation of creators}: OpenAI.
    \item \textbf{Domain}: linguistics.
    \item \textbf{Tasks in fairness literature}: data bias evaluation \citep{tan2019assessing}.
    \item \textbf{Data spec}: text.
    \item \textbf{Sample size}: $\sim$ 8M documents.
    \item \textbf{Year}: 2019.
    \item \textbf{Sensitive features}: textual references to people and their demographics.
    \item \textbf{Link}: \url{https://github.com/openai/gpt-2-output-dataset} (partial)
    \item \textbf{Further info}: \citet{radford2019language}
\end{itemize}

\subsection{Wholesale}
\begin{itemize}
    \item \textbf{Description}: this dataset represents Portuguese businesses from the catering industry purchasing goods from the same wholesaler. The businesses are located in Lisbon, Oporto, and a third undisclosed area; 298 are from the Horeca (Hotel/Restaurant/Café) channel and 142 from the Retail channel. Each data point comprises this information along with yearly expenditures on different categories of products (e.g. milk, frozen goods, delicatessen). Collection of this data was presumably carried out by the wholesaler in a business intelligence initiative primarily aimed at customer segmentation and targeted marketing.
    \item \textbf{Affiliation of creators}: Université Pierre et Marie Curie; University Institute of Lisbon; INRIA.
    \item \textbf{Domain}: marketing.
    \item \textbf{Tasks in fairness literature}: fair data summarization \citep{jones2020fair}.
    \item \textbf{Data spec}: tabular data.
    \item \textbf{Sample size}: $\sim400$ businesses.
    \item \textbf{Year}: 2014.
    \item \textbf{Sensitive features}: geography.
    \item \textbf{Link}: \url{http://archive.ics.uci.edu/ml/datasets/wholesale+customers}
    \item \textbf{Further info}: \citet{baudry2015enhancing}
\end{itemize}

\subsection{Wikidata}
\begin{itemize}
    \item \textbf{Description}: founded in 2012, Wikidata is a free, collaborative, multilingual knowledge base, maintained by editors and partly automated. It consists of items linked by properties. The most common items include humans, administrative territorial entities, architectural structures, chemical compounds, films, and scholarly articles.
    \item \textbf{Affiliation of creators}: Wikimedia Foundation.
    \item \textbf{Domain}: information systems.
    \item \textbf{Tasks in fairness literature}: fairness evaluation in graph mining \citep{fisher2020measuring}.
    \item \textbf{Data spec}: item-property-value triples.
    \item \textbf{Sample size}: $\sim90$M items.
    \item \textbf{Year}: present.
    \item \textbf{Sensitive features}: demographics of people featured in entities (age, sex, geography) and their relations.
    \item \textbf{Link}: \url{https://www.wikidata.org/wiki/Wikidata:Data_access}
    \item \textbf{Further info}: \url{https://www.wikidata.org/wiki/Wikidata:Main_Page}
\end{itemize}

\subsection{Wikipedia dumps}
\begin{itemize}
    \item \textbf{Description}: Wikipedia dumps are maintained and updated regularly by the Wikimedia Foundation. Typically, they contain every article available in a language at a given time. As a large source of curated text, they have often been used by the natural language processing and computational linguistics communities to extract models of human language. We find usage of German, English, Mandarin Chinese, Spanish, Arabic, French, Farsi, Urdu, and Wolof dumps in the surveyed articles.
    \item \textbf{Affiliation of creators}: Wikimedia Foundation.
    \item \textbf{Domain}: linguistics.
    \item \textbf{Tasks in fairness literature}: bias evaluation in WEs \citep{liang2020artificial,papakyriakopulos2020bias,brunet2019understanding,chen2021gender}.
    \item \textbf{Data spec}: text.
    \item \textbf{Sample size}: $\sim6$M articles (EN), $\sim3$M articles (DE) as of May 2021.
    \item \textbf{Year}: present.
    \item \textbf{Sensitive features}: textual references to people and their demographics.
    \item \textbf{Link}: \url{https://dumps.wikimedia.org/enwiki/}; \url{https://dumps.wikimedia.org/dewiki/}
    \item \textbf{Further info}: \url{https://meta.wikimedia.org/wiki/Data_dumps}
\end{itemize}

\subsection{Wikipedia Toxic Comments}
\begin{itemize}
    \item \textbf{Description}: this dataset was developed as a resource to analyze discourse and personal attacks on Wikipedia talk pages, which are used by editors to discuss improvements. It is aimed at using ML for better online conversations and flag posts that are likely to make other participants leave. The data consists of Wikipedia comments labelled by 5,000 crowd-workers according to their toxicity level (toxic, severe\_toxic) and type (obscene, threat, insult, identity\_hate). This resource powers a public Kaggle competition.
    \item \textbf{Affiliation of creators}: Wikimedia foundation; Google.
    \item \textbf{Domain}: social media.
    \item \textbf{Tasks in fairness literature}: fair classification, \citep{garg2019counterfactual,shah2021rawslian}, fairness evaluation \citep{dixon2018measuring}.
    \item \textbf{Data spec}: text.
    \item \textbf{Sample size}: $\sim160$K comments.
    \item \textbf{Year}: 2017.
    \item \textbf{Sensitive features}: textual reference to people and their demographics.
    \item \textbf{Link}: \url{https://www.kaggle.com/c/jigsaw-toxic-comment-classification-challenge}
    \item \textbf{Further info}: \url{https://www.perspectiveapi.com/research/}
\end{itemize}

\subsection{Willingness-to-Pay for Vaccine}
\begin{itemize}
    \item \textbf{Description}: this dataset resulted from a study of willingness to pay for a vaccine against tick-borne encephalitis in Sweden. Thousands of citizens from different areas of the country filled in a survey about exposure, risk perception, knowledge, and protective behavior related to ticks and tick-borne diseases, along with socioeconomic information. The central question of the survey asks how much respondents would be willing to pay for a vaccine that provides a three-year protection against tick-borne encephalitis.
    \item \textbf{Affiliation of creators}: University of Gothenburg.
    \item \textbf{Domain}: public health.
    \item \textbf{Tasks in fairness literature}: fair pricing evaluation \citep{kallus2021fairness}.
    \item \textbf{Data spec}: tabular data.
    \item \textbf{Sample size}: $\sim2$K respondents.
    \item \textbf{Year}: 2015.
    \item \textbf{Sensitive features}: age, gender, geography.
    \item \textbf{Link}: \url{https://snd.gu.se/sv/catalogue/study/snd0987/1#dataset}
    \item \textbf{Further info}: \citet{slunge2015willingness}
\end{itemize}

\subsection{Winobias}
\begin{itemize}
    \item \textbf{Description}: similarly to Winogender, this benchmark was built to study coreference resolution and gender bias, focusing on words that relate to professions with diverse gender representation. Example: ``The physician hired the secretary because he (she) was overwhelmed with clients''. The correct pronoun resolution is clear from the syntax or semantics of the sentence and can be either stereotypical or counter-stereotypical. The accuracy of biased coreference resolution systems will vary accordingly.
    \item \textbf{Affiliation of creators}: University of California Los Angeles;  University of Virginia; Allen Institute for Artificial Intelligence.
    \item \textbf{Domain}: linguistics.
    \item \textbf{Tasks in fairness literature}: fair entity resolution evaluation. \citep{vig2020investigating}.
    \item \textbf{Data spec}: text.
    \item \textbf{Sample size}: $\sim3$K sentences.
    \item \textbf{Year}: 2020.
    \item \textbf{Sensitive features}: gender.
    \item \textbf{Link}: \url{https://github.com/uclanlp/corefBias/tree/master/WinoBias/wino}
    \item \textbf{Further info}: \citet{zhao2018gender}
\end{itemize}

\subsection{Winogender}
\begin{itemize}
    \item \textbf{Description}: this dataset was crafted to systematically study gender bias in systems for coreference resolution, the task of resolving whom pronouns refer to in a sentence. This resource follows the Winograd schemas, with sentence templates mentioning a profession (nurse), a participant (patient), and a pronoun referring to either one of them: ``The nurse notified the patient that her/his/their shift would be ending in an hour.'' Sentence templates have been crafted so that the pronoun resolution can be done unambiguously based on contextual information, hence unbiased systems should display similar error rates, regardless of gender concentrations in different professions. The ground truth for each sentence has been validated by workers on Mechanical Turk with accuracy over 99\%.
    \item \textbf{Affiliation of creators}: Johns Hopkins University.
    \item \textbf{Domain}: linguistics.
    \item \textbf{Tasks in fairness literature}: fair entity resolution evaluation \citep{vig2020investigating}, fairness evaluation in entity recognition \citep{mishra2020assessing}.
    \item \textbf{Data spec}: text.
    \item \textbf{Sample size}: $\sim700$ sentences.
    \item \textbf{Year}: 2018.
    \item \textbf{Sensitive features}: gender.
    \item \textbf{Link}: \url{https://github.com/rudinger/winogender-schemas}
    \item \textbf{Further info}: \citet{rudinger2018gender}
    \item \textbf{Variants}: Winogender-NER \citep{mishra2020assessing} is a modified version of the template appropriate for named entity recognition.
\end{itemize}

\subsection{Word Embedding Association Test (WEAT)}
\begin{itemize}
    \item \textbf{Description}: this resource was created to audit biases in English WEs. Following the Implicit Association Test (IAT) from social psychology \citep{greenwald1998measuring}, this dataset defines two groups of target words, relating e.g. to flowers and insects, and two groups of attribute words, relating e.g. to pleasantness and unpleasantness. The dataset can be used to measure biased associations between the target words and the attribute words represented by a set of WEs. WEAT comprises ten tests across different word categories. The most salient for the purposes of algorithmic fairness support tests of associations between race and pleasantness, age and pleasantness, gender and career (vs family), gender and propensity to math (vs arts). Race-related words are first names predominantly associated with African American or European American individuals. Gender is encoded in a similar fashion, or with intrinsically gendered words (e.g. mother).
    \item \textbf{Affiliation of creators}: Princeton University; University of Bath.
    \item \textbf{Domain}: linguistics.
    \item \textbf{Tasks in fairness literature}: bias evaluation in WEs \citep{brunet2019understanding,guo2021detecting}.
    \item \textbf{Data spec}: text.
    \item \textbf{Sample size}: $\sim$10 groups of words, with $\sim$10-60 words in each group.
    \item \textbf{Year}: 2017.
    \item \textbf{Sensitive features}: race, gender.
    \item \textbf{Link}: \url{https://arxiv.org/pdf/1608.07187.pdf}
    \item \textbf{Further info}: \citet{caliksan2017semantics}
\end{itemize}

\subsection{Yahoo! A1 Search Marketing}
\begin{itemize}
    \item \textbf{Description}: this dataset contains bids from all advertisers who participated in Yahoo! Search Marketing auctions for the top 1000 search queries from June 15, 2002, to June 14, 2003. The identities of advertisers and the queries they target are anonymized for confidentiality reasons.
    \item \textbf{Affiliation of creators}: Yahoo! Labs.
    \item \textbf{Domain}: marketing.
    \item \textbf{Tasks in fairness literature}: fair advertising \citep{celis2019toward,nasr2020bidding}.
    \item \textbf{Data spec}: advertiser-keyword pairs.
    \item \textbf{Sample size}: $\sim20$M bids by $\sim10$K advertisers over $\sim1$K search queries.
    \item \textbf{Year}: after $2003$.
    \item \textbf{Sensitive features}: none.
    \item \textbf{Link}: \url{https://webscope.sandbox.yahoo.com/catalog.php?datatype=a}
    \item \textbf{Further info}: 
\end{itemize}

\subsection{Yahoo! c14B Learning to Rank}
\begin{itemize}
    \item \textbf{Description}: this resource consists of 2 datasets which encode the interactions of Yahoo! users with the search engine in the US and an unknown Asian country. This data is a subset of the entire training set used internally to train the ranking functions of the Yahoo! search engine. Textual features are deliberately obfuscated and the final data consists of numerical features which encode query-document pairs. Query-document pairs are assigned multigraded relevance judgements by a professional editor.
    \item \textbf{Affiliation of creators}: Yahoo! Labs.
    \item \textbf{Domain}: information systems.
    \item \textbf{Tasks in fairness literature}: fair ranking \citep{singh2019policy}.
    \item \textbf{Data spec}: query-document pairs.
    \item \textbf{Sample size}: $\sim40$K queries, $\sim900$K documents.
    \item \textbf{Year}: 2011.
    \item \textbf{Sensitive features}: none.
    \item \textbf{Link}: \url{https://webscope.sandbox.yahoo.com/catalog.php?datatype=c}
    \item \textbf{Further info}: \citet{chapelle2010yahoo}
\end{itemize}

\subsection{YouTube Dialect Accuracy}
\begin{itemize}
    \item \textbf{Description}: this dataset was curated to audit the accuracy of YouTube's automated captioning system across two genders and five dialects of English. Eighty speakers were sampled from videos matching the query ``accent challenge $<$region$>$'' or ``accent tag $<$region$>$'', where $<$region$>$ is one of five areas selected for geographic separation and distinct local dialects: California, Georgia, New England, New Zealand and Scotland. This curation choice targets a popular internet phenomenon (called ``accent tag'', ``dialect meme'' or ``accent challenge'') consisting of videos of people from different areas presenting themselves and their linguistic background, subsequently reading a list of words designed to elicit pronounciation differences dependent on dialect. This resource focuses only on the word portion of these videos, with a ``phonetically-trained listener familiar with the dialects'' performing the annotation for word caption accuracy.
    \item \textbf{Affiliation of creators}: University of Washington.
    \item \textbf{Domain}: social media.
    \item \textbf{Tasks in fairness literature}: fairness evaluation of speech recognition \citep{tatman2017:gender}.
    \item \textbf{Data spec}: tabular data.
    \item \textbf{Sample size}: $\sim100$ speakers.
    \item \textbf{Year}: 2016.
    \item \textbf{Sensitive features}: gender, geography.
    \item \textbf{Link}: \url{https://github.com/rctatman/youtubeDialectAccuracy}
    \item \textbf{Further info}: \citet{tatman2017:gender}
\end{itemize}

\subsection{Yow news}
\begin{itemize}
    \item \textbf{Description}: this dataset was collected to support research on personalized information integration and retrieval. The data, consisting of implicit and explicit user feedback stored in interaction logs, was gathered in a user study via a special browser accessing a web-based news story filtering system. The task associated with this resource is personalized news recommendation.
    \item \textbf{Affiliation of creators}: Carnegie Mellon University.
    \item \textbf{Domain}: news, information systems.
    \item \textbf{Tasks in fairness literature}: fair ranking \citep{singh2018fairness}.
    \item \textbf{Data spec}: user-story pairs.
    \item \textbf{Sample size}: $\sim10$K interaction logs.
    \item \textbf{Year}: 2009.
    \item \textbf{Sensitive features}: news provider.
    \item \textbf{Link}: \url{https://users.soe.ucsc.edu/~yiz/papers/data/YOWStudy/}
    \item \textbf{Further info}: \citet{zhang2005bayesian}; \url{https://users.soe.ucsc.edu/~yiz/piir/}
\end{itemize}

\subsection{Zillow Searches}
\begin{itemize}
    \item \textbf{Description}: this is a proprietary dataset from Zillow, a famous real estate marketplace. It consists of a random sample of over 13,000 search sessions covering more than 36,000 property listings. Each listing consists of several features, some of which are considered salient by the creators and a sensible target for fair ranking algorithms. Among these are the ownership of the house (Zillow, independent realtor, new construction listed by builders) and the availability of 3D/video tours of the property. This dataset was collected internally to study the problem of fair recommendation and ranking on Zillow data.
    \item \textbf{Affiliation of creators}: Boston University; Zillow Group.
    \item \textbf{Domain}: information systems.
    \item \textbf{Tasks in fairness literature}: fair ranking \citep{chaudhari2020general}.
    \item \textbf{Data spec}: unknown.
    \item \textbf{Sample size}: $\sim10$K search sessions featuring $\sim40$K property listings.
    \item \textbf{Year}: 2020.
    \item \textbf{Sensitive features}: ownership, tour availability.
    \item \textbf{Link}: not available
    \item \textbf{Further info}: \citet{chaudhari2020general}
\end{itemize}

\section{Adult}
\label{sec:adult}

Key references include \citet{cohany1994revisions,kohavi1996scaling,kohavi1994adult_uci,usdeptcomm1995current,hardt2021facing,mckenna2019:history,mckenna2019:history_drb}.



\subsection{Datasheet}

\subsubsection{Motivation}
\begin{itemize}
    \item \que{For what purpose was the dataset created?}
    
    \ans{The Adult dataset was created as a resource to benchmark the performance of machine learning algorithms. Rather than powering a specific task or application, the dataset was likely chosen as a real-world source of socially relevant data \citep{kohavi1996scaling}.}
    \item \que{Who created the dataset?}
    
    \ans{\textbf{Barry Becker} extracted this dataset from the 1994 Census database. Ronny Kohavi and Barry Becker donated it to UCI Machine Learning Repository in 1996. At that time, both were working for Silicon Graphics Inc \citep{kohavi1994adult_uci}}
    \item \que{Who funded the creation of the dataset?}
    
    \ans{The underlying database is a product of the Current Population Survey (CPS) of March 1994, a joint effort by the US Census Bureau and the US Bureau of Labor Statistics (BLS), funded by the US federal government. The extraction of Adult from the larger database was plausibly part of work remunerated by Silicon Graphics.}
\end{itemize}
\subsubsection{Composition}
\begin{itemize}
    \item \que{What do the instances that comprise the dataset represent?}
    
    \ans{Each instance is a \textbf{March 1994 CPS respondent}, represented along demographic and socio-economic dimensions.}
    \item \que{How many instances are there in total?}
    
    \ans{The dataset consists of \textbf{48,842 instances}.}
    \item \que{Does the dataset contain all possible instances or is it a sample (not necessarily random) of instances from a larger set?}
    
    \ans{Adult contains individuals from a \textbf{sample} of US households, extracted from the 1994 Annual Social and Economic Supplement (ASEC) of the CPS with the following query: 
    \begin{equation*}
        (AAGE>16) \&\& (AGI>100) \&\& (AFNLWGT>1)\&\& (HRSWK>0).
    \end{equation*}
    This means Adult focuses on a subset of ASEC respondents aged 17 or older, whose income is above \$100, working at least 1 hour per week. While these were conceived as conditions to filter out noisy records \citep{kohavi1994adult_uci}, they may introduce sampling effects. Moreover, the 1994 CPS data was itself a sample, selected according to Census Bureau best practices, reaching over 70,000 households in nearly 2,000 US counties. The March 1994 CPS sample aimed at obtaining more reliable information on the Hispanic population, and was hence extended to an additional 2,500 eligible housing units.}
    \item \que{What data does each instance consist of?}
    
    \ans{Each instance consists of a combination of nominal, ordinal and continuous attributes, denominated age, workclass, fnlwgt, education, education-num, marital-status, occupation, relationship, race, sex, capital-gain, capital-loss, hours-per-week, native-country. See Table \ref{tab:adult_meta1} for a detailed explanation of features and their values.}
    \item \que{Is there a label or target associated with each instance?}
    
    \ans{\textbf{Yes}. Each person instance comes with a binary label encoding whether their income is above a $50,000$ threhsold.}
    \item \que{Is any information missing from individual instances?}
    
    \ans{\textbf{Yes}. Over $7\%$ of the instances have missing values. This is likely due to issues with data recording and coding or respondents' inability to recall information.}
    \item \que{Are relationships between individual instances made explicit e.g., users’ movie ratings, social network links)?}
    
    \ans{\textbf{No}. Some instances are related persons from the same household \citep{usdeptcomm1995current} but this information is not reported in the dataset.}
    \item \que{Are there recommended data splits?}
    
    \ans{\textbf{Yes}. The dataset comes with a specified train/test split made using MLC++ GenCVFiles, resulting in a 2/3--1/3 random split \citep{kohavi1994adult_uci}. The training set consists of 32561 instances, the test set of 16281 instances.}
    \item \que{Are there any errors, sources of noise, or redundancies in the dataset?}
    
    \ans{\textbf{Yes}. Sources of error include definitional difficulties, differences in interpretation of questions, respondents inability or unwillingness to provide correct information, errors made during data collection, data processing or missing value imputation. The tendency in household surveys for respondents to under-report their income was an explicit concern. Finally, noise infusion such as topcoding (saturation to \$99,999) was applied to avoid re-identification of certain individuals \citep{usdeptcomm1995current}.}
    \item \que{Is the dataset self-contained, or does it link to or otherwise rely on external resources?}
    
    \ans{The dataset is \textbf{self-contained}.}
    \item \que{Does the dataset contain data that might be considered confidential?}
    
    \ans{\textbf{Yes}. The data is protected by Title 13 of the United States Code, protecting individuals against identification from Census data.\footnote{\url{https://www.census.gov/about/policies/privacy/data_stewardship/title_13_-_protection_of_confidential_information.html}}}
    \item \que{Does the dataset contain data that, if viewed directly, might be offensive, insulting, threatening, or might otherwise cause anxiety?}
    
    \ans{\textbf{No}, not strictly. Interpreting the question more broadly, however, the envisioned racial and sexual categories may be deemed inadequate.}
    
    \item \que{Does the dataset identify any subpopulations (e.g., by age, gender)?}
    
    \ans{\textbf{Yes}. The dataset provides information on sex, age and race of respondents. These were self-reported, although self-identification was bounded by envisioned categories. These are (female, male) for sex and (White, Black, American Indian/Aleut Eskimo, Asian or Pacific Islander, Other) for race. Table \ref{tab:adult_demographics} summarizes the marginal distribution of the Adult dataset across these subpopulations.}
    \begin{table}[h] 
    \centering
    \begin{tabular}{||c c||}
        \hline 
        \textbf{Demographic Caracteristic} & \qquad \textbf{Values} \\ [1ex]
        \hline
        Percentage of male subjects & \qquad 66.85\% \\[1ex]
        Percentage of female subjects & \qquad 33.15\% \\[1ex]
        \hline
        Percentage of White subjects & \qquad 85.50\% \\[1ex]
        Percentage of Black subjects & \qquad 9.60\% \\[1ex]
        Percentage of Asian-Pac-Islander subjects & \qquad 3.11\% \\[1ex]
        Percentage of Amer-Indian-Eskimo subjects & \qquad 0.96\% \\[1ex]
        Percentage of people belonging to other races & \qquad 0.83\% \\[1ex]
        \hline
        Percentage of people between 16-19 years old & \qquad 5.14\% \\[1ex]
        Percentage of people between 20-29 years old & \qquad 24.58\% \\[1ex]
        Percentage of people between 30-39 years old & \qquad 26.47\% \\[1ex]
        Percentage of people between 40-49 years old & \qquad 21.95\% \\[1ex]
        Percentage of people between 50-59 years old & \qquad 13.55\% \\[1ex]
        Percentage of people between 60-69 years old & \qquad 6.25\% \\[1ex]
        Percentage of people between 70-79 years old & \qquad 1.67\% \\[1ex]
        Percentage of people between 80-89 years old & \qquad 0.27\% \\[1ex]
        Percentage of people between 90-99 years old & \qquad 0.11\% \\[1ex]
        \hline
        \end{tabular}
    \caption{Demographic Characteristics of the Adult dataset.} \label{tab:adult_demographics}
    \end{table}
    \item \que{Is it possible to identify individuals (i.e., one or more natural persons), either directly or indirectly (i.e., in combination with other data) from the dataset?} 
    
    \ans{\textbf{Unknown}. Important variables for data re-identification, such as birth date or ZIP code, are absent from the Adult dataset. However, instances in this dataset may be linked to the original CPS 1994 data \citep{hardt2021facing}. Moreover, re-identification studies internal to the Census Bureau pointed to combinations of variables that could potentially be used to re-identify respondents from Census microdata \citep{mckenna2019:history_drb}.}
    \item \que{Does the dataset contain data that might be considered sensitive in any way?}
    
    \ans{\textbf{Yes}. This dataset contains sensitive data, such as sex, race, native country and financial situation of respondents.}
    \item \que{Any other comments?}
    
    \ans{A precise definition for the variable called fnlwgt is unknown. It was used by Census Bureau statisticians to obtain population-level estimates from the CPS sample. For this reason, its use in classification tasks would be unusual.}
\end{itemize}
\subsubsection{Collection process}
\begin{itemize}
    \item \que{How was the data associated with each instance acquired?}
    
    \ans{Trained interviewers asked questions directly to respondents \citep{usdeptcomm1995current}. The data was made available through US Census data products which were used by Barry Becker to extract the Adult dataset.}
    \item \que{What mechanisms or procedures were used to collect the data?}
    
    \ans{Interviewers conducted the survey either in person at the respondent's home or by phone. They used laptop computers with ad-hoc software to prompt questions and record answers. At the end of each day, interviewers transmitted the collected data via modem to the Bureau headquarters \citep{usdeptcomm1995current}}.
    \item \que{If the dataset is a sample from a larger set, what was the sampling strategy?}
    
    \ans{A probabilistic sample was selected according to US Census Bureau best practice, with a muti-stage stratified design. The US territory was divided into strata, from which one county (or group of counties) was selected. From each selected county a sample of addresses was later obtained and added to the sample \citep{usdeptcomm1978current}. Barry Becker extracted a ``set of reasonably clean records'' using the following conditions:
    \begin{equation*}
        (AAGE>16) \&\& (AGI>100) \&\& (AFNLWGT>1)\&\& (HRSWK>0).
    \end{equation*}
    }
    \item \que{Who was involved in the data collection process and how were they compensated?}
    
    \ans{Interviewers trained by the US Census Bureau were involved in the data collection process. Data extraction was later performed by Barry Becker while affiliated with Silicon Graphics. Their compensation is unknown.}
    \item \que{Over what timeframe was the data collected?}
    
    \ans{Respondents were interviewed in March 1994, while the Adult dataset was donated to UCI ML Repository in May 1996.}
    \item \que{Were any ethical review processes conducted?}
    
    \ans{The Microdata Review Panel likely reviewed this data for compliance with Title 13 \citep{mckenna2019:history_drb} and authorized its publication.}

    \item \que{Was the data collected from the individuals in question directly, or obtain it via third parties or other sources?} 
    
    \ans{\textbf{Directly}. US Census Bureau interviewers collected the data through interviews, conducted in person or over the phone. Danny Kohavi and Barry Becker later processed this data, obtaining it from the Census Bureau website.}
    \item \que{Were the individuals in question notified about the data collection?}
    
    \ans{\textbf{Yes}. Individuals knew they were part of a sample chosen by the Census Bureau chosen for statistical analysis. They were not notified about their data being included in the Adult dataset.}
    \item \que{Did the individuals in question consent to the collection and use of their data?}
    
    \ans{\textbf{Yes}. For the CPS, participation is voluntary. A recent version of the information provided to respondents before interviews is available on the US Census Website.\footnote{\url{https://www2.census.gov/programs-surveys/cps/advance_letter.pdf}}}
    \item \que{If consent was obtained, were the consenting individuals provided with a mechanism to revoke their consent in the future or for certain uses?}
    
    \ans{\textbf{Unknown}.}
    \item \que{Has an analysis of the potential impact of the dataset and its use on data subjects been conducted?}
    
    \ans{\textbf{Yes}. Re-identification studies have been conducted both internally \citep{mckenna2019:history_drb} and externally \citep{rocher2019estimating} on Census Bureau data. \citet{mckenna2019:history_drb} mention finding combinations of variables on Census files that can lead to successful re-identification, which were subsequently removed or protected with noise injection. \citet{rocher2019estimating} demonstrate on the Adult dataset that the likelihood of a specific individual to have been correctly re-identified can be estimated with high accuracy. We are unaware of studies about the potential impact of successful re-identification on respondents.}
\end{itemize}
\subsubsection{Preprocessing/cleaning/labelling}
\begin{itemize}
    \item \que{Was any preprocessing/cleaning/labeling of the data done?}
    
    \ans{\textbf{Yes}. Preprocessing operations by the Census Bureau include missing value imputation and topcoding. Furthermore, Barry Becker and Ron Kohavi binarized the income variable ($>\$50$K) and discarded several CPS respondents who are not included in the Adult dataset.}
    \item \que{Was the “raw” data saved in addition to the preprocessed/cleaned/labeled data?}
    
    \ans{\textbf{Unknown}.}
    \item \que{Is the software used to preprocess/clean/label the instances available?}
    
    \ans{\textbf{Likely no}. It seems unlikely for the code to be available 25 years after its last known use.}
\end{itemize}
\subsubsection{Uses}
\begin{itemize}
    \item \que{For what tasks has the dataset been used?}
    
    \ans{This dataset probably owes its status in the ML community to an early position of publicly-available and interesting resource based on real-world data. For this reason, rather than powering specific applications, Adult is used as a benchmark for classifiers in many fields of machine learning. Due to its encoding of sensitive attributes, it has also become the most used dataset in the fair ML literature.}
    \item \que{Is there a repository that links to any or all papers or systems that use the dataset?}
    
    \ans{\textbf{Yes}. A selection of early works (pre-2005) using this dataset can be found in \citet{kohavi1994adult_uci}. A more recent list is available under the beta version of the UCI ML Repository.\footnote{\url{https://archive-beta.ics.uci.edu/ml/datasets/2}} See Appendix \ref{sec:adult_db} for a (non-exhaustive) list of algorithmic fairness works using this resource.}
    \item \que{What (other) tasks could the dataset be used for?}
    
    \ans{The Adult dataset is used in tasks where data of social significance is deemed important, for example privacy-preserving ML.}
    \item \que{Is there anything about the composition of the dataset or the way it was collected and preprocessed/cleaned/labeled that might impact future uses?}
    
    \ans{\textbf{Yes}. The threshold used to quantize income for a binary classification task is very high (\$50K). As a result a trivial rejector achieves very large accuracy on the black subpopulation (93\%). For the same reason, models are often more accurate for the female subpopulation than for the male one \citep{hardt2021facing}. Some numerical results on Adult may be an artifact of this threshold choice.}
    \item \que{Are there tasks for which the dataset should not be used?}
    
    \ans{Based on the previous answer, we caution against drawing overarching conclusions based on experimental results obtained on this dataset alone.}
\end{itemize}
\subsubsection{Distribution}
\begin{itemize}
    \item \que{Is the dataset distributed to third parties outside of the entity on behalf of which the dataset was created?}
    
    \ans{\textbf{Yes}. The dataset is publicly available \citep{kohavi1994adult_uci}.}
    \item \que{How is the dataset distributed?}
    
    \ans{The dataset is available as a \textbf{csv file}.}
    \item \que{When was the dataset distributed?}
    
    \ans{The dataset was released on the UCI ML Repository in \textbf{May 1996}.}
    \item \que{Is the dataset distributed under a copyright or other intellectual property (IP) license, and/or under applicable terms of use (ToU)?}
    
    \ans{\textbf{Yes}. The UCI ML repository has a citation policy. Terms of Use concerning the privacy of CPS respondents are likely to apply.}
    \item \que{Have any third parties imposed IP-based or other restrictions on the data associated with the instances?}
    
    \ans{\textbf{Likely no}. We are unaware of any IP-based restrictions.}
    \item \que{Do any export controls or other regulatory restrictions apply to the dataset or to individual instances?}
    
    \ans{\textbf{Likely no}.}
\end{itemize}
\subsubsection{Maintenance}
\begin{itemize}
    \item \que{Who is supporting/hosting/maintaining the dataset?}
    
    \ans{The dataset is hosted and maintained by the \textbf{UCI Machine Learning Repository} \citep{kohavi1994adult_uci}.}
    \item \que{How can the owner/curator/manager of the dataset be contacted?}
    
    \ans{Comments and inquiries may be directed at ml-repository@ics.uci.edu. Ronny Kohavi is the primary contact for this specific resource, available at ronnyk@live.com.}
    \item \que{Is there an erratum?}
    
    \ans{\textbf{Likely no}. We are unaware of any erratum.}
    \item \que{Will the dataset be updated?}
    
    \ans{A superset of the dataset without quantization of the target income variable is available \citep{hardt2021facing}.}
    \item \que{If the dataset relates to people, are there applicable limits on the retention of the data associated with the instances?}
    
    \ans{\textbf{Unknown}.}
    \item \que{Will older versions of the dataset continue to be supported/hosted/maintained?}
    
    \ans{Unless otherwise indicated, the Adult dataset will remain hosted on the UCI ML Repository in its current version.}
    \item\que{ If others want to extend/augment/build on/contribute to the dataset, is there a mechanism for them to do so?}
    
    \ans{\textbf{Unknown}.}
\end{itemize}

\clearpage
\subsection{Data Nutrition Label}

\ifconf\else{\subsubsection{Metadata}}\fi

\begin{table}[h]
\centering
\begin{tabular}{||r l||}
 \hline 
 \multicolumn{2}{|c|}{\textbf{METADATA}}\\
 \hline
 \textbf{Filenames} & \qquad adult  \\ 
 \textbf{Format} & \qquad csv \\
 \textbf{Url} & \qquad \url{https://archive.ics.uci.edu/ml/datasets/adult}  \\
 \textbf{Domain} & \qquad Economics  \\
 \textbf{Keywords} & \qquad US census, income  \\
 \textbf{Type} & \qquad Tabular  \\
 \textbf{Rows} & \qquad 48842 \\ 
 \textbf{Columns} & \qquad 14  \\
  \textbf{\% of missing cells} & \qquad 0.9\%  \\
 \textbf{Rows with missing cells} & \qquad 7\%  \\
 \textbf{License} & \qquad UCI Repository citation policy  \\
 \textbf{Released} & \qquad May 1996  \\
 \textbf{Range} & \qquad 1994 \\ [3pt]
 \textbf{Description} & \qquad \multirow{3}{20em}{A benchmark for classifiers tasked with predicting whether individual income exceeds \$50K/yr based on demographic and socio-economic information. Also known as ``Census Income'' dataset.}  \\[12ex] 
 \hline
\end{tabular}
\caption{Metadata of the Adult dataset}
\end{table}

\ifconf\else{\subsubsection{Provenance}}\fi

\begin{table}[h]
\centering
\begin{tabular}{||r l||}
 \hline
 \multicolumn{2}{|c|}{\textbf{PROVENANCE}}\\
 \hline
 \textbf{Source} & \\
 \qquad Name & \qquad U. S. Census Bureau  \\
 \qquad Url & \qquad \url{https://www.census.gov/en.html}  \\
 \qquad email & \qquad //  \\ [2ex]
 \textbf{Authors} & \\
 \qquad Names & \qquad Ronny Kohavi and Barry Becker   \\
 \qquad Url & \qquad \url{https://archive.ics.uci.edu/ml/datasets/}  \\
 \qquad email & \qquad ronnyk@live.com  \\
 \hline
\end{tabular}
\caption{Provenance of the Adult dataset}
\end{table}

\ifconf\else{\subsubsection{Variables}}\fi

\begin{table}[h]
\centering
\begin{tabular}{||r l||}
 \hline 
 \multicolumn{2}{|c|}{\textbf{VARIABLES}}\\
 \hline
 \textbf{age} & \qquad \multirow{1}{20em}{Respondent's age.} \\ [2ex]
 \textbf{workclass} & \qquad \multirow{2}{20em}{Broad classification of employment, with following envisioned classes. \\Private\\ Self-emp-not-inc (Self employed not-incorporated) \\Self-emp-inc (Self employed incorporated) \\Federal-gov \\Local-gov \\State-gov \\Without-pay (Without pay in family business) \\Never-worked} \\ [32ex]
 \textbf{fnlwgt} &  \qquad \multirow{1}{20em}{Variable used to produce population estimates from the CPS sample.} \\[7ex]
 \textbf{education} & \qquad \multirow{1}{20em}{Educational attainment of respondent. \\Preschool \\ 1st-4th\\ 5th-6th \\ 7th-8th\\ 9th \\ 10th\\ 11th\\ 12th (no diploma) \\ HS-grad (High school graduation) \\ Some-college (no degree) \\ Assoc-voc (associate degree in college, vocation program) \\ Assoc-acdm (associate degree in college, academic program)\\ Bachelors \\ Masters \\ Prof-school (professional school) \\ Doctorate} \\ [54ex]
 \textbf{education-num} & \qquad \multirow{1}{20em}{Ordinal encoding of previous variable.} \\ [2ex]
 \hline
\end{tabular}
\caption{Variables of the Adult dataset (1/3).} \label{tab:adult_meta1}
\end{table}

\begin{table}[h]
\centering
\begin{tabular}{||r l||}
 \hline 
 \multicolumn{2}{|c|}{\textbf{VARIABLES}}\\
 \hline
 \textbf{marital-status} & \qquad \multirow{1}{20em}{Respondent's marital status, with following envisioned classes. \\ Married-civ-spouse (married, civilian spouse present) \\ Divorced \\Never-married\\ Separated\\ Widowed\\ Married-spouse-absent\\ Married-AF-spouse (married, armed force spouse)} \\ [28ex]
 \textbf{occupation} & \qquad \multirow{1}{20em}{Job of respondent. \\ Tech-support (Technical, sales, and administrative support)\\ Craft-repair (Precision production, craft, and repair)\\ Other-service\\ Sales\\ Exec-managerial (Managerial and professional speciality)\\ Prof-specialty (Professional speciality) \\ Handlers-cleaners (Handlers, equipment cleaners, helpers, and laborers)\\ Machine-op-inspct (Operators, fabricators, and laborers)\\ Adm-clerical (Administrative support occupations, including clerical) \\ Farming-fishing (Farming, forestry, and fishing) \\ Transport-moving (Transportation and material moving)\\ Priv-house-serv (Private household service, e.g. cooks, cleaners) \\ Protective-serv (Protective service, e.g. firefighters, police) \\ Armed-Forces} \\ [70ex]
 \textbf{relationship} & \qquad \multirow{2}{20em}{Familial role wihtin household. \\ Wife \\ Own-child \\ Husband \\Not-in-family \\ Other-relative \\Unmarried} \\ [20ex]
 \hline
\end{tabular}
\caption{Variables of the Adult dataset (2/3).}
\end{table}

\begin{table}[h]
\centering
\begin{tabular}{||r l||}
 \hline 
 \multicolumn{2}{|c|}{\textbf{VARIABLES}}\\
 \hline
 \textbf{race} & \qquad \multirow{1}{20em}{Respondent's race.  \\ Amer-Indian-Eskimo \\ Asian-Pac-Islander\\ Black \\ White \\ Other \\} \\ [16ex]
 \textbf{sex} & \qquad \multirow{1}{20em}{Respondent's sex. \\ Female \\ Male} \\ [8ex]
 \textbf{capital-gain} & \qquad \multirow{1}{20em}{Profits from sale of assets.} \\ [2ex]
 \textbf{capital-loss} & \qquad \multirow{1}{20em}{Losses from sale of assets.} \\ [2ex]
 \textbf{hours-per-week} & \qquad \multirow{1}{20em}{Average hours of work per week.}  \\[2ex]
 \textbf{native-country} & \qquad \multirow{1}{20em}{Native Country of respondent}  \\[2ex]
 \textbf{target variable} & \qquad \multirow{1}{20em}{Does respondent's income exceed \$50,000?}  \\[4ex]
 \hline
\end{tabular}
\caption{Variables of the Adult dataset (3/3).}
\end{table}

\clearpage

\ifconf\else{\subsubsection{Statistics}}\fi

\begin{table}[h]
\begin{tabular}{||lllllll||}
\hline
\multicolumn{7}{|c|}{\textbf{STATISTICS}} \\ \hline
\multicolumn{7}{|l|}{\textbf{Ordinal}} \\ \hline
\multicolumn{1}{|c}{name} & \multicolumn{1}{c}{type} & \multicolumn{1}{c}{count} & \multicolumn{1}{c}{uniqueEntries} & \multicolumn{1}{c}{mostFrequent} & \multicolumn{1}{c}{leastFrequent} & \multicolumn{1}{c|}{missing} \\
\hline
education-num & int & 48842 & 16 & 9 & 1 & 0 \\\hline
\end{tabular}
\caption{Ordinal variables statistics of the Adult dataset}
\end{table}

\begin{table}[h]
\begin{tabular}{||lllllll||}
\hline
\multicolumn{7}{|l|}{\textbf{Categorical}} \\ \hline
\multicolumn{1}{|c}{name} & \multicolumn{1}{c}{type} & \multicolumn{1}{c}{count} & \multicolumn{1}{c}{uniqueEntries} & \multicolumn{1}{c}{mostFrequent} & \multicolumn{1}{c}{leastFrequent} & \multicolumn{1}{c|}{missing} \\
\hline
workclass & string & 48842 & 8 & Private & Never-worked & 2799 \\
education & string & 48842 & 16 & HS-grad & Preschool & 0 \\
marital-status & string & 48842 & 7 & Married-civ-spouse & Married-AF-spouse & 0 \\
occupation & string & 48842 & 14 & Prof-specialty & Armed-Forces & 2809  \\
relationship & string & 48842 & 6 & Husband & Other-relative & 0 \\
race & string & 48842 & 5 & White & Other & 0 \\
sex & string & 48842 & 2 & Male & Female & 0 \\
native-country & string & 48842 & 41 & United-States & Holand-Netherlands & 857 \\ 
target variable & string & 48842 & 2 & $<=50$K & $>50$K & 0 \\ \hline
\end{tabular}
\caption{Categorical variables statistics of the Adult dataset}
\end{table}

\begin{table}[h]
\begin{tabular}{||llllllllll||}
\hline
\multicolumn{10}{|l|}{\textbf{Quantitative}} \\ \hline
\multicolumn{1}{|c}{name} & \multicolumn{1}{c}{type} & \multicolumn{1}{c}{count} & \multicolumn{1}{c}{min} & \multicolumn{1}{c}{median} & \multicolumn{1}{c}{max} & \multicolumn{1}{c}{mean} & stdDev & miss & \multicolumn{1}{c|}{zeros} \\\hline
age & int & 48842 & 17 & 37 & 90 & 38.64 & 13.71 & 0 & 0 \\
fnlwgt & int & 48842 & 12285 & 178144.5 & 1490400 & 189664.13 & 105604.03 & 0 & 0 \\
capital-gain & int & 48842 & 0 & 0 & 99999 & 1079.07 & 7452.02 & 0 & 44807 \\
capital-loss & int & 48842 & 0 & 0 & 4356 & 87.50 & 403 & 0 & 46560 \\
hours-per-week & int & 48842 & 1 & 40 & 99 & 40.42 & 12.39 & 0 & 0 \\
\hline
\end{tabular}
\caption{Quantitative variables statistics of the Adult dataset.}
\end{table}

\clearpage
\section{COMPAS}
\label{sec:compas}

Key references include \citet{angwin2016machine,larson2016how,dieterich2016compas,propublica2016compas,equivant2019compas,brennan2009evaluating,bao2021COMPASlicated,barenstein2019propublica}.



\subsection{Datasheet}

\subsubsection{Motivation}
\begin{itemize}
    \item \que{For what purpose was the dataset created?}
    
    \ans{This dataset was created for an external audit of racial biases in the Correctional Offender Management Profiling for Alternative Sanctions (COMPAS) risk assessment tool developed by Northpointe (now Equivant), which estimates the likelihood of a defendant becoming a recidivist.}
    \item \que{Who created the dataset and on behalf of which entity?}
    
    \ans{The dataset was created by Julia Angwin (senior reporter), Jeff Larson (data editor), Surya Mattu (contributing researcher), Lauren Kirchner (senior reporting fellow). All four contributors were affiliated with ProPublica at the time.}
    \item \que{Who funded the creation of the dataset?} 
    
    \ans{The dataset curation work was likely remunerated by ProPublica.}
\end{itemize}
\subsubsection{Composition}
\begin{itemize}
    \item \que{What do the instances that comprise the dataset represent?}
    
    \ans{Each instance is a person that was scored for risk of recidivism by the COMPAS system in Broward County, Florida, between 2013--2014. In other words, instances are \textbf{defendants}.}
    \item \que{How many instances are there in total?}
    
    \ans{The COMPAS dataset \citep{propublica2016compas} consists of \textbf{11,757} defendants assessed at the pretrial stage (\texttt{compas-scores.csv}). A separate dataset is released for a subset of 7,214 defendants that were observed for two years after screening (\texttt{compas-scores-two-years.csv}). Finally a smaller subset of 4,743 defendants focuses on violent recidivism (\texttt{compas-scores-two-years-violent.csv}).}
    \item \que{Does the dataset contain all possible instances or is it a sample of instances from a larger set?}
    
    \ans{The dataset represents a \textbf{convenience sample} of all individuals that were scored by the COMPAS tool. It concentrates on defendants in Broward County, as it is a large jurisdiction in a state with strong open-records laws \citep{larson2016how}. Moreover, due to Broward County using COMPAS primarily in release/detain decisions prior to a defendant's trial, scores assessed at parole, probation or other stages were discarded. A notable anomaly in the sample is the low amount of defendants screened between June and July 2013 compared to the remaining time span of the COMPAS dataset \citep{barenstein2019propublica}.}
    \item \que{What data does each instance consist of?}
    
    \ans{Instances represent Broward County defendants scored with COMPAS for risk of recidivism. For each defendant the data provided by ProPublica includes tens of variables ($\sim50$) summarizing their demographics, criminal record, custody and COMPAS scores.}
    \item \que{Is there a label or target associated with each instance?}
    
    \ans{\textbf{Yes}. Instances are associated with two target variables (is\_recid and is\_violent\_recid), indicating whether defendants were booked in jail with a criminal offense (potentially violent) that took place after their COMPAS screening but within two years. The definition of recidivism and the two-year cutoff were selected by ProPublica staff to align their audit with definitions by Northpointe \citep{brennan2009evaluating,angwin2016machine}.}
    \item \que{Is any information missing from individual instances?}
    
    \ans{\textbf{Yes}. There are several columns where data is missing for one or more instances, including dates when defendants committed the offense (c\_offense\_date) were incarcerated (c\_jail\_in) or released (c\_jail\_out). Missingness in this dataset is not surprising as its curation was a complex endeavour that required cross-referencing information from three separate sources, namely Broward County Sheriff’s Office, Broward County Clerk’s Office and Florida Department of Corrections. Moreover, Northpointe's response to the ProPublica's study points out important risk factors considered by the COMPAS algorithm that are not present in the dataset, among which the criminal involvement scale, drug problems sub-scale, age at first adjudication, arrest rate and vocational educational scale \citep{dieterich2016compas}. Finally, a clear indication of whether defendants were released or detained pretrial seems to be missing.
    }
    \item \que{Are relationships between individual instances made explicit?}
    
    \ans{\textbf{No}. While it is plausile for some Broward County defendants to be connected, this information is not available.}
    \item \que{Are there recommended data splits?}
    
    \ans{\textbf{No}.}
    \item \que{Are there any errors, sources of noise, or redundancies in the dataset?}
    
    \ans{\textbf{Yes}. Clerical errors in records caused incorrect matches between individuals' COMPAS scores and their criminal records, leading to an error rate close to 4\% \citep{larson2016how}. Moreover, an important temporal trend was spuriously introduced by ProPublica's preprocessing in \texttt{compas-scores-two-years.csv} and \texttt{compas-scores-two-years-violent.csv}, due to which defendants with a screening date after April 2014 are all recidivists \citep{barenstein2019propublica}. In terms of redundancies, \texttt{compas-scores.csv} contains two identical columns (called decile\_score and decile\_score.1).}
    \item \que{Is the dataset self-contained, or does it link to or otherwise rely on external resources?}
    
    \ans{The dataset is \textbf{self-contained}.}
    \item \que{Does the dataset contain data that might be considered confidential?} 
    
    \ans{\textbf{No}. However it does contains first names and last names of defendants, connecting them to their criminal history.}
    \item\que{Does the dataset contain data that, if viewed directly, might be offensive, insulting, threatening, or might otherwise cause anxiety?}
    
    \ans{\textbf{Yes}. The column vr\_charge\_desc describing violent recidivism charges is one such example.}

    \item \que{Does the dataset identify any subpopulations (e.g., by age, gender)?}
    
    \ans{\textbf{Yes}. The dataset identifies population by age, sex and race.  The curators of the COMPAS dataset maintained the race classifications used by the Broward County Sheriff’s Office, identifying individuals as Asian, Black, Hispanic, Native American and White \citep{larson2016how}. Age is reported as an integer, sex as either Male or Female. A distribution along these dimensions is reported in Table \ref{tab:compas_demographics_2y} which summarizes data in \texttt{compas-scores-two-years.csv}. Distributions in remaining files are similar.}
    \begin{table}[h]
    \centering
    \begin{tabular}{||c c||}
        \hline
        \multicolumn{2}{||c||}{\textbf{compas-scores-two-years}}\\
        \hline 
        \textbf{Demographic Caracteristic} & \qquad \textbf{Values} \\ [1ex]
        \hline
        Percentage of male subjects & \qquad 80.83\% \\[1ex]
        Percentage of female subjects & \qquad 19.17\% \\[1ex]
        \hline
        Percentage of African-American subjects & \qquad 51.46\% \\[1ex]
        Percentage of Caucasian subjects & \qquad 33.63\% \\[1ex]
        Percentage of Hispanic subjects & \qquad 8.67\% \\[1ex]
        Percentage of Asian subjects & \qquad 0.48\% \\[1ex]
        Percentage of Native American subjects & \qquad 0.20\% \\[1ex]
        Percentage of people belonging to other races & \qquad 5.56\% \\[1ex]
        \hline
        Percentage of people under-19 years old & \qquad 0.42\% \\[1ex]
        Percentage of people between 20-29 years old & \qquad 42.41\% \\[1ex]
        Percentage of people between 30-39 years old & \qquad 28.04\% \\[1ex]
        Percentage of people between 40-49 years old & \qquad 14.60\% \\[1ex]
        Percentage of people between 50-59 years old & \qquad 11.00\% \\[1ex]
        Percentage of people between 60-69 years old & \qquad 3.01\% \\[1ex]
        Percentage of people over-70 years old & \qquad 0.51\% \\[1ex]
        \hline
        \end{tabular}
    \caption{Demographic Characteristics of compas-scores-two-years.} \label{tab:compas_demographics_2y}
    \end{table}
    
    \item \que{Is it possible to identify individuals , either directly or indirectly from the dataset?}
    
    \ans{\textbf{Yes}. The dataset reports defendants' first name, last name and date of birth.}
    \item \que{Does the dataset contain data that might be considered sensitive in any way?} 
    
    \ans{\textbf{Yes}. The COMPAS dataset reports individuals' race, criminal history, full name and date of birth.}
\end{itemize}
\subsubsection{Collection process}
\begin{itemize}
    \item \que{How was the data associated with each instance acquired?} 
    
    \ans{The data was obtained cross-referencing three sources. From the Broward County Sheriff’s Office in Florida, ProPublica obtained COMPAS scores associated with all 18,610 people scored in 2013 and 2014. Defendants' public criminal records were obtained from the Broward County Clerk’s Office website matching them based on date of birth, first and last names. The dataset was augmented with jail records provided by the Broward County Sheriff’s Office. Finally public incarceration records were downloaded from the Florida Department of Corrections website.}
    \item \que{What mechanisms or procedures were used to collect the data?}
    
    \ans{The original data was plausibly recorded by employees of the Broward County Sheriff's Office, Broward County Clerk's Office, and Florida Department of Corrections. The curators of the COMPAS dataset obtained records from the County Sheriff's Office through a public records request, while data from the County Clerk's Office and the Florida Department of Correction was downloaded from their official website, matching the methodology of a COMPAS validation study \citep{larson2016how}.}
    \item \que{If the dataset is a sample from a larger set, what was the sampling strategy?}
    
    \ans{In terms of auditing the COMPAS risk assessment tool, this dataset represents a \textbf{convenience sample}, focused on a single county and scoring period 2013--2014. Considering a single county in a state with strong open-records laws reduced the data cross-referencing overhead. Concentrating on recent scores predating the study by 2--3 years kept the study timely and permitted a measurement of recidivism aligned with the one by Northpointe. The fact that Northpointe's response to the ProPublica study only contains minor criticism of the sample (concerning the definition of pretrial defendants \citep{dieterich2016compas}) may be interpreted as testimony to its overall quality.  More broadly and beyond the COMPAS audit, arrest data as a proxy for crime brings about specific sampling effects, inevitably mediated by law enforcement practices \cite{xie2012racial,malcolm2008:minority}}.
    \item \que{Who was involved in the data collection process and how were they compensated?}
    
    \ans{The original data was plausibly recorded by Broward County and Florida Department of Corrections employees. On ProPublica's side, we assume that key curation choices were made and implemented by four employees credited in the article \citep{angwin2016machine} and accompanying technical report \citep{larson2016how}, namely Julia Angwin, Jeff Larson, Surya Mattu and Lauren Kirchner. Given the focus on arrest data, the Broward County law enforcement community is also important in the data sampling process.}
    \item \que{Over what timeframe was the data collected?}
    
    \ans{COMPAS scores are from 2013 and 2014, while jail records cover the period from January 2013 to April 2016. The dataset was first released by ProPublica in May 2016 \citep{propublica2016compas}.} 
    \item \que{Were any ethical review processes conducted?}
    
    \ans{\textbf{Unknown}.}
    \item \que{Was the data collected from the individuals in question directly, or obtained via third parties or other sources?}
    
    \ans{The data was obtained \textbf{via third parties}, namely the Broward County Sheriff’s Office in Florida through a public records request, from the Broward County Clerk’s Office through the official website and through the Florida Department of Corrections through the official website. Collection from interested individuals would not have been viable.}
    \item \que{Were the individuals in question notified about the data collection?}
    
    \ans{\textbf{Likely no}. Most of the COMPAS data was publicly available and downloaded from the official websites of Broward County Clerk’s Office and the Florida Department of Corrections.}
    \item \que{Did the individuals in question consent to the collection and use of their data?}
    
    \ans{\textbf{Likely no}. Public availability of arrest/conviction records is associated with collateral consequences that typically damage subjects socially and financially \citep{pinard2010collateral,angwin2016machine}.}
    \item \que{If consent was obtained, were the consenting individuals provided with a mechanism to revoke their consent in the future or for certain uses?}
    
    \ans{\textbf{Likely no}.}
    \item \que{Has an analysis of the potential impact of the dataset and its use on data subjects been conducted?}
    
    \ans{\textbf{Likely no}. We are unaware of analyses specifically focused on the COMPAS dataset. More broadly, public availability of criminal records is related to studies on the employability of offenders \citep{graffam2008perceived}.}
\end{itemize}
\subsubsection{Preprocessing/cleaning/labelling}
\begin{itemize}
    \item \que{Was any preprocessing/cleaning/labeling of the data done?}
    
    \ans{\textbf{Yes}. Instances were discarded if assessed with COMPAS at parole, probation or other stages in the criminal justice system. This data is unavailable. Moreover, ProPublica published its datasets with accompanying preprocessing code which has become standard \citep{propublica2016compas}. The standard preprocessing removes instances for which (1) arrest dates or charge dates are not within 30 days of the COMPAS assessment, (2) true recidivism cannot be decided, (3) charge degree is not defined as misdemeanor or felony, (4) the COMPAS score is not clearly defined. The remaining COMPAS scores were bucketed into low, medium and high risk.}
    \item \que{Was the “raw” data saved in addition to the preprocessed/cleaned/labeled data?}
    
    \ans{\textbf{Yes}. The data is available in the official ProPublica github repository \citep{propublica2016compas}. This is an intermediate data artifact, already cross-referenced by ProPublica across three separate sources.}
    
    \item \que{Is the software used to preprocess/clean/label the instances available?}
    
    \ans{\textbf{Yes}. The standard preprocessing software can be found in the official ProPublica github repository \citep{propublica2016compas}. The software used to cross-reference data from separate sources is not publicly available.}
    
    \end{itemize}
\subsubsection{Uses}
\begin{itemize}
    \item \que{For what tasks has the dataset been used?}
    
    \ans{The creators used this dataset to audit the COMPAS tool for racial bias. In the literature it has also been used to evaluate the fairness and accuracy of different algorithms and, more broadly, to study definitions of algorithmic fairness.}
    \item \que{Is there a repository that links to any or all papers or systems that use the dataset?}
    
    \ans{See Appendix \ref{sec:compas_db} for a (non-exhaustive) list of algorithmic fairness works using this resource.}
    \item \que{What (other) tasks could the dataset be used for?}
    
    \ans{In terms of immediate applications, the dataset could be used to train novel recidivism risk assessment tools. From a methodological perspective, COMPAS may be used in high-stakes domains connected with decision-making about human subjects, including explainable and privacy-preserving ML.}
    \item \que{Is there anything about the composition of the dataset or the way it was collected and preprocessed/cleaned/labeled that might impact future uses?}
    
    \ans{From a very narrow perspective, the fact that all defendants with a screening date after April 2014 are recidivists introduces artificially inflated recidivism base rates \citep{barenstein2019propublica}, which would likely be inherited by tools trained on the COMPAS dataset. Moreover, the dataset contains no clear indication concerning pretrial detention or release of defendants. Therefore, researchers must come up with subjective criteria to label individuals as detained or released if they are interested in studying pretrial detention as an intervention deviating from a default course of action \citep{mishler2021fairness}. From a broader perspective, the data is likely influenced by historical biases in criminal justice, with differential impact on different communities \citep{malcolm2008:minority,xie2012racial,angwin2016machine}. Zooming out further, the use of automated risk assessment tools in pretrial decisions is the subject of controversial debate \citep{barabas2019problems} which cannot be overlooked.}
    \item \que{Are there tasks for which the dataset should not be used?}
    
    \ans{Given the above considerations and the narrow geographical scope of the dataset, COMPAS should not be used to train and deploy risk assessment tools for the judicial system. In research settings, users should exercise care in selecting both rows and columns. \citet{bao2021COMPASlicated} suggest avoiding the use of COMPAS 
    to demonstrate novel approaches in algorithmic fairness, as considering data without proper context may bring to misleading conclusions which could misguidedly enter the broader debate on criminal justice.}
\end{itemize}

\subsubsection{Distribution}
\begin{itemize}
    \item \que{Is the dataset distributed to third parties outside of the entity on behalf of which the dataset was created?}
    
    \ans{\textbf{Yes}. The COMPAS dataset is publicly available.}
    \item \que{How is the dataset distributed?}
    
    \ans{The dataset is hosted on ProPublica's official github repository \citep{propublica2016compas}.}
    \item \que{When was the dataset distributed?}
    
    \ans{Since \textbf{May 2016}.}
    \item \que{Is the dataset distributed under a copyright or other intellectual property (IP) license, and/or under applicable terms of use (ToU)?}
    
    \ans{As of June 2021 the COMPAS dataset is freely distributed under ProPublica's standard ToU \citep{propublica2021:tou}. The dataset cannot be republished in its entirety, it cannot be sold, and can only be used for publication if ProPublica's work is properly referenced.}
    \item \que{Have any third parties imposed IP-based or other restrictions on the data associated with the instances?}
    
    \ans{\textbf{Likely no}.}
    \item \que{Do any export controls or other regulatory restrictions apply to the dataset or to individual instances?}
    
    \ans{\textbf{Unknown}.}
\end{itemize}

\subsubsection{Maintenance}
\begin{itemize}
    \item \que{Who is supporting/hosting/maintaining the dataset?}
    
    \ans{The dataset is currently hosted and maintained by \textbf{ProPublica} on github.}
    \item \que{How can the owner/curator/manager of the dataset be contacted?}
    
    \ans{The contact for ProPublica's data store is data.store@propublica.org.}
    \item \que{Is there an erratum?}
    
    \ans{\textbf{No}. There is no official erratum. An external report highlighting anomalies in the data is available \citep{barenstein2019propublica}.}
    \item \que{Will the dataset be updated?}
    
    \ans{\textbf{Likely no}. In the event of an update, ProPublica's data store ToU specifies users are solely responsible for checking their sites for updates \citep{propublica2021:tou}}
    
    \item \que{If the dataset relates to people, are there applicable limits on the retention of the data associated with the instances?}
    
    \ans{\textbf{Unknown}.}
    \item \que{Will older versions of the dataset continue to be supported/hosted/maintained?}
    
    \ans{\textbf{Unknown}.}
    \item \que{If others want to extend/augment/build on/contribute to the dataset, is there a mechanism for them to do so?}
    
    \ans{\textbf{Likely no}.}
\end{itemize}

\clearpage
\subsection{Data Nutrition Label}

The following analysis refers to compas-scores-two-years.csv after applying the standard COMPAS preprocessing \citep{propublica2016compas}.

\ifconf\else{\subsubsection{Metadata}}\fi

\begin{table}[h]
\centering
\begin{tabular}{||r l||}
 \hline 
 \multicolumn{2}{|c|}{\textbf{METADATA}}\\
 \hline
 \textbf{Filenames} & \qquad compas-scores-two-years  \\ 
 \textbf{Format} & \qquad csv \\
 \textbf{Url} & \qquad \multirow{2}{26em}{\url{https://www.propublica.org/datastore/dataset/compas-recidivism-risk-score-data-and-analysis}}  \\[3ex] 
 \textbf{Domain} & \qquad Law  \\
 \textbf{Keywords} & \qquad  risk assessment, pretrial, recidivism\\
 \textbf{Type} & \qquad Tabular \\
 \textbf{Rows} & \qquad 6,172 \\ 
 \textbf{Columns} & \qquad 57 \\
 \textbf{\% missing cells} & \qquad 5\% \\
 \textbf{Rows with missing cells} & \qquad 100\% \\
 \textbf{License} & \qquad ProPublica's ToU \citep{propublica2021:tou} \\
 \textbf{Released} & \qquad May 2016  \\
 \textbf{Range} & \qquad \multirow{2}{20em}{2013-2014 for COMPAS scores, 2013-2016 for arrest and detention history.} \\ [10pt]
 \textbf{Description} & \qquad \multirow{2}{20em}{Dataset curated by ProPublica to audit COMPAS software for racial biases, focusing on Broward County 2013--2014.}  \\[8ex] 
 \hline
\end{tabular}
\caption{Metadata of COMPAS dataset.}
\end{table}

\ifconf\else{\subsubsection{Provenance}}\fi

\begin{table}[h]
\centering
\begin{tabular}{||r l||}
 \hline
 \multicolumn{2}{|c|}{\textbf{PROVENANCE}}\\
 \hline
 \textbf{Source} & \\
 \qquad Name & \qquad Broward County Sheriff's Office  \\
 \qquad Url & \qquad \url{http://www.sheriff.org/}  \\
 \qquad email & \qquad //  \\ [2ex]
 \qquad Name & \qquad Broward County Clerk’s Office  \\
 \qquad Url & \qquad \url{https://www.browardclerk.org}  \\
 \qquad email & \qquad Eclerk@browardclerk.org \\ [2ex]
 \qquad Name & \qquad Florida Department of Corrections  \\
 \qquad Url & \qquad \url{http://www.dc.state.fl.us/}  \\
 \qquad email & \qquad FDCCitizenServices@fdc.myflorida.com  \\ [2ex]
 \textbf{Authors} & \\ 
 \qquad Names & \qquad Julia Angwin, Jeff Larson, Surya Mattu and Lauren Kirchner\\
 \qquad Url & \qquad \multirow{2}{25em}{\url{https://www.propublica.org/datastore/dataset/compas-recidivism-risk-score-data-and-analysis}}  \\[3.2ex]
 \qquad email & \qquad data.store@propublica.org \\
 \hline
 \end{tabular}
\caption{Provenance of COMPAS dataset.}
\end{table}

\clearpage

\ifconf\else{\subsubsection{Variables}}\fi

\begin{table}[h]
\centering
\begin{tabular}{||r l||}
 \hline 
 \multicolumn{2}{|c|}{\textbf{VARIABLES}}\\
 \hline
 \textbf{id} & \qquad \multirow{1}{20em}{Unique identifier assigned by the authors} \\ [1ex]
 \textbf{name} & \qquad \multirow{1}{20em}{Defendant's first and last name} \\ [1ex]
 \textbf{first} & \qquad \multirow{1}{20em}{Defendant's first name} \\ [1ex]
 \textbf{last} & \qquad \multirow{1}{20em}{Defendant's last name} \\ [1ex]
 \textbf{compas\_screening\_date} & \qquad \multirow{1}{20em}{Day defendant was scored by COMPAS} \\ [1ex]
 \textbf{sex} & \qquad \multirow{1}{20em}{Defendant's sex} \\ [1ex]
 \textbf{dob} & \qquad \multirow{1}{20em}{Defendant's date of birth} \\ [1ex]
 \textbf{age} & \qquad \multirow{1}{20em}{Defendant's age} \\ [1ex]
 \textbf{age\_cat} &  \qquad \multirow{1}{20em}{Age quantization: \\ less than 25 \\ 25-45 \\ greater than 45} \\[9ex]
 \textbf{race} & \qquad \multirow{1}{20em}{Defendant's race: \\ African-American \\ Asian \\ Caucasian \\ Hispanic \\ Native American \\ Other } \\ [16ex]
 \textbf{juv\_fel\_count} & \qquad \multirow{1}{20em}{Number of juvenile felonies} \\ [1ex]
 \textbf{decile\_score} & \qquad \multirow{1}{20em}{COMPAS recidivism score (10-point scale)} \\ [1ex]
 \textbf{juv\_misd\_count} & \qquad \multirow{1}{20em}{Number of juvenile misdemeanors} \\ [1ex]
 \textbf{juv\_other\_count} & \qquad \multirow{1}{20em}{Number of other juvenile convictions (not considering misdemeanor and felonies)} \\ [4ex]
 \textbf{priors\_count} & \qquad \multirow{1}{20em}{Number of prior crimes} \\ [1ex]
 \textbf{days\_b\_screening\_arrest} & \qquad \multirow{1}{20em}{Days between imprisonment (c\_jail\_in) and COMPAS screening (compas\_screening\_date) } \\ [4ex]
 \hline
\end{tabular}
\caption{Variables of COMPAS dataset (1/3).}
\end{table}

\begin{table}[h]
\centering
\begin{tabular}{||r l||}
 \hline 
 \multicolumn{2}{|c|}{\textbf{VARIABLES}}\\
 \hline
 \textbf{c\_jail\_in} & \qquad \multirow{1}{20em}{Date of imprisonment} \\ [1ex]
 \textbf{c\_jail\_out} & \qquad \multirow{1}{20em}{Date of release} \\ [1ex]
 \textbf{c\_case\_number} & \qquad \multirow{1}{20em}{Alpha-numeric case identifier} \\ [1ex]
 \textbf{c\_offense\_date} & \qquad \multirow{1}{20em}{Date on which the offense was committed} \\[1ex]
 \textbf{c\_arrest\_date} & \qquad \multirow{1}{20em}{Date on which defendant was arrested} \\ [1ex]
 \textbf{c\_days\_from\_compas} & \qquad \multirow{1}{20em}{Days elapsed between offense/arrest and the date of COMPAS screening} \\ [3ex]
 \textbf{c\_charge\_degree} & \qquad \multirow{1}{20em}{Degree of charge: \\ F (felony) \\ M (misdemeanor)} \\ [6ex]
 \textbf{c\_charge\_desc} & \qquad \multirow{1}{20em}{Textual description of charge} \\ [1ex]
 \textbf{is\_recid} & \qquad \multirow{1}{20em}{Binary indication of recidivism.} \\ [1ex]
 \textbf{r\_case\_number} & \qquad \multirow{1}{20em}{Alpha-numeric case identifier for recidivist offense} \\ [3ex]
 \textbf{r\_charge\_degree} & \qquad \multirow{1}{20em}{Degree of recidivist charge} \\ [1ex]
 \textbf{r\_days\_from\_arrest} & \qquad \multirow{1}{20em}{Days elapsed between date of recidivist offense (r\_offense\_date) and date of recidivist incarceration (r\_jail\_in)} \\ [6ex]
 \textbf{r\_offense\_date} & \qquad \multirow{1}{20em}{Date of recidivist offense} \\ [1ex]
 \textbf{r\_charge\_desc} & \qquad \multirow{1}{20em}{Textual description of recidivist charge} \\ [1ex]
 \textbf{r\_jail\_in} & \qquad \multirow{1}{20em}{Date of incarceration for recidivist offense} \\ [1ex]
 \textbf{r\_jail\_out} & \qquad \multirow{1}{20em}{Date of release for recidivist offense} \\ [2ex] 
 \hline
\end{tabular}
\caption{Variables of COMPAS dataset (2/3).}
\end{table}

\begin{table}[h]
\centering
\begin{tabular}{||r l||}
 \hline 
 \multicolumn{2}{|c|}{\textbf{VARIABLES}}\\
 \hline
 \textbf{violent\_recid} & \qquad \multirow{1}{20em}{Unknown; all nan} \\ [1ex]
 \textbf{is\_violent\_recid} & \qquad \multirow{1}{20em}{Binary indication of violent recidivism. If true, then is\_recid is true.} \\ [3ex]
 \textbf{vr\_case\_number} & \qquad \multirow{1}{20em}{Alpha-numeric case identifier for violent recidivist offense} \\ [3ex]
 \textbf{vr\_charge\_degree} & \qquad \multirow{1}{20em}{Degree of violent recidivist offense} \\ [1ex]
 \textbf{vr\_offense\_date} & \qquad \multirow{1}{20em}{Date of violent recidivist offense} \\ [1ex]
 \textbf{vr\_charge\_desc} & \qquad \multirow{1}{20em}{Textual description of the violent recidivist charge} \\ [3ex]
 \textbf{type\_of\_assessment} & \qquad \multirow{1}{20em}{Type of COMPAS assessment - all 'Risk of Recidivism'.} \\ [3ex]
 \textbf{decile\_score\_1} & \qquad \multirow{1}{20em}{Identical to decile\_score} \\ [1ex]
 \textbf{score\_text} & \qquad \multirow{1}{20em}{Quantization of decile\_score: \\ LOW (1-4) \\ MEDIUM (5-7) \\ HIGH (8-10).} \\ [8ex]
 \textbf{screening\_date} & \qquad \multirow{1}{20em}{Identical to compas\_screening\_date} \\ [1ex]
 \textbf{v\_type\_of\_assessment} & \qquad \multirow{1}{20em}{Type of COMPAS violent assessment - all 'Risk of Violence'.} \\ [3ex]
 \textbf{v\_decile\_score} & \qquad \multirow{1}{20em}{COMPAS violent recidivism score (10-point scale)} \\ [3ex]
 \textbf{v\_score\_text} & \qquad \multirow{1}{20em}{Quantization of v\_decile\_score: \\ LOW (1-4) \\ MEDIUM (5-7) \\ HIGH (8-10).} \\ [8ex]
 \textbf{v\_screening\_date} & \qquad \multirow{1}{20em}{Identical to compas\_screening\_date.} \\ [1ex]
 \textbf{in\_custody} & \qquad \multirow{1}{20em}{Unknown} \\ [1ex]
 \textbf{out\_custody} & \qquad \multirow{1}{20em}{Unknown} \\ [1ex]
 \textbf{priors\_count.1} & \qquad \multirow{1}{20em}{Identical to priors\_count.} \\ [1ex]
 \textbf{start} & \qquad \multirow{1}{20em}{Unknown} \\ [1ex]
 \textbf{end} & \qquad \multirow{1}{20em}{Unknown} \\ [1ex]
 \textbf{event} & \qquad \multirow{1}{20em}{Unknown} \\ [1ex]
 \textbf{two\_year\_recid} & \qquad \multirow{1}{20em}{Unknown} \\ [1ex]
 \hline
\end{tabular}
\caption{Variables of COMPAS dataset (3/3).}
\end{table}
\FloatBarrier

\ifconf\else{\subsubsection{Statistics}}\fi

\begin{table}[h]
\begin{tabular}{||lllllll||}
\hline
\multicolumn{7}{|c|}{\textbf{STATISTICS}} \\ \hline
\multicolumn{7}{|l|}{\textbf{Ordinal}} \\ \hline
\multicolumn{1}{|c}{name} & \multicolumn{1}{c}{type} & \multicolumn{1}{c}{count} & \multicolumn{1}{c}{uniqueEntries} & \multicolumn{1}{c}{mostFrequent} & \multicolumn{1}{c}{leastFrequent} & \multicolumn{1}{c|}{missing} \\
\hline
id & int & 6,172 & 6,172 & multiple & multiple & 0 \\
compas\_screening\_date & date & 6,172 & 685 & 2013-04-20 & multiple & 0\\ 
dob & date & 6,172 & 4,830 & multiple & multiple & 0 \\ 
age\_cat & string & 6,172 & 3 & 25 - 45 & Greater than 45 & 0 \\
c\_jail\_in & date & 6,172 & 6,172 & multiple & multiple & 433\\ 
c\_jail\_out & date & 6,172 & 6,161 & 2013-09-14 05:58:00 & multiple & 433\\ 
c\_offense\_date & date & 6,172 & 737 & multiple & multiple & 1388\\ 
c\_arrest\_date & date & 6,172 & 417 & 2013-02-06 & multiple & 8425\\
r\_offense\_date & date & 6,172 & 1,041 & 2014-12-08 & multiple & 3,182\\ 
r\_jail\_in & date & 6,172 & 928 & multiple & multiple & 4,175\\ 
r\_jail\_out & date & 6,172 & 893 & multiple & multiple & 4,175\\ 
vr\_offense\_date & date & 6,172 & 505 & 2015-08-15 & multiple & 5,480\\ 
v\_score\_text & string & 6,172 & 3 & Low & High & 0 \\
v\_screening\_date & date & 6,172 & 685 & 2013-04-20 & multiple & 0\\
score\_text & string & 6,172 & 3 & Low & High & 0 \\
screening\_date & date & 6,172 & 685 & 2013-04-20 & multiple & 0\\ 
in\_custody & date & 6,172 & 1,087 & multiple & multiple & 0\\
out\_custody & date & 6,172 & 1,097 & 2020-01-01 & multiple & 0\\ \hline
\end{tabular}
\caption{Ordinal variables statistics of COMPAS dataset}
\end{table}

\begin{table}[h]
\begin{tabular}{||lllllll||}
\hline
\multicolumn{7}{|l|}{\textbf{Categorical}} \\ \hline
\multicolumn{1}{|c}{name} & \multicolumn{1}{c}{type} & \multicolumn{1}{c}{count} & \multicolumn{1}{c}{uniqueEntries} & \multicolumn{1}{c}{mostFrequent} & \multicolumn{1}{c}{leastFrequent} & \multicolumn{1}{c|}{missing} \\
\hline
name & string & 6,172 & 9,128 & mutiple & multiple & 0 \\
first & string & 6,172 & 2,493 & michael & multiple & 0 \\
last & string & 6,172 & 3,465 & williams & multiple & 0 \\
sex & string & 6,172 & 2 & Male & Female & 0 \\
race & string & 6,172 & 6 & African-American & Native American & 0 \\
c\_case\_number & string & 6,172 & 6,172 & multiple & multiple & 0 \\
c\_charge\_desc & string & 6,172 & 390 & Battery & multiple & 5 \\
c\_charge\_degree & string & 6,172 & 2 & F & M & 0 \\ 
r\_case\_number & string & 6,172 & 2,991 & multiple & multiple & 3,182 \\ 
r\_charge\_desc & string & 6,172 & 319 & \multirow{2}{15em}{Possess Cannabis/ \\ 20 Grams Or Less} & multiple & 3,228 \\ [3ex] 
r\_charge\_degree & string & 6,172 & 11 & (M1) & (F5) & 0 \\ 
vr\_case\_number & string & 6,172 & 693 & multiple & multiple & 5,480 \\ 
vr\_charge\_desc & string & 6,172 & 82 & Battery & multiple & 5,480 \\
vr\_charge\_degree & string & 6,172 & 10 & (M1) & (F5) & 5,480 \\ 
type\_of\_assessment & string & 6,172 & 1 & Risk of Recidivism & Risk of Recidivism & 0 \\
v\_type\_of\_assessment & string & 6,172 & 1 & Risk of Violence & Risk of Violence & 0 \\
is\_recid & binary & 6,172 & 2 & 0 & 1 & 0 \\
is\_violent\_recid & binary & 6,172 & 2 & 0 & 1 & 0 \\
event & binary & 6,172 & 2 & 0 & 1 & 0 \\ 
two\_year\_recid & binary & 6,172 & 2 & 0 & 1 & 0 \\ 
\hline
\end{tabular}
\caption{Categorical variables statistics of COMPAS dataset}
\end{table}

\begin{table}[h]
\begin{tabular}{||llllllllll||}
\hline
\multicolumn{10}{|l|}{\textbf{Quantitative}} \\ \hline
\multicolumn{1}{|c}{name} & \multicolumn{1}{c}{type} & \multicolumn{1}{c}{count} & \multicolumn{1}{c}{min} & \multicolumn{1}{c}{median} & \multicolumn{1}{c}{max} & \multicolumn{1}{c}{mean} & stdDev & miss & \multicolumn{1}{c|}{zeros} \\\hline

age & int & 6,172 & 18 & 31 & 96 & 34.53 & 11.73 & 0 & 0 \\
juv\_fel\_count & int & 6,172 & 0 & 0 & 20 & 0.06 & 0.46 & 0 & 5,964\\
juv\_misd\_count & int & 6,172 & 0 & 0 & 13 & 0.09 & 0.50 & 0 & 5,820\\
juv\_other\_count & int & 6,172 & 0 & 0 & 9 & 0.11 & 0.47 & 0 & 5,711\\
priors\_count & int & 6,172 & 0 & 1 & 38 & 3.25 & 4.74 & 0 & 2,085\\
days\_b\_screening\_arrest & int & 6,172 & -30.0 & -1 & 30.0 & -1.74 & 5.08 & 0 & 1,379\\
c\_days\_from\_compas & int & 6,172 & 0 & 1 & 9,485 & 24.90 & 276.81 & 0 & 869\\
r\_days\_from\_arrest & int & 6,172 & -1 & 0 & 993 & 20.10 & 76.54 & 4,175 & 1,452\\
decile\_score & int & 6,172 & 1 & 4 & 10 & 4.42 & 2.84 & 0 & 0\\
v\_decile\_score & int & 6,172 & 1 & 3 & 10 & 3.64 & 2.49 & 0 & 0\\
start & int & 6,172 & 0 & 0 & 937 & 13.32 & 50.14 & 0 & 3,485\\
end & int & 6,172 & 0 & 539 & 1,186 & 555.05 & 400.26 & 0 & 1\\
\hline
\end{tabular}
\caption{Quantitative variables statistics of COMPAS dataset. }
\end{table}




\FloatBarrier

\clearpage

\section{German Credit}
\label{sec:german}

Key references include  \citet{haussler1979empirische,hofmann1994:sg,gromping2019:sg,gromping2019:sg2}.



\subsection{Datasheet}
\subsubsection{Motivation}
\begin{itemize}
    \item \que{For what purpose was the dataset created?} 
    
    \ans{This dataset was created to study the problem of automated credit decisions at a regional Bank in southern Germany.}
    \item \que{Who created the dataset and on behalf of which entity?}
    
    \ans{The dataset was created at a regional Bank of southern Germany (most likely Hypo Bank) and first used by Walter Häußler in the late 1970s as part of his PhD thesis. Hans Hofmann, affiliated with Universität Hamburg at the time, is credited as dataset source \citep{hofmann1994:sg}. Presumably, he donated the dataset to the European Statlog project and a representative of Strathclyde University donated it to UCI \citep{gromping2019:sg}.}
    \item \que{Who funded the creation of the dataset?}
    
    \ans{The first known work using the dataset describes it as originating from a regional Bank of southern Germany \citep{haussler1979empirische}. Given the affiliation of the author is Hypo Bank, which fit the description at the time, we assume the dataset was collected, curated and funded at Hypo Bank.}
\end{itemize}
\subsubsection{Composition}
\begin{itemize}
    \item \que{What do the instances that comprise the dataset represent?}
    
    \ans{Instances represent Hypo bank \textbf{loan recipients} from 1973--1975.}
    \item \que{How many instances are there in total?}
    
    \ans{The dataset consists of \textbf{1,000} instances.}
    \item \que{Does the dataset contain all possible instances or is it a sample of instances from a larger set?}
    
    \ans{In principle this is a \textbf{convenience sample}, consisting of people who were deemed creditworthy by a bank clerk. A representative sample stemming from indiscriminate credit grants would not have been viable \citep{haussler1979empirische}. However, if the envisioned application was \emph{post-screening} credit decisions, the influence of this selection bias would be reduced. Finally loan recipients associated with delayed payment or loan default (``bad credit'') are oversampled (30\%).}
    
    \item \que{What data does each instance consist of?}
    
    \ans{For each instance, 13 categorical and 7 quantitative variables are provided, summarizing their financial situation, credit history, and personal situation, including housing, number of liable people, and a mixed variable encoding marital status and sex. A more through description is deferred to Tables \ref{tab:german_meta1}-\ref{tab:german_meta3}.
    }
    \item \que{Is there a label or target associated with each instance?}
    
    \ans{\textbf{Yes}. A binary label encodes whether loan recipients punctually payed each installment (``good credit'') or not (``bad credit''). The latter label includes a range of situations from delayed payment up to loan default.}
    \item \que{Is any information missing from individual instances?}
    
    \ans{\textbf{No}. No cell is missing, however the variable ``property'' has a level jointly encoding the conditions ``no property'' and ``unknown''. A similar joint encoding exists for ``savings'', so some values may actually be deemed missing for these variables.}
    \item \que{Are relationships between individual instances made explicit?}
    
    \ans{\textbf{No}. There are no known relationships between instances.}
    \item \que{Are there recommended data splits?}
    
    \ans{\textbf{No}.}
    \item \que{Are there any errors, sources of noise, or redundancies in the dataset?} 
    
    \ans{\textbf{Yes}. The dataset documentation is filled with errors, so that several levels of categorical variables do not correspond to what they should according to the official documentation from \citet{hofmann1994:sg}. This is not necessarily an issue if one is purely interested in the evaluation of a method. For example, according to the official documentation, a majority of loan recipients are foreign workers, while in reality this should appear rather strange and indeed is not true \citep{gromping2019:sg}. Computationally, this will make no difference, as the input to a machine learning method will remain the same. However if one is interested to the context surrounding the data, as should be the case with fairness research, the wrong encoding poses several problems. The most significant problem is the impression that one can retrieve people's sex from the joint sex-marital-status encoding, which is simply false as a single level corresponds to both single males and divorced/separated/married females \citep{gromping2019:sg}. Despite this information being available since 2019, the fairness community does not seem to have taken notice. Several experiments of algorithmic fairness on this dataset consider the protected attribute ``sex'' (sometimes even called ``gender''). These experiments are part of work recently published in the most reputable venues for fairness research (Appendix \ref{sec:german_db}). More mistakes in the documentation of eight variables and the relative errata are outlined in \citet{gromping2019:sg}. A clean version of the dataset is available at \citet{gromping2019:sg2}.} 
    \item \que{Is the dataset self-contained, or does it link to or otherwise rely on external resources?}
    
    \ans{The dataset is \textbf{self-contained}.}
    \item \que{Does the dataset contain data that might be considered confidential?}
    
    \ans{\textbf{Yes}. The dataset summarizes customers' financial and personal situation, including past credit history.}
    \item \que{Does the dataset contain data that, if viewed directly, might be offensive, insulting, threatening, or might otherwise cause anxiety?}
    
    \ans{\textbf{No}.}

    \item \que{Does the dataset identify any subpopulations?}
    
    \ans{\textbf{Yes}. The dataset identifies subpopulation by age and sex. Sex is jointly encoded with marital status and cannot be retrieved, contrary to documentation accompanying the dataset \citep{hofmann1994:sg}. A summary based on amended documentation \citep{gromping2019:sg} is presented in Table \ref{tab:German_demographics}.}
    
    \begin{table}[h]
    \centering
    \begin{tabular}{||c c||}
        \hline
        \textbf{Demographic Caracteristic} & \qquad \textbf{Values} \\ [1ex]
        \hline
        Percentage of people under-19 years old & \qquad 0.20\% \\[1ex]
        Percentage of people between 20-29 years old & \qquad 36.70\% \\[1ex]
        Percentage of people between 30-39 years old & \qquad 33.20\% \\[1ex]
        Percentage of people between 40-49 years old & \qquad 17.60\% \\[1ex]
        Percentage of people between 50-59 years old & \qquad 7.20\% \\[1ex]
        Percentage of people between 60-69 years old & \qquad 4.40\% \\[1ex]
        Percentage of people over-70 years old & \qquad 0.70\% \\[1ex]
        \hline
        Percentage of people who are male : divorced/separated & \qquad 5.00\% \\[1ex]
        Percentage of people who are female : non-single or male : single & \qquad 31.00\% \\[1ex]
        Percentage of people who are male : married/widowed & \qquad 54.80\% \\[1ex]
        Percentage of people who are female : single & \qquad 9.20\% \\[1ex]
        \hline
        \end{tabular}
    \caption{Demographic characteristics of the German credit dataset.} \label{tab:German_demographics}
    \end{table}

    \item \que{Is it possible to identify individuals, either directly or indirectly, from the dataset?}
    
    \ans{\textbf{Likely no}, especially given the fact that these records data back to almost 50 years ago. Also, important variables for re-identification, such as ZIP cose and date of birth are missing and many other variables are bucketed.}
    
    \item \que{Does the dataset contain data that might be considered sensitive in any way?}
    
    \ans{\textbf{Yes}. For each instance, the dataset encodes sex, marital status and financial situation.}
\end{itemize}
\subsubsection{Collection process}
\begin{itemize}
    \item \que{How was the data associated with each instance acquired?}
    
    \ans{The data was collected by Hypo bank clerks. Some variables were observable (e.g. credit history with the bank), other variables were reported by subjects (e.g. loan purpose).}
    \item \que{What mechanisms or procedures were used to collect the data?}
    
    \ans{\textbf{Unknown}.}
    \item \que{If the dataset is a sample from a larger set, what was the sampling strategy?}
    
    \ans{The so-called ``bad credits'' are heavily oversampled to make the classification problem more balanced. A natural selection bias is present in the data, as it only consist of applicants who were deemed creditworthy and were thus granted a loan.}
    \item \que{Who was involved in the data collection process and how were they compensated?}
    
    \ans{The data was likely collected by Hypo bank clerks. Walter Häußler was likely involved in sample selection.}
    \item \que{Over what timeframe was the data collected?}
    
    \ans{The dataset covers loans granted in the period \textbf{1973--1975}. Its first publicly-known use dates back to 1979 \citep{haussler1979empirische}. It became publicly available in November 1994 \citep{hofmann1994:sg}.}
    \item \que{Were any ethical review processes conducted?}
    
    \ans{\textbf{Unknown}.}
    \item \que{Was the data collected from the individuals in question directly, or obtained via third parties or other sources?}
    
    \ans{\textbf{Likely both}. Some variables were necessarily collected from loan applicants (e.g. loan purpose), while other variables were likely available from bank records (e.g. credit history with the bank).}
    \item \que{Were the individuals in question notified about the data collection?}
    
    \ans{Individuals provided some of this data as part of a loan application. Collection and notification practices for variables like credit history are unclear.}
    \item \que{Did the individuals in question consent to the collection and use of their data?}
    
    \ans{\textbf{Likely yes}, for the purposes of the immediate credit decision. However it seems implausible they agreed to their data becoming publicly available in an anonymized fashion.}
    \item \que{If consent was obtained, were the consenting individuals provided with a mechanism to revoke their consent in the future or for certain uses?}
    
    \ans{\textbf{Likely no}.}
    \item \que{Has an analysis of the potential impact of the dataset and its use on data subjects been conducted?}
    
    \ans{\textbf{Unknown}.}
\end{itemize}
\subsubsection{Preprocessing/cleaning/labelling}
\begin{itemize}
    \item \que{Was any preprocessing/cleaning/labeling of the data done?}
    
    \ans{\textbf{Yes}. Some instances were discarded. Remaining instances were associated with a binary label according to compliance with the contract. Bucketing took place on several variables, including balance on checking and savings account (A1, A6) and duration of current employment (A7). Sex and marital status were jointly coded (A9).}
    \item \que{Was the “raw” data saved in addition to the preprocessed/cleaned/labeled data?}
    
    \ans{\textbf{Unknown}.}
    \item \que{Is the software used to preprocess/clean/label the instances available?}
    
    \ans{\textbf{Likely no}.}
\end{itemize}
\subsubsection{Uses}
\begin{itemize}
    \item \que{For what tasks has the dataset been used?}
    
    \ans{The dataset was originally used to study the problem of automated credit scoring \citep{haussler1979empirische}. Similarly to the Adult dataset, since becoming publicly available it has been used as a benchmark in various machine learning fields.}
    \item \que{Is there a repository that links to any or all papers or systems that use the dataset?}
    
    \ans{\textbf{Yes}. A selection of early works (pre-2005) using this dataset can be found in \citet{hofmann1994:sg}. A more recent list is available under the beta version of the UCI ML Repository.\footnote{\url{https://archive-beta.ics.uci.edu/ml/datasets/144}} See Appendix \ref{sec:german_db} for a (non-exhaustive) list of algorithmic fairness works using this resource.} 
    \item \que{What (other) tasks could the dataset be used for?}
    
    \ans{The German Credit could be used in fields that concentrate on socially relevant goals and require socially relevant data, such as privacy and explainability. The task at hand is always credit scoring.}
    \item \que{Is there anything about the composition of the dataset or the way it was collected and preprocessed/cleaned/labeled that might impact future uses?}
    
    \ans{Contrary to documentation accompanying the dataset \citep{hofmann1994:sg}, the sex of loan recipients cannot be reliably retrieved. Works of algorithmic fairness should not use this feature.}
    \item \que{Are there tasks for which the dataset should not be used?} 
    
    \ans{In its most common version \citep{hofmann1994:sg} the German Credit dataset should not be used in works of explainability/interpretability as the incorrect documentation would result in counter-intuitive explanations. The 2019 version \citep{gromping2019:sg2} associated with the erratum \citep{gromping2019:sg} is recommended.}
\end{itemize}
\subsubsection{Distribution}
\begin{itemize}
    \item \que{Is the dataset distributed to third parties outside of the entity on behalf of which the dataset was created?}
    
    \ans{\textbf{Yes}. The dataset is publicly available \citep{hofmann1994:sg}}
    \item \que{How is the dataset distributed?} 
    
    \ans{The dataset is available as a \textbf{csv file}.}
    \item \que{When was the dataset distributed?}
    
    \ans{The dataset was released to the UCI ML Repository in \textbf{November 1994}.}
    \item \que{Is the dataset distributed under a copyright or other intellectual property (IP) license, and/or under applicable terms of use (ToU)?} 
    
    \ans{\textbf{Yes}. The UCI ML repository has a citation policy.}
    \item \que{Have any third parties imposed IP-based or other restrictions on the data associated with the instances?}
    
    \ans{\textbf{Likely no}. We are unaware of any IP-based restrictions.}
    \item \que{Do any export controls or other regulatory restrictions apply to the dataset or to individual instances?}
    
    \ans{\textbf{Unknown}.}
\end{itemize}
\subsubsection{Maintenance}
\begin{itemize}
    \item \que{Who is supporting/hosting/maintaining the dataset?}
    
    \ans{The dataset is hosted and maintained by the \textbf{UCI Machine Learning Repository} \citep{hofmann1994:sg}. A clean and well-documented version of the same dataset donated by Ulrike Gromping \citep{gromping2019:sg2} is also available on the same repository.}
    \item \que{How can the owner/curator/manager of the dataset be contacted?} 
    
    \ans{The dataset donor, Hans Hofmann retired in 2008. Comments and inquiries for UCI may be sent to ml-repository@ics.uci.edu.}
    \item \que{Is there an erratum?}
    
    \ans{\textbf{Yes}. A clean data release \citep{gromping2019:sg2} and accompanying report \citep{gromping2019:sg} are available online.}
    \item \que{Will the dataset be updated?}
    
    \ans{\textbf{Likely no}. The recently released South German Credit Data Set \citep{gromping2019:sg2} may be considered an update.}
    \item \que{If the dataset relates to people, are there applicable limits on the retention of the data associated with the instances?}
    
    \ans{\textbf{Unknown}.}
    \item \que{Will older versions of the dataset continue to be supported/hosted/maintained?} 
    
    \ans{Unless otherwise indicated, both the new \citep{gromping2019:sg2} and the old version \citep{hofmann1994:sg} of the German Credit dataset will remain hosted on the UCI ML Repository in its current version.}
    \item\que{ If others want to extend/augment/build on/contribute to the dataset, is there a mechanism for them to do so?}
    
    \ans{\textbf{Unknown}.}
\end{itemize}

\clearpage
\subsection{Data Nutrition Label}

For the sake of correctness, we report redacted information based on the new South German Credit Data Set \citep{gromping2019:sg2} and accompanying documentation \citep{gromping2019:sg}.

\ifconf\else{\subsubsection{Metadata}}\fi 

\begin{table}[h]
\centering
\begin{tabular}{||r l||}
 \hline 
 \multicolumn{2}{|c|}{\textbf{METADATA}}\\
 \hline
 \textbf{Filenames} & \qquad SouthGermanCredit  \\ 
 \textbf{Format} & \qquad .asc \\
 \textbf{Url} & \qquad \url{https://archive.ics.uci.edu/ml/datasets/South+German+Credit}  \\
 \textbf{Domain} & \qquad Economics  \\
 \textbf{Keywords} & \qquad credit scoring, Germany, loan, classification  \\
 \textbf{Type} & \qquad Tabular  \\
 \textbf{Rows} & \qquad 1000 \\ 
 \textbf{Columns} & \qquad 21  \\
 \textbf{\% missing cells} & \qquad 0\% \\
 \textbf{Rows with missing cells} & \qquad 0\% \\
 \textbf{License} & \qquad UCI Repository citation policy  \\
 \textbf{Released} & \qquad November 2019  \\
 \textbf{Range} & \qquad 1973-1975 \\ [3pt]
 \textbf{Description} & \qquad \multirow{3}{30em}{This dataset encodes socio-economical features of loan recipients from a bank in southern Germany, along with binary variable encoding whether they punctually payed every installment, which is he target of a classification task.}  \\[9ex] 
 \hline
\end{tabular}
\caption{Metadata of South German Credit dataset.}
\end{table}

\ifconf\else{\subsubsection{Provenance}}\fi 

\begin{table}[h]
\centering
\begin{tabular}{||r l||}
 \hline
 \multicolumn{2}{|c|}{\textbf{PROVENANCE}}\\
 \hline
 \textbf{Source} & \\
 \qquad Name & \qquad Walter Häußler  \\
 \qquad Url & \qquad \multirow{2}{20em}{\url{https://archive.ics.uci.edu/ml/datasets/Statlog+\%28German+Credit+Data\%29}} \\ [3.2ex]
 \qquad email & \qquad//  \\ [2ex]
 \textbf{Authors} & \\
 \qquad Names & \qquad Ulrike Grömping\\
 \qquad Url & \qquad \url{https://archive.ics.uci.edu/ml/datasets/South+German+Credit}\\
 \qquad email & \qquad \url{groemping@bht-berlin.de}  \\
 \hline
\end{tabular}
\caption{Provenance of South German Credit dataset}
\end{table}

\ifconf\else{\subsubsection{Variables}}\fi

\begin{table}[h]
\centering
\begin{tabular}{||r l||}
 \hline 
 \multicolumn{2}{|c|}{\textbf{VARIABLES}}\\
 \hline
 \textbf{status} & \qquad \multirow{1}{20em}{Checking account balance (in Deutsche Mark) \\ 1 (no checking account) \\ 2 ($<$ 0 DM)\\ 3 (0 $\leq$ ... $<$ 200 DM)\\ 4 ($\geq$ 200 DM)} \\ [17ex]
 \textbf{duration} & \qquad \multirow{1}{20em}{Credit duration (in months)} \\ [3ex]
 \textbf{credit\_history} &  \qquad \multirow{3}{20em}{Applicant's credit history \\0 (delay in past payments)\\1 (critical account/other credits elsewhere)\\ 2 (no credits taken/all credits paid back duly)\\ 3 (existing credits paid back duly till now)\\ 4 (all credits at this bank paid back duly)} \\[18ex]
 \textbf{purpose} & \qquad \multirow{1}{20em}{Purpose of loan \\0 (other)\\ 1 (new car) \\ 2 (used car)\\ 3 (furniture/equipment) \\ 4 (radio/television) \\ 5 (domestic appliances) \\ 6 (repairs)\\ 7 (education) \\ 8 (vacation)\\ 9 (retraining) \\10 (business)}\\[34ex]
 \textbf{amount} & \qquad \multirow{1}{20em}{Credit amount (result of unknown monotonic transformation)} \\[3ex]
 \hline
\end{tabular}
\caption{Variables of South German Credit dataset (1/3).}
\label{tab:german_meta1}
\end{table}

\begin{table}[h]
\centering
\begin{tabular}{||r l||}
 \hline 
 \multicolumn{2}{|c|}{\textbf{VARIABLES}}\\
 \hline
 \textbf{savings} & \qquad \multirow{1}{20em}{Savings account balance (in Deutsche Mark) \\ 1 (unknown/ no savings account)\\ 2($<$ 100 DM)\\ 3 (100 $\leq$ ... $<$ 500 DM)\\ 4 (500 $\geq$ ... $<$ 1000 DM) \\ 5 ($\geq$ 1000 DM)} \\ [17ex]
 \textbf{employment\_duration} & \qquad \multirow{1}{20em}{Duration of applicant's current employment\\ 1 (unemployed) \\ 2 ($<$ 1 year)\\ 3 (1 $\leq$ ... $<$ 4 years)\\4 (4 $\leq$ ... $<$ 7 years) \\5 ($\geq$ 7 years)} \\ [18ex]
 \textbf{installment\_rate} & \qquad \multirow{2}{20em}{Installment amount to disposable income ratio $[\%]$ \\ 1 ($\geq$35) \\ 2 (25 $\leq$ ... $<$ 35)\\ 3 (20 $\leq$ ... $<$ 25)\\4 ($<$ 20)} \\ [17ex]
 \textbf{personal\_status\_sex} & \qquad \multirow{1}{20em}{Joint encoding of sex and marital status of applicant \\ 1 (male - divorced/separated) \\2 (female - non single or male - single) \\ 3 (male - married/widowed) \\4 (female - single)} \\ [18ex]
 \textbf{other\_debtors} & \qquad \multirow{1}{20em}{Presence of co-debtor or guarantor \\1 (none) \\ 2 (co-applicant)\\ 3 (guarantor)} \\ [14ex]
 \textbf{present\_residence} & \qquad \multirow{2}{20em}{Years living at current address  \\ 1 ($<$ 1 year)\\ 2 (1 $\leq$ ... $<$ 4 years)\\ 3 (4 $\leq$ ... $<$ 7 years)\\ 4 ($\geq$ 7 years)} \\ [14ex]
 \textbf{property} & \qquad \multirow{2}{20em}{Applicant's most valuable property \\1 (unknown / no property) \\ 2 (car or other)\\ 3 (building soc. savings agr / life insurance) \\ 4 (real estate)} \\ [14ex]
 \hline
\end{tabular}
\caption{Variables of South German Credit dataset (2/3).}
\label{tab:german_meta2}
\end{table}

\begin{table}[h]
\centering
\begin{tabular}{||r l||}
 \hline 
 \multicolumn{2}{|c|}{\textbf{VARIABLES}}\\
 \hline
 \textbf{age} & \qquad \multirow{1}{20em}{Applicant's age (years)} \\ [2ex]
 \textbf{other\_installment\_plans} & \qquad \multirow{1}{20em}{Installment plans with other banks \\ 1 (bank)\\  2 (stores) \\ 3 (none)} \\ [11ex]
 \textbf{housing} & \qquad \multirow{1}{20em}{Type of housing \\ 1 (for free) \\ 2 (rent) \\ 3 (own)} \\ [12ex]
 \textbf{number\_credits} & \qquad \multirow{1}{20em}{Number of credits (ongoing or past, including current) with this bank \\ 1 (1) \\ 2 (2-3) \\ 3 (4-5) \\ 4($\geq$ 6)} \\ [16ex]
 \textbf{job} & \qquad \multirow{1}{20em}{Applicant's job and emplyability \\ 1 (unemployed/ unskilled - non-resident)\\ 2 (unskilled - resident) \\ 3 (skilled employee / official) \\ 4 (manager / self-empl. / highly qualif. employee)} \\ [16ex]
\textbf{people\_liable} & \qquad \multirow{1}{20em}{Number of people who financially depend on the applicant \\ 1 (3 or more) \\ 2 (0 to 2)} \\ [10ex]
 \textbf{telephone} & \qquad \multirow{1}{20em}{Presence of telephone landline registered under applicant's name (2) or not (1)}  \\[4ex]
 \textbf{foreign\_worker} & \qquad \multirow{1}{20em}{Foreign worker (1) or not (2)} \\[3ex]
 \textbf{credit\_risk} & \qquad \multirow{1}{20em}{Punctually payed back every installment (1) or not (2)} \\[3ex]
 \hline
\end{tabular}
\caption{Variables of South German Credit dataset (3/3).}
\label{tab:german_meta3}
\end{table}

\clearpage
\ifconf\else{\subsubsection{Statistics}}\fi 

\begin{table}[h]
\begin{tabular}{||lllllll||}
\hline
\multicolumn{7}{|c|}{\textbf{STATISTICS}} \\ \hline
\multicolumn{7}{|l|}{\textbf{Ordinal}} \\ \hline
\multicolumn{1}{|c}{name} & \multicolumn{1}{c}{type} & \multicolumn{1}{c}{count} & \multicolumn{1}{c}{unique} & \multicolumn{1}{c}{mostFrequent} & \multicolumn{1}{c}{leastFrequent} & \multicolumn{1}{c|}{missing} \\
\hline
status & string & 1000 & 4 & 4 ($\geq$200) & 3 (0$\leq$...$<$200) & 0 \\
savings & string & 1000 & 5 & 1 (unknown/no savings) & 4 (500$\leq$...$<$1000) & 0 \\
employment\_duration & string & 1000 & 5 & 3 (1$\leq$...$<$4) & 1 (unemployed) & 0 \\
installment\_rate & string & 1000 & 4 & 4 ($<$20) & 1 $\geq$35 & 0 \\
present\_residence & string & 1000 & 4 & 4 ($\geq$7 yrs) & 1 ($<$ 1 yr) & 0 \\
number\_credits & string & 1000 & 4 & 1 (1) & 4 ($\geq$6) & 0  \\
people liable & string & 1000 & 2 & 2 (0 to 2) & 1 (3 or more) & 0 \\
\hline
\end{tabular}
\caption{Ordinal variables statistics of South German Credit dataset}
\end{table}

\begin{table}[h]
\begin{tabular}{||lllllll||}
\hline
\multicolumn{7}{|l|}{\textbf{Categorical}} \\ \hline
\multicolumn{1}{|c}{name} & \multicolumn{1}{c}{type} & \multicolumn{1}{c}{count} & \multicolumn{1}{c}{uniqueEntries} & \multicolumn{1}{c}{mostFrequent} & \multicolumn{1}{c}{leastFrequent} & \multicolumn{1}{c|}{missing} \\
\hline
credit\_history & string & 1000 & 5 & 2 (no credits taken) & 0 (delay in paying off)  & 0 \\
purpose & string & 1000 & 11 & 3 (furniture/equipment) & 8 (vacation) & 0 \\
status\_sex & string & 1000 & 4 & 3 (male-marr/widow) & 1 (male-divorc/separ) & 0 \\
other\_debtors & string & 1000 & 3 & 1 (none) & 2 (co-appliant) & 0 \\
property & string & 1000 & 4 & 3 (building soc. savings) & 4 (real estate) & 0 \\
other\_plans & string & 1000 & 3 & 3 (none) & 2 (stores) & 0 \\
housing & string & 1000 & 3 & 2 (rent) & 3 (own) & 0 \\
job & string & 1000 & 4 & 3 (skilled empl/offic) & 1 (unempl/unsk non-res) & 0 \\
telephone & string & 1000 & 2 & 1 (no) & 2 (yes) & 0 \\
foreign\_worker & string & 1000 & 2 & 2 (no) & 1 (yes) & 0 \\
credit\_risk & string & 1000 & 2 & 1 (good) & 0 (bad) & 0 \\
\hline
\end{tabular}
\caption{Categorical variables statistics of South German Credit dataset}
\end{table}

\begin{table}[h]
\begin{tabular}{||llllllllll||}
\hline
\multicolumn{10}{|l|}{\textbf{Quantitative}} \\ \hline
\multicolumn{1}{|c}{name} & \multicolumn{1}{c}{type} & \multicolumn{1}{c}{count} & \multicolumn{1}{c}{min} & \multicolumn{1}{c}{median} & \multicolumn{1}{c}{max} & \multicolumn{1}{c}{mean} & stdDev & miss & \multicolumn{1}{c|}{zeros} \\\hline

duration & number & 1000 & 4 & 18 & 72 & 20.90 & 12.06 & 0 & 0 \\
amount & number & 1000 & 250 & 2319.50 & 18424 & 3271.25 & 2822.75 & 0 & 0 \\
age & number & 1000 & 19 & 33 & 75 & 35.54 & 11.35 & 0 & 0 \\ \hline
\end{tabular}
\caption{Quantitative variables statistics of South German Credit dataset.}
\end{table}

\end{appendices}

\end{document}